\documentclass[11pt,english,appendixprefix=true]{article}
\usepackage[T1]{fontenc}
\usepackage[latin9]{inputenc}
\usepackage[paperwidth=24cm]{geometry}
\usepackage{wrapfig}
\usepackage{calc}
\usepackage{textcomp}
\usepackage{amsmath}
\usepackage{amssymb}
\usepackage{graphicx}
\usepackage{setspace}
\usepackage{esint}
\makeatletter

\newcommand{\noun}[1]{\textsc{#1}}

\numberwithin{figure}{section}
\numberwithin{equation}{section}

\makeatother

\usepackage{babel}
\begin{document}
\thispagestyle{empty} 

\noindent \begin{center}
\textbf{\Huge{}From Field Theory to the Hydrodynamics of Relativistic
Superfluids}
\par\end{center}{\Huge \par}

~

~

~

~

~

~

~

~

\begin{center}
{\LARGE{}A Dissertation Submitted to the Faculty of Physics at the}
\par\end{center}{\LARGE \par}

~

\begin{center}
{\huge{}Vienna University of Technology}
\par\end{center}{\huge \par}

~

\begin{center}
{\Large{}by}
\par\end{center}{\Large \par}

~

~

\begin{center}
{\Huge{}Stephan Stetina}
\par\end{center}{\Huge \par}

~

~

~

~

~

~

~

~

{\Large{}Vienna, September 2014}\newpage{}

~

\thispagestyle{empty} 

\newpage{}

\thispagestyle{empty}

\part*{Abstract}

~

~

~

~

The hydrodynamic description of a superfluid is usually based on a
two-fluid picture. In this thesis, basic properties of such a relativistic
two-fluid system are derived from the underlying microscopic physics
of a complex scalar quantum field theory. To obtain analytic results
of all non-dissipative hydrodynamic quantities in terms of field theoretic
variables, calculations are first carried out in a low-temperature
and weak-coupling approximation. In a second step, the 2-particle-irreducible
formalism is applied: This formalism allows for a numerical evaluation
of the hydrodynamic parameters for all temperatures below the critical
temperature. In addition, a system of two coupled superfluids is studied.
As an application, the velocities of first and second sound in the
presence of a superflow are calculated. The results show that first
(second) sound evolves from a density (temperature) wave at low temperatures
to a temperature (density) wave at high temperatures. This role reversal
is investigated for ultra-relativistic and near-nonrelativistic systems
for zero and nonzero superflow. The studies carried out in this thesis
are of a very general nature as one does not have to specify the system
for which the microscopic field theory is an effective description.
As a particular example, superfluidity in dense quark and nuclear
matter in compact stars are discussed. 

\newpage{}

~

\thispagestyle{empty} 

\noindent \textit{\large{}Dedicated to my parents with deep and profound
gratitude. I would not be the person I am today without their constant
support and guidance.}\newpage{}

\thispagestyle{empty} 

\newpage{}

\part*{Preface}

~

~

~

~

\noindent Superfluidity is a very general phenomenon. Critical temperatures
of known superfluids extend over 17 orders of magnitude (from about
200 nK for cold atomic gases to up to $10^{10}$ K for nuclear and
quark matter). In the frame of this thesis, I will be most interested
in the upper end of this scale and discuss relativistic superfluids
which presumably exist in dense nuclear or quark matter in the interior
of compact stars. Due to its interdisciplinarity, the study of this
field is as fascinating as it is challenging. Concepts of many different
branches such as particle physics, solid state physics or astrophysics
have to be taken into account and I will devote a rather large part
of this thesis to describe the rich spectrum of physics to which the
research presented in this thesis is relevant. 

\noindent The basic idea is to relate the effective hydrodynamic description
of a superfluid to the underlying microscopic theory. The relevant
microscopic physics are determined by a relativistic quantum field
theory to which a Bose-Einstein condensate is introduced. Even though
it is to some extent obvious that the hydrodynamics of a superfluid
evolve from such a microscopic background, this relation has never
been made explicit before. The existence of such a ``gap'' in our
perception of superfluidity has led the existence of two separate
communities (a ``phenomenological'' and a ``microscopic'' one),
each equipped with a terminology of its own. This difference in terminology
is a frequent source of confusion and I hope the results of this work
can help to eliminate at least some of this confusion. In the second
part of this thesis, I shall demonstrate explicitly that it is indeed
possible to the derive the hydrodynamic description and calculate
all non-dissipative hydrodynamic parameters purely in terms of field
theoretic variables. It shall also be indicated how the obtained results
are related to recent research outside of high energy physics, in
particular to liquid helium and cold atoms. In the third part I will
use this microscopic framework to study the sound excitations of a
superfluid. In part IV, the more complicated but phenomenologically
important scenario of two coupled superfluids will be discussed.\newpage{}

\noindent Most of the results were obtained in a collaboration with
my thesis advisor Andreas Schmitt%
\footnote{Univ.Ass. Dr. Andreas Schmitt, Institute for Theoretical Physic, Technical
University Vienna

~~andreas.schmitt@tuwien.ac.at

~%
} as well as Mark Alford %
\footnote{Prof. Mark G. Alford, Physics Department, Washington University of
St. Louis, alford@wuphys.wustl.edu

~%
} and Kumar Mallavarapu %
\footnote{S. Kumar Mallavarapu, Physics Department, Washington University of
St. Louis, kumar.s@go.wustl.edu%
} from the Washington University in St. Louis. In particular the results
presented in sections 8 and 10,11 and 12-15 of this thesis were published
in the following two articles
\begin{itemize}
\item M. G. Alford, S. K. Mallavarapu, A. Schmitt, S. Stetina, Phys. Rev.
D 87, 065001 (2013)
\item M. G. Alford, S. K. Mallavarapu, A. Schmitt, S. Stetina, Phys.Rev.
D89, 085005 (2014)
\end{itemize}
\newpage{}

\part*{Acknowledgments}

~

\noindent I would like to use this opportunity to express my deepest
gratitude towards my thesis advisor Andreas Schmitt. Over the past
years, I have come to value him not only as an excellent and dedicated
teacher, but also as a great mentor. In addition to uncountable discussions
in frame of the research projects we carried out together, his counseling
in academic affairs in general and in preparation of conference talks
or writing scientific articles in particular have been invaluable
to me. I find it hard to believe that the quality of my PhD training
could have been any better. 

\noindent I would also like to thank Anton Rebhan for many valuable
comments and suggestions and in particular for granting me additional
financial support. 

\noindent I thank Mark Alford, Karl Landsteiner and Eduardo Fraga
for inspiring and valuable discussions. 

\noindent Finally I would like to thank the Austrian Science Fund
FWF for the financial support I have received over the past three
years. 

\newpage{}

\tableofcontents{}

\newpage{}

\part{Introduction \label{part:Introduction}}

\noindent The discovery of superfluidity originated from the study
of the low-temperature dynamics of helium. Helium exhibits the rather
peculiar feature of remaining liquid - even when the limit of absolute
zero temperature is approached. While we are not aiming to describe
the properties of helium as accurately as possible, no introduction
to superfluidity can be complete without mentioning some of the spectacular
experimental results of these studies. Most of the discussed properties
are not exclusive for helium but are properties of superfluids in
general. In particular, we shall explain how the theoretical attempts
to describe the observed phenomena ultimately led to the powerful
two-fluid formalism which will be introduced in section \ref{sec:The-two-fluid-model}.
A relativistic version of this two-fluid formalism is widely and successfully
used today to describe the phenomenology of superfluids in compact
stars (see also section \ref{sec:Relativistic-thermodynamics-and-hydro}).
After a short historical outline of the discovery of compact stars
we will discuss the properties of matter which presumably exists in
compact stars from a microscopic point of view. In particular, we
shall explain in which phases of dense matter we expect to encounter
superfluidity and how the occurrence of superfluidity influences observable
phenomena of compact stars. The microscopic terminology used in section
\ref{sec:Compact-stars-from-a-modern-POV} will inevitably be very
different from the phenomenological one used in sections \ref{sec:Early-experiments}
and \ref{sec:The-two-fluid-model}. It shall be the goal of second
part of this thesis to make the relationship between both approaches
as explicit as possible. A review of the discovery of superfluidity
containing many interesting historical details can be found in \cite{historic}.
In addition, reference \cite{Donelli} provides a good introduction
to the phenomenology of superfluidity.

\section{Early experiments with liquid helium \label{sec:Early-experiments}}

~

\begin{onehalfspace}
\noindent The experimental discovery of superfluidity dates back to
the year 1927 when Keesom, Wolfke and Clusius discovered an anomaly
in the properties of liquid $\textrm{\ensuremath{^{4}}He}$ %
\footnote{This is by far the most abundant isotope of helium and used in most
laboratory experiments. Each pair of neutrons, protons and electrons
occupies a 1s orbital, none of them possesses orbital angular momentum
and the spin of each pair adds up to zero. In this configuration,
helium is extremely stable. %
}: the specific heat as a function of temperature shows a sharp maximum
at 2.17 K \cite{Lambda1,Lambda2}. According to the shape of this
diagram this point was called ``lambda point'' and marks the boundary
of two different liquid states referred to as helium I (above $\textrm{\ensuremath{T_{\lambda}}}$)
and helium II (below $\textrm{\ensuremath{T_{\lambda}}}$). Shortly
after the discovery of this novel He II phase, experimental efforts
focused on the study of heat-flow and viscosity of helium II which
seemed to show a rather unusual behavior: 
\begin{figure}[t]
\fbox{\begin{minipage}[t]{1\columnwidth}%
~~~~\includegraphics[scale=0.56]{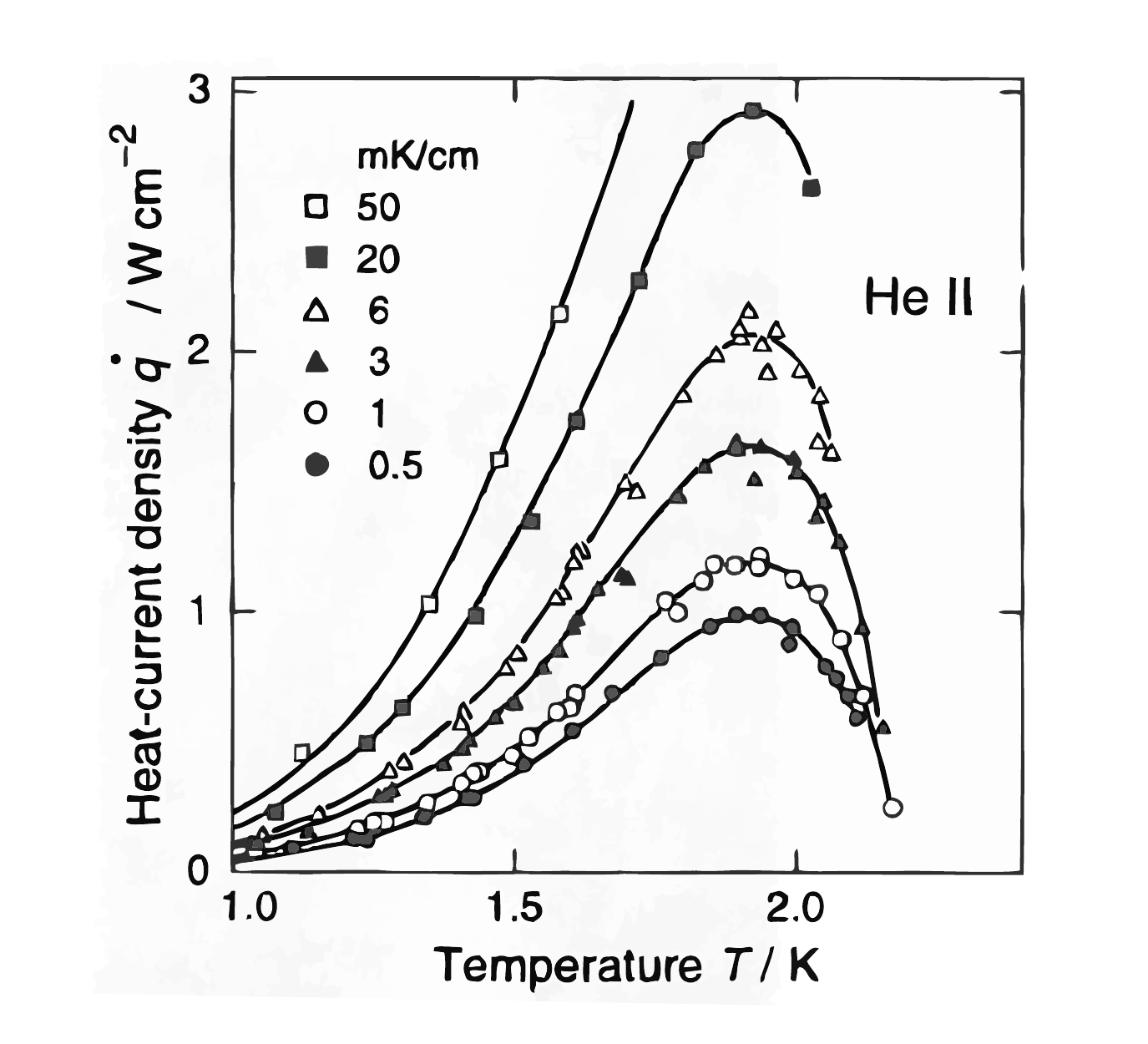}~~~~~~~\includegraphics[bb=0bp 13bp 396bp 324bp,scale=0.65]{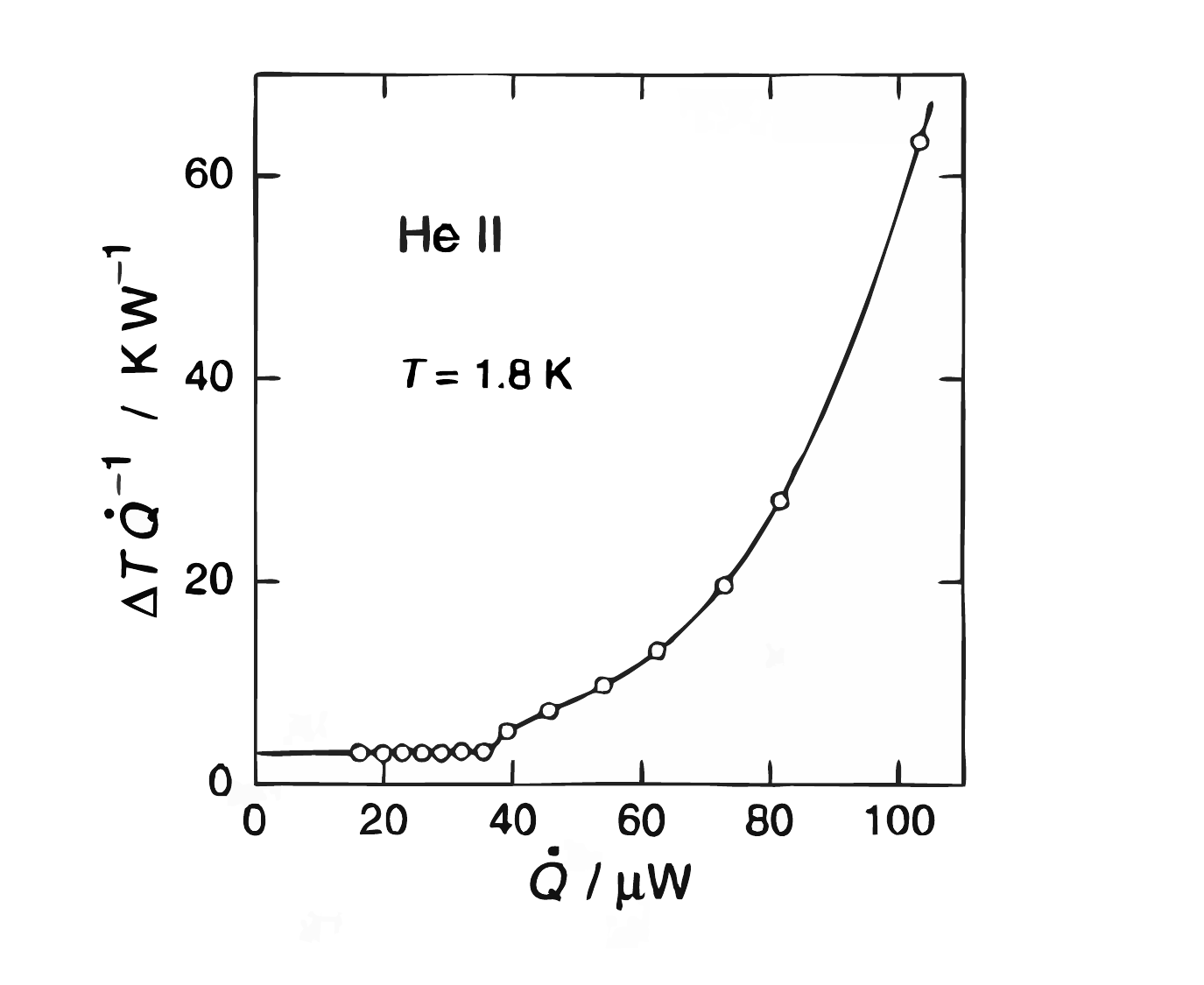}%
\end{minipage}}\protect\caption{Left panel: the heat-flow of He II as a function of the temperature
for different temperature gradients taken from reference \cite{Keesom1970}.
The general spectrum shows a sharp maximum around 2 Kelvin and a strong
decrease as the critical temperature for the phase transition to He
I is approached. Right Panel: thermal resistance as a function of
the heat flow taken from reference \cite{Brewer1961}. The onset of
the critical velocity is clearly visible. \label{fig:Left-panel:Heatflow}}
\end{figure}
 Rollin in Oxford realized \cite{Rollin} that the \textit{heat transport}
in Helium II is up to 5 times as effective as in Helium I and shows
a sharp peak at a temperature around 2 K (a typical plot is shown
in \ref{fig:Left-panel:Heatflow}).
\end{onehalfspace}

\noindent It was conjectured that convection (due to the bulk motion
of the fluid as opposed to heat conduction due to excitations of molecules)
could be responsible for the anomalously large value of heat transport
provided that the viscosity is small enough. However, the usual convection
law $\partial_{t}\vec{q}=-\Lambda\vec{\nabla}T$ (i.e. a linear relation
between heat-flow and temperature gradient where $\Lambda$ denotes
the thermal conductivity) was only recovered for sufficiently small
temperature gradients. In addition, Brewer and Edwards measured the
\textit{thermal resistance} as function of the heat flow \cite{Brewer1961}.
This relation remains constant up to a critical value of the heat
flux at which the thermal resistance suddenly increases rapidly (see
right panel of figure \ref{fig:Left-panel:Heatflow}). This critical
value of the heat flow corresponds precisely to the critical value
of the temperature gradient at which nonlinear deviations from the
convection law become measurable which suggests a deeper connection
between both phenomena. In fact, both can be explained by the concept
of a \textit{critical velocity}: as we will discuss in the paragraph
below, a heat flux induces a \textit{counter-flow} of mass flux (and
vice versa). The mass flow however is limited by a certain critical
velocity at which turbulence arises (see discussion in section \ref{sub:The-critical-velocity}).
This in turn is the origin of the sudden increase in thermal resistivity
and for the onset of a nonlinear regime of the heat flow.

\noindent The \textit{counter-flow mechanism} was discovered by two
spectacular experiments which showed the so called ``thermomechanical
effect'' and its inverse, the ``fountain effect'' \cite{Fountain}:
in case of the thermomechanical effect, two containers were filled
with helium II and connected by a very thin capillary - a so called
``superleak'' which should be thin enough to block any viscous fluid.
Increasing the pressure \begin{wrapfigure}{o}{0.3\columnwidth}%
~~~\includegraphics[scale=0.35]{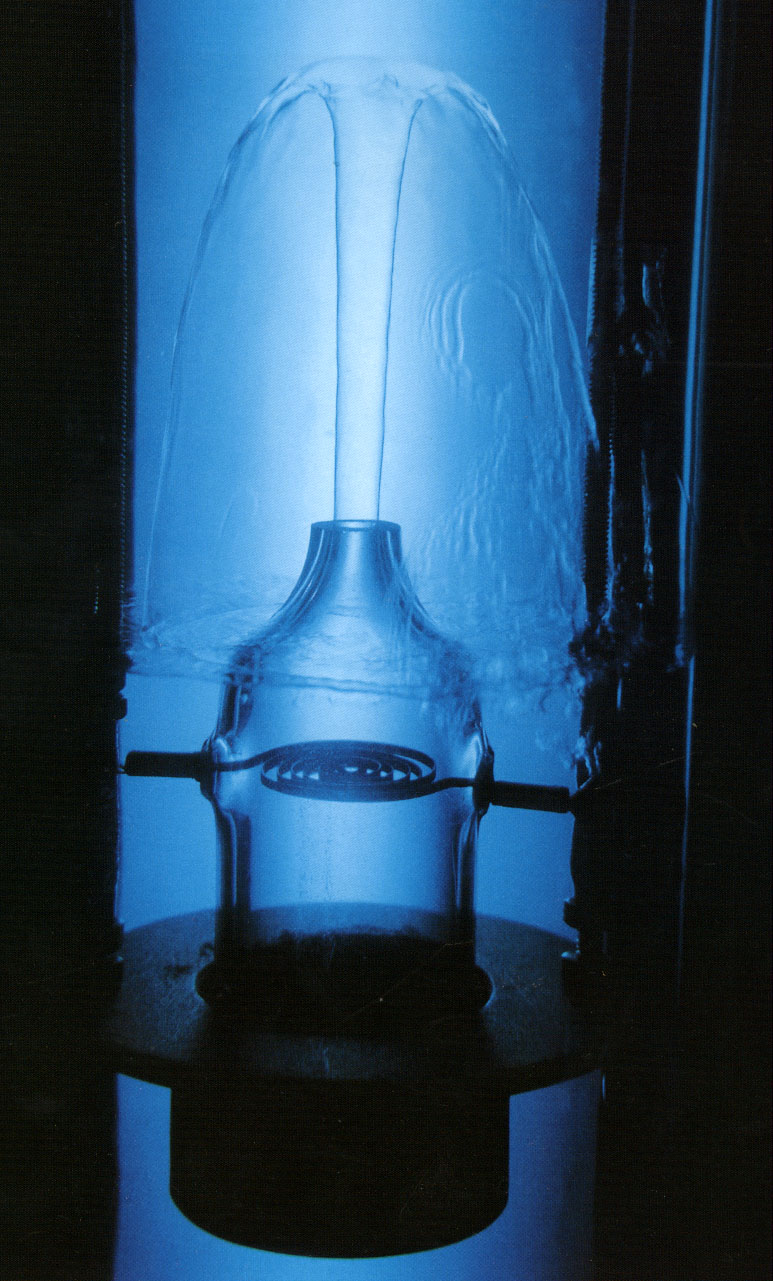}

\protect\caption{Photograph of an experiment which demonstrates the fountain effect
taken from reference \cite{FountainURL}.}
\end{wrapfigure}%
 in container A leads to a flow of helium towards container B. Surprisingly,
this induces a temperature difference in both containers: while the
temperature increases in A, it decreases in B. This shows that mass
flow and heat flow are not just interconnected, but even pointed in
opposite directions. The fountain effect was discovered, when a flask
(open at the bottom) with a thin neck was lowered into a bath of helium
II. Additionally, the lower part of the flask is filled with a fine
compact powder - again with the purpose of preventing any viscous
liquid from escaping through the bottom. As soon as the helium in
the flask is heated up, a fountain of liquid helium sprays out at
the top. Such a process can in principle go on as long as the heat
supply as well as the cooling of the bath are provided. 

\noindent The \textit{absence of viscosity} in helium II was shown
in 1938 independently by two groups, Kapitza \cite{Kapitza} in Moscow
and Allen and Misener \cite{AllenMisener} in Cambridge by measuring
the flow velocity of helium through thin capillaries. The published
articles of both groups include remarkable statements about the nature
of superfluidity: Kapitza proposed \cite{Kapitza} that \textit{``by
analogy with superconductors {[}...{]} helium below the $\lambda$
point enters a special state which might be called superfluid''}.
This is the first time the term superfluidity appears in literature.
Furthermore, superfluidity is related to superconductivity long before
the microscopic theory of the latter phenomenon was established. Kapitza
received the Nobel price for his discovery in 1978. Allen and Misener
on the other hand claimed \cite{AllenMisener} that\textit{ ``the
observed type of flow{[}...{]} cannot be treated as laminar or even
as ordinary turbulent flow''.} This statement implied that helium
II requires an entirely new fluid-dynamical description and was in
contradiction to the common view of that time that liquid helium is
an ``ordinary'' fluid with very small viscosity (i.e. an ideal fluid
describable by Euler`s equations of motion).

\begin{onehalfspace}
\noindent The pioneers in setting up this entirely new theory of fluidity
were Fritz London and Laszlo Tisza. London argued \cite{London} that
since $\textrm{\ensuremath{^{4}}He}$ atoms were Bose particles, they
should undergo Bose-Einstein condensation, a rather new concept at
that time. He then calculated the transition temperature of an ideal
Bose gas with the density of liquid $\textrm{\ensuremath{^{4}}He}$
and arrived at a value of 3.1 K, quite close to $\textrm{\ensuremath{T_{\lambda}}}$.
London also explained, why Helium II remains liquid when the temperature
approaches absolute zero: even at very low pressure, the quantum kinetic
energy of helium atoms is large compared to their binding energies
due to Van der Waals forces. In addition, helium atoms are particularly
light. As a net result, helium atoms do not remain ``frozen'' at
fixed lattice positions even at very low temperatures. Shortly after
Tisza learned about London\textquoteright s ideas, he proposed a two-fluid
model \cite{Tisza} consisting of a superfluid that would have zero
entropy and viscosity and a viscous normal fluid which carries entropy.
With the aid of such a model he was able to provide an intriguingly
simple explanation of the thermomechanical effect: heating in terms
of the two fluids means converting superfluid into normal fluid at
a rate sufficient to absorb the applied energy. Near the heater, this
results in an excess of the normal fluid and a deficiency of the superfluid.
Convection then leads to a counterflow of both components: while the
superfluid is drawn towards the heater (where it will be transformed),
the normal fluid flows away from the heater. Any temperature inhomogeneity
in a superfluid is efficiently smoothened out by this counterflow
mechanism. This effect is directly visible in experiments: \begin{wrapfigure}{o}{0.7\columnwidth}%
\includegraphics[scale=0.6]{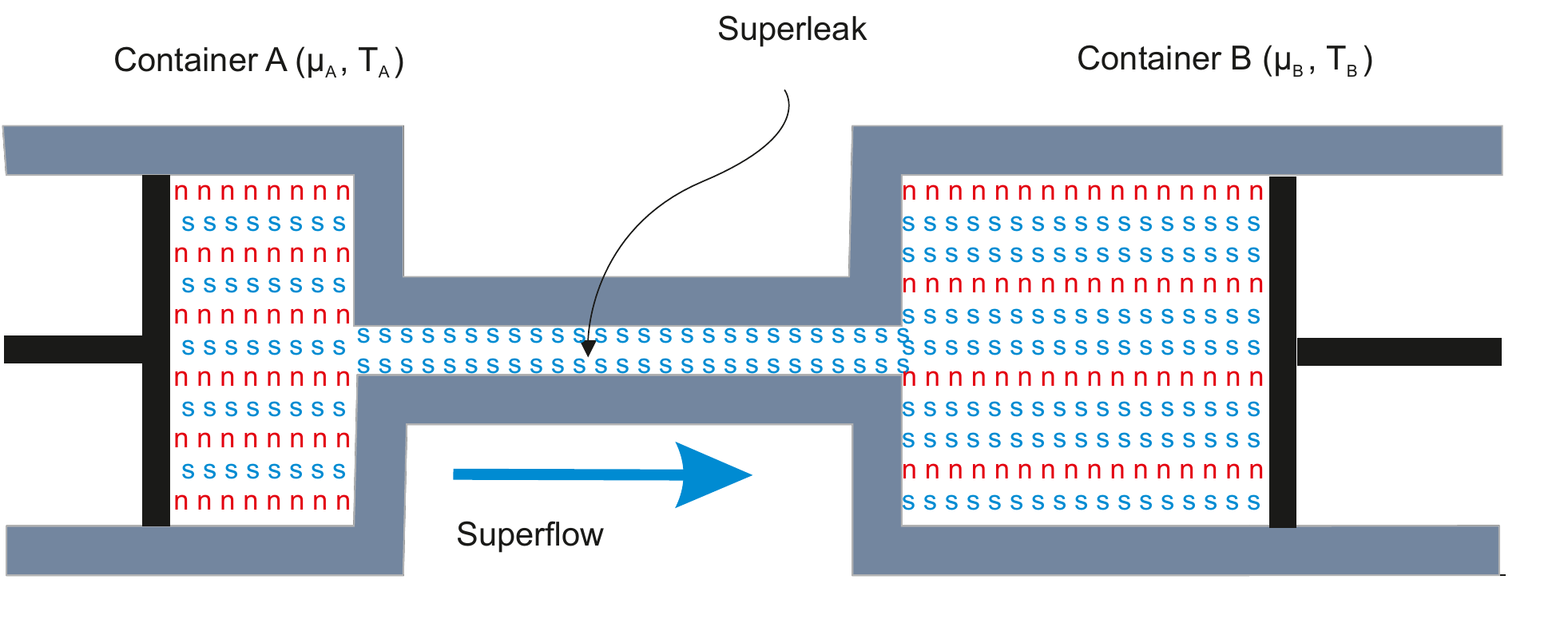}\protect\caption{Illustration of the normal-fluid and superfluid components in a superleak.
As long as the chemical potential $\mu_{A}$ is larger than $\mu_{B}$
, the superfluid (s) will propagate towards container B. The counterflow
is blocked by the narrow capillary.\label{fig:Superleak} }
\end{wrapfigure}%
when pressure is reduced below vapor pressure, we expect that boiling
of helium II sets in. Usually boiling becomes visible by the onset
of bubbles, which represent local hot spots in a liquid. However,
in boiling helium II one cannot observe any bubbles%
\footnote{Once again, this effect is limited by the critical velocity. If the
induced counterflow becomes too large an onset of bubbles is indeed
visible. %
}.

\noindent In case of the two containers which are connected by a small
capillary, only the superfluid can pass (the viscosity of the normal
fluid prevents it from creating a counterflow through the narrow capillary
such that no equilibrium between the two containers can be achieved).
This situation is illustrated in figure \ref{fig:Superleak}. Since
there is now more mass (but the same entropy!) in container B, the
temperature \textit{decreases}. By the same argument, the temperature
\textit{increases} in container A. From the same experiment, it was
deduced that the charge is carried by the superflow: a steady state
(when there is no more flow through the superleak) is only achieved
when the chemical potentials are equal in both containers $\mu_{A}(T_{A},P_{A})=\mu_{B}(T_{B},P_{B})$
. According to the thermodynamic relation, a pressure gradient corresponds
to gradients in temperature and chemical potential $dP=nd\mu+sdT$,
but only a gradient in the chemical potential will induce a superflow.
Tisza was also able to explain the fountain experiment: heating creates
a temperature difference between the helium within the flask and the
helium bath below causing the superfluid to enter the flask. The viscous
normal fluid on the other hand is prevented from leaving the flask
due to the powder at the bottom. The volume of the liquid in the flask
thus increases rapidly resulting in a fountain shooting out at the
top.

\noindent Finally it was Landau, who perfect the two-fluid model.
At this point, it is interesting to mention that Landau, despite his
certainly outstanding intuition refused to believe in the relevance
of Bose-Einstein condensation. Nevertheless, Landau`s theory is a
remarkable success. The main advance compared to Tisza`s model lies
in the definition of the normal component. While Tisza believed that
the normal fluid consists of uncondensed atoms, Landau proposed that
it is made of quasiparticles of the quantum fluid. In a manner of
speaking, both Landau and Tisza were correct about a different half
of the two-fluid model. Landau divided all quasiparticle excitations
into two groups which he termed ``phonons'' and ``rotons''. While
the ``phonon'' has a dispersion relation linear in momentum and
determines the low-temperature properties of the fluid, the ``roton''
dispersion is quadratic in momentum and can only be excited after
a certain energy gap $\Delta$ has been reached%
\footnote{The microscopic nature of the roton excitations is still heavily debated
today. Different microscopic models aim to reproduce the roton spectrum
with different levels of success.%
}. After fitting model parameters such as $\Delta$ to reproduce the
experimental value of the specific heat, he calculates a transition
temperature of about 2.3 K in very good agreement with the experimental
value. 
\end{onehalfspace}

\noindent Perhaps the most striking prediction of the two-fluid model
is the existence of an additional sound mode: Landau proposed that
heat should propagate as what he calls ``second sound'' rather than
diffuse as in an ordinary fluid. Even though the idea of temperature
waves also originated from Tisza, the results differed in both models:
Landau predicted that second sound approaches a value of $c/\sqrt{3}$
with $c$ being the velocity of ordinary sound waves in the zero temperature
limit while in Tisza`s model the velocity of second sound approaches
zero. The experimental confirmation of Landau`s result ultimately
marked the success of his model which we shall briefly review in the
next section. 

\newpage{}

\section{The two-fluid model\label{sec:The-two-fluid-model}}

~

\noindent Landau essentially constructed his famous two-fluid model
based on the experimental results reviewed in the last section. We
will now discuss the non-dissipative version of this model, i.e. we
will consider a mixture of a superfluid and an ideal fluid (the normal
component). A generalization which takes into account the viscosity
of the normal fluid is discussed for example in \cite{Khalatnikov}.
The basic equations of motion of an ideal fluid are built on the fact
that the motion of a fluid within a given volume element can be described
by the mass density $\rho(\vec{r},t)$, the entropy density $s(\vec{r},t)$
and the fluid velocity $\vec{v}(\vec{r},t)$. The corresponding equations
read:

\begin{eqnarray}
\frac{\partial\rho}{\partial t}+\vec{\nabla}\cdot\rho\vec{v} & = & 0,\,\,\,\,\,\,\textrm{\,\,\,\,\,\,\,\ conservation of mass}\\
\frac{\partial s}{\partial t}+\vec{\nabla}\cdot s\vec{v} & = & 0,\,\,\,\,\,\,\textrm{\,\,\,\,\,\,\,\ conservation of entropy}\label{eq:IdealFluid1}\\
\frac{\partial\vec{v}}{\partial t}+\left(\vec{v}\cdot\vec{\nabla}\right)\vec{v} & = & -\frac{\vec{\nabla}P}{\rho}\,.\,\,\,\textrm{Euler equation}
\end{eqnarray}

\noindent Equivalently to the Euler equation the conservation of momentum
can be used 

\begin{equation}
\partial_{t}\, g_{j}+\partial_{i}\Pi_{ij}=0\,,\label{eq:IdealFluid2}
\end{equation}

\noindent where the momentum density is given by $g_{i}=\rho v_{i}$
and the stress tensor of an ideal fluid by

\begin{equation}
\Pi_{ij}=P\delta_{ij}+\rho v_{i}v_{j}\,.\label{eq:IdealFluid3}
\end{equation}

\noindent These equations can be solved for $\rho$, $s$ and $\vec{v}$
provided that the equation of state $P=P(\rho,s)$ is supplemented. 

\noindent To explain the experimental fact that entropy does not flow
with the center mass velocity $\vec{v}$, it is obviously necessary
to go beyond the Euler set of equations and introduce an independent
velocity field associated with the entropy flow. This velocity is
denoted by $\vec{v}_{n}$, reflecting that only the normal fluid carries
entropy. The velocity of the superfluid on the other hand is constrained
by the condition that no turbulence occurs (at least not for flow
velocities below the critical velocity). In mathematical terms, the
superfluid velocity is assumed to be \textit{irrotational,} $\vec{\nabla}\times\vec{v}_{s}=\vec{0}$.
The mass density is divided into superfluid and normal-fluid density%
\footnote{It should be noted that this is merely an interpretation, the two
densities cannot be physically separated. It is not possible to determine
which helium atoms belong to the normal fluid and which to the superfluid.%
} $\rho=\rho_{n}+\rho_{s}$ where the normal-fluid density vanishes
at zero temperature and the superfluid density at the critical temperature.
By definition, the two fluid components interpenetrate each other
without mutual friction. The total mass flow adds up to:

\begin{equation}
\vec{g}=\rho_{n}\vec{v}_{n}+\rho_{s}\vec{v}_{s}\,.\label{eq:2fluid1}
\end{equation}

\noindent The conservation of mass remains unchanged 

\begin{equation}
\frac{\partial\rho}{\partial t}+\vec{\nabla}\cdot\vec{g}=0,\,\,\,\,\,\,\,\,\,\,\rho=\rho_{n}+\rho_{s},\label{eq:2fluid2}
\end{equation}

\noindent whereas the conservation of entropy is now given by\smallskip{}

\noindent 
\begin{equation}
\frac{\partial s}{\partial t}+\vec{\nabla}\cdot s\vec{v}_{n}=0\,\,.\label{eq:2fluid3}
\end{equation}

\noindent Momentum conservation can be written in the form of (\ref{eq:IdealFluid2})
using equation (\ref{eq:2fluid1}) and the stress tensor of the two-fluid
system

\smallskip{}

\noindent 
\begin{equation}
\Pi_{ij}=(P_{n}+P_{s})\delta_{ij}+\rho_{n}v_{ni}v_{nj}+\rho_{s}v_{si}v_{sj}\,.\label{eq:2fluid4}
\end{equation}

\noindent It remains to find the corresponding Euler equation for
the superfluid. As demonstrated by Khalatnikov \cite{Khalatnikov},
this equation can in principle be derived from conservation equations
postulated above, the irrotationality of the superflow and the principle
of Galilean invariance (see also reference \cite{Putterman}). However,
such a derivation is tedious and we shall follow Landau`s original
approach: in order to explain why the chemical potentials on both
sides of a superleak are equal in the steady state Landau \textit{postulated}
that the chemical potential acts as the potential energy of the superfluid
component and $-\vec{\nabla}\mu$ as the corresponding force. The
superfluid Euler equation then reads
\begin{equation}
\frac{\partial\vec{v_{s}}}{\partial t}+\left(\vec{v}_{s}\cdot\vec{\nabla}\right)\vec{v}_{s}=-\vec{\nabla}\mu.\label{eq:2fluid5}
\end{equation}

\noindent Equations (\ref{eq:2fluid2}, \ref{eq:IdealFluid3}) and
momentum conservation form a complete set of eight independent hydrodynamic
equations describing the motion of a superfluid in terms of the eight
variables $\rho$, $s$ ,$\vec{v}_{s}$ and $\vec{v}_{n}$ provided
that the equations of state $p=p[\rho,\, s,\,(\vec{v}_{n}-\vec{v}_{s})^{2}]$,
$\rho_{n}=\rho_{n}[\rho,\, s,\,(\vec{v}_{n}-\vec{v}_{s})^{2}]$ and
$\mu=\mu[\rho,\, s,\,(\vec{v}_{n}-\vec{v}_{s})^{2}]$ are supplemented.
Galilean invariance requires the equations of state to depend only
on the difference of $\vec{v}_{n}$ and $\vec{v}_{s}$ and to be invariant
under rotations. A few concluding remarks about the two fluid equations
are in order:
\begin{itemize}
\item In the strict mathematical derivation of the two-fluid equations,
frame dependence plays an important role. While in the single fluid
case, one can always define a local rest frame of the fluid, this
is no longer possible in the two-fluid case. In the way listed above,
the two-fluid equations are obviously given in a lab frame where $\vec{v}_{s}$
and $\vec{v}_{n}$ are both nonzero. When we consider relativistic
superfluids, the Galilean transformation connecting two frames of
reference will have to be replaced by a Lorentz transformation. Especially
when deriving the corresponding hydrodynamic equations from microscopic
physics, frame dependence will be a non-trivial issue. 
\item The irrotationality condition of the superfluid implies that the superflow
can be expressed as the gradient of a scalar potential $\vec{v}_{s}=-\vec{\nabla}\psi(\vec{x},t)$.
When terms quadratic in the velocities are neglected, equation (\ref{eq:2fluid5})
turns into $\partial\vec{v}_{s}/\partial t=-\vec{\nabla}\mu$. This
is an important relation indicating that velocity and chemical potential
of a superfluid can be obtained as time and spatial derivatives of
the same scalar field $\psi(\vec{x},t)$. The relation $\partial_{t}\psi=\mu$
follows after taking the time derivative 
\begin{equation}
\vec{v}_{s}=-\vec{\nabla}\psi,\,\rightarrow\,\partial_{t}\vec{v_{s}}=-\vec{\nabla}\partial_{t}\psi=-\vec{\nabla}\mu\,,\label{eq:PotentFlow}
\end{equation}
In a relativistic context, it seems natural to unite chemical potential
and superflow using the four gradient $\partial_{\mu}\psi$. In a
microscopic context, we will later identify the field $\psi$ as the
phase of a Bose-Einstein condensate. 
\item The two-fluid framework can be extended to include dissipation. The
equation of motion for the normal component is then to be replaced
by a Navier-Stokes equation. In the simplest case, linear deviations
from equilibrium in the hydrodynamic parameters are considered. While
the conservation equations of mass and momentum can be extended to
include dissipative terms, entropy is no longer conserved. One rather
has: 
\begin{equation}
\frac{\partial s}{\partial t}+\vec{\nabla}\cdot\left(s\vec{v}_{n}+\frac{1}{T}\vec{s}_{diss}\right)=\frac{1}{T}\Sigma\,.
\end{equation}
Here, $\vec{s}_{diss}$ denotes the dissipative entropy flux and $\Sigma$
is a positive definite quantity (the positive entropy production due
to dissipation will drive the system back into equilibrium after some
time). These assumption result in a modified Navier-Stokes equation
which is rather complicated (an explicit expression can be found in
\cite{Khalatnikov}). The two-fluid Navier-Stokes equation differs
from the regular one in the number of viscosity coefficients: in addition
to the shear viscosity, three bulk viscosity coefficients rather than
one are present. It should be noted that the concept of two fluid
components which interpenetrate each other without mutual friction
becomes problematic in the presence of dissipation. A detailed discussion
of the validity of the two-fluid picture including viscosity can be
found in \cite{Putterman}. \newpage{}
\end{itemize}

\section{The discovery of compact stars\label{sec:The-discovery-of-CS}}

~

\noindent The discovery of Sirius B by Walter Adams in 1915 is regarded
as the first discovery of a compact star. Using stellar spectroscopy,
he was able to deduce \cite{Whitedwarf} that despite of its size
which is roughly about the size of the earth, Sirius B had a mass
that is comparable to that of the sun. In reference to their hot temperature
and small size, such objects were called ``white dwarfs''. Due to
the high densities present in a white dwarf, atoms are fully ionized,
all electrons are free and form a degenerate gas. It was soon realized
that relativistic effects are important for a realistic description
of such an electron gas \cite{WhiteDwarf2}. 

\noindent In February 1932, James Chadwick discovered the neutron
\cite{chadwick} after only two weeks of experimentation. It is often
mentioned in literature that this discovery served as a motivation
for Landau to speculate about the existence of neutron stars. This
however seems not to be the case (see reference \cite{histNeutronStar}
for a historical review). The submission of Landau`s first publication
on compact stars \cite{LandauStar} dates back to January 1932 - one
month before Chadwick`s discovery. While it is true that Landau predicted
the existence of stars with the structure of ``gigantic atomic nuclei'',
he was obviously unaware of the existence of neutrons at that time
as he describes the atomic nucleus as being made of protons only (Landau
would not consider neutrons for another six years). The first model
of a nucleus made of protons and neutrons was suggested by Ivanenko
in April 1932 \cite{Iwanenko}. 

\noindent The term neutron star was first introduced in 1933 by Walter
Baade an Fritz Zwicky \cite{firstNeutronStar} at Caltech in an attempt
to explain the enormous amount of energy released in supernova explosions
(the gravitational collapse of the core of a massive star). As they
correctly explained, such a supernova explosion represents the transition
of an ordinary star into a neutron star - an object made up of closely
packed neutrons with a very small radius and extremely high density.
In 1939, George Gamov realized \cite{Gamow} that white dwarfs are
analogous to neutron stars: both represent the final evolutionary
state of a star, but a white dwarf is the supernova remnant of a star
whose mass was not large enough to become a neutron star (which is
the case for over 97\% of the stars in our galaxy). However, the existence
of neutron stars remained controversial. In 1939, Robert Oppenheimer
and George Volkov found analytic solutions to the Einstein equations
of general relativity for the special case of static and spherical
stars made of isotropic matter. Based on the resulting equation (also
called Tolman-Oppenheimer-Volkov equation), they calculated an upper
limit for the mass of neutron star to roughly 0.7 solar masses \cite{TOV}
- smaller than the mass of stellar cores that could collapse into
neutron stars. The crucial ingredient is the equation of state: while
Oppenheimer and Volkov used an equation of state for a degenerate
non-interacting neutron gas, a similar calculation based on Skyrme-model
effective nucleon interactions resulted in a maximum mass as high
as two times the solar mass \cite{Cameron} and general interest in
neutron stars raised again. \begin{wrapfigure}{o}{0.5\columnwidth}%
~~~~\includegraphics[scale=0.9]{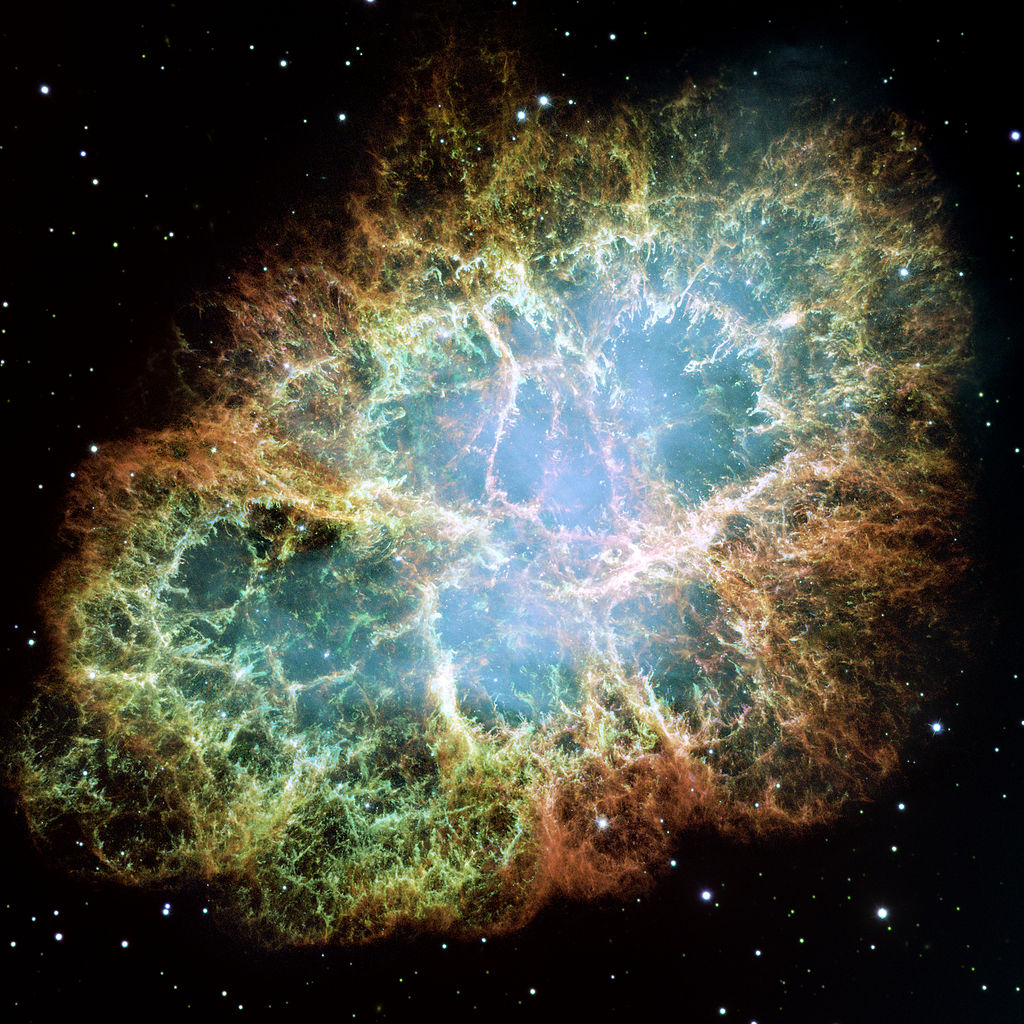}

\protect\caption{The crab nebular - remnant of a gigantic supernova explosion - photographed
by the Hubble space telescope in 2005. }
\end{wrapfigure}%
 In the meantime a consistent microscopic theory of Cooper pairing
was established and possible applications in nuclear matter where
studied by Nicolay Bogolyubov in 1958 \cite{Bogolyubov}. Only one
year later, Arkady Midgal suggested that superfluidity might be present
in neutron stars - a first study was carried out by Vitaly Ginzburg
and David Kirzhnits in 1964 \cite{Ginzburg}. Since then, a great
number of possible phases of matter (including superfluid ones) in
compact stars that go beyond ordinary nuclear matter have been suggested
and we will review some of them in the next chapter. It was not until
1967 that the first direct observation of a neutron star finally took
place. Jocelyn Bell Burnell and Antony Hewish measured radio emissions
originating from a fast rotating neutron star, a so called pulsar
\cite{Jocelyn}, located in the ``crab-nebular'' - a remnant of
a gigantic supernova bright enough to be directly observed by Chinese
astronomers in 1054. Hewish was awarded the Nobel price for this discovery
in 1974. 

\newpage{}

\section{Compact stars from a microscopic point of view\label{sec:Compact-stars-from-a-modern-POV}}

~

\noindent In the frame of this thesis, we will reserve the word compact
star for objects which are dense enough to support nuclear and/or
deconfined quark matter in their inner layers. The physics of white
dwarfs mentioned in the last chapter will not be considered here.
Nucleons and quarks are strongly interacting particles and therefore,
the dynamics of matter inside a compact star are mainly determined
by quantum chromodynamics (QCD). However, electroweak interactions
induce further constraints on the composition of matter. To describe
dense matter in compact stars, we first need to discuss the phase
diagram of QCD and clarify where in this diagram compact stars exist.
Since we are ultimately interested in superfluidity (i.e. a low-temperature
phenomenon), we shall discuss in which sense matter inside a compact
star can be considered as cold. In order to review superfluidity in
dense and strongly interacting matter, we will make use of a modern
microscopic picture, in which superfluidity is the result of spontaneous
symmetry breaking. We will not elaborate further why a system subject
to spontaneous symmetry breaking exhibits superfluidity at this point
as this will be discussed in great detail in part II. The impact of
superfluidity on the phenomenology of a compact star will be explained
in section \ref{sub:Phenomenology-of-compact-stars}.

~

\subsection{Cold and dense nuclear and quark matter\label{sub:Cold-and-dense-nuclear-matter}}

~

\noindent In what follows, only three quark flavors (up, down and
strange) will be considered. Quark chemical potentials in a compact
star can reach values up to about 500 MeV - by far not enough to excite
heavier quarks. (If not explicitly stated otherwise, the symbol $\mu$
will refer to the quark chemical potential.) The presence of electrons
might be required in compact star matter to achieve electric neutrality
and the corresponding chemical potential will be denoted as $\mu_{e}$.
If we were to consider purely strong interactions, we could assign
a chemical potential to each separately conserved quark flavor. However,
weak interactions violate flavor symmetries and we shall see that
as a result the number of independent chemical potentials is reduced
from four ($\mu_{u},\,\mu_{d},\,\mu_{s},\,\mu_{e}$) to two ($\mu_{e}$
and $\mu$). Despite the high densities of matter inside a compact
star, the mean free path of neutrinos is still large enough to allow
them to escape. Lepton number is thus not conserved and no chemical
potential can be assigned to it. Finally, any cluster of matter is
formally required to be a color singlet and thus there is no net chemical
potential for color charges in a compact star (with the exception
of non-uniform phases in which sub domains with positive and negative
charge can in principle exist \cite{CSreview}). In the high energy
limit, quark masses can be neglected and the overall symmetry group
of QCD including color gauge group as well as left and right-handed
flavor groups reads:

\begin{equation}
G_{QCD}=SU(3)_{C}\times SU(3)_{L}\times SU(3)_{R}\times U(1)_{L}\times U(1)_{R}\label{eq:SymFullQCD}
\end{equation}

~

\begin{figure}
~~~~~~~~~~~~~~~~~~~~~~~~~\includegraphics[scale=0.55]{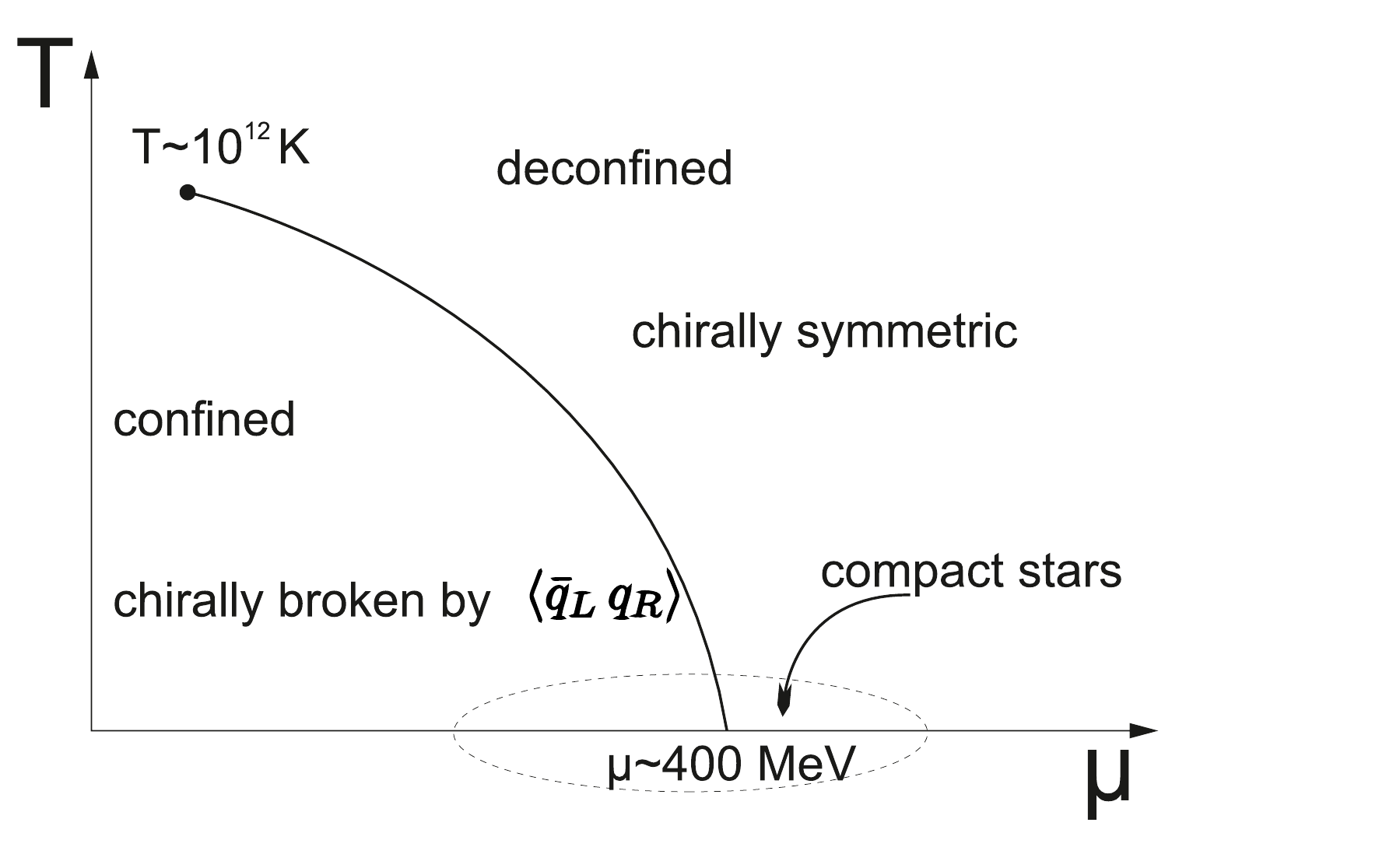}\protect\caption{A crude first approximation of the QCD phase diagram. A single phase
transition line separates confined and chirally broken from deconfined
and chirally symmetric matter. \label{fig:QCDdiagSimple}}
\end{figure}

\noindent While the current quark masses of up and down quark of about
5 MeV are negligible, the strange quark mass of about 90 MeV will
certainly have a strong impact on the composition of matter inside
a compact star. The following features of QCD are particularly important
for our understanding of the phase structure of strong interactions:
\begin{itemize}
\item QCD is an \textit{asymptotically free} theory which means that the
coupling strength between quarks \textit{decreases with increasing
momentum transfer}. At sufficiently high energies and/or densities,
QCD thus behaves like a free field theory. This behavior can effectively
be described by a running coupling $\alpha_{QCD}=\alpha_{QCD}(q/\Lambda_{QCD})$
where the characteristic energy scale $\textrm{\ensuremath{\Lambda_{QCD}}}$
is experimentally determined to a value of $\textrm{\ensuremath{\Lambda_{QCD}}}$
\ensuremath{\approx} 200 MeV. Only when the momentum transfer is larger
than this value, say above $q=1\textrm{GeV}$, perturbative calculations
are valid. 
\item At sufficiently low temperatures and/or densities, quarks are \textit{confined}
into color-neutral composite particles (hadrons). Critical temperature
and density of the deconfinement transition can vaguely be related
to the characteristic energy scale: $\textrm{\ensuremath{\Lambda_{QCD}}}$
corresponds to a temperature of the order of $10^{12}$ K at which
hadrons are melted into their constituent quarks. Furthermore, the
size of a light hadron measures about 1 fm which roughly corresponds
to $\Lambda_{QCD}^{^{-1}}$ . If the average separation distance of
quarks is below 1 fm (at a chemical potential $\mu$ of around 400
MeV) deconfinement sets in. It should be noted that even though asymptotic
freedom and confinement can at least vaguely be related to one energy
scale $\Lambda_{QCD}$, they should be treated as independent: while
confinement is the dominant characteristic of the theory at low energies,
asymptotic freedom becomes dominant at high energies. 
\item At low temperatures and densities\textit{, chiral symmetry }(the symmetry
of independent left and right handed flavor rotations) is \textit{spontaneously
broken} by a color-neutral quark/anti-quark condensate $\phi\thicksim\left\langle \bar{q}_{L}\, q_{R}\right\rangle .$
The resulting ground state is only invariant under simultaneous rotations
of left and right handed quark flavors (i.e. vector rotations)%
\footnote{\noindent The full symmetry group $G_{QCD}$ can be decomposed in
vector and axial-vector symmetries which correspond to simultaneous
($V=L+R$) and opposite ($A=L-R$) rotations of left and right handed
flavors. However, the $U(1)_{A}$ symmetry is violated in any region
of the phase diagram by quantum effects (axial anomaly) and reduced
to the descrete group $Z(6)$. We will surpress this residual group
in the breaking patterns. The vector symmetry $U(1)_{V}$ corresponds
to baryon number conservation and will from now on be denoted as $U(1)_{B}$. 

~%
}
\begin{equation}
SU(3)_{C}\times SU(3)_{L}\times SU(3)_{R}\times U(1)_{B}\rightarrow SU(3)_{C}\times SU(3)_{V}\times U(1)_{B}
\end{equation}
The above pattern indicates that the axial symmetry $SU(3)_{A}$ is
maximally broken $SU(3)_{A}\rightarrow1$ resulting in $N_{f}^{2}-1=8$
Goldstone bosons - the pseudoscalar meson%
\footnote{\noindent Pseudoscalar particles are characterized by zero total spin
and odd parity, usually denoted as $J^{P}=0^{-}$. %
} octet. It should be noted that chirality is an approximate symmetry
valid only at asymptotically high energies and its breaking is a complicated
dynamical matter: instead of exactly massless Goldstone bosons one
obtains pseudo-Goldstone modes with small masses.
\end{itemize}
\noindent For the sake of completeness, it should be mentioned that
chiral symmetry breaking and confinement not necessarily share a common
phase transition line and in principle a confined but chirally symmetric
phase might exist. We will ignore this possibility here and project
a crude first version of the phase diagram with a single phase boundary
separating confined and deconfined quark matter in figure \ref{fig:QCDdiagSimple}. 

\noindent From these generic features of QCD, we can deduce that compact
stars are located in an area of cold and dense matter in the QCD phase
diagram: shortly after their creation in a supernova explosion, the
temperature of compact stars is of the order of 10 MeV ( roughly $10^{11}\, K$).
During the evolution of a compact star, it further cools down to temperatures
in the keV range which is small compared to scale set by $\Lambda_{QCD}$.
Chemical potentials of compact stars on the other hand can become
as large as $\mu\leq500$ MeV. We are thus particularly interested
in a region of $T\ll\Lambda_{QCD}$ and $T\ll\mu$. At very high temperatures
$T\gg\mu$ a plasma of asymptotically free quarks and gluons is realized.
Entropy prohibits a well ordered ground state and there is no spontaneous
symmetry breaking in this region of the phase diagram - in other words
all symmetries of the group $G_{QCD}$ are effectively restored. Experimental
data of the transition to this state matter can be obtained from relativistic
heavy ion colliders. A powerful theoretical tool to probe this transition
is lattice QCD which, at least in the vicinity of the temperature
axis, predicts a smooth crossover from hadronic matter to quark-gluon
plasma. In the limit of $\mu\gg T$ , a rich phase structure due to
a large variety of symmetry breaking patterns is anticipated. Because
of asymptotic freedom, it seems reasonable to begin a discussion of
cold and dense quark matter at very high densities where properties
of the ground state can be deduced from first principles (i.e. QCD)
and then investigate what happens once we progress downwards in density. 

\noindent 

\subsubsection{Highest densities, Color flavor locking \label{sub:Highest-densities-CFL}}

~

\noindent A comprehensive introduction to the physics of high density
quark matter can be found for example in references \cite{CSreview}
and \cite{SchmittBook}. At asymptotically high densities, quark masses
can be neglected and the quark Fermi momenta $p_{F}=\sqrt{\mu^{2}-m_{q}^{2}}\backsim\mu$
become large. Due to Pauli blocking, only states in the vicinity of
the Fermi sphere are modified by interactions. Such interactions then
involve large momentum transfer and are governed by weak coupling.
As a result, we expect to find a Fermi liquid of weakly interacting
quarks and quark-holes. However, in contrast to Coulomb forces acting
between electrons, interactions between quarks are certainly attractive
in some channel which can be deduced from the existence of baryons
which are bound states of quarks. Such attractive interactions between
quark quasi-particles will render the ground state unstable with respect
to Cooper pairing%
\footnote{\noindent As discussed for example in \cite{CSreview}, the energy
scale at weak coupling $g$ below which the quasiparticle picture
of quarks breaks down is parametrically of order $e^{-const/g^{2}}$
while the BCS order parameter (the energy gap in the excitation spectrum
of the quasiparticles) is parametrically larger of order $e^{-const/g}$.
In other words, pairing occurs in a region of the phase diagram where
the quasiparticle picture of the quarks and thus also the BCS argument
remains rigorously valid. 

\noindent ~%
}and these pairs, possessing bosonic quantum numbers, will undergo
Bose Einstein condensation. This argument, as originally presented
by Bardeen, Cooper and Schrieffer (BCS) \cite{BCS} holds true for
quarks in quark matter in the same way as it does for electrons in
a solid%
\footnote{\noindent In some sense, the pairing mechanism is even simpler in
quark matter as an attractive interaction in QCD is directly provided
by single gluon exchange (which is the dominant process at weak coupling)
whereas in a solid, a complicated framework of electron - phonon interactions
is required. %
}. So far, we can make two important observations on high density quark
matter:
\begin{itemize}
\item The ground state of high density quark matter spontaneously breaks
baryon conservation, and therefore is a baryon superfluid. Since baryon
conservation is an exact symmetry of QCD for any given density, the
spontaneous breaking of $U(1)_{B}$ is also the origin of superfluidity
in nuclear matter. 
\item Since the order parameter $\Delta$ of Cooper pairing is a di-fermion
condensate $\Delta\sim\left\langle qq\right\rangle $, it cannot be
a color singlet but rather breaks the $SU(3)_{C}$ symmetry. Thus,
quark matter at highest densities is not only a superfluid but also
a color superconductor. 
\end{itemize}
\noindent It remains to determine the structure of the BCS order parameter
in color, flavor and spin space. Interactions between quarks can be
decomposed into a symmetric sextet as well as an antisymmetric anti-triplet
channel:

\noindent 
\begin{equation}
\left[3\right]\otimes\left[3\right]=\bar{\left[3\right]}^{A}\oplus\left[6\right]^{S}
\end{equation}

\noindent This holds true for color as well as flavor degrees of freedom.
In color space, quarks must be in the anti-triplet representation
as this channel provides attractive interactions while interactions
in the sextet channel are repulsive. Since pairing is preferred in
the antisymmetric spin zero channel and the overall wave function
of a Cooper pair has to be antisymmetric, we can conclude that quarks
pair in an anti-triplet \textit{flavor} channel. The color and flavor
structure of the Cooper pair is thus given by:

\smallskip{}

\noindent 
\begin{equation}
\left\langle q\, q\right\rangle \in\bar{\left[3\right]}_{c}^{A}\otimes\bar{\left[3\right]}_{f}^{A}\label{eq:AntiTrip}
\end{equation}

\begin{figure}
~~~~~~~~~~~~~~~ \includegraphics[scale=0.6]{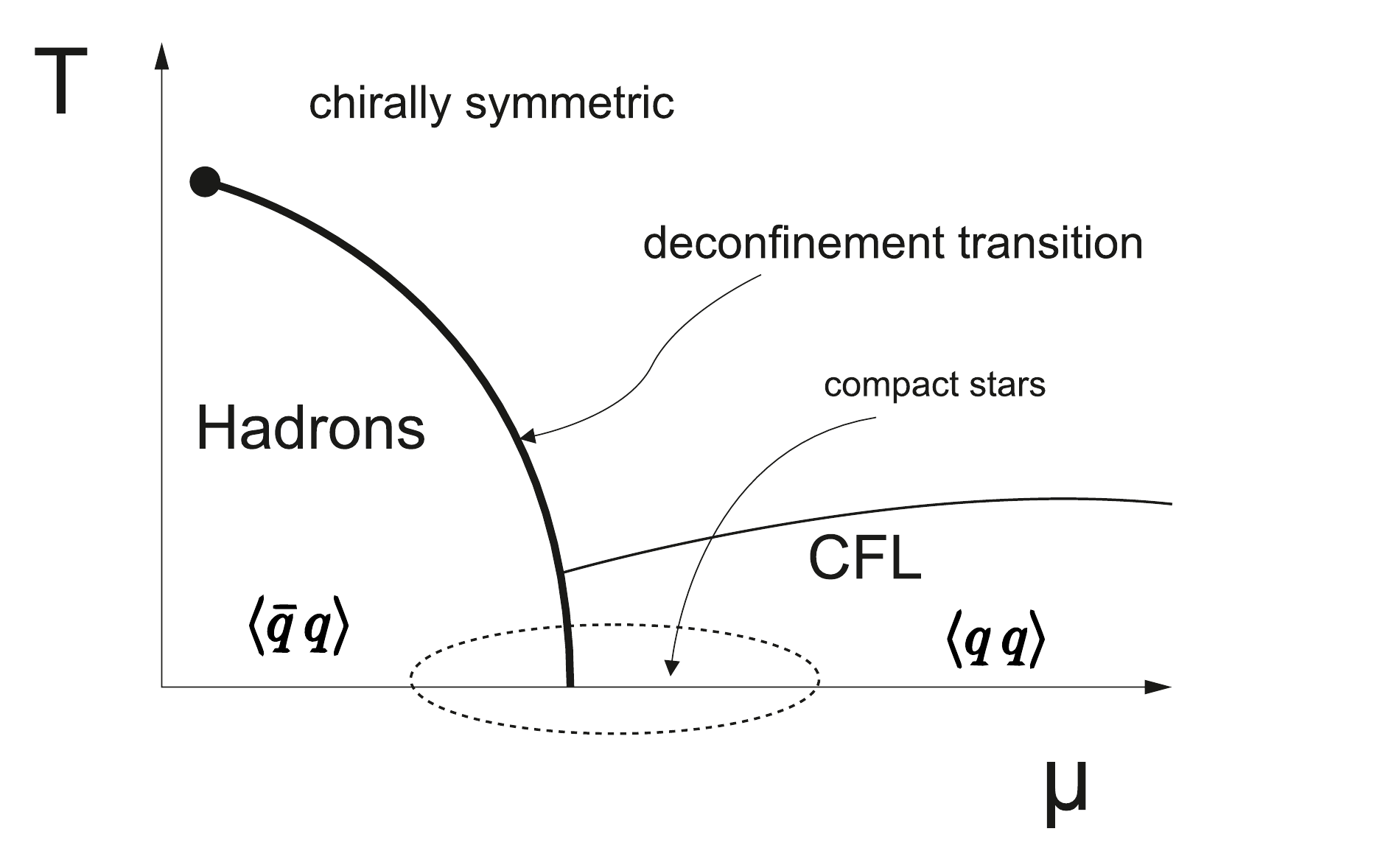}\protect\caption{Phase diagram of QCD including the color-flavor locked phase. Both,
the hadronic and the CFL phase spontaneously break chiral symmetry.
In case of hadronic phase, the condensate is a quark-antiquark bound
state whereas in CFL it is a Cooper pair of quarks. Only at a chemical
potential as large as $\mu\sim10^{8}\,\textrm{MeV}$ the existence
of the CFL phase can be taken for granted and it seems unlikely that
``pure'' CFL will survive all the way down to the phase boundary
of hadronic matter. Therefore, this diagram still represents a strong
simplification. \label{fig:QCDDiagCFL}}
\end{figure}

\noindent Expanding in an antisymmetric color and flavor basis, we
can write: 

\smallskip{}

\noindent 
\begin{equation}
\left\langle q\, q\right\rangle \propto\epsilon^{\alpha\beta a}\epsilon_{ijb}\Phi_{a}^{b}
\end{equation}
The $3\times3$ matrix $\Phi_{a}^{b}$ now determines the specific
color and flavor structure of the Cooper pair within the antisymmetric
basis. To maximize the condensation energy, quarks of all colors and
flavors are required to contribute to the Cooper pairing. This allows
for an unique determination of the order parameter and one obtains
$\Phi_{a}^{b}=\delta_{a}^{b}$ \cite{CFL}. This diquark order parameter
breaks the symmetry group $G_{QCD}$ down to simultaneous rotations
of color and flavor degrees of freedom:

\smallskip{}

\noindent 
\begin{equation}
SU(3)_{c}\times SU(3)_{L}\times SU(3)_{R}\times U(1)_{B}\rightarrow SU(3)_{c+L+R}\times Z(2)\label{eq:SymmCFL}
\end{equation}

\noindent and related to this breaking pattern, the corresponding
ground state has been termed color-flavor locking (CFL) \cite{CFL}.
The breaking of baryon conservation to the discrete subgroup $Z(2)$
reflects the cooper pair nature of the ground state. To complete the
discussion of CFL, we list all elementary excitations of this phase:
\begin{itemize}
\item The spontaneous breaking of baryon conservation $U(1)_{B}$ gives
rise to a discrete massless Goldstone mode as well as a gapped mode
with a finite spectral weight \cite{Fukushima2005}. We shall see
that the Goldstone mode is crucial in the discussion of superfluidity,
the massive mode becomes relevant only at higher energies. 
\item CFL breaks chiral symmetry as can be seen in the breaking pattern
(\ref{eq:SymmCFL}). The low energy spectrum of CFL hence contains
8 light (pseudo)-Goldstone modes with quantum numbers identical to
those of the meson octet resulting from chiral symmetry breaking in
low-density QCD (\ref{eq:SymFullQCD}). The corresponding excitations
can therefore be considered to be the high density analogues of pions,
kaons and the $\eta$ particle%
\footnote{At this point, it is interesting to mention that the symmetry properties
of CFL and hadronic matter in principle allow for the intriguing possibility
of a quark-hadron continuity, see references \cite{continuity}, \cite{Yamamoto2007}
and \cite{Schmitt2011}.

~%
}. The meson octet is complemented by a singlet state $\eta^{\prime}$
resulting from the breaking of $U(1)_{A}$. Due to the axial anomaly,
the $\eta^{\prime}$ particle mass becomes large at lower densities
whereas at large densities, the effect of the anomaly becomes arbitrarily
small (see also discussion in section \ref{sub:intermediate-densities}).
The (pseudo) Goldstone modes together with the (exact) Goldstone mode
resulting from the breaking of $U(1)_{B}$ determine the dynamics
of CFL at low energies $\epsilon<\Delta$. 
\item The $8\oplus1$ (pseudo) Goldstone modes in CFL are accompanied by
nine gapped excitations (i.e. the quark-quasiparticles) where one
of them has a gap of magnitude $2\Delta$ and the remaining eight
of magnitude $\Delta$ \cite{SchmittBook}. 
\item The color gauge group $SU(3)_{c}$ is completely broken resulting
in Meissner masses for all glouns. The generator of the electromagnetic
charge on the other hand is contained in the flavor group $T{}_{em}\subset SU(3)_{L+R}$%
\footnote{\noindent This is visible for example in the covariant derivative.
To couple electric charges to the three quark flavors, the charge
generator $Q=diag(2/3,\,-1/3,$~-1/3) is coupled to the electron
chemical potential $\mu_{e}$.%
} and due to symmetry breaking in CFL, only the residual generator
$\tilde{T}_{em}\subset SU(3)_{c+L+R}$ remains unbroken. In other
words, all diquark condensates carry zero net $U(1)_{\tilde{T}_{em}}$-charge.
This phenomenon is called rotated electromagnetism. $\tilde{T}_{em}$
is a linear combination of the generator of the original electromagnetic
charge and the gluon generator $T_{8}$ and the new gauge field reads
$\tilde{A}_{\mu}=cos(\theta)A_{\mu}-sin(\theta)G_{\mu}^{8}$. However,
since the mixing angle $\theta$ is very small, one may say that the
(original) photon does not acquire a Meissner mass and CFL is not
an electromagnetic superconductor. 
\end{itemize}
\noindent A phase diagram including the CFL phase is shown in figure
\ref{fig:QCDDiagCFL}. In summary, the properties of the CFL phase
at highest densities allow for a rigorous theoretical treatment from
first principles: QCD is weakly coupled at high densities and infrared
divergences are are cut off by the Meissner masses of the gluons.
Furthermore, magnetic interactions in QCD are screened by Landau damping
\cite{CSreview}. 

\noindent It remains to provide a quantitative estimate in which density
regime one can expect CFL to represent the ground state. Calculations
of the order parameter $\Delta$ of CFL in the frame of BCS theory
are reliable at a chemical potential of the order of about $10^{8}$
MeV%
\footnote{It should be emphasized that this magnitude of the chemical potential
specifically describes the limit at which the BCS gap equation is
valid. It should not be confused with a threshold at which perturbative
calculations become applicable - the gap equation is derived under
the assumption of weak coupling, but it is still non perturbative. %
} \cite{PertMu} (roughly 15 orders of magnitude larger than the maximum
value for chemical potentials inside compact stars). Parametrically,
one obtains the following result \cite{Son19999}: 

\begin{onehalfspace}
\begin{equation}
\Delta\propto\mu e^{-c/g}\,.
\end{equation}

\end{onehalfspace}

\noindent This shows that the gap in CFL is parametrically larger
than the standard BCS result for the gap which is proportional to
$e^{-c/g^{2}}$. This deviation results from the fact that the point
like four-fermion interaction has been replaced with the long-range
gluon interaction. As $\mu$ increases faster than $exp(-c/g)$ decreases
\cite{CSreview}, one can conclude that the gap increases for asymptotically
large $\mu$. The critical temperature of CFL deviates by a factor
of $2^{1/3}$ from the standard BCS result

\smallskip{}

\begin{onehalfspace}
\noindent 
\begin{equation}
T_{C}\simeq2^{1/3}\cdot0.57\,\Delta_{T=0}\,.
\end{equation}

\end{onehalfspace}

\noindent As in standard BCS theory, the critical temperature is of
the same order of magnitude as the zero temperature gap. With these
results at hand, it is tempting to try a bold extrapolation to densities
existing inside compact stars. According to the QCD beta function,
a chemical potential of about 400 MeV corresponds to a coupling of
$g\sim3.5$ (of course we can only rely on the two loop approximation
of the beta function which strictly speaking is not valid at all at
lower densities). This results in a gap (and thus also in a critical
temperature) of the order of 10 MeV - above temperatures of a compact
star except for the first minutes after their creation. Even though
such an extrapolation seems of course unreliable, a comparison with
models specifically designed to describe an intermediate density region
such as the NJL (Nambo Jona Lasinio) model shows surprisingly good
qualitative agreement \cite{NJLGap}. This suggests that color superconductors
are at least strong candidates for the ground state of matter inside
a compact star. What really happens to the ground state of QCD once
we leave the save grounds of asymptotically high densities is very
hard to determine as our current theoretical control over the region
of intermediate densities is very limited. Some insights can be obtained
by extrapolations from nuclear theory (upwards in density) or, as
we will discuss in the next section, from CFL (downwards in density).
Another possibility is to use effective models for this region of
the phase diagram. More powerful and reliable methods such as lattice
QCD or experimental insights from heavy ion collisions are limited
to lower densities (one should not however, that future accelerator
facilities such as NICA might provide some insight \cite{NICA} and
also in lattice QCD some progress in the effort to extend calculations
to higher densities has been made \cite{Seiler2013}). From this point
of view, the study of compact stars as the only ``laboratory'' where
such intermediate densities are realized in nature could prove to
be invaluable. 

~\newpage{}

\subsubsection{High, but not asymptotically high densities \label{sub:intermediate-densities}}

~

\noindent Two effects become important once we leave the realm of
asymptotically high densities: 
\begin{itemize}
\item The coupling strength increases. This effect renders calculations
from first principles outside the asymptotic density region very complicated
and it is very challenging to include higher order effects in the
coupling constant. Approaches to resolve this issue include the construction
of an effective field theory for quasi-quarks and gluons near the
Fermi surface \cite{EffQuark} or renormalization group theory \cite{RenoGroup}.
In the effective field theory approach, strong coupling makes it necessary
to include non Fermi liquid effects at energy scales above the gap
\cite{SchwenzerSchaefer}, \cite{Gerhold2004}. 
\item The mass of the strange quark increases and the Fermi momentum $k_{F,s}=\sqrt{\mu_{s}^{2}-m_{s}^{2}}$
decreases. The separation of the Fermi momenta of different quark
flavors eventually leads to the breakdown of Cooper pairing%
\footnote{\noindent For two fermion species with chemical potentials $\mu_{1}$
and $\mu_{2}$, a first order transition to the unpaired phase sets
in at $\delta\mu=\frac{1}{2}(\mu_{1}-\mu_{2})=\Delta/\sqrt{2}$, the
so called Chandrasekhar-Clogston point.

~%
}. To obtain a quantitative estimate, when this happens for 3-flavor
CFL, one has to take into account that matter inside a compact star
is constrained by charge neutrality and beta equilibrium%
\footnote{\noindent In quark matter, $\beta$ decay and electron capture are
represented by $d\rightarrow u+e+\bar{\nu}_{e}$, $s\rightarrow u+e+\bar{\nu}_{e}$
and $u+e\rightarrow d+\nu_{e}$, $u+e\rightarrow s+\nu_{e}$ respectively
and an additional non-leptonic process is given by $s+u\longleftrightarrow d+u$
. These constrain the chemical potentials to $\mu_{d}=\mu_{e}+\mu_{u}$
and $\mu_{s}=\mu_{e}+\mu_{u}$ (as stated before, there is no chemical
potential for neutrinos). %
} , which couples the chemical potentials. In \textit{unpaired quark
matter}, the lack of negative electric charge due to the reduced number
of strange quarks (charge $-1/3$) is compensated by lowering the
up-quark (charge $+2/3$) Fermi momentum and increasing the down-quark
(charge$-1/3$) Fermi momentum (the electron contribution to the charge
density is parametrically negligibly compared to the quark contributions,
for a more detailed discussion see \cite{SchmittBook}) resulting
in the ordering $k_{F,s}<k_{F,u}<k_{F,d}$. To leading order in $m_{s}$
one finds an equidistant separation of $\delta k_{F}=m_{s}^{2}/4\mu$
between all quark flavors. In \textit{CFL quark matter}, the pairing
locks the Fermi momenta together as long as the energy cost of enforcing
the pairing is compensated by the energy released from the condensation
of Cooper pairs. As the cost of maintaining a common Fermi surface
is parametrically $\mu^{2}\delta k_{F}^{2}\propto m_{s}^{4}$ and
the gain of condensation energy is $\Delta^{2}\mu^{2}$, we can expect
paired quark matter to remain stable as long as $\Delta\gtrsim m_{s}^{2}/\mu$.
It should be noted that such a limit is not exclusive to the pairing
pattern of CFL: less symmetric pairing patterns which might appear
as $m_{s}$ increases (for example patterns in which only two flavors
contribute to the pairing) suffer the same fate of stressed pairing
\cite{stressedpair}. From these simple estimates, it is of course
not possible to decide, whether CFL is robust enough to extend all
the way down to the phase boundary of nuclear matter or not. 
\end{itemize}
Systematic studies show that as long as densities are still high enough
and the stress on the pairing pattern is not too large, CFL will most
likely react with the development of a kaon condensate \cite{SchaeferBedaque}.
To understand why kaon condensation is particularly important in this
context, we will take a closer look at the effective theory for mesons
in CFL first derived in reference \cite{SchwenzerSchaefer}. The construction
of this effective theory works analogously to chiral perturbation
theory in nuclear matter: the chiral group $G_{\chi}$ is assumed
to be intact which is an appropriate approximation as long as quark
masses are small compared to the specific scale of chiral symmetry
breaking. This scale is set by the high density analogue of the pion
decay constant $f_{\pi}$ which was calculated \cite{piondecay} to
\begin{equation}
f_{\pi}^{2}=\frac{21-8\,\textrm{log}\,2}{18}\frac{\mu^{2}}{2\pi^{2}}\,.
\end{equation}

\noindent It should be noted that in contrast to the chiral effective
theory in vacuum, the finite chemical potential in the high density
effective theory explicitly breaks Lorentz invariance (see also discussion
in section \ref{sub:Temperature-and-chemical-pot}). The meson fields
$\theta_{a}$ appear to all orders in the chiral field $\Sigma\in SU(3)$,

\medskip{}

\noindent 
\begin{equation}
\Sigma=\textrm{exp}(i\theta_{a}\lambda^{a}/f_{\pi})\,,
\end{equation}

\noindent where $\lambda_{a}$ are the Gellmann matrices. $\theta_{a}$
contains an octet of mesons $(\pi^{\pm}\,,\pi^{0}\,,K^{\pm}\,,K^{0}\,,\bar{K}^{0}\,,\eta\,)$
under the unbroken $SU(3)_{c+L+R}$ symmetry. The subscripts $\pm$
and $0$ now correspond to the $\tilde{Q}$ charges which are attributed
in the same way as (regular) electric charges are attributed to vacuum
mesons. There is however an important difference to vacuum mesons:
the Cooper pair nature of the ground state in CFL leads to mesons
which are given by a $\bar{q}\bar{q}qq$ condensate rather than by
quark/anti-quark bound states. This can be deduced from the fact that
quark flavors in CFL are paired in the anti-triplet representation,
see (\ref{eq:AntiTrip}). Replacing quarks with anti-quarks (and vice
versa) while preserving their flavor quantum numbers results in the
identification $(u\rightarrow\bar{d}\bar{s},\, d\rightarrow\bar{u}\bar{s},\, s\rightarrow\bar{u}\bar{d})$
or $(\bar{u}\rightarrow ds,\,\bar{d}\rightarrow us,\,\bar{s}\rightarrow ud)$.
In order to reproduce the flavor quantum number of, say, a neutral
kaon, one then has to replace $K_{0}\propto\bar{s}d$ with $K_{0}\propto\bar{u}\bar{s}du$.
Obviously, as the quark content of the ``CFL-mesons'' differs from
the vacuum mesons, so will their mass ordering. 

\noindent Since the effect of the axial anomaly becomes arbitrarily
small at high densities, the overall symmetry group under of the effective
theory is given by $SU(3)_{L}\times SU(3)_{R}\times U(1)_{A}$. The
chiral field and mass matrix transform under chiral rotations $(L,R)\in SU(3)_{L}\times SU(3)_{R}$
as $\Sigma\rightarrow L\Sigma R^{\dagger}$ and $M\rightarrow LMR^{\dagger}$.
The somewhat peculiar transformation property of $M$ is related to
the explicit breaking of chiral symmetry induced by finite quark masses:
in order to recover a chirally symmetric theory, it is necessary to
require that the mass matrix is not passive under chiral transformations
but transforms as $LMR^{\dagger}$. Under $U(1)_{A}$ transformations,
the chiral field $\Sigma$ transforms as $exp(-i4\eta_{a})\Sigma$.
Once again, to enforce invariance under axial transformations, $M$
is required to transform as $exp(-2i\eta_{a})\, M$. Collecting all
possible mass contributions up to the second order, the effective
Lagrangian for mesons in CFL reads

\medskip{}

\noindent 
\begin{equation}
\mathcal{L}_{eff}=\frac{f_{\pi}^{2}}{4}Tr\left[D_{0}\Sigma D_{0}\Sigma^{\dagger}-v_{\pi}^{2}\partial_{i}\Sigma\partial_{i}\Sigma^{\dagger}\right]+a\frac{f_{\pi}^{2}}{2}det\, M\, Tr\left[M^{^{-}1}(\Sigma+\Sigma^{\dagger})\right]\,,\label{eq:EffMesonCFL}
\end{equation}

\noindent where $v_{\pi}=1/3$ and the constant $a$ can be obtained
from weak-coupling calculations. Remember that the mesons fields are
contained in the exponent of $\Sigma$ and are thus present to any
order. It can be shown \cite{SchaeferBedaque} that $\mu_{R}=M^{\dagger}M/(2k_{F})$
and $\mu_{L}=MM^{\dagger}/(2k_{F})$ act as effective chemical potentials
for right-handed and left-handed fields. The corresponding symmetry
group $SU(3)_{L}\times SU(3)_{R}$ can formally be treated as a gauge
symmetry, which allows us to introduce a covariant derivative $D_{0}\Sigma$.
The corresponding mass terms then enter the theory in the usual way
of a chemical potential as the zeroth component of the covariant derivative: 

\medskip{}

\noindent 
\begin{equation}
D_{0}\Sigma=\partial_{0}\Sigma+i\mu_{L}\Sigma-i\mu_{R}\Sigma\,.
\end{equation}

\noindent If we where only to consider the $SU(3)_{L}\times SU(3)_{R}$
symmetry, these terms would cover all possible mass contributions.
Due to the additional requirement of $U(1)_{A}$ invariance, also
the second term in $\mathcal{L}_{eff}$ proportional to $a$ is allowed
up to second order in $M$ (for more details on the construction of
mass terms see also \cite{SonSteph}). Terms linear in $M$ are in
principle forbidden by the symmetry group of $\mathcal{L}_{eff}$
as they break the axial symmetry $U(1)_{A}$. In vacuum chiral perturbation
theory where the effect of the axial anomaly is strong, a linear term
of the form $B\left[M\Sigma^{\dagger}+M^{\dagger}\Sigma\right]$ is
included instead of the $U(1)_{A}$ invariant term proportional to
$a$ in (\ref{eq:EffMesonCFL}). Weak-coupling results in the high
density regime for $a$ and $B$ indeed show that $B$ is suppressed
for large $\mu$ and the term proportional to $a$ is dominant while
at lower densities the situation is reversed. 

\noindent Diagonalization of the mass terms of (\ref{eq:EffMesonCFL})
leads to the result that the mass ordering of mesons in CFL is reversed%
\footnote{In particular, the $\eta^{\prime}$ meson which is the Goldstone boson
corresponding to $U(1)_{A}$ breaking is now the lightest meson since
the effect of the anomaly is suppressed.

~%
} as compared to vacuum mesons \cite{SonSteph} . This is one of the
reasons why in CFL kaon condensation is favored over pion condensation.
It is worth emphasizing that this effective theory is constructed
on symmetry properties only. If the scaling of the CFL gap and the
quark masses with the chemical potential were known, the effective
theory would be applicable far outside the weak-coupling regime and
represent a powerful tool to calculate properties of matter inside
a compact star. 

\noindent To study kaon condensation, we can set all meson fields
except the neutral kaons in the exponent of $\Sigma$ to zero and
denote the fields in $\theta_{a}\lambda^{a}$ which correspond to
the neutral kaons by the complex field $\varphi_{K^{0}}=(K^{0},\bar{K}^{0})$.
Then we separate the vacuum expectation value $\Phi_{K^{0}}:=\left\langle \varphi_{K^{0}}\right\rangle $
from fluctuations

~

\noindent $\varphi_{K^{0}}$ and expand up to fourth order in the
fields\cite{CritTKaon}: 

\medskip{}

\noindent 
\begin{equation}
\mathcal{L}_{K^{0}}\rightarrow-U_{K^{0}}\left(\Phi_{K^{0}}\right)+\mathcal{L}_{\varphi}^{(2)}+\mathcal{L}_{\varphi}^{(3)}+\mathcal{L}_{\varphi}^{(4)}\,.
\end{equation}

\noindent As a result, one obtains a complex scalar $\varphi^{4}$
theory. We shall use such a theory as a starting point for the derivation
of superfluid hydrodynamics in part II. For the current discussion
we can neglect fluctuations and consider only the potential: 

\medskip{}

\noindent 
\begin{equation}
U_{K^{0}}=\frac{m_{K^{0}}^{2}-\mu_{K^{0}}^{2}}{2}\Phi_{K^{0}}^{2}+\frac{1}{4}\lambda_{K^{0}}\Phi_{K^{0}}^{4}
\end{equation}

\noindent where the effective mass, chemical potential and coupling
are obtained in terms of the parameters of the high density effective
theory as:

\medskip{}

\noindent 
\begin{equation}
m_{K^{0}}^{2}=a\, m_{u}\,\left(m_{s}+m_{d}\right),\,\,\,\,\,\,\,\,\,\mu_{K^{0}}=\frac{m_{s}^{2}-m_{d}^{2}}{2\mu},\,\,\,\,\,\,\,\,\,\lambda_{K^{0}}=\frac{4\mu_{K^{0}}^{2}-m_{K^{0}}^{2}}{6f_{\pi}^{2}}\,.
\end{equation}

\noindent The mass depends on the coefficient $a$ which in turn is
proportional to $a\propto\Delta^{2}/f_{\pi}^{2}\propto exp(-const/g)$
. A similar derivation of the effective neutral pion mass results
in $m_{\pi^{0}}^{2}\propto m_{s}\left(m_{u}+m_{d}\right)$ which is
indeed larger than the kaon mass. From these results it becomes obvious
why CFL with (neutral) kaon condensation ($\textrm{CFL-\ensuremath{K^{0}}}$)
is a reasonable candidate to succeed (pure) CFL once we progress downwards
in density. We have seen that the stress induced by a finite strange
quark mass on unpaired quark matter in $\beta$ equilibrium is compensated
by converting strange quarks into mostly down quarks. In CFL matter
where all quasi-quarks are gapped, the system rather counteracts the
lack of strangeness with the population of mesons that contain down
quarks and strange holes (i.e. kaons). Bose-Einstein condensation
of these particles will occur if the effective chemical potential
becomes larger than the effective mass. Since $m_{K^{0}}^{2}\propto m_{s}m_{u}\, exp(-const/g)$
and $\mu_{K^{0}}\propto m_{s}^{2}/\mu$, the condensation of neutral
kaons is likely to occur in a region of decreasing $\mu$ and increasing
$m_{s}$ and $g$. Similar arguments also apply to positively charged
kaons $K^{+}$. However, the positive charge requires the presence
of electrons to achieve electric neutrality which disfavors a $K^{+}$
condensed phase. A calculation of the critical temperature of kaon
condensation \cite{CritTKaon} extrapolated to densities of about
$\mu=500$ MeV yields $T_{C}\simeq60$ MeV. This is of the order of
or even larger than the critical temperature of CFL. In other words,
in a region of the phase diagram, where parameters are such that $\mu_{K^{0}}>m_{K^{0}}$
is guaranteed, CFL quark matter should develop a kaon condensate.
Since the kaon condensate spontaneously breaks conservation of strangeness

\smallskip{}

\noindent 
\[
SU(3)_{L+R+C}\supseteq U(1)_{s}\rightarrow1,
\]
we expect this phase to be a kaon superfluid. It should be noted that
weak interactions violate the conservation of strangeness and $U(1)_{s}$
is not an exact symmetry to begin with. However, this violation is
relatively mild as the finite mass of the (pseudo) Goldstone boson
of 50 keV \cite{SchmittBook} is small compared to the critical temperature
of kaon condensation. 

\noindent Given the rather exotic value of $\mu\backsimeq10^{8}$
MeV at which CFL can be taken for granted, it is unknown whether this
phase (with or without meson condensation) really extends all the
way down to phase boundaries of nuclear matter. Other candidate ground
states such as superconductors with two-flavor or single flavor pairings,
Spin-1 superconductors or crystalline phases have been discussed in
great detail in literature (see reference \cite{CSreview} for a review).
As discussed in the end of section \ref{sub:Highest-densities-CFL},
the absence of reliable theoretical or experimental tools in this
region of the phase diagram leaves us behind with a formidable challenge. 

\noindent 
\begin{figure}
~~~~~~~~~~~~~~~~\includegraphics[scale=0.65]{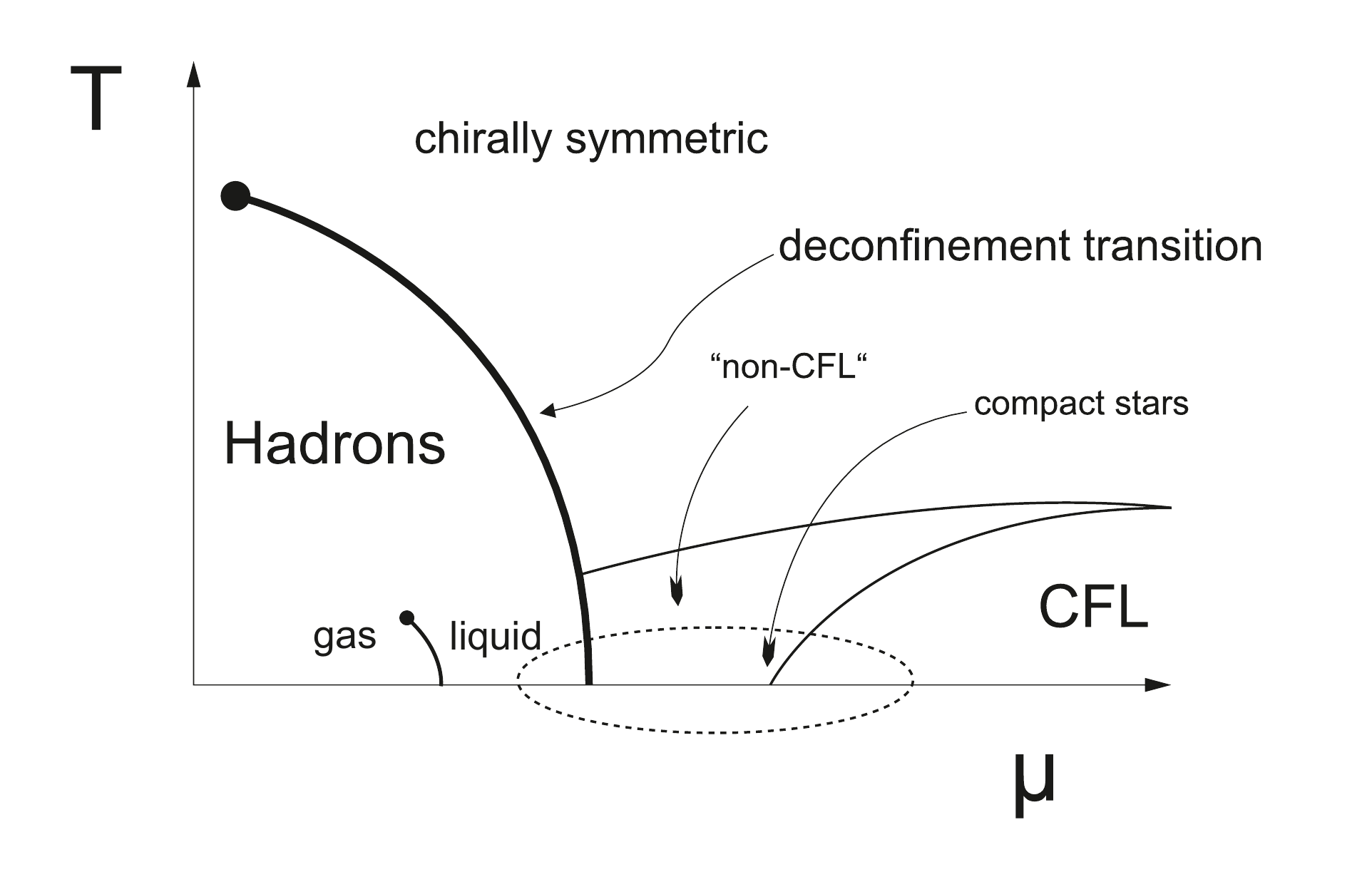}\protect\caption{Phase diagram of QCD. It seems unlikely that CFL will persist all
the way down to the phase boundary of nuclear matter and very little
is known about the region of cold matter at intermediate densities.
This region has been termed ``non-CFL''. One possible approach to
decide which among the candidate ground states for this region is
realized in nature is the study of compact stars which are presumably
located in this area of the phase diagram. \label{fig:PhaseDiagNoCFL} }

\end{figure}

\noindent Finally, after crossing this unknown region we reach the
hadronic phase which is relatively well describable by means of effective
models which are - at least at sufficiently low densities - well constrained
by nuclear scattering data. The hadronic phase is subdivided in a
gaseous and a liquid phase at a chemical potential of $\mu\simeq300$
MeV. The phase structure is summarized in figure \ref{fig:PhaseDiagNoCFL}.
At higher densities and low temperatures, nuclear matter might be
superfluid. In fact it is the spontaneous breaking of the same symmetry
group $U(1)_{B}$ that leads to superfluidity in dense nuclear matter
as baryon number is always an exact symmetry in any density regime
of the QCD phase diagram. Since we are now in the confined phase of
QCD, we cannot deduce the properties of the ground state from first
principles but have to rely on an effective microscopic description.
The attractive interaction between protons and neutrons necessary
for the formation of Cooper pairs can be described by the exchange
of mesons instead of gluons%
\footnote{An appropriate model to describe such interactions is for example
the Walecka model. In its simplest version, protons and neutrons interact
via the exchange of the scalar $\sigma$ and the vector $\omega$
meson (superfluidity in such a model is discussed for example in reference
\cite{ComerJoynt}). 

~%
}. In hadronic matter at low densities, protons and neutrons pair in
the $^{1}S_{0}$ channel, where we have used the spectroscopic notation
$^{2S+1}L_{J}$ to specify total spin $S$, angular momentum $L$
and total angular momentum $J$. At higher densities, medium effects
as well as three-body interactions of the nuclear forces become important
and pairing most likely happens in the $^{3}P_{2}$ channel (see reference
\cite{NuclPair} for a recent review on pairing in nuclear matter).

\noindent Also meson condensation is considered in nuclear matter
where again the focus lies on the condensation of (in this case negatively
charged) kaons. $K^{-}$ condensation is motivated by the fact that
medium effects in nuclear matter lead to an increase of the pion mass
whereas the effective kaon mass is reduced (for a review of strangeness
in neutron stars see reference \cite{Strangeness}). Furthermore,
as soon as densities are large enough for kaons to condense, it becomes
favorable to create a new Fermi sphere for negatively charged kaons
instead of adding additional electrons at large momenta in order to
achieve electric neutrality. The conversion of electrons into kaons
is subdivided into electron capture ($p+e^{-}\rightarrow n+\nu_{e}$)
followed by neutron decay ($n\rightarrow K^{-}+p$) with the neutrinos
escaping the star. As a net result, a large number of neutrons is
converted into protons and matter becomes more and more isospin symmetric.

\noindent Hyperons are estimated to appear at densities of about two
times nuclear saturation density%
\footnote{Nuclear saturation density is defined is the density of nucleons in
an infinite volume at zero pressure. It should be emphasized that
the term nuclear matter does not address matter inside a nucleus but
rather an idealized state of matter of a huge number of neutrons and
protons interacting via strong forces only. In the absence of external
forces, the nucleons will then arrange themselves in a preferred density
of $n_{0}=0.153\,\textrm{f\ensuremath{m^{-3}}}$ (If for instance,
nucleons are added to a very large nucleus, the density of the nucleons
will remain approximately constant at the value of $n_{0}$). The
corresponding binding energy per nucleon is $E_{0}=-16.3\,\textrm{MeV}$.
In a compact star, densities are typically as large as several times
nuclear saturation density. 

~%
} $\textrm{\ensuremath{\rho_{0}}}$ \cite{Strangeness} and depending
on critical temperatures and densities, they might constitute additional
superfluid components. In particular, the $\Lambda$ and $\Sigma^{-}$
particle are often considered in literature as they are conjectured
to appear first with increasing density. It should be mentioned that
the recent discovery of a two-solar-mass neutron star \cite{2SolarNeutron}
challenges the hypothesis of hyperonic matter and/or meson condensation
in neutron stars since the equation of state including hyperons seems
to limit the maximum mass of a neutron star to lower values. This
is all the more surprising since at a certain density threshold, the
onset of hyperons seems unavoidable. This so called ``hyperon puzzle''
is currently among the most heavily debated issues in compact star
physics. 

\noindent ~

\noindent In summary, one can see that superfluidity appears in many
different spots and at various densities in cold strongly interacting
matter :
\begin{itemize}
\item \noindent CFL is a fermionic superfluid since it spontaneously breaks
baryon conservation $U(1)_{B}$. The scalar field theory analyzed
in part II can be seen as a low-temperature approximation to such
a system for energies smaller than the magnitude of the superconductive
gap. Expected temperatures in the interior of compact stars suggest
that this is a reasonable approximation. 
\item \noindent At high but not asymptotically high densities, systematic
studies show that CFL will most likely develop a kaon condensate.
The corresponding ground state $\mathrm{CFL-K^{0}}$ is a bosonic
superfluid since it spontaneously breaks conservation of strangeness
$U(1)_{s}$. The effective theory for kaons in CFL (\ref{eq:EffMesonCFL})
can even be more directly related to a complex scalar field theory
as we have discussed above. However, it is still an approximation
since explicit symmetry breaking effects due to weak interactions
are neglected. It is important to realize that the spontaneous breaking
of $U(1)_{s}$ happens ``on top'' of the symmetry breaking pattern
of CFL. A hydrodynamic description of CFL and kaon condensation is
therefore highly non-trivial: in addition to two superfluid components
(a quark and a kaon superfluid), a normal fluid is present. In part
\ref{sec:A-mixture-of-two-SF}, we will discuss how such a coupled
system of superfluids can effectively be described in terms of coupled
scalar fields. 
\item \noindent Finally we encounter superfluidity in compact stars at much
lower densities, where proton superconductivity and neutron superfluidity
might coexist. Again, one might use an effective bosonic description
to model the low-temperature dynamics. However, in order the describe
proton superconductivity, the global $U(1)$ symmetry would have to
be replaced by a local gauge symmetry. If coexisting hyperon superfluidity
is taken into account, one has to handle an even more complicated
mixture of different fluid components (a mixture of nucleon-hyperon
superfluids has for example been considered in \cite{HypSuperfl}). 
\end{itemize}
\newpage{}

\subsection{Phenomenology of compact stars\label{sub:Phenomenology-of-compact-stars}}

~

\noindent We have seen that nuclear or quark matter at intermediate
densities is notoriously hard to tackle and the study of compact stars
perhaps the only available way to gain further insights. Naturally,
the question arises whether the microscopic composition of matter
inside a compact star can be related to macroscopic effects observable
to astrophysicists. This could lead to a fruitful symbiosis: a profound
understanding of dense matter from first principles allows for a more
precise modeling of compact stars whereas on the other hand observations
of compact stars can help to constrain microscopic models. Superfluidity,
being a macroscopic quantum phenomenon, is certainly of great interest
in this respect. 

\begin{figure}[t]
\noindent \centering{}\includegraphics[scale=0.45]{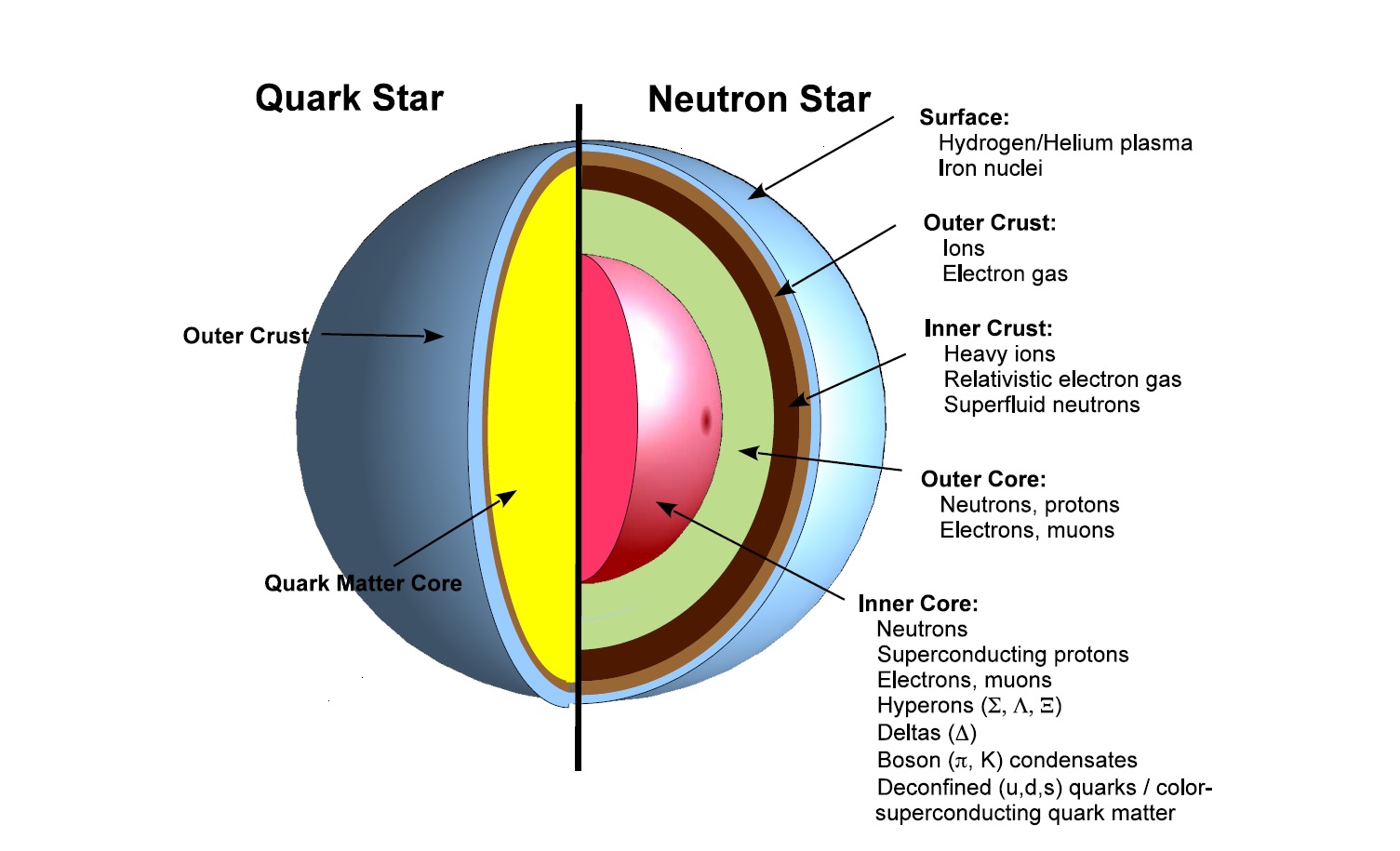}\protect\caption{Conjectured structure of a neutron star with quark core taken from
reference \cite{CompStarStruct}. This figure also shows the configuration
of a speculative bare quark star. A bare quark star might have an
extremely thin surface layer of roughly about 1 fm and an out-flowing
$e^{+}/e^{-}$ plasma of very high luminosity (see \cite{CompStarStruct}
and references therein). \label{fig:CSstruc}}
\end{figure}

\noindent To the best of current knowledge, a compact star can be
subdivided into 3 different regions \cite{NS1}, see figure \ref{fig:CSstruc};
Atmosphere, crust, and core. Nuclear and quark matter are most likely
limited to the core with a radius of several kilometers which in turn
is subdivided into inner and outer core: in the outer core, densities
can reach up to two times $\rho_{0}$ and matter is most likely composed
of a large fraction of neutrons accompanied by a small admixture of
protons as well as electrons and possibly muons constrained by the
condition of electric neutrality. While electrons and muons presumably
form an ideal Fermi liquid, neutrons and protons constitute a strongly
interacting Fermi liquid and are most likely in a superfluid (superconducting)
state. In the inner core of a compact star, densities can reach up
to 15 times $\rho_{0}$. Among the candidates for the ground state
of matter are CFL and CFL-$\textrm{\ensuremath{K^{0}}}$ as well as
nuclear matter including pion or kaon condensation or hyperons. The
crust is again subdivided into outer and inner crust. The outer crust
is a very thin surface layer with a radius of a few hundred meters
and densities below $0.5\,\rho_{0}$ consisting of ionized atoms and
an electron gas. In deeper layers, this electron gas becomes strongly
degenerate and ultrarelativistic while the ions constitute a strongly
coupled Coulomb system (a liquid or a solid). At the boundary to the
inner crust, neutrons start to drip out from the nuclei. Matter inside
the inner crust thus consists of electrons, free neutrons and and
neutron rich atomic nuclei. In analogy to semiconductors, it has been
suggested that neutrons in the crust can be divided into ``conduction''
neutrons and neutrons which are effectively bound to nuclei. A speculative
band structure of these conduction neutrons has been investigated
for example in \cite{Chamel2012}, \cite{LivingCrust}. The fraction
of free neutrons increases with the density until nuclei completely
disappear at the interface to the core. Finally, the outermost layer
(atmosphere) is hypothesized to be a thin plasma layer (at most several
micrometers thick) and its dynamics are assumed to be fully controlled
by the star's magnetic field. \newpage{}

\noindent We shall now discuss observables where superfluidity acts
as an important link between microscopic and macroscopic physics.
Some of these observables such as pulsar glitches explained below
are best described in a hydrodynamic framework. For two reasons, such
an effective hydrodynamic description should be consistent with the
principles of special%
\footnote{We shall ignore effects of general relativity in the frame of this
work. While they are certainly important to model the structure compact
star on a large scale, they can be neglected when we discuss microscopic
properties of matter inside a compact star.

~%
} relativity: first of all, due to the high densities in a compact
star, Fermi momenta are much larger than the masses of particles%
\footnote{This is certainly true for deconfined quark matter. Nucleons in the
crust are assumed to be at the ``borderline'' at which relativistic
effects become important while nucleons in the core definitely have
to be treated relativistically.%
}. Secondly, compact stars can reach rotation frequencies up to $\textrm{f\,\ensuremath{\lesssim}1\,\ m\ensuremath{s^{-1}}}$.
This corresponds to a point on the equator moving with a velocities
up to 20 percent of the speed of light. A relativistic version of
the two-fluid formalism will be introduced in section \ref{sec:Relativistic-thermodynamics-and-hydro}.

\newpage{}
\begin{itemize}
\item Pulsar glitches are sudden increases in the rotational frequency $\Omega$
of a pulsar. The rotation of a compact star is expect to decrease
slowly with time due to the loss of rotational energy to electromagnetic
radiation (dipole radiation or radiation due to electron-positron
winds). They are observed in various pulsars at intervals from days
to years and magnitudes of $\Delta\Omega/\Omega=10^{-6}-10^{-8}$
. The most popular explanation of this phenomenon is related to the
superfluid nature of matter in the crust of a compact star: the angular
momentum of a superfluid is quantized by the formation of vortex lines
(singular regions in which the superfluid density vanishes). The loss
of angular momentum of spinning pulsar thus corresponds to a decrease
in the density of vortex lines (the vortices ``move apart'') just
as the increase of angular momentum corresponds to the creation of
new lines. In the inner regions of the crust close to the neutron
drip density, free neutrons might be superfluid and coexist with the
crust (see reference \cite{LivingCrust} for a theoretical modeling
of the superfluid/crust interaction). If however the vortices are
immobilized because they are ``pinned'' to the rigid structure of
the crust, then after some time the superfluid component will move
faster than the rest of the star. This differential motion will result
in a rising tension and at some critical value, a sudden transfer
of angular momentum from the superfluid to the crust as a result of
the collective ``unpinning'' of several vortex lines might take
place. The vortex lines then move outwards as the angular momentum
of superfluid is decreased and ``re-pin''. There is however doubt
that neutron vortices really posses the ability to ``re-pin'' (see
\cite{CSreview} and references therein). 
\item A compact star has a very rich and complex structure of pulsation
modes which can be classified in terms of their respective main restoring
force (see \cite{rmodes} for a review). Of special interest are so
called r-modes (or Rossby modes) whose restoring force is due to the
Coriolis effect. R-modes are known to become generically unstable
at a certain critical frequency above which this mode grows exponentially.
In other words, if a neutron star is spun up by accretion of surrounding
matter, its spin will be limited by a value slightly above this critical
frequency at which the torque due to accretion is balanced by gravitational
radiation emission which is coupled to the r-mode. Fast spinning stars
are nevertheless observed in nature, which indicates that the r-mode
instability is effectively damped by some mechanism. One such mechanism
is viscous damping, in particular viscosity effects at the boundary
of crust and core are suspected to provide an efficient enough suppression
of the instability. The description of r-modes requires a hydrodynamic
framework which properly takes into account the microscopic composition
of a star. To take into account possible superfluid phases in a compact
star, a two-fluid model must be used which results in a distinction
of ``ordinary'' and superfluid r-modes \cite{2fluidRmode}.
\end{itemize}
\newpage{}

\noindent Another observable which sensitive to the energy gap $\Delta$
but does not require a hydrodynamic description is the cooling of
a star. One minute after the creation of compact star in a supernova,
it becomes transparent to neutrino emission. Neutrinos will then dominate
the cooling for millions of years. Three ingredients are necessary
to gain an understanding of the cooling behavior of a star: the emission
rate of the neutrinos, the specific heat and the heat transport properties
of matter inside a compact star. We will restrict this discussion
to the low-temperature properties where superfluidity is important
(that is we consider only temperatures small compared to the critical
temperature of superfluidity or superconductivity). When compact stars
are formed, their interior temperatures are of the order of $10^{11}$
K. Within days, the star cools down to less than $10^{10}$ K and
during most of its existence, it will sustain a temperature in between
$10^{7}$K and $10^{9}$K. In case of CFL, we have discussed that
critical temperatures in a compact star can roughly be extrapolated
to a value of 10 MeV $\sim\,10^{11}$K, which would mean that for
(almost) any evolutionary state, the low-temperature approximation
is justified. In nuclear matter on the other hand, recent measurements
\cite{Cooling} indicate that the critical temperature of neutron
superfluidity is of the order of $5.5\cdot10^{8}$ K which means that
one needs to go beyond the low-temperature description - at least
in the early evolutionary stages of the star. 

\noindent Most matter inside a compact star transports heat very efficiently
and can thus be assumed to be isothermal to a good approximation.
As a result, the cooling of a compact star should be dominated by
the layer that provides the highest emission rate. As heat transport
in superfluids is particularly large, it is probably the most important
channel to distribute thermal energy in a compact star. However, it
should be emphasized that the convective heat transporting mechanism
that we have discussed in the context of pure liquid helium might
be suppressed in nuclear matter in the presence of electrons or muons
due to entrainment (see discussion in section \ref{sub:Relativistic-thermodynamics-and-entrain}).
In CFL on the other hand, no additional leptons are required to achieve
electric neutrality and therefore the convective counterflow might
indeed be the dominant process.

\noindent The most efficient neutrino emissivity process is the so
called ``direct Urca'' process where in the case of nuclear matter
the emission of neutrinos originates directly from neutron decay and
electron capture reactions. From the principle of momentum conservation,
it can be shown \cite{SchmittBook} that both processes will only
take place if the proton fraction is larger than 10 percent of the
overall baryon density. At least in the case of non-interacting nuclear
matter, this condition would rule out the Urca process. This situation
can change significantly in interacting nuclear matter as we have
discussed before and therefore, the cooling curve might provide a
rough estimate of the proton fraction in a compact star. If the proton
fraction is not large enough, a modified and much less efficient version
of the direct Urca process will most likely take over in which case
a spectator proton or neutron is added to ensure the conservation
of angular momentum (the neutron decay process is then for example
modified to $\textrm{N+n\ensuremath{\rightarrow}N+p+e+\ensuremath{\bar{\nu}_{e}}}$
where N denotes either a neutron or a proton). In case of quark matter,
the corresponding weak processes for the direct Urca process involve
single quarks (see section \ref{sub:Highest-densities-CFL}). In both
cases, this means that one has to come up with the necessary energy
to break Cooper pairs. As a result, specific heat and Urca process
are exponentially suppressed by a factor of $\textrm{exp}(-\Delta/T)$
which might provide a way to determine whether or not superconducting
or superfluid matter exists in a compact star: any shell of a compact
star in which matter is not superfluid, will dominate the cooling.
If all fermionic modes in a compact star are gapped, less efficient
cooling mechanisms originating from the Goldstone mode should become
dominant. Comparing with experimental data, it seems unlikely that
this is the case. 

\newpage{}

\part{Superfluidity from field theory\label{part:Superfluidity-from-field-theory}}

~

~

\section{Relativistic thermodynamics and hydrodynamics \label{sec:Relativistic-thermodynamics-and-hydro}}

~

\noindent To introduce superfluid hydrodynamics for relativistic systems,
it is necessary to replace Galilean invariance, which was the guiding
principle in the construction of Landau`s non-relativistic two-fluid
model, with Lorentz invariance. This concerns thermodynamics as well
as hydrodynamics. A relativistic generalization of thermodynamics
requires to answer questions such as: ``How does temperature or chemical
potential transform under Lorentz transformations?''. We shall not
attempt to find the most general answer to these questions%
\footnote{There has been a rather long debate how to set up a consistent relativistic
description of thermodynamics, see for instance reference \cite{Callen1971}. %
}, but rather search for the correct transformation properties within
the frame of the two-fluid model. The relativistic invariance in this
model is implemented by requiring that the central quantity - the
so called ``master function'' - is built from Lorentz scalars similar
to the Lagrangian of a relativistic field theory in vacuum. However,
the introduction of finite temperature and chemical potential in a
field theory goes hand in hand with the introduction of boundary conditions
which explicitly break Lorentz invariance. Furthermore, performing
calculations within a theory - even if a perfectly invariant one -
can obviously lead to results which are manifestly not invariant:
in the calculation of dispersion relations for example, the zeroth
component of the four-vector $k^{\mu}$ is expressed in terms of its
spatial components $\epsilon_{k}=k_{0}=f(\vec{k})$. We therefore
cannot expect to be able to write down results covariantly at any
intermediate stage of a calculation but we will at least be able to
\textit{reformulate }the final results in terms of invariants - within
certain limits as we shall see in section \ref{sub:Generalized-thermodynamics-from-field-theory}.
We will now review, how to construct relativistic hydrodynamics on
the basis of Lorentz invariance and then in section \ref{sub:Temperature-and-chemical-pot}
discuss in which sense temperature and chemical potential violate
Lorentz invariance in a microscopic approach.

\newpage{}

\subsection{Relativistic thermodynamics and entrainment\label{sub:Relativistic-thermodynamics-and-entrain}}

~

\noindent A generalization of hydrodynamics and thermodynamics for
relativistic superfluids was introduced by Lebedev and Khalatnikov
\cite{LandauLebedev,LebedevKhalatnikov} and Carter\cite{CarterBook}.
Their models - termed ``potential'' and ``convective'' variational
approaches, respectively - differ in formulation but are equivalent
and can be translated into each other \cite{CarterKhala,LReviewsAnderssonComer}.
As a starting point to set up a relativistic generalization of thermodynamics,
we consider the thermodynamic relation between pressure and energy
density,
\begin{equation}
\epsilon+P=\mu n+Ts\,.\label{eq:TDrel}
\end{equation}
The right hand side couples extensive variables (variables that are
proportional to the size of a system such as charge and entropy densities)
to their conjugate intensive variables (variables that describe bulk
properties of matter and do not scale with the size of the system
such as chemical potential or temperature). The relativistic generalization
of the first group of variables is given in terms of the two four-vectors
$j^{\mu}$ and $s^{\mu}$ which contain $n$ and $s$ respectively
in their zeroth component. In the same manner we can define a relativistic
generalization for the second group of variables by introducing two
independent four-vectors for the conjugate momenta which include $\mu$
and $T$ in their zeroth component. Motivated by the non-relativistic
two-fluid formalism (in particular equation (\ref{eq:PotentFlow})),
we introduce a four-gradient $\partial^{\mu}\psi$ as the conjugate
momentum to $j^{\mu}$. We shall see later that the chemical potential
in the superfluid case is indeed proportional to the time derivative
of the phase of the superfluid condensate. For the current purpose,
the symbol $\partial^{\mu}\psi$ denotes some gradient field which
reflects the potential flow of a superfluid. The conjugate momentum
to $s^{\mu}$ is usually denoted by $\Theta^{\mu}$. The convective
approach uses the two four-currents as basic hydrodynamic variables,
whereas the potential approach uses the two conjugate momenta. The
straightforward relativistic generalization of equation (\ref{eq:TDrel})
then reads:

\begin{equation}
\Lambda+\Psi=j\cdot\partial\psi+s\cdot\Theta\,,\label{eq:RelTDrel}
\end{equation}

\noindent where by $j\cdot\partial\psi$ we denote the invariant contraction
$j_{\mu}\partial^{\mu}\psi$ of the two four-vectors. $\Lambda$ and
$\Psi$ denote generalized versions of energy (Hamilton) and pressure
(Lagrange) density (in hydrodynamic literature $\Lambda$ is often
referred to as ``master function''). Both $\Lambda$ and $\Psi$
are connected by the covariant version of a Legendre transform (\ref{eq:RelTDrel}).
To obtain some intuition about the microscopic origin of the conjugate
momenta, we consider the case of zero temperature where there is no
entropy current and where the Legendre transform reduces to $\Lambda+\Psi=j\cdot\partial\psi$.
In this case, the \textit{conjugate momentum} is a direct covariant
generalization of the \textit{canonically conjugate momentum }in the
sense of Hamiltonian mechanics. For a Lagrangian depending on the
gradients of the field $\psi$, we can at tree level identify $\Psi_{T=0}=\mathcal{L}_{T=0}$
(this identification will be made explicit in section \ref{sub:Zero-temperature-hydrodynamics:-generalized})
and the transition from canonical to generalized mechanics is simply
given by 

\noindent 
\begin{eqnarray*}
\pi^{0} & = & \partial\mathcal{L}/\partial\left(\partial_{0}\psi\right)\,\,\,\rightarrow\,\,\pi^{\mu}=\partial\mathcal{L}/\partial\left(\partial_{\mu}\psi\right)\,,\\
\mathcal{H} & = & \pi_{0}\partial_{0}\psi-\mathcal{L}\,\,\,\,\,\,\rightarrow\,\,\,\Lambda=\pi^{\mu}\partial_{\mu}\psi-\mathcal{L}.
\end{eqnarray*}

\noindent This analogy is less obvious at finite temperature as there
is no field theoretic prescription how to construct the conjugate
momentum $\Theta^{\mu}$ to the entropy current. 

\noindent To ensure Lorentz invariance, the only allowed building
blocks to construct $\Lambda$ or $\Psi$ are contractions of the
currents or the conjugate momenta respectively:

\begin{equation}
\Lambda=\Lambda[j^{2}\,,s^{2}\,,j\cdot s],\,\,\,\,\,\,\,\,\,\,\,\,\Psi=\Psi[\partial\psi^{2}\,,\Theta^{2}\,,\partial\psi\cdot\Theta]\,.\label{eq:InvariantMaster}
\end{equation}

\noindent Variations of $\Lambda$ correspond to variations of the
conjugate momenta while variations of $\Psi$ correspond to variations
of the currents:

\medskip{}

\noindent 
\begin{equation}
d\Lambda=\partial_{\mu}\psi dj^{\mu}+\Theta_{\mu}ds^{\mu},\,\,\,\,\, d\Psi=j_{\mu}d\left(\partial^{\mu}\psi\right)+s_{\mu}d\Theta^{\mu}\,.\label{eq:VariationalMaster}
\end{equation}
 Starting from $\Lambda$ and applying the chain rule, the conjugate
momenta are then obtained from

\noindent 
\begin{eqnarray}
\partial^{\mu}\psi & = & \frac{\partial\Lambda}{\partial j_{\mu}}={\cal B}\, j^{\mu}+{\cal A}\, s^{\mu}\,,\label{eq:Entrainment1}\\
\Theta^{\mu} & = & \frac{\partial\Lambda}{\partial s_{\mu}}={\cal A}\, j^{\mu}+{\cal C}\, s^{u}\,,\label{eq:Entrainment2}
\end{eqnarray}

\noindent with: 

\noindent 
\begin{equation}
{\cal A}=\frac{\partial\Lambda}{\partial\left(j\cdot s\right)},\,\,\,\,\,\,\,\,{\cal \,\, B}=\frac{\partial\Lambda}{\partial j^{2}},\,\,\,\,\,\,\,\,\,\,\,\,\,\,\,{\cal C}=\frac{\partial\Lambda}{\partial s^{2}}\,.\label{eq:LambdaDeriv}
\end{equation}

\smallskip{}

\noindent In the same way, starting from $\Psi$ we obtain the currents
as:

\noindent 
\begin{eqnarray}
j^{\mu} & = & \frac{\partial\Psi}{\partial\left(\partial_{\mu}\psi\right)}=\bar{{\cal B}}\,\partial^{\mu}\psi+\bar{{\cal A}}\,\Theta^{\mu}\,,\label{eq:Entrainment3}\\
s^{\mu} & = & \,\,\,\frac{\partial\Psi}{\partial\Theta^{\mu}}\,\,\,\,=\bar{{\cal A}}\,\partial^{\mu}\psi+\bar{{\cal C}}\,\Theta^{\mu}\,,\label{eq:Entrainment4}
\end{eqnarray}

\noindent with:

\noindent 
\begin{equation}
\bar{{\cal A}}=\frac{\partial\Psi}{\partial\left(\Theta\cdot\partial\psi\right)},\,\,\,\,\,\,\,\,\,\,\,\,\,\,\bar{{\cal B}}=\frac{\partial\Psi}{\partial\left(\partial\psi^{2}\right)},\,\,\,\,\,\,\,\,\,\,\,\,\,\bar{{\cal C}}=\frac{\partial\Psi}{\partial\Theta^{2}}\,.\label{eq:PsiDeriv}
\end{equation}

\smallskip{}

\newpage{}

~

\noindent Obviously, the coefficients are related by a simple matrix
inversion,

\bigskip{}

\noindent 
\begin{equation}
\overline{{\cal C}}=\frac{{\cal B}}{{\cal B}{\cal C}-{\cal A}^{2}}\,,\,\,\,\qquad\overline{{\cal B}}=\frac{{\cal C}}{{\cal B}{\cal C}-{\cal A}^{2}}\,,\,\,\,\qquad\overline{{\cal A}}=-\frac{{\cal A}}{{\cal B}{\cal C}-{\cal A}^{2}}\,.\label{eq:AAbar}
\end{equation}

\medskip{}

\noindent As we can see, only two out of the four different four-vectors
are independent and it is up to us which two we are going to use to
set up our hydrodynamics. Even a ``mixed form'' based on one momentum
and one current is possible. (As we will demonstrate in the next section,
Landau`s two-fluid formalism is precisely such a mixed form!) Furthermore,
these relations reveal a very important intrinsic feature of multi-component
fluids: currents are not necessarily aligned with their respective
conjugate momenta. One rather finds that a conjugate momentum is given
as a linear combination of all available currents (and vice versa).
This effect is termed \textit{entrainment} and as we can see, it manifests
itself in the appearance of the coefficient ${\cal A}$ (or $\bar{{\cal A}}$)
which is therefore often called entrainment coefficient%
\footnote{The letter ${\cal A}$ originally referred to ``anomalous''. It
has since been realized that entrainment is a key feature of most
multi-fluid systems. ${\cal B}$ and ${\cal C}$ are called bulk and
caloric coefficients respectively. %
}. Currents and momenta can be viewed as the skeleton of hydrodynamics,
the microscopic physics enter through the coefficients. Entrainment
is by no means a finite temperature effect, it rather concerns any
system in which more than one (particle or heat) current is present.
It was first discovered by Andreev and Bashkin (especially in non-relativistic
literature it is still often referred to as the Andreev-Bashkin effect)
who developed a hydrodynamic description of a mixtures of liquid $\textrm{\ensuremath{^{4}}He}$
and $\textrm{\ensuremath{^{3}}He}$ \cite{AndreevBashkin}. $\textrm{\ensuremath{^{3}}He}$
atoms which obey Fermi statistics need to undergo Cooper pairing first
before they can condense and dissolve in the surrounding $\textrm{\ensuremath{^{4}}He}$
condensate. Such a system therefore consists of two very different
kinds of condensates and correspondingly of two different kinds of
superfluid flow. Entrainment becomes evident since the effective mass
of a $\textrm{\ensuremath{^{3}}He}$ atom due to strong interactions
with the surrounding $\textrm{\ensuremath{^{4}}He}$ becomes more
than twice as large as the $\textrm{\ensuremath{^{3}}He}$ mass itself.
The flow of $\textrm{\ensuremath{^{3}}He}$ quasiparticles therefore
transports a significant fraction of $\textrm{\ensuremath{^{4}}He}$
atoms. It should be noted that entrainment is a non-dissipative effect
owing to the microscopic interactions between particles. The reason
why it is usually not observed in a mixture of ordinary fluids is
viscosity, which tends to equalize velocities. A similar situation
exists in a mixture of neutron and proton superfluids (see reference
\cite{Sauls} or \cite{sjoeberg,Borumand1996,Gusakov2005}) where
the strong interactions due to nuclear forces lead to a coupling between
both fluids (again this is a non-dissipative effect, distinct from
the scattering of neutron and proton quasiparticles which leads to
dissipation): the neutron and proton quasiparticles are dressed 

\noindent \begin{flushleft}
\newpage{}by a common polarization cloud, leading to effective masses
of the form
\par\end{flushleft}

\vspace{-1cm}

\begin{eqnarray}
m_{n}^{*} & = & m_{n}+\delta m_{nn}^{*}+\delta m_{np}^{*}\,,\label{eq:EntrainExample1}\\
m_{p}^{*} & = & m_{p}+\delta m_{pp}^{*}+\delta m_{pn}^{*}\,,\nonumber 
\end{eqnarray}

\noindent with $\delta m_{np}^{*}=\delta m_{pn}^{*}\sim0.5\, m_{p}$
being the entrainment contribution. The corresponding relations between
the conserved momenta and velocity-fields

\vspace{-1cm}

\begin{eqnarray}
\vec{g}_{n} & = & \rho_{nn}\vec{v}_{n}+\rho_{np}\vec{v}_{p}\,,\label{eq:Entrainexample2}\\
\vec{g}_{p} & = & \rho_{pp}\vec{v}_{p}+\rho_{np}\vec{v}_{n}\,,\nonumber 
\end{eqnarray}

\noindent are a non-relativistic equivalent of equations (\ref{eq:Entrainment1})
and (\ref{eq:Entrainment2}). An entrainment matrix for a mixture
of nucleons and hyperons has been calculated in reference \cite{HypSuperfl}.
We shall construct a relativistic effective theory for entrainment
at zero temperature in part \ref{sec:A-mixture-of-two-SF}. 

\noindent Entrainment effects are suspected to play a key role in
many observables of neutron stars \cite{LivingCrust}: they are important
in the determination of frequency and damping of oscillation modes
of neutron star cores composed of neutron and proton superfluids.
Furthermore, they might have a strong influence on the heat transport
mechanism in neutron stars. As we have argued in the context of helium,
the dominant heat transport mechanism in a superfluid is convection
- at least in the absence of other charged particles. In the interior
of a rotating neutron star, entrainment effects induce a flow of protons
around vortices of the neutron superfluity resulting in huge magnetic
fields of about $10^{14}$ G for each vortex line. Electrons then
scatter on these large magnetic fields and thereby induce a mutual
friction between the neutron superfluid and the electrons. This might
effectively damp the counterflow mechanism and favor heat conduction
over convection. Finally, entrainment effects between a lattice of
ionized nucleons and superfluid neutrons at the inner crust of a neutron
star most likely have a strong influence on pulsar glitches \cite{EntrainmentGlitches}. 

\noindent For relativistic systems, the generalized Legendre transform
which takes into account superfluids originating from several particles
species denoted by the chemical index $x$ is then given by:

\begin{equation}
\Lambda=\Psi-s\cdot\Theta-\sum_{x}j_{x}\cdot\partial\psi_{x}\,.\label{eq:RelTDrel2}
\end{equation}

\noindent From this general discussion of entrainment now back to
the two-fluid formalism. If one of the two

\noindent \begin{flushleft}
\newpage{}central quantities $\Lambda$ and $\Psi$, say the generalized
energy density is known, the equations of motion can be obtained by
applying the variational principle %
\footnote{To extend this treatment according to the principles of general relativity,
the Einstein-Hilbert action can be added to the fluid action. One
then not only performes variations of $\Lambda$ with respect to the
currents but also with respect to the metric $g$. %
} \cite{LReviewsAnderssonComer}:
\par\end{flushleft}

\begin{equation}
\delta I_{fluid}=\delta\int d^{4}x\,\sqrt{-g}\Lambda=0\,.
\end{equation}

\noindent In the perfect heat conducting case, one obtains the conservation
equations for charge and entropy:

\medskip{}

\noindent 
\begin{equation}
\partial_{\mu}j^{\mu}=0,\,\,\,\,\,\,\,\,\partial_{\mu}s^{\mu}=0\,,\label{eq:ConCurrents}
\end{equation}

\noindent as well as two additional Euler equations:

\medskip{}

\noindent 
\begin{equation}
j^{\mu}\left(\partial_{\mu}p_{\nu}-\partial_{\nu}p_{\mu}\right)=0,\,\,\,\,\,\,\,\, s^{\mu}\left(\partial_{\mu}\Theta_{\nu}-\partial_{\nu}\Theta_{\mu}\right)=0\,.\label{eq:RelEuler}
\end{equation}

\noindent To demonstrate the general structure of these equations,
the conjugate momentum to $j_{\mu}$ has been denoted as $p_{\mu}$.
In the superfluid case, $p_{\mu}=\partial_{\mu}\psi$ and the equation
to the left is trivially fulfilled. The second equation is called
the vorticity equation. We will encounter this equation later and
use it to derive the wave equations which determine the speeds of
sound. While this mathematically very elegant and rigorous formalism
is often applied in relativistic astrophysics, its applicability in
field theory is very limited since here we do not possess the master
function from the beginning and therefore cannot use it to obtain
the correct equations of motion. Nevertheless, we can use the knowledge
of the structure of these equations to correctly construct the corresponding
fluid variables from field theory and then also construct the masterfunction
\textit{a posteriori}. 

\noindent It remains to construct a generalized version of the stress-energy
tensor:

\medskip{}

\noindent 
\begin{equation}
T^{\mu\nu}=-\Psi g^{\mu\nu}+\frac{\partial\Psi}{\partial\left(\partial_{\mu}\psi\right)}\partial^{\nu}\psi+\frac{\partial\Psi}{\partial\Theta_{\mu}}\Theta^{\nu}=-\Psi g^{\mu\nu}+j^{\mu}\partial^{\nu}\psi+s^{\mu}\Theta^{\nu}\,.\label{eq:RelGenT}
\end{equation}

\noindent Again, one can see that in the zero-temperature case, (tree
level) field theoretic and effective description coincide (simply
replace $\Psi$ with $\mathcal{L}$) whereas the finite temperature
generalization is more involved in field theory. $T^{\mu\nu}$ is
manifestly symmetric in the Lorentz indices even in the presence of
entrainment as can easily be checked by eliminating the conserved
currents by their conjugate momenta from equations (\ref{eq:Entrainment3}),(\ref{eq:Entrainment4})
(or the other way around). With the aid of $T^{\mu\nu}$, we can also
formulate the relation between pressure and energy density as

\medskip{}

\noindent 
\begin{equation}
\Lambda=T_{\,\mu}^{\mu}+3\Psi\,.\label{eq:LambdaPsiT}
\end{equation}

\newpage{}

\noindent The conservation of the stress-energy tensor provides us
with an alternative way to obtain the equations of motion

\medskip{}

\noindent 
\begin{equation}
0=\partial_{\mu}T^{\mu\nu}=-\partial^{\nu}\Psi+\left(\partial_{\mu}j^{\mu}\right)\partial^{\nu}\psi+\left(\partial_{\mu}s^{\mu}\right)\Theta{}^{\nu}+j_{\mu}\partial^{\mu}\partial^{\nu}\psi+s_{\mu}\partial^{\mu}\Theta^{\nu}\,.\label{eq:conservT}
\end{equation}

\noindent With $\partial^{\nu}\Psi=j_{\mu}\partial^{\nu}\partial^{\mu}\psi+s_{\mu}\partial^{\nu}\theta^{\mu}$
from equation (\ref{eq:VariationalMaster}), we find:

\bigskip{}

\noindent 
\begin{equation}
0=\partial_{\mu}T^{\mu\nu}=\left(\partial_{\mu}j^{\mu}\right)\partial^{\nu}\psi+j_{\mu}(\partial^{\mu}\partial^{\nu}\psi-\partial^{\nu}\partial^{\mu}\psi)+\left(\partial_{\mu}s^{\mu}\right)\Theta^{\nu}+s_{\mu}(\partial^{\mu}\Theta^{\nu}-\partial^{\nu}\Theta^{\mu})\,.\label{eq:conservT2}
\end{equation}

\noindent Using charge and entropy conservation as well as the fact
that the second term of equation (\ref{eq:conservT2}) is zero by
construction, we find that the vorticity equations follows directly
from the conservation of $T^{\mu\nu}$. 

\noindent Finally, we shall demonstrate how to derive useful expressions
for the coefficients ${\cal A}$, ${\cal B}$, ${\cal C}$ in terms
of various contractions of $s^{\mu}$, $j^{\mu}$, $T^{\mu\nu}$ and
the Lorentz scalar $\Lambda$ \cite{ComerJoynt}. To do so, we first
contract equations (\ref{eq:Entrainment1}-\ref{eq:Entrainment4})
with $j^{\mu}$ and $s^{\mu}$ resulting in:

\noindent 
\begin{eqnarray}
\partial\psi\cdot s & = & {\cal B}\, j\cdot s+{\cal A}\, s^{2},\,\,\,\,\,\,\,\,\,\,\,\,\,\,\,\partial\psi\cdot j={\cal B}\, j^{2}+{\cal A}\, s\cdot j\,,\\
\theta\cdot s & = & {\cal A}\, j\cdot s+{\cal C}\, s^{2},\,\,\,\,\,\,\,\,\,\,\,\,\,\,\,\,\,\,\,\,\theta\cdot j={\cal A}\, j^{2}+{\cal C}\, s\cdot j\,.\nonumber 
\end{eqnarray}

\noindent The equations in the first line can be used to eliminate
${\cal A}$ in favor of ${\cal B}$: 

\bigskip{}

\noindent 
\[
{\cal B}=\frac{\left(j\cdot s\right)\partial\psi\cdot s-s^{2}\partial\psi\cdot j}{\left(j\cdot s\right)^{2}-j^{2}s^{2}}\,.
\]

\noindent In the numerator we still have mixed terms in momenta and
currents. However, it is easy to check that the expression in the
numerator can be rewritten as $s_{\mu}s_{\nu}T^{\mu\nu}-s^{2}\Lambda$.
In an analogous way we obtain expressions for the remaining two coefficients:

\bigskip{}

\noindent 
\begin{equation}
{\cal B}=\frac{s_{\mu}s_{\nu}T^{\mu\nu}-s^{2}\Lambda}{\left(j\cdot s\right)^{2}-j^{2}s^{2}},\,\,\,\,\,\,\,\,\,{\cal A}=-\frac{j_{\mu}s_{\nu}T^{\mu\nu}-\left(j\cdot s\right)\Lambda}{\left(j\cdot s\right)^{2}-j^{2}s^{2}},\,\,\,\,\,\,\,\,\,{\cal C}=\frac{j_{\mu}j_{\nu}T^{\mu\nu}-j^{2}\Lambda}{\left(j\cdot s\right)^{2}-j^{2}s^{2}}\,.\label{eq:ABC1}
\end{equation}

\noindent The inverse relations are:

\bigskip{}

\noindent 
\begin{equation}
\bar{{\cal B}}=\frac{\Theta_{\mu}\Theta_{\nu}T^{\mu\nu}-\Theta^{2}\Lambda}{\left(\partial\psi\cdot\Theta\right)^{2}-\sigma^{2}\Theta^{2}},\,\,\,\,\,\,\,\,\bar{{\cal A}}=-\frac{\partial_{\mu}\psi s_{\nu}T^{\mu\nu}-\sigma^{2}\Lambda}{(\partial\psi\cdot\Theta)^{2}-\sigma^{2}\Theta^{2}},\,\,\,\,\,\,\,\,\bar{{\cal C}}=\frac{\partial^{\mu}\psi\partial^{\nu}\psi T_{\mu\nu}-\sigma^{2}\Lambda}{(\partial\psi\cdot\Theta)^{2}-\sigma^{2}\Theta^{2}}\,.\label{eq:ABC2}
\end{equation}

\newpage{}

\subsection{\noindent A relativistic version of Landau`s two-fluid formalism\label{sub:A-relativistic-version-of-Landau}}

~

\noindent From the choice of variables, it becomes evident that the
original two-fluid formalism of Landau introduced in section \ref{sec:The-two-fluid-model}
differs from the generalized hydrodynamic formalism discussed above.
Instead of setting up hydrodynamics in terms of the conserved currents
$j^{\mu}$ and $s^{\mu}$, the current $j^{\mu}$ is decomposed in
a normal fluid and a superfluid contribution

\medskip{}

\noindent 
\begin{equation}
j^{\mu}=j_{n}^{\mu}+j_{s}^{\mu}=n_{s}v_{s}^{\mu}+n_{n}u^{\mu}\,,\label{eq:RelLandauCurrent}
\end{equation}

\noindent where neither $j_{s}^{\mu}$ nor $j_{n}^{\mu}$ is conserved
on its own. The velocity of the normal fluid is usually denoted by
$u^{\mu}$. By definition, only the normal fluid carries entropy.
Therefore, we define the normal-fluid velocity as:
\begin{equation}
u^{\mu}=\frac{1}{s}s^{\mu},\,\,\,\,\,\,\,\, s=\sqrt{s_{\mu}s^{\mu}}\,.\label{eq:norFluid}
\end{equation}

\noindent In other words, the normal-fluid rest frame is defined by
$u^{\mu}=(1,\vec{0})$ or $\vec{s}=\vec{0}$. The superfluid velocity
on the other hand is as we know constructed from the gradient field
$\partial^{\mu}\psi$:

\medskip{}

\noindent 
\begin{equation}
v_{s}^{\mu}=\frac{1}{\sigma}\partial^{\mu}\psi,\,\,\,\,\,\,\,\,\sigma=\sqrt{\partial_{\mu}\psi\partial^{\mu}\psi}.\label{eq:SuperFl}
\end{equation}
In both cases, the normalization factors guarantee that the condition
$v_{\mu}v^{\mu}=1$ is fulfilled. The basic fluid variables now include
one current and one conjugate momentum. In the strict sense of generalized
hydrodynamics, Landau`s formalism is to be regarded as a mixed form.
This means that in contrast to the rest frame of the normal fluid
defined by $\vec{s}=\vec{0}$, the rest frame of the superfluid is
not defined by a vanishing three-current $\vec{j}$ but rather by
a vanishing three-momentum $\vec{\nabla}\psi=\vec{0}$. Only at zero
temperature, where $\vec{j}=\vec{j}_{s}=n_{s}\vec{\nabla}\psi/\sigma$
both rest frames coincide. In this limit, we will introduce the chemical
potential in our field-theoretic calculation (see section \ref{sub:Zero-temperature-hydrodynamics:-Landau}).
Landau`s mixed form is a frequent source of confusion and we shall
devote the rest of the section to analyze, how pure and mixed form
are related to each other. 

\noindent In an analogous way to $j^{\mu}$ from equation (\ref{eq:RelLandauCurrent}),
we can construct the stress-energy tensor by adding contributions
from two ideal-fluid tensors:

\medskip{}

\noindent 
\begin{equation}
T^{\mu\nu}=\left(\epsilon_{s}+P_{s}\right)v_{s}^{\mu}v_{s}^{\nu}-g^{\mu\nu}P_{s}+\left(\epsilon_{n}+P_{n}\right)u^{\mu}u^{\nu}-g^{\mu\nu}P_{n}\,.\label{eq:RelTLandau}
\end{equation}

\noindent One should note that while in the single fluid case $T^{\mu\nu}$
is constructed such that $\epsilon$ and $P$ are measured in the
rest frame of the fluid, this is in general no longer possible in
the presence of two fluids. In particular $T^{ij}$ is always anisotropic.
In section \ref{sub:From-generalized-to-frames} we will relate the
generalized pressure $\Psi$ which appears in the thermodynamic relation
to the components of $T^{ij}$ in the superfluid and the normal-fluid
rest frames. The decomposition in terms of superfluid and normal components
in the form (\ref{eq:RelLandauCurrent}) and (\ref{eq:RelTLandau})
can be found for instance in \cite{SonTwoFl,GusakovAndersson,GusakovTwoFl,HerzigKovtun,HerzogYaron}.
To translate between the mixed and the pure form (see also appendix
of \cite{HerzigKovtun}), we rewrite the current and stress-energy
tensor with the help of equations (\ref{eq:Entrainment1}), (\ref{eq:Entrainment2}),
(\ref{eq:Entrainment3}) and (\ref{eq:Entrainment4}):

\noindent 
\begin{eqnarray}
j^{\mu} & = & \frac{1}{{\cal B}}\,\partial^{\mu}\psi-\frac{{\cal A}}{{\cal B}}\, s^{\mu}\,,\label{eq:translate1}\\
T^{\mu\nu} & = & -g^{\mu\nu}\Psi+\frac{1}{{\cal B}}\,\partial^{\mu}\psi\partial^{\nu}\psi+\frac{{\cal B}{\cal C}-{\cal A}^{2}}{{\cal B}}\, s^{\mu}s^{\nu}\,,\label{eq:translate2}
\end{eqnarray}

\noindent which shows the stress-energy tensor in its manifestly symmetric
from. Comparing this with equations (\ref{eq:RelLandauCurrent}),
(\ref{eq:RelTLandau}), we can identify: 
\begin{equation}
n_{s}=\frac{\sigma}{{\cal B}}\,,\qquad n_{n}=-\frac{{\cal A}s}{{\cal B}}\,,\qquad\epsilon_{s}+P_{s}=\frac{\sigma^{2}}{{\cal B}}\,,\qquad\epsilon_{n}+P_{n}=\frac{{\cal B}{\cal C}-{\cal A}^{2}}{{\cal B}}\, s^{2}\,.\label{eq:LandauABC}
\end{equation}

\noindent Finally, we show how the invariant quantities $\Psi$ and
$\Lambda$ relate to $\epsilon_{s},\,\epsilon_{n},\, P_{s},\,\textrm{and}\, P_{n}$.
From (\ref{eq:RelGenT}) one can derive the following expression for
the generalized pressure \cite{ComerJoynt} 

\bigskip{}

\noindent 
\begin{equation}
\Psi=\frac{1}{2}\left[\frac{s\cdot\partial\psi(s_{\mu}\partial_{\nu}\psi+s_{\nu}\partial_{\mu}\psi)-s^{2}\partial_{\mu}\psi\partial_{\nu}\psi-\sigma^{2}s_{\mu}s_{\nu}}{(s\cdot\partial\psi)^{2}-s^{2}\sigma^{2}}-g_{\mu\nu}\right]T^{\mu\nu}\,.
\end{equation}

\noindent Plugging in the stress-energy tensor in its mixed form from
(\ref{eq:RelTLandau}) on the right-hand side of this equation, we
find

\noindent 
\begin{equation}
\Psi=P_{s}+P_{n}\,.\label{eq:PsiLandau}
\end{equation}
Consequently, the generalized pressure is the sum of the pressures
of the superfluid and normal components, each measured in their respective
rest frames. Analogously, we find for $\Lambda$

\begin{equation}
\Lambda=\epsilon_{s}+\epsilon_{p}\,.\label{eq:LambdaLandau}
\end{equation}

\subsection{Temperature and chemical potential in field theory \label{sub:Temperature-and-chemical-pot}}

~

\noindent As elegant as the way how temperature and chemical potential
in the covariant formalism were introduced may be - as soon as we
try to derive the corresponding hydrodynamics from field theory, we
are confronted with conceptual difficulties. In field theory, Lorentz
invariance can explicitly be violated as soon as we impose boundary
conditions on a partition function which is otherwise perfectly Lorentz
invariant. In our microscopic calculations, we will work in the Matsubara
formalism in which temperature is associated with imaginary time $\tau=i\, t$.
The boundary conditions now require the quantum fields $\varphi(\tau,\vec{x})$
to vary periodically with $\tau$, i.e. $\varphi(0,\vec{x})=\varphi(\beta,\vec{x})$
where $\beta$ is the inverse temperature. In momentum space, this
condition translates into $exp(i\omega_{n}\beta)=1$ which leads to
the definition of the (bosonic) Matsubara frequencies $\omega_{n}=2in\pi/T$.
It is important to keep in mind that we haven\textquoteright t made
any identification yet how the variable $T$ which appears in the
field-theoretic calculations relates to the temperature variable which
appears in the hydrodynamic two-fluid model. A more physical way to
say why Lorentz invariance appears to be broken at finite temperature
is to point out that finite temperature introduces a preferred rest
frame, the rest frame of the heat bath. The heat bath is characterized
as the frame in which the averaged kinetic energy of an ensemble particles
vanishes. In such a frame, one can measure a net entropy $s_{0}$
but no entropy flow $\vec{s}$. At this point one should remember
that, in the terminology of the two-fluid formalism, the rest frame
of the heat bath characterized by $\vec{s}=\vec{0}$ corresponds the
the rest frame of the normal fluid. Usually one can switch to a covariant
formulation by introducing the velocity four-vector of the heat bath
(see for example the discussion in references \cite{Israel1981} or
\cite{Kraemmer2004}). However, as we will argue in section \ref{sub:Generalized-thermodynamics-from-field-theory}
this is complicated in our case. 

\noindent In the same way, finite density breaks Lorentz invariance
(again the concept of finite density is linked to the preferred rest
frame of the heat bath in which the medium is a rest). In field theory,
the chemical potential couples to the zeroth component of the charge
(Noether) current $j^{\mu}$ in the partition function. In the particular
case of a complex scalar $\varphi^{4}$ theory, we have:

\begin{equation}
Z=\int_{periodic}D\varphi D\varphi^{*}\int D\pi D\pi^{*}exp\int_{0}^{\beta}d\tau\int d^{3}x\left[\pi^{*}\partial_{0}\varphi+\pi\partial_{0}\varphi^{*}-\mathcal{H}+\mu j^{0}\right].\label{eq:Partition}
\end{equation}

\noindent Here, $\pi$ denotes the canonically conjugate momentum
$\pi=\partial\mathcal{L}/\partial(\partial_{0}\varphi)$. After performing
a shift in the conjugate momentum, the new variable $\tilde{\pi}$
appears only quadratically and can be integrated out. This redefinition
leads to a new Lagrangian in which the chemical potential appears
similar to the temporal component of a gauge field (i.e. the time
derivative $\partial_{0}\varphi$ is replaced by $\left(\partial_{0}-i\mu\right)\varphi$).
The details of this calculation can be found for example in \cite{Kapusta}.
Again we haven\textquoteright t made any statement how the variable
$\mu$ is connected to the chemical potential in the two-fluid model.
In an effective theory for two coupled superfluids with chemical potentials
$\mu_{1}$ and $\mu_{2}$, we can still introduce a chemical potential
to the Lagrangian by replacing $\partial_{0}\varphi_{k}$ with $\left(\partial_{0}-i\mu_{k}\right)\varphi_{k}$
where $k=1,2$ (the prove of this statement is somewhat more involved
and carried out later).

\noindent From what we just discussed, it might seem natural to identify
$T=\Theta^{0}$ and $\mu=\partial^{0}\psi$ and assume that $T$ and
$\mu$ are both measured in the normal-fluid rest frame. A formal
prove of these relations will have to wait until section \ref{sub:Generalized-thermodynamics-from-field-theory}.

\noindent The effect of finite density on our microscopic calculation
will be in some sense less severe compared to the effects of finite
temperature: It will always be possible to calculate, how the spatial
components $\vec{j}$ of the charge current (conjugate to $\mu$)
enter our field theoretic calculations. On the other hand we will
not be able to obtain any information how the spatial components of
the entropy current $\vec{s}$ (conjugate to $T$) enter our field
theoretic calculations as these have been projected out. This limitation
also implies that explicit expressions of all hydrodynamic parameters
in the superfluid rest frame are impossible to obtain. We can still
consider the case $\vec{v}_{s}=\vec{0}$ , but this simply means that
now \textit{both fluids} share a common rest frame. The only exception
is the zero-temperature limit discussed in section \ref{sec:Zero-temperature:-Single-fluid},
at which the flow of the normal fluid is zero by definition. In this
case, the only velocity dependence is due to $\vec{v}_{s}$ and $\vec{v}_{s}=\vec{0}$
really corresponds to the superfluid rest frame. 

~

\subsection{From generalized to frame dependent thermo- and hydrodynamics. \label{sub:From-generalized-to-frames}}

~

\noindent Since we anticipate that our microscopic calculations will
be tied to the rest frame of the normal fluid, it is a useful preparation
to calculate frame dependent expressions for $T^{\mu\nu}$ and $j^{\mu}$
which we will later calculate from field theory. Before we do so,
we have to clarify how temperature and chemical potentials in these
frames are related to the generalized four-vectors $\partial^{\mu}\psi$
and $\Theta^{\mu}$ . To obtain, say, the temperature in an arbitrary
frame moving with velocity $v^{\mu}$ we have to evaluate the contraction
$v^{\mu}\Theta_{\mu}$. This leads in the particular cases of superfluid
and normal-fluid rest frames to the following definitions:
\begin{eqnarray}
\textrm{superframe}:\,\,\partial_{\mu}\psi & = & \left(\partial_{0}\psi,\vec{0}\right),\,\,\, T_{s}=\frac{1}{\sigma}\partial_{\mu}\psi\Theta^{\mu},\,\,\,\,\mu_{s}=\frac{1}{\sigma}\partial_{\mu}\psi\partial^{\mu}\psi=\sigma\,,\label{eq:Framedef}\\
\textrm{normal\,\ frame:}\,\, s_{\mu} & = & \left(s_{0},\vec{0}\right),\,\,\,\,\,\,\, T_{n}=\frac{1}{s}s_{\mu}\Theta^{\mu},\,\,\,\,\,\,\,\,\mu_{n}=\frac{1}{s}s_{\mu}\partial^{\mu}\psi\,.
\end{eqnarray}

\noindent To extract the explicit form of $T^{00}$ in either rest
frame, we eliminate $\Psi$ from equations (\ref{eq:RelTDrel}), (\ref{eq:RelGenT})
and find

\noindent 
\begin{equation}
T^{00}=\Lambda-\vec{j}\cdot\vec{\nabla}\psi+\vec{s}\cdot\vec{\Theta}\,,
\end{equation}

\noindent from which the normal-fluid and superfluid rest frame expressions
can easily be read off. For $T^{i0}$ we find from equation (\ref{eq:RelGenT})
\begin{equation}
T^{i0}=j^{i}\partial^{0}\psi+s^{i}\Theta^{0}\,.
\end{equation}

\noindent Remember, that the stress energy tensor as given in equation
(\ref{eq:RelGenT}) is not manifestly symmetric. It is obtained in
its symmetric form when relations (\ref{eq:Entrainment1}), (\ref{eq:Entrainment2})
or (\ref{eq:Entrainment3}), (\ref{eq:Entrainment4}) are inserted.
Finally, the spatial components can be obtained from proper projections
on transverse and longitudinal directions 

~

\noindent with respect to the fluid component in motion: 

\bigskip{}

\noindent 
\begin{eqnarray}
\textrm{superframe}:\,\, T_{\parallel} & = & \frac{s_{i}s_{j}}{\vec{s}^{2}}T^{ij},\,\,\,\,\,\, T_{\perp}=\frac{1}{2}\left(\delta_{ij}-\frac{s_{i}s_{j}}{\vec{\nabla}s^{2}}\right)T^{ij}\,,\label{eq:TsuperParPerp}\\
\nonumber \\
\textrm{normal\,\ frame:}\,\, T_{\parallel} & = & \frac{\partial_{i}\partial_{j}}{\vec{\nabla}\psi^{2}}T^{ij},\,\,\, T_{\perp}=\frac{1}{2}\left(\delta_{ij}-\frac{\partial_{i}\partial_{j}}{\vec{\nabla}\psi^{2}}\right)T^{ij}\,.\label{eq:TnorParPerp}
\end{eqnarray}

\noindent \begin{table*}[t] \begin{tabular}{|c||c|c|} \hline \rule[-1.5ex]{0em}{5ex} & $\;\;$ normal-fluid rest frame $\;\;$ & superfluid rest frame \\[1ex] \hline\hline \rule[-1.5ex]{0em}{6ex} 
charge density $j^0$ & $\;\;\displaystyle{n_n+n_s\frac{\partial^0\psi}{\sigma}}\;\;$ & $\displaystyle{\;\;n_s+n_n\frac{s^0}{s}\;\;}$ \\[2ex] \hline \rule[-1.5ex]{0em}{6ex} 
spatial current ${\bf j}$ & $\;\;\displaystyle{\frac{\partial^0\psi}{\sigma}n_s {\bf v}_s }\;\;$ & $\displaystyle{\;\;\frac{s^0}{s}n_n{\bf v}_n\;\;}$ \\[2ex] \hline \rule[-1.5ex]{0em}{6ex} 
energy density $T^{00}$ & $\;\;\displaystyle{\Lambda-{\bf j}\cdot\nabla\psi}\;\;$ & $\displaystyle{\;\;\Lambda+{\bf s}\cdot{\bf \Theta}\;\;}$ \\[1ex] \hline \rule[-1.5ex]{0em}{5ex} 
$\;\;$momentum density $T^{0i}$$\;\;$ & $\;\;\displaystyle{j^i\partial^0\psi}\;\;$ & $\displaystyle{\;\;\Theta^i s^0\;\;}$ \\[1ex] \hline \rule[-1.5ex]{0em}{5ex} 
long.\ pressure $T_{||}$ & $\;\;\displaystyle{\Psi-{\bf j}\cdot\nabla\psi}\;\;$ & $\displaystyle{\;\;\Psi+{\bf s}\cdot{\bf \Theta}\;\;}$ \\[1ex] \hline \rule[-1.5ex]{0em}{5ex} 
transv.\ pressure $T_{\perp}$ & $\;\;\displaystyle{\Psi}\;\;$ & $\displaystyle{\;\;\Psi\;\;}$ \\[1ex] \hline \rule[-1.5ex]{0em}{5ex} 
$T^{00}+T_\perp-T_{||}$ & $\;\;\displaystyle{\Lambda}\;\;$ & $\displaystyle{\Lambda}$ \\[1ex] \hline 
\end{tabular} \caption{Components of the current and the stress-energy tensor in the normal and superfluid rest frames. In each frame, $\partial^0\psi$ is the chemical potential, $s^0$ the entropy, and $\Theta^0$ the temperature, while $n_n$ and $n_s$ are the normal and superfluid number densities, measured in their respective rest frames and ${\vec{v}}_n={\vec{s}}/s^0$ (${\vec{v}}_s=-\vec{\nabla}\psi/\partial^0\psi$) the three-velocities of the normal (superfluid) component, measured in the superfluid (normal) rest frame. Longitudinal and transverse pressures are defined with respect to the three-direction of the velocity of the other fluid component.}  \end{table*}The rest frame expressions of the current can be obtained directly
from (\ref{eq:RelLandauCurrent}). The various components are listed
in table 1 and can be interpreted as follows: the components of the
current are written in terms of $n_{n}$ and $n_{s}$. (Alternatively,
by means of the translation given in equation (\ref{eq:translate1}),
we could have written them in terms of the coefficients ${\cal A}$
and ${\cal B}$.) Since $n_{n}$ and $n_{s}$ are measured in their
respective rest frames, the charge density of the other fluid component
contains an explicit Lorentz factor, i.e., $n_{s}$ is multiplied
by $\partial^{0}\psi/\sigma$ (see also the discussion at the end
of section \ref{sub:Zero-temperature-hydrodynamics:-Landau}) in the
normal-fluid rest frame, and $n_{n}$ is multiplied by $s^{0}/s$
in the superfluid rest frame. The spatial components of the currents
are given by the respective number densities times the three-velocities
of the other fluid. The components of the stress-energy tensor are
written in terms of the Lorentz scalars $\Psi=P_{s}+P_{n}$ and $\Lambda=\epsilon_{s}+\epsilon_{n}$
. The last two lines of the table illustrate the meaning of these
quantities. The transverse pressure, i.e., the pressure measured in
the spatially orthogonal direction with respect to the fluid velocity
of the other current, is identical to the generalized pressure $\Psi$.
The energy density $T^{00}$ contains the kinetic energy from the
other fluid component. This is exactly the term that distinguishes
the transverse from the longitudinal pressure. Therefore, the combination
of the frame-dependent quantities $T^{00}+T_{\perp}-T_{||}$ is identical
to the generalized energy density $\Lambda$ . 

\newpage{}

\section{The microscopic point of view - what makes a superfluid? \label{sec:The-microscopic-point-of-view}}

~

\noindent Now that we reviewed the hydrodynamic framework which we
ultimately want to derive, we can finally turn to the discussion of
the microscopic foundations of superfluidity. Our understanding of
the microscopic nature of superfluidity is inevitably connected with
the question what happens to matter at very low temperature (``low
temperature'' is of course a relative term, depending on the corresponding
critical temperature). This leads us directly to the third law of
thermodynamics: at absolute zero temperature, the entropy of any substance
must go to zero. In terms of Boltzmann`s relation $S=k_{B}\,\textrm{ln}\,\Omega$,
where $\Omega$ is the number of available quantum states of a system,
this implies $\Omega=1$ (i.e. the system must have reached its ground
state and this ground state is non-degenerate). An obvious condition
for the existence of superfluidity is that matter doesn\textquoteright t
solidify at very low temperatures. As mentioned in the introduction,
it was London who realized that quantum mechanics are responsible
for this phenomenon. As a consequence of Heisenbergs Uncertainty Principle,
the lowest energy state of a system is not zero and as a result, the
molecules of a solid are not entirely fixed to their lattice sites
- even at the absolute zero temperature. If in addition these molecules
are loosely bound by weak van der Walls forces, than even the zero
point motion is enough to render the solid unstable. In other words,
the solid melts under its own zero point energy! Naturally, the approach
to zero temperature is very different for bosons and fermions: in
case of fermions, the Pauli principle forbids that any two particles
can occupy the same state. The result is that even in the ground state,
states up to a very high energy (the Fermi energy) are occupied. For
bosons, there is no such exclusion principle which leads to the conclusion
that all particles occupy the same lowest energy state as we approach
zero temperature. What is special about Bose gases is the fact that
the transition into this ordered state suddenly sets in at a critical
temperature $T_{c}>0$. In case of $\textrm{\ensuremath{^{4}}He}$,
this transition temperature is located at a value of about $2.17$
K at vapor pressure. The fraction of particles which are in the ground
state is called the Bose-Einstein condensate. All particles within
the condensate can be described by \textit{one coherent }wave function
(i.e. the complex phase of the wave function is fixed to one specific
value for all particles in the ground state) which renders this state
particularly robust. The fundamental mechanism behind this transition
is spontaneous symmetry breaking. In the ground state, the phase of
the condensate randomly assumes one specific value. The theory itself
is invariant under global changes of the phase and all values of the
phase lead to the exact same ground state energy. In field theoretical
terms this means that the Lagrangian which describes our system is
invariant under global transformations of the phase, but the ground
state is not. For completeness, it should be mentioned that interactions
slightly modify this simple picture as quantum scattering will drive
a small number of particles out of the ground state even at zero temperature.
Provided that interactions are sufficiently small, this effect can
be neglected. 

\noindent As we shall see, the condensate alone is not enough the
explain superfluidity. Of crucial importance is the nature of the
available excitations. Above absolute zero, all sorts of excitations
including vibrations or spin and orbital degrees of freedom may contribute
to the entropy of a given substance. Such excitations can be described
as a gas of quasiparticles and we will call them \textit{elementary
}in the sense that they correspond to the natural spectrum of excited
states of the substance under consideration%
\footnote{In some sense, a quasiparticle can of course also be considered a
\textit{collective excitation }since, without the presence of a surrounding
medium the whole quasiparticle picture doesn\textquoteright t make
any sense. We shall however reserve the term collective excitation
for sound waves which involve oscillations in the density of the quasi-particles
(i.e. in the density of the \textit{elementary excitations}). %
}. In substances which exhibit superfluidity at low temperatures, the
quasiparticle picture typically remains valid up to very high wave
numbers. Usually these excitations simply ``freeze out'' as we approach
zero temperature because their dispersion relations are gapped. In
a superfluid however, there is a second kind of excitation which is
massless and therefore present for any given temperature. Quite confusingly,
this excitation has been named \textit{superfluid phonon }even though
the term phonon usually describes an elementary excitation of a lattice
of atoms in condensed matter theory. In field theoretic terminology,
such an excitation is called \textit{Goldstone mode. }According to
Goldstone`s theorem, it is a direct consequence of the spontaneous
breaking of a continuous global symmetry. Corresponding to this symmetry,
there is a conserved charge which is carried by the superflow. We
will not consider complications such as spin, our system simply consists
of self-interacting spin-0 particles. The quasiparticle spectrum will
include the superfluid phonon as well as a gapped massive mode which
is not present at low temperatures. We shall explicitly calculate
the dispersion relations of both excitations. The presence of a Goldstone
mode which can be excited for arbitrarily small energies raises the
question why the superfluid ground state does not immediately dissipate.
The answer to this question is Landau`s critical velocity which we
will introduce now. 

~

\subsection{The critical velocity of a superfluid\label{sub:The-critical-velocity}}

~

\noindent To explain under which conditions a fluid can propagate
without friction, Landau considered a superfluid flowing with velocity
$\vec{v}$ through a capillary which defines the rest frame of the
laboratory. Dissipation involves the creation of quasiparticles which,
in the rest frame of the fluid are characterized by an energy $\epsilon_{p}$
and a momentum $\vec{p}$. In the frame of the laboratory, we therefore
have 

\begin{equation}
E=E_{kin}+\epsilon_{p}+\vec{p}\cdot\vec{v}\,.
\end{equation}

\noindent Landau`s original argument is non-relativistic and therefore
both frames are connected by a Galilean transformation. Dissipation
now means that the fluid loses energy which translates into the condition

\medskip{}

\noindent 
\begin{equation}
\epsilon_{p}+\vec{p}\cdot\vec{v}<0\,.
\end{equation}
The minimum value of the expression on the left hand side is obtained
when $\vec{p}$ and $\vec{v}$ are anti-parallel. One then finds: 

\noindent 
\begin{equation}
\left|v_{c}\right|=min\frac{\epsilon_{p}}{\left|p\right|}\,.
\end{equation}

\noindent This minimum condition is identical to\emph{
\begin{equation}
0=\frac{\partial(\epsilon_{p}/p)}{\partial p}\rightarrow\frac{\partial\epsilon_{p}}{\partial p}=\frac{\epsilon_{p}}{p}\,.
\end{equation}
}

\noindent This leads to a very simple geometric interpretation of
the critical velocity: first, one needs to plot the dispersions of
all elementary excitations in an $\epsilon_{p}-p$ plane. In the next
step, one considers a horizontal line through the origin in this plane
and rotates it upwards. 
\begin{figure}[t]
\fbox{\begin{minipage}[t]{1\columnwidth}%
~\includegraphics[scale=0.6]{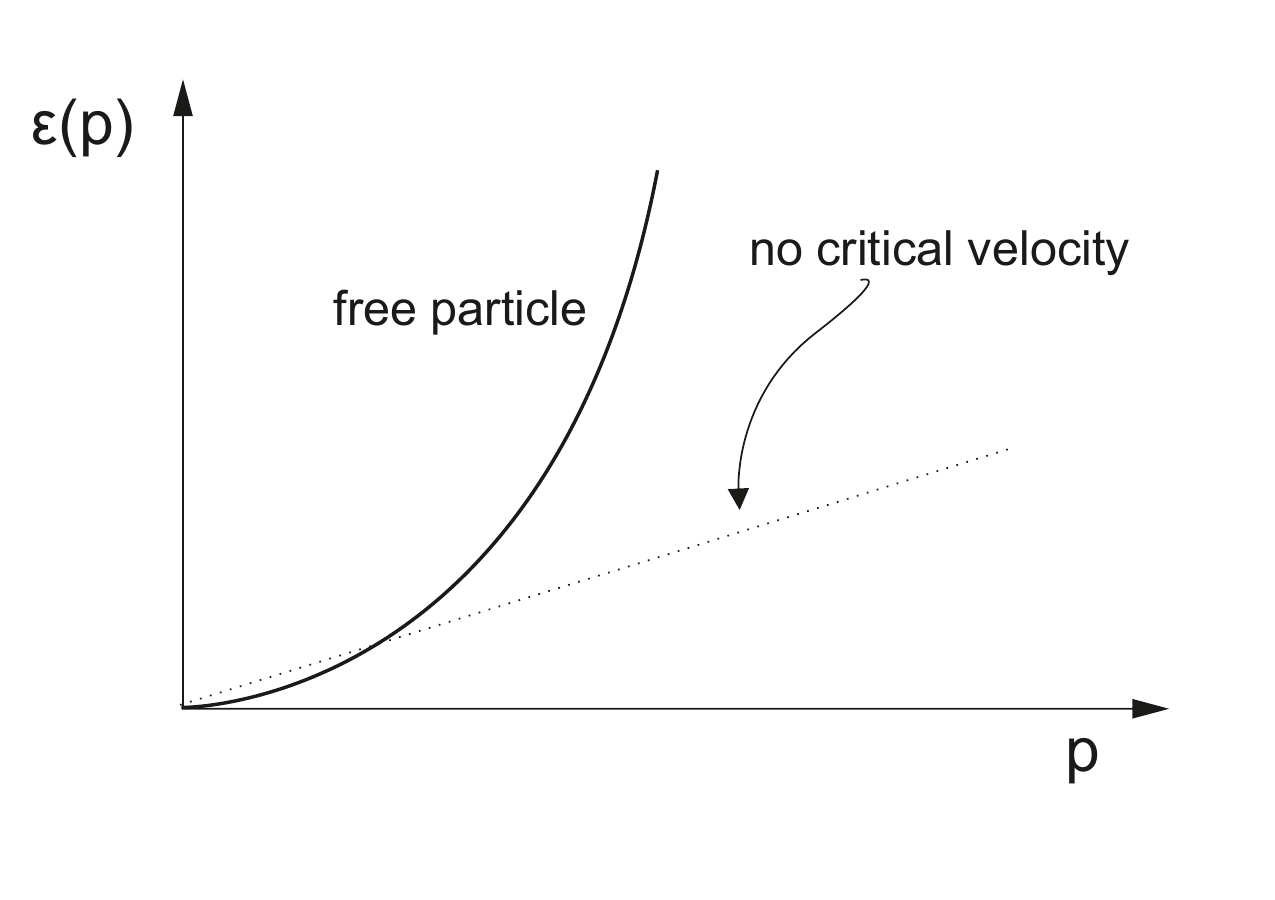}~~~\includegraphics[scale=0.6]{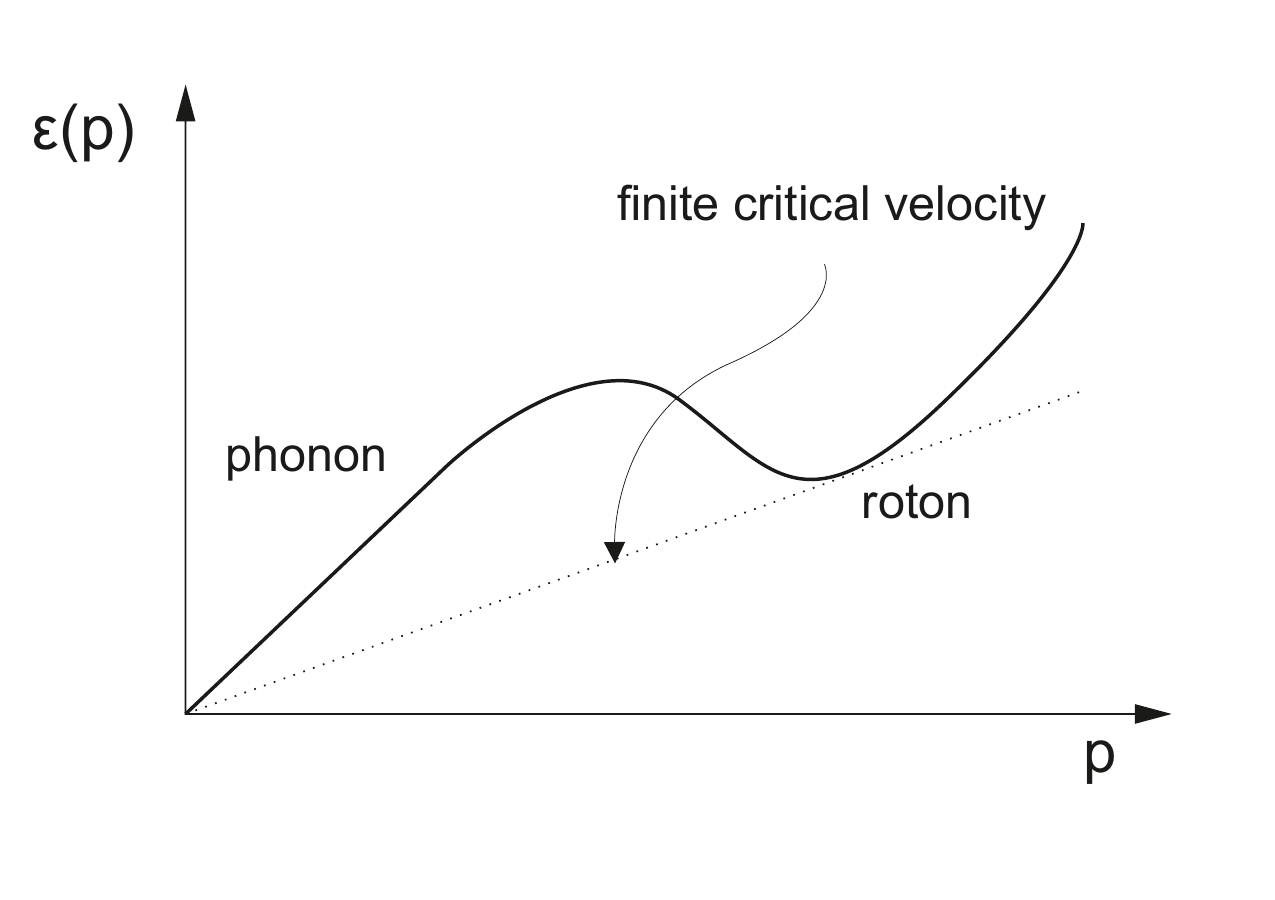}%
\end{minipage}}\protect\caption{Comparison of a free particle and a quasi-particle dispersion relation.
Obviously an ensemble of non-interacting Bose particles does not exhibit
superfluidity.\label{fig:CritVComp} }
\end{figure}
If one can do so without intersecting the curve of the dispersion
relation, superfluidity is supported. In an ideal Bose gas, the particle
dispersions are given by $\epsilon_{p}=p^{2}/2m$ and therefore superfluidity
is not supported. This lead Landau to his famous conclusion that the
elementary excitations in $\textrm{\ensuremath{^{4}}He}$ must be
of an entirely different nature consisting of phonons and rotons.
The two scenarios are compared in figure \ref{fig:CritVComp}. A similar
argument applies for fermionic systems such as $\textrm{\ensuremath{^{3}}He}$.
Excitations are generated as particle-hole pairs and usually measured
relative to the Fermi surface. As properly described by BCS theory,
Cooper pairing leads to an energy gap $\Delta$ in the excitation
spectrum of the fermion quasi-particles. There is however also a Goldstone
mode in $\textrm{\ensuremath{^{3}}He}$ from which the critical velocity
has to be determined. 

\noindent The above argument holds for zero temperature. At finite
temperature, the situation is more complicated as the elementary excitations
will be thermally populated and therefore thermal excitations of the
Goldstone mode are present for any superfluid velocity. It is clear
that a proper derivation of the hydrodynamics of a superfluid requires
both ingredients, condensate and excitations and they will also appear
coupled to each other. This derivation will be carried out in three
steps: we shall begin at zero temperature, where the relation between
hydrodynamics and field theory is well established, review it and
make it as explicit as possible. In the second step, we derive hydrodynamics
in a low-temperature regime. In this approximation, we will still
be able to obtain analytic results for all hydrodynamic parameters.
Finally we study the full temperature range up the the critical temperature
numerically in a self-consistent formalism. We express the basic currents
and momenta of the two-fluid formalism in terms of field-theoretic
quantities, and calculate the generalized pressure $\Psi$ at nonzero
temperature in the presence of a superflow. The microscopic model
will be a relativistic $\varphi^{4}$ field theory (see section \ref{sec:Lagrangian})
with a U(1) symmetry that is spontaneously broken by a Bose-Einstein
condensate of the fundamental scalars. We have discussed how such
a model can in principle be related to physics relevant to compact
stars in section \ref{sub:intermediate-densities}. Nevertheless,
the study which we will carry out is very general since we do not
have to specify the system for which our microscopic theory is an
effective description.

\noindent Several existing studies in the literature are related to
this approach. For instance, in \cite{CarterLanglois} the two-fluid
formalism is connected with a simple statistical approach to the phonon
contribution, \cite{ComerJoynt} connects it with a Walecka model
describing nuclear matter in a neutron star, and \cite{Nicolis} makes
the connection to a very general effective field theory. In \cite{Son1},
a hydrodynamic interpretation of a field-theoretic effective action
was discussed for zero temperature. For the simplified case of a dissipationless,
homogeneous fluid, our study is a generalization of this work to nonzero
temperatures. In that reference, an effective Lagrangian for the superfluid
phonons is formulated which has been employed to study transport properties
of quark matter, see for instance \cite{Manuel1}. 

\newpage{}

\subsection{Spontaneous symmetry breaking in field theory\label{sub:Spontaneous-symmetry-breaking}}

~

\noindent Before we begin the translation from field theory to hydrodynamics,
we shall discuss how spontaneous symmetry breaking manifests itself
in field theory which is a useful preparation for the chapters to
come. We use the same model which will also serve as a basis for the
derivation of the two-fluid equations: a complex scalar $\varphi^{4}$
model describing (repulsively) interacting spin-0 bosons with mass
$m$ and coupling $\lambda>0$ in the presence of a finite chemical
potential $\mu$ (see also section \ref{sub:Temperature-and-chemical-pot}) 

\begin{equation}
\mathcal{L}=\left|(\partial_{0}-i\mu)\varphi\right|^{2}-\left|\vec{\nabla}\varphi\right|^{2}-m^{2}\left|\varphi\right|^{2}-\lambda\left|\varphi\right|^{4}\,.\label{eq:LagrangianClassic}
\end{equation}

\noindent This model is invariant under global $\textrm{U(1)}$ transformations
$\varphi\rightarrow e^{i\alpha}\varphi$ (note that a \textit{complex
}scalar field theory is necessary to introduce a U(1) symmetry). The
complex field can be written as $\varphi=\frac{1}{\sqrt{2}}\left(\varphi_{1}+i\varphi_{2}\right)$.
To allow for Bose-Einstein condensation, we separate the zero momentum
mode $\varphi_{i}\rightarrow\varphi_{i}+\phi_{i}$ where for symmetry
reasons, we can set $\phi_{1}=0$ and define $\phi_{2}:=\phi$. 
\begin{figure}[t]
\fbox{\begin{minipage}[t]{1\columnwidth}%
~\includegraphics[scale=0.55]{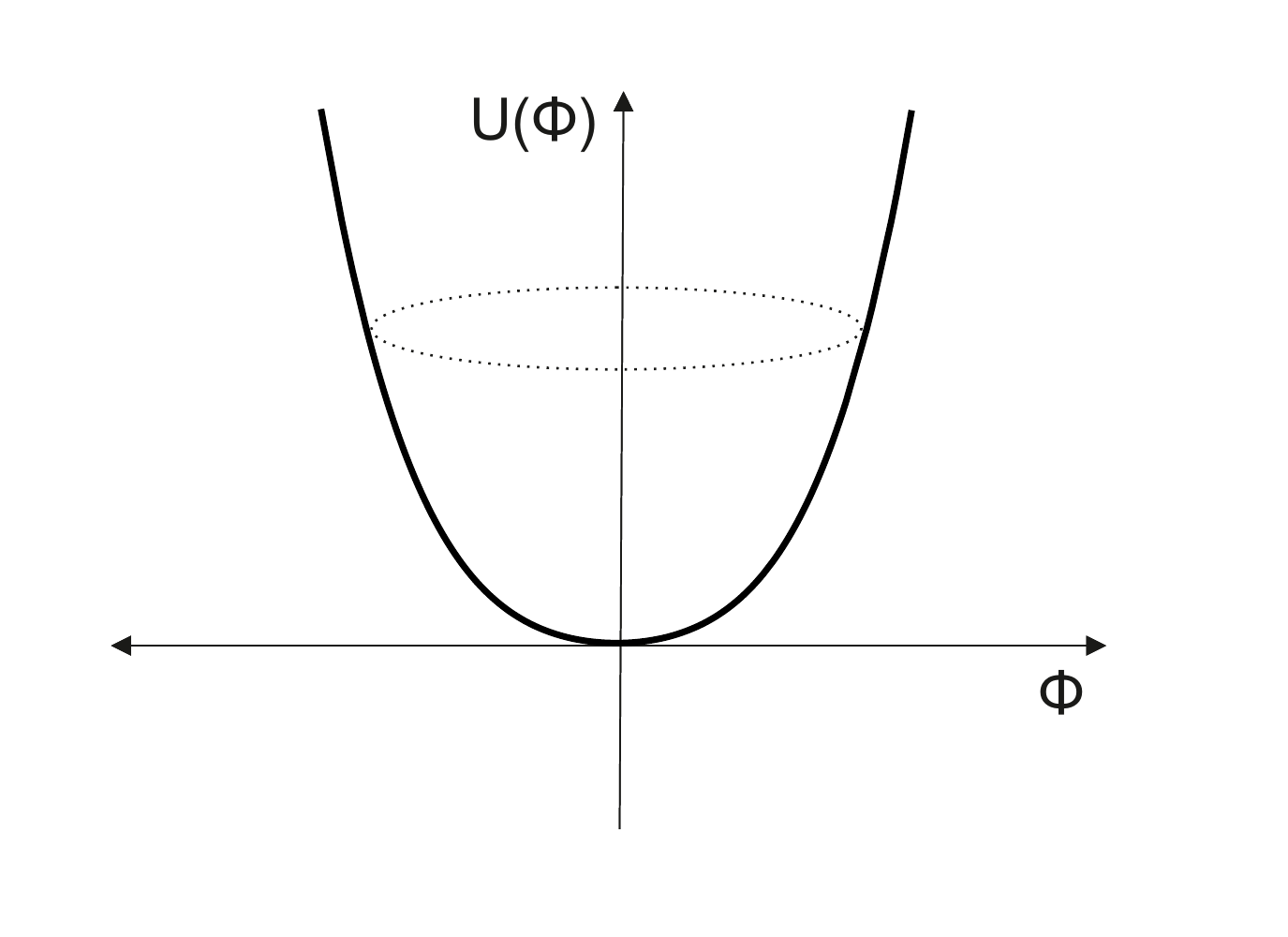}~~~\includegraphics[scale=0.55]{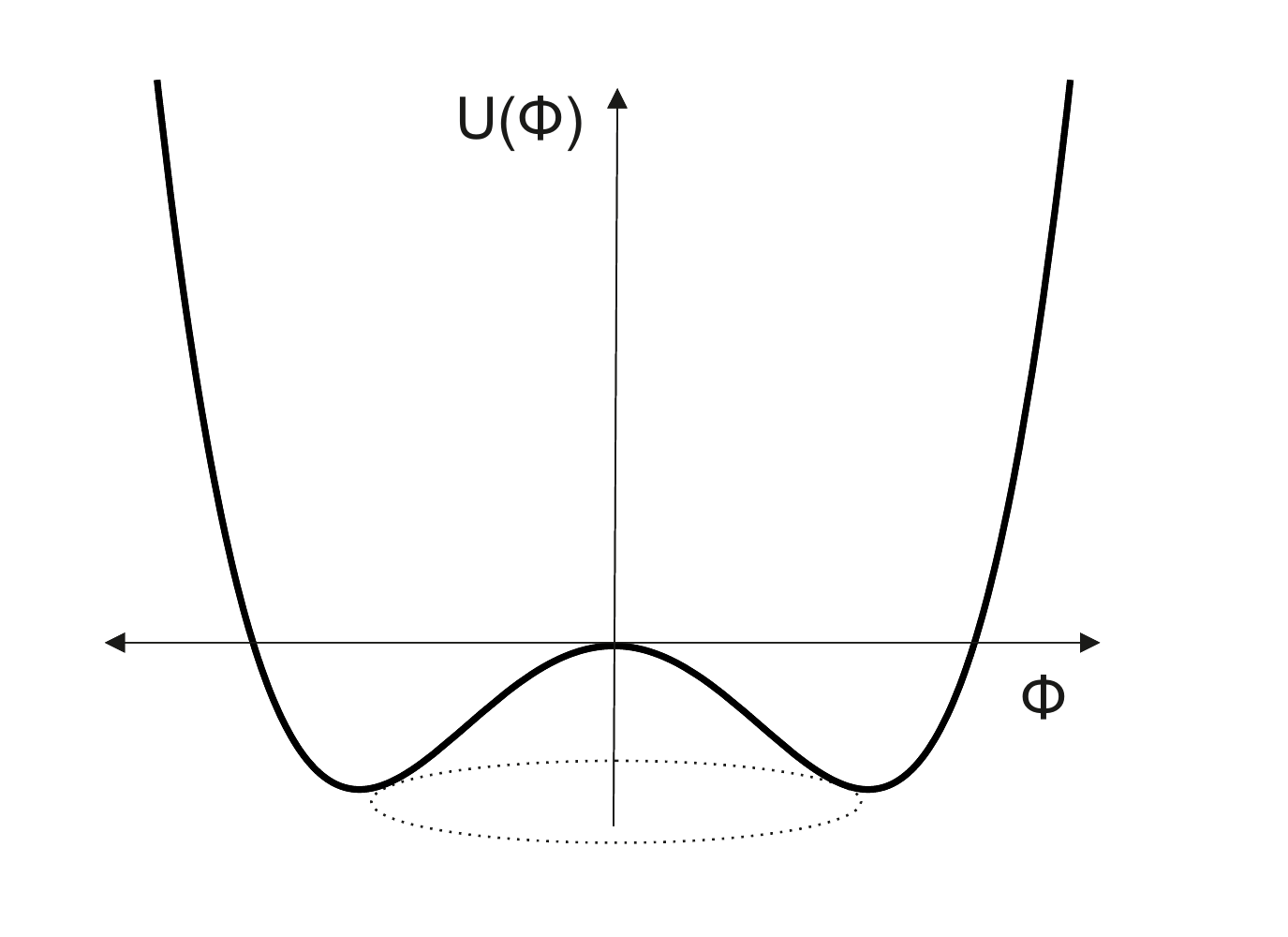}%
\end{minipage}}\protect\caption{Tree level potential for $\mu<m$ (left) and $\mu>m$ (right). \label{fig:SSB}}
\end{figure}
 To discuss spontaneous symmetry breaking, it is sufficient to consider
the tree level potential $U(\phi)$. From (\ref{eq:LagrangianClassic})
we obtain
\begin{equation}
U(\phi)=\frac{m^{2}-\mu^{2}}{2}\phi^{2}+\frac{\lambda}{4}\phi^{4}\,.
\end{equation}

\noindent As we can see, a positive coupling $\lambda>0$ is necessary
to ensure the stability of the potential. Minimization with respect
to $\phi$ yields the ground state (i.e. the state with the largest
pressure). For $\left|\mu\right|<m$, the only minimum is given by
the trivial solution $\phi=0$. If on the other hand $\left|\mu\right|>m$,
we 

\noindent \begin{flushleft}
\newpage{}find a minimum located at nonzero $\phi$ (see figure \ref{fig:SSB}),
\par\end{flushleft}

\noindent 
\begin{equation}
\phi^{2}=\frac{\mu^{2}-m^{2}}{\lambda}\,.\label{eq:CondPrev}
\end{equation}
Such a ground state can no longer be $U(1)$ invariant. In field theoretic
terminology the ground state \textit{spontaneously breaks }the $U(1)$
symmetry of the Lagrangian. One should keep in mind that the ground
state itself is degenerate and by choosing $\phi_{1}=0$ we have chosen
a specific direction in the $\phi_{1}$- $\phi_{2}$ plane. 

\noindent To obtain the dispersion relations of the elementary excitations,
we need to calculate the propagator (i.e. the term quadratic in the
fluctuations $\varphi(x)$, we will demonstrate this in a more complicated
case in the next section)\bigskip{}

\noindent 
\begin{equation}
S_{0}^{-1}=\left(\begin{array}{cc}
-k^{2}+m^{2}+3\lambda\phi^{2}-\mu^{2} & -2ik_{0}\mu\\
2ik_{0}\mu & -k^{2}+m^{2}+\lambda\phi^{2}-\mu^{2}
\end{array}\right)\,.\label{eq:TestProp}
\end{equation}

\noindent The excitations are given by the poles of $S_{0}$ (or the
roots of det $S_{0}^{-1}$)
\begin{equation}
\epsilon_{k}^{\pm}=\sqrt{\vec{k}^{2}+m^{2}+2\lambda\phi^{2}+\mu^{2}\mp\sqrt{4\mu^{2}(\vec{k}^{2}+m^{2}+2\lambda\phi^{2})+\lambda^{2}\phi^{4}}}\,.\label{eq:DispPrev}
\end{equation}

\noindent \begin{wrapfigure}{o}{0.5\columnwidth}%

\includegraphics[scale=0.5]{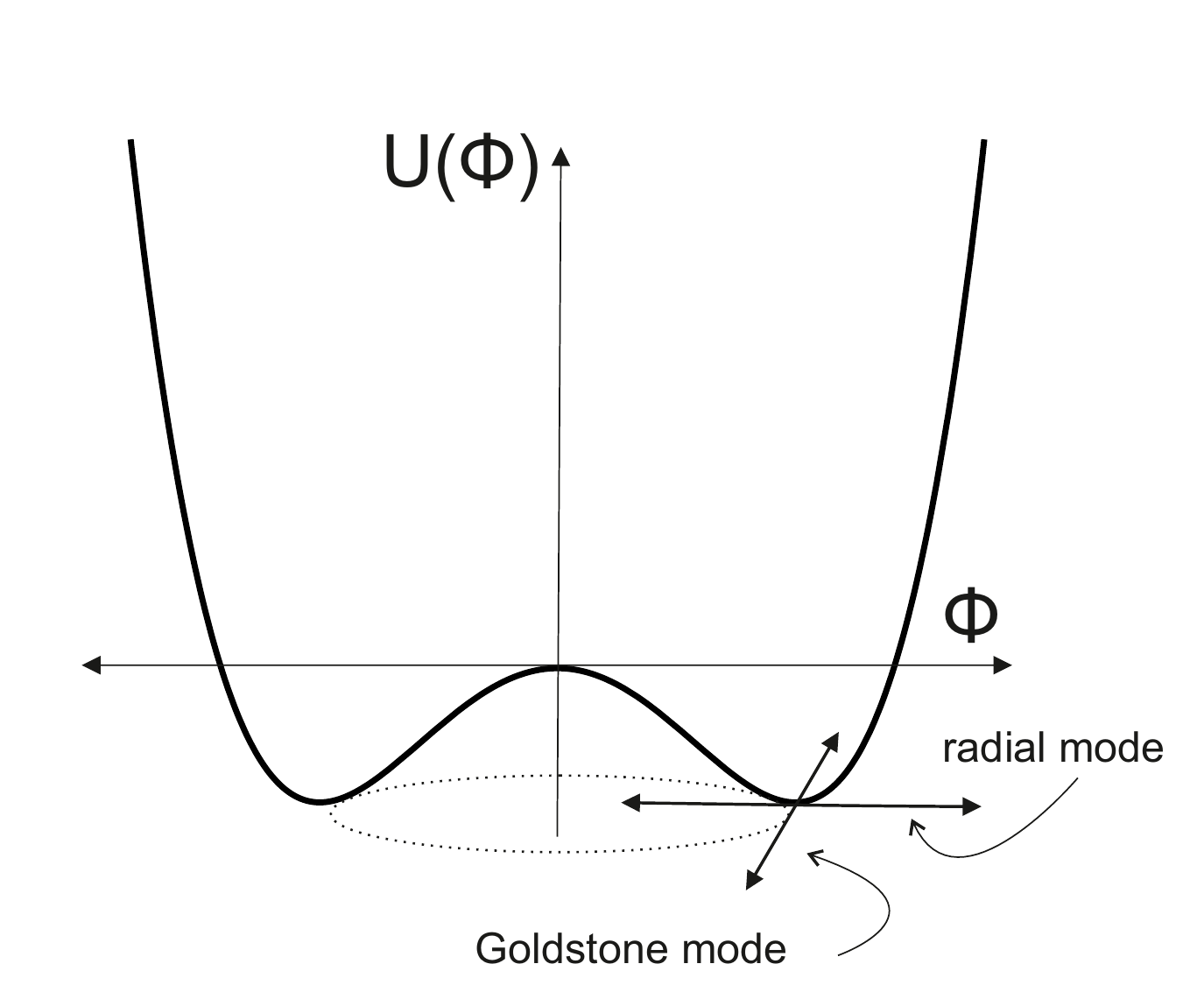}\protect\caption{Elementary excitations in a complex $\varphi^{4}$ model. The spectrum
includes a massless Goldstone mode and a massive radial mode which
is orthogonal to the Goldstone mode.\label{fig:Elementary-excitations-in} }
\end{wrapfigure}%
As expected, the spectrum includes a gapless Goldstone mode which
is linear to lowest order in an expansion in $k$

\bigskip{}

\noindent 
\begin{equation}
\epsilon_{\vec{k}}^{+}\simeq\sqrt{\frac{\mu^{2}-m^{2}}{3\mu^{2}-m^{2}}}\,|\vec{k}|,\,\,\,\,\,\epsilon_{|\vec{k}|=0}=0\,,
\end{equation}

\noindent as well as a massive mode $\epsilon^{-}$. To obtain this
result, we have inserted the condensate from equation (\ref{eq:CondPrev}).
As pointed out, the nature of these excitations is a key ingredient
to understand superfluidity and we shall discuss their properties
in detail in the chapters to come. By now, we have discussed all microscopic
concepts necessary to describe superfluidity. To derive superfluid
properties from field theory, it will be very important to consider
modulus and phase of the condensate $\phi$ separately. The general
setup for the upcoming calculations is described in the next chapter. 

\newpage{}

\section{Lagrangian and formalism \label{sec:Lagrangian}}

~

\noindent Our starting point is the Lagrangian 
\begin{equation}
{\cal L}=\partial_{\mu}\varphi\partial^{\mu}\varphi^{*}-m^{2}|\varphi|^{2}-\lambda|\varphi|^{4}\,,\label{eq:Lagrange}
\end{equation}
with the same properties as discussed in the last section. Note however
that the chemical potential $\mu$ is set to zero. We shall see later
that it can be obtained from the (time dependency of the) phase of
the Bose-Einstein condensate. For simplicity, we might also set the
mass $m$ to zero in some of the results.  We can now allow for Bose-Einstein
condensation in the usual way by separating the condensate,

\begin{equation}
\varphi(x)=\frac{e^{i\psi(x)}}{\sqrt{2}}\left[\rho(x)+\varphi'_{1}(x)+i\varphi'_{2}(x)\right]\,.\label{eq:field}
\end{equation}
Here, $\rho(x)$ is the modulus and $\psi(x)$ the phase of the condensate

\[
\phi(x)=\frac{1}{\sqrt{2}}\rho(x)e^{i\psi(x)}\,.
\]
For convenience, we have introduced the transformed fluctuation field
$\varphi'(x)$, which we have written in terms of its real and imaginary
parts. Inserting this into the Lagrangian yields 

\begin{equation}
{\cal L}=-U+{\cal L}^{(1)}+{\cal L}^{(2)}+{\cal L}^{(3)}+{\cal L}^{(4)}\,,
\end{equation}
with the tree-level potential 

\begin{equation}
U=-\frac{1}{2}\partial_{\mu}\rho\,\partial^{\mu}\rho-\frac{\rho^{2}}{2}(\partial_{\mu}\psi\partial^{\mu}\psi-m^{2})+\frac{\lambda}{4}\rho^{4}\label{eq:potential}
\end{equation}
and the fluctuation terms, listed by their order in the fluctuation
from linear to quartic, 

\begin{eqnarray}
{\cal L}^{(1)} & = & \partial_{\mu}\psi(\rho\partial^{\mu}\varphi_{2}'-\varphi_{2}'\partial^{\mu}\rho)+\rho(\partial_{\mu}\psi\partial^{\mu}\psi-m^{2}-\lambda\rho^{2})\varphi_{1}'+\partial_{\mu}\varphi_{1}'\partial^{\mu}\rho\,,\label{eq:fluct1}\\
{\cal L}^{(2)} & = & \frac{1}{2}[\partial_{\mu}\varphi_{1}'\partial^{\mu}\varphi_{1}'+\partial_{\mu}\varphi_{2}'\partial^{\mu}\varphi_{2}'+(\varphi_{1}'^{2}+\varphi_{2}'^{2})(\partial_{\mu}\psi\partial^{\mu}\psi-m^{2})\label{eq:fluct2}\\
 &  & +2\partial_{\mu}\psi(\varphi_{1}'\partial^{\mu}\varphi_{2}'-\varphi_{2}'\partial^{\mu}\varphi_{1}')-\lambda\rho^{2}(3\varphi_{1}'^{2}+\varphi_{2}'^{2})]\,,\nonumber 
\end{eqnarray}

\begin{eqnarray}
{\cal L}^{(3)} & = & -\lambda\rho\varphi_{1}'(\varphi_{1}'^{2}+\varphi_{2}'^{2})\,,\label{eq:fluct3}\\
{\cal L}^{(4)} & = & -\frac{\lambda}{4}(\varphi_{1}'^{2}+\varphi_{2}'^{2})^{2}\,.\label{eq:fluct4}
\end{eqnarray}

\begin{figure}
~~~~~~~~~~~~~~~~~\includegraphics[scale=0.8]{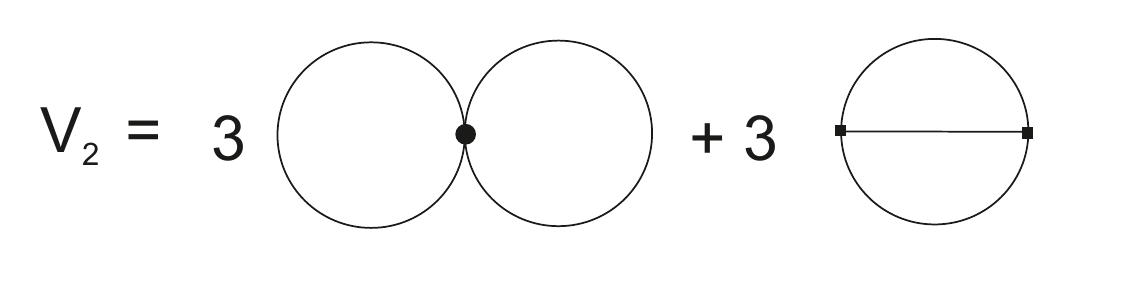}

\protect\caption{The 2PI potential $V_{2}$ is constructed from 2PI diagrams which
can be obtained from $\mathcal{L}^{(3)}$ and $\mathcal{L}^{(4)}$.
The black square at each vertex in the second diagram denotes a multiplication
with the condensate $\rho$. Both contributions are of order $\lambda$
though one might naively expect the second contribution to be of order
$\lambda^{2}$. The second diagram has a combinatorial factor of 3!
which is multiplied by an additional factor of 1/2 because this diagram
originates from a second order contribution of $\mathcal{L}^{(3)}$.
\label{fig:Two-particle-irreducible} }
\end{figure}

In the following sections, we will perform calculations in three steps:
\begin{itemize}
\item At strictly zero temperature, we can restrict ourselves to the tree
level potential $U$ given by equation (\ref{eq:potential}). In terms
of the two-fluid model, this scenario corresponds to a single fluid
case where only the superfluid is present. The (static) zero temperature
value of the condensate is obtained from $\partial U/\partial\rho=0$.
This scenario is discussed in section \ref{sec:Zero-temperature:-Single-fluid}.
\item For low-temperature approximations, we shall need the tree-level potential
and the terms quadratic in the fluctuations. The linear terms ${\cal L}^{(1)}$
can be rewritten such that they are - up to a total derivative term
- proportional to the equations of motion (\ref{eq:EOMRho}) and (\ref{eq:EOMPsi}),
and thus do not contribute to the on-shell action. From the contributions
of ${\cal L}^{(2)}$, we can construct the one-particle irreducible
(1PI) effective action $\Gamma[\rho]$ and we will use the one-loop
approximation for this quantity. While it is of course possible to
construct 1PI diagrams from $\mathcal{L}^{(3)}$ and $\mathcal{L}^{(4)}$,
they would include external legs. The 1PI effective action includes
only ``legless'' 1PI diagrams, the so-called ``proper vertices'',
as we will explain in the next section. The temperature dependent
condensate $\rho(T)$ can in principle be obtained from the stationary
equation $\delta\Gamma[\rho]/\delta\rho=0$. However, we will approximate
$\rho$ by its tree-level value. The low-temperature approximation
to the two-fluid model is discussed in section \ref{sec:Finite-temperature:-Two-fluid}. 
\item To go beyond the low-temperature limit, we will use a two-particle
irreducible (2PI) formalism. In the 2PI framework, contributions from
$\mathcal{L}^{(3)}$ and $\mathcal{L}^{(4)}$ obviously cannot be
neglected. All diagrams which are two-particle irreducible and can
be constructed from the interaction terms above are collected in the
potential $V_{2}$ (see figure \ref{fig:Two-particle-irreducible}).
The ``double-bubble'' diagram can be directly constructed from $\mathcal{L}^{(4)}$.
From $\mathcal{L}^{(3)}$ alone it is not possible to construct a
2PI diagram. We can see however that every vertex in $\mathcal{L}^{(3)}$
contains a condensate, and as we know, each condensate comes with
a factor of $\lambda^{-1/2}$(see equation (\ref{eq:CondPrev})).
Therefore, to order $\lambda$ we have to include second order contributions
from $\mathcal{L}^{(3)}$ which are given by the second diagram in
figure \ref{fig:Two-particle-irreducible}. The one-particle and two-particle
irreducible effective action will be derived in section \ref{sec:Preface-1PI2PI}.
The 2PI effective action is a functional of the condensate $\rho$
as well as the full propagator $S$. The stationarity equations $\delta\Gamma[\rho,S]/\delta\rho=0$
and $\delta\Gamma[\rho,S]/\delta S=0$ not only allow for a self consistent
determination for the condensate $\rho$ but also of the mass parameter.
This is important for the following reason (see also discussion in
reference \cite{CritTKaon}): consider the simplest case with no condensation
(and no superflow). The dispersion relations (\ref{eq:DispPrev})
then reduce to the simple form
\[
\epsilon_{\vec{k}}^{\pm}=\sqrt{\vec{k}^{2}+m^{2}}\mp\mu\,.
\]
Condensation occurs for $m<\mu$ in which case $\epsilon^{+}$ acquires
an unphysical negative value. A non-vanishing condensate $\phi$ leads
to corrections proportional to $\phi^{2}$ in the propagator, see
(\ref{eq:TestProp}). The resulting dispersions (\ref{eq:DispPrev})
are then positive for all three-momenta $\vec{k}$ and can be used
to calculate thermal integrals in the low-temperature approximation.
However, as the condensate is expected to melt away at the critical
temperature, this problem will indefinitely occur for sufficiently
large temperatures if the mass remains fixed. Therefore, self-consistency
equations for mass \textit{and} condensate are required. Calculations
for arbitrary temperatures are discussed in section \ref{sec:The-two-fluid-model-at-arbitrary-T}.
\end{itemize}

\newpage{}

\section{Zero temperature: Single fluid formalism\label{sec:Zero-temperature:-Single-fluid}}

~

\noindent In a first step, the translation of field theory into hydrodynamics
will be carried out at tree-level, ignoring all quantum fluctuations.
This allows us to present the translation into hydrodynamic equations
in the simplest case, based on reference \cite{Son1}. We begin with
a discussion of the conservation equations as well as the equations
of motion and solutions thereof. 

~

\subsection{Equations of motion and conservation equations\label{sub:Zero-temperature-hydrodynamics}}

~

\noindent The classical equations of motion for $\rho$ and $\psi$
are 

\noindent 
\begin{eqnarray}
\square\rho & = & \rho(\sigma^{2}-m^{2}-\lambda\rho^{2})\,,\label{eq:EOMRho}\\
0 & = & \partial_{\mu}(\rho^{2}\partial^{\mu}\psi)\,.\label{eq:EOMPsi}
\end{eqnarray}
The field-theoretic expressions for the conserved charge current and
the stress-energy tensor are

\begin{eqnarray}
j^{\mu} & = & \frac{\partial{\cal L}}{\partial(\partial_{\mu}\psi)}=\rho^{2}\partial^{\mu}\psi\,,\label{eq:microCurrent}\\
T^{\mu\nu} & = & \frac{2}{\sqrt{-g}}\frac{\delta(\sqrt{-g}\,{\cal L})}{\delta g_{\mu\nu}}=2\frac{\partial{\cal L}}{\partial g_{\mu\nu}}-g^{\mu\nu}{\cal L}=\partial^{\mu}\rho\,\partial^{\nu}\rho+\rho^{2}\partial^{\mu}\psi\partial^{\nu}\psi-g^{\mu\nu}{\cal L}\,.\label{eq:microT}
\end{eqnarray}
For the stress-energy tensor, we have used the gravitational definition
by formally introducing a general metric $g$ which, after taking
the derivatives, we set to be the metric of flat Minkowski space.
This definition guarantees that $T^{\mu\nu}$ is symmetric and conserved.
Since $j^{\mu}$ is also conserved, we have

\begin{equation}
\partial_{\mu}j^{\mu}=0\,,\qquad\partial_{\mu}T^{\mu\nu}=0\,,
\end{equation}
which are the hydrodynamic equations. As we expect from Noether's
theorem, current conservation is identical to the equation of motion
for $\psi$. The reason is that $\psi$ only appears through its derivatives
in the Lagrangian, as it should be for the radial mode in the presence
of an exact U(1) symmetry. 

~

~

~

\noindent As an ansatz for the solution of the equations of motion
we may choose

\begin{eqnarray}
\psi & = & p_{\mu}x^{\mu}+{\rm Re}\sum_{\vec{k}}\delta\psi_{\vec{k}}\, e^{i(\omega t-\vec{k}\cdot\vec{x})}\,,\label{eq:AnsatzPsi}\\
\rho & = & \sqrt{\frac{p^{2}-m^{2}}{\lambda}}+{\rm Re}\sum_{\vec{k}}\delta\rho_{\vec{k}}\, e^{i(\omega t-\vec{k}\cdot\vec{x})}\,.\label{eq:AnsatzRho}
\end{eqnarray}
In this ansatz, we assume $\partial^{\mu}\psi$ and $\rho$ are each
composed of a static part plus small oscillations around it. The static
part of the solution is the superfluid mode, corresponding to an infinite
and uniformly flowing superfluid. The density and flow are specified
by the values of the components of $p_{\mu}$, which are pure numbers,
not functions of $x$, and are \textit{not }constrained by the equation
of motion. The value of $p_{\mu}$ is determined by the boundary conditions,
which specify the topology of the field configuration, namely the
number of times the phase winds around as we traverse the space-time
region in which the superfluid resides%
\footnote{Usually, Bose-Einstein condensation is introduced by separating the
zero momentum mode in the expansion of 
\[
\varphi(x^{\mu})=\varphi(0)+\sum_{k^{\mu}\neq0}\varphi(k^{\mu})e^{ik\cdot x}\,.
\]

This case however corresponds to the separation of the $k^{\mu}=p^{\mu}$
mode resulting in

\[
\varphi(x^{\mu})=\varphi_{k^{\mu}=p^{\mu}}e^{ip\cdot x}+\sum_{k^{\mu}\neq p^{\mu}}\varphi(k^{\mu})e^{ik\cdot x}\,.
\]

$\varphi_{k^{\mu}=p^{\mu}}$ is given by $\rho$ from equation (\ref{eq:rhomin}). %
}. The oscillations around the static solution yield two modes whose
dispersion relations $\omega(\vec{k})$ are determined by the equations
of motion. As expected, one of the modes is the massless Goldstone
mode, the other is a massive ``radial'' mode (see also figure \ref{fig:Elementary-excitations-in}).
The oscillatory modes can be thermally populated to yield the normal
fluid. The boundary conditions place no constraint on the amplitudes
of the oscillatory modes because they are topologically trivial. In
this section we will assume uniform density and flow of the superfluid,
but in general one could obtain a space-time-dependent superfluid
flow by giving nonzero classical background values to the $\delta\psi_{\vec{k}}$
and $\delta\rho_{\vec{k}}$. The normal fluid would then consist of
a thermal population on top of that background. We can use the poles
of the tree-level propagator (\ref{eq:Propagator}) to determine the
dispersion relation of the oscillatory modes. This yields the same
result as solving the equations of motion (\ref{eq:EOMRho}), (\ref{eq:EOMPsi})
with the ansatz (\ref{eq:AnsatzPsi}), (\ref{eq:AnsatzRho}). Oscillations
in $\psi$ will also be relevant to the calculation of the sound velocities.
In fact, for low temperatures we will demonstrate that the coefficient
of the linear term in the massless mode $\omega(\vec{k})$ is nothing
but the velocity of first sound. For now, we set $\delta\psi_{\vec{k}}=\delta\rho_{\vec{k}}=0$.
In this case, $p^{\mu}=\partial^{\mu}\psi$, and we can write 
\begin{equation}
\rho=\sqrt{\frac{\sigma^{2}-m^{2}}{\lambda}}\,.\label{eq:rhomin}
\end{equation}
Inserting this solution back into the Lagrangian gives

\begin{equation}
{\cal L}=-U=\frac{(\sigma^{2}-m^{2})^{2}}{4\lambda}\,.\label{eq:PotentialMin}
\end{equation}
It is clear from the usual definition of the pressure in terms of
the partition function that, in our current tree-level treatment,
this Lagrangian is identical to the pressure. Bose-Einstein condensation
obviously appears for $\left|\sigma\right|>m$ which is a first hint
that $\sigma$ plays the role of a chemical potential. It is important
to keep in mind that we will keep the approximation of uniform density
and flow of the superfluid throughout the remainder of this work,
including the finite temperature calculations of sections \ref{sec:Finite-temperature:-Two-fluid}
and \ref{sec:The-two-fluid-model-at-arbitrary-T}.

~

\subsection{Zero-temperature hydrodynamics: Landau`s formalism\label{sub:Zero-temperature-hydrodynamics:-Landau}}

~

\noindent We now explore the connection between field-theoretic quantities
and hydrodynamics. We recall that the hydrodynamic expressions for
current and stress-energy tensor at zero temperature are:

\vspace{-0.5cm}

\begin{eqnarray}
j^{\mu} & = & n_{s}v^{\mu}\,,\label{eq:T02luid1}\\
T^{\mu\nu} & = & (\epsilon_{s}+P_{s})v^{\mu}v^{\nu}-g^{\mu\nu}P_{s}\,,\label{eq:T02fluid2}
\end{eqnarray}
where $v^{\mu}$ is the velocity of the superfluid which fulfills
$v_{\mu}v^{\mu}=1$. To identify $n_{s}$, $\epsilon_{s}$, and $P_{s}$
we go to the superfluid rest frame $v^{\mu}$ = (1,0,0,0), where there
is no flow of charge, energy, or momentum: $j^{i}=T^{0i}=0$. Then
we have $j^{0}=n_{s}$, $T^{00}=\epsilon_{s}$, $T^{ij}=\delta^{ij}P_{s}$,
so $n_{s}$, $\epsilon_{s}$, and $P_{s}$ are the charge density,
energy density, and pressure in the superfluid rest frame; the pressure
is isotropic. (We will discuss the superfluid density $\rho_{s}$
below). They can be expressed covariantly via appropriate contractions,

\begin{equation}
n_{s}=\sqrt{j^{\mu}j_{\mu}}=v^{\mu}j_{\mu}\,,\qquad\epsilon_{s}=v_{\mu}v_{\nu}T^{\mu\nu}\,,\qquad P_{s}=-\frac{1}{3}(g_{\mu\nu}-v_{\mu}v_{\nu})T^{\mu\nu}\,.\label{eq:macroProjection}
\end{equation}

\newpage{}

\noindent To relate these hydrodynamical quantities to the microscopic
physics we can use equations (\ref{eq:microCurrent}), (\ref{eq:microT}).
For $n_{s}$ and $v^{\mu}$ we obtain immediately 

\bigskip{}

\noindent 
\begin{equation}
n_{s}=\sigma\frac{\sigma^{2}-m^{2}}{\lambda}\,,\qquad v^{\mu}=\frac{\partial^{\mu}\psi}{\sigma}=\gamma\,(1,\vec{v}_{s})\ ,\label{eq:microNsV}
\end{equation}
where $\vec{v}_{s}$ is the superfluid 3-velocity with Lorentz factor
$\gamma=\partial_{0}\psi/\sigma=1/\sqrt{1-\vec{v}_{s}^{2}}$. This
is precisely the factor multiplying $n_{s}$ when measured in the
normal-fluid rest frame (see table 1 in section \ref{sub:From-generalized-to-frames}).
The velocity is given by $\vec{v}_{s}=-\vec{\nabla}\psi/\partial_{0}\psi$%
\footnote{\noindent The minus sign appears because the 3-velocity $\vec{v}_{s}$
corresponds to the spatial components of the contravariant 4-vector
$v^{\mu}$, while the operator $\vec{\nabla}$ corresponds to the
spatial components of the covariant 4-vector $\partial_{\mu}$, i.e.,$\partial^{\mu}=(\partial_{t},-\vec{\nabla})$%
}. We have thus recovered the expected irrotational flow of the superfluid.
The difference between expressions (\ref{eq:microNsV}) and (\ref{eq:SuperFl})
is that we have identified the scalar potential $\psi$ as the phase
of the condensate. For $\epsilon_{s}$ and $P_{s}$ we first use (\ref{eq:macroProjection})
and (\ref{eq:microT}) to obtain partly microscopic expressions

\medskip{}

\noindent 
\begin{equation}
\epsilon_{s}=v_{\mu}\partial^{\mu}\psi\, n_{s}-{\cal L}\,,\qquad P_{s}={\cal L}+(v_{\mu}\partial^{\mu}\psi-\sigma)n_{s}\,.\label{eq:EPproject}
\end{equation}
From contracting the relation for $v^{\mu}$ in (\ref{eq:microNsV})
with $\partial_{\mu}\psi$ we know that $v^{\mu}\partial_{\mu}\psi=\sigma$;
as a consequence, we obtain the expected relation $P_{s}={\cal L}$.
We can now identify the physical meaning of $\sigma$: using the zero-temperature
thermodynamic relation in the superfluid rest frame $\epsilon_{s}+P_{s}=\mu_{s}n_{s}$,
we find

\medskip{}

\noindent 
\begin{equation}
\mu_{s}=\sigma=v^{\mu}\partial_{\mu}\psi\,,\label{eq:microMuS}
\end{equation}
 so $\sigma$ is identified with $\mu_{s}$ \textit{the chemical potential
in the superfluid rest frame}. Going back to the solution for the
modulus $\rho$ (\ref{eq:rhomin}) we see that, as expected, Bose
condensation only occurs for $\mu_{s}^{2}>m^{2}$. Using (\ref{eq:PotentialMin})
and (\ref{eq:EPproject}) we finally obtain fully microscopic expressions
for the energy density and the pressure, 

\noindent 
\begin{equation}
P_{s}=\frac{(\sigma^{2}-m^{2})^{2}}{4\lambda}\,,\qquad\epsilon_{s}=\frac{(3\sigma^{2}+m^{2})(\sigma^{2}-m^{2})}{4\lambda}\,.\label{eq:microEP}
\end{equation}
Note that $m$ is the only mass scale in our Lagrangian. The trace
of the stress-energy tensor $T^{\mu}{}_{\mu}=\epsilon_{s}-3P_{s}=m^{2}\rho^{2}$
vanishes for $m=0$. Finally, the \textit{superfluid density} $\rho_{s}$
is defined via the expansion in small three-velocities of the momentum
and energy densities 

\bigskip{}

\noindent 
\begin{equation}
T^{0i}=\rho_{s}v_{si}+{\cal O}(|\vec{v}_{s}|^{3})\,,\qquad T^{00}=\frac{\epsilon_{s}+P_{s}\vec{v}_{s}^{2}}{1-\vec{v}_{s}^{2}}=\epsilon_{s}+\rho_{s}\vec{v}_{s}^{2}+{\cal O}(|\vec{v}_{s}|^{4})\,,
\end{equation}
 From these expansions we obtain its microscopic form 

\bigskip{}

\noindent 
\begin{equation}
\rho_{s}=\epsilon_{s}+P_{s}=\sigma^{2}\frac{\sigma^{2}-m^{2}}{\lambda}\,.\label{eq:microRho}
\end{equation}

\newpage{}

\subsubsection{Generalized single fluid formalism \label{sub:Zero-temperature-hydrodynamics:-generalized} }

~

\noindent We can now repeat the above translation in the frame of
the generalized hydrodynamic formalism. Although this might seem to
be an unnecessary complication for the isotropic zero-temperature
case, it is a useful preparation for the nonzero temperature case.
The basic variables are the conserved current density $j^{\mu}$ and
its conjugate momentum. We will now confirm that this conjugate momentum
is given by the gradient of the phase $\partial^{\mu}\psi$. Recall
that in the single fluid case the stress-energy tensor is given by
\begin{equation}
T^{\mu\nu}=-g^{\mu\nu}\Psi+j^{\mu}\partial^{\nu}\psi\,,\label{eq:GentT0}
\end{equation}

\noindent with the generalized pressure $\Psi$ and the generalized
energy density is given by 

\bigskip{}

\noindent 
\begin{equation}
\Lambda\equiv T_{\;\;\mu}^{\mu}+3\Psi=-\Psi+j^{\mu}\partial_{\mu}\psi\,.\label{eq:LambdaPsiT0}
\end{equation}
This is the Legendre transform in the single fluid case where $\Psi=\Psi[\sigma^{2}]$,
$\Lambda=\Lambda[j^{2}]$ and $j^{2}=j_{\mu}j^{\mu}=n_{s}^{2}$. By
comparison with the expressions from section \ref{sub:Zero-temperature-hydrodynamics:-Landau},
it is easy to see that in the single-fluid case $\Psi$ and $\Lambda$
are simply pressure and energy density in the fluid rest frame (see
also discussion in section \ref{sec:Relativistic-thermodynamics-and-hydro}),
$\Psi=P_{s}={\cal L}$, $\Lambda=\epsilon_{s}$. In the single fluid
case, the current is indeed proportional to the conjugate momentum
with a coefficient given by the underlying microscopic physics

\bigskip{}

\noindent 
\begin{equation}
\partial_{\mu}\psi=\frac{\partial\Lambda}{\partial j^{\mu}}={\cal B}\, j_{\mu}\,,\qquad\,\,{\cal B}\equiv2\frac{\partial\Lambda}{\partial j^{2}}\,,\label{eq:Entrainment1T0}
\end{equation}
 and 

\noindent 
\begin{equation}
j^{\mu}=\frac{\partial\Psi}{\partial(\partial_{\mu}\psi)}=\overline{{\cal B}}\,\partial^{\mu}\psi\,,\qquad\,\,\overline{{\cal B}}\equiv2\frac{\partial\Psi}{\partial\sigma^{2}}\,,\label{eq:Entrainment2T0}
\end{equation}

\noindent where we can simply read off

~
\begin{equation}
\bar{{\cal B}}={\cal B}^{-1}=\frac{\sigma^{2}-m^{2}}{\lambda}\,.\label{eq:BBarT0}
\end{equation}

\noindent In summary, it is important to keep in mind that the connection
from macroscopic to microscopic physics is made by identifying the
conjugate momentum $\partial^{\mu}\psi$ as the four-gradient of the\textit{
phase of the condensate}. In addition, in equations (\ref{eq:microNsV})
-(\ref{eq:microEP}) we express all thermodynamic and hydrodynamic
parameters of the Landau model in terms of field theoretic variables
while the microscopic physics enter the generalized hydrodynamic formalism
via the coefficient ${\cal B}$ from equation (\ref{eq:BBarT0}).
In addition, from equation (\ref{eq:microNsV}) and (\ref{eq:microMuS})
we draw the following important conclusions:
\begin{itemize}
\item \noindent The rotation of the phase (at the bottom of the ``Mexican
hat'' potential) generates the chemical potential of the superfluid.
\item \noindent The number of rotations per unit length determines the superfluid
velocity.
\item \noindent $\sigma$ represents the invariant expression for the chemical
potential $\sigma=\left(\partial_{\mu}\psi\partial^{\mu}\psi\right)^{1/2}$.
In case $\vec{v}_{s}=\vec{0}$ this relation simplifies to $\sigma=\partial_{0}\psi$.
In case $\vec{v}_{s}\neq0$ we have $\sigma=\left(\partial_{o}\psi\right)^{\prime}\sqrt{1-\vec{v}_{s}^{2}}$
. Here we have indicated that $\partial_{0}\psi$ alone is \textit{not
}an invariant and therefore changes its value to $\left(\partial_{0}\psi\right)^{\prime}$when
measured from outside the superfluid rest frame. The Lorentz factor
in $\sigma$ compensates this change such that $\mathbb{\sigma}$
is indeed invariant. 
\end{itemize}
~

\subsubsection{\noindent The non-relativistic limit\label{sub:The-non-relativistic-limit}}

~

\noindent The relativistic two-fluid equations incorporate the correct
non-relativistic limit of section \ref{sec:The-two-fluid-model} (for
a general discussion, see for example \cite{CarterKhala}). In this
section we investigate this limit in the single fluid case at zero
temperature. According to Landau and Lifschitz \cite{LandauLifshitzFluid}
we need to write the relativistic energy density as the sum of the
rest energy of particles that constitute the fluid and the non-relativistic
energy density $\epsilon=mn_{0}+\epsilon^{\prime}$. Here $n_{0}$
is the particle number per volume in the rest frame of the fluid.
To transform into the laboratory frame (i.e. to eliminate $n_{0}$
in favor of $n$), we write the energy density as %
\footnote{For notational consistency, we still omit factors of $c$ which are
usually taken into account in the non-relativistic literature. %
}:

\noindent 
\begin{equation}
\epsilon=mn\sqrt{1-\vec{v}^{2}}+\epsilon^{\prime}\simeq mn-mn\vec{v}^{2}/2+\epsilon^{\prime}\,.\label{eq:epsnonrel}
\end{equation}

\noindent In the single fluid case, the components of $T^{\mu\nu}$
can be written as

\bigskip{}

\noindent 
\begin{equation}
T_{00}=\frac{\epsilon+P\,\vec{v}^{2}}{1-\vec{v}^{2}},\,\,\,\,\,\, T^{ij}=\frac{P\left[\left(1-\vec{v}^{2}\right)\delta^{ij}+v^{i}v^{j}\right]+\epsilon v^{i}v^{j}}{1-\vec{v}^{2}},\,\,\,\,\,\, T^{0i}=-\frac{\epsilon+P}{1-\vec{v}^{2}}v^{i}\,.\label{eq:TSingleFl}
\end{equation}

\noindent Here, we have inserted the $\gamma$ factor contained in
definition of $v^{\mu}$. Expanding for small velocities and inserting
$\epsilon$ from equation (\ref{eq:epsnonrel}), $T_{00}$ reads

\noindent \bigskip{}

\noindent 
\begin{equation}
T_{00}=mn+\epsilon^{\prime}+\left(\frac{mn}{2}+\epsilon^{\prime}+P\right)\vec{v}^{2}+\mathcal{O}(v^{4})\simeq mn+\epsilon^{\prime}+\frac{mn}{2}\vec{v}^{2}+\mathcal{O}(\vec{v}^{4})\,.
\end{equation}

\medskip{}

\noindent Here, we have used that the non-relativistic pressure and
energy density are much smaller than the rest energy: $\epsilon^{\prime},\, P\,\ll\rho=mn$
. After subtraction of the rest energy $\rho$, we obtain the expected
result

\bigskip{}

\noindent 
\begin{equation}
T_{00}^{\prime}=\epsilon^{\prime}+\frac{\rho}{2}\vec{v}^{2}\,.
\end{equation}

\noindent Next, we expand $T^{0i}$ up to the third order in $v$
resulting in \bigskip{}

\noindent 
\begin{equation}
T_{0i}=-mnv_{i}+\left[\epsilon^{\prime}+P+\left(\frac{mn}{2}+\epsilon^{\prime}+P\right)\vec{v}^{2}\right]v_{i}\simeq-mnv_{i}-\left(\epsilon^{\prime}+P+\frac{mn}{2}\vec{v}^{2}\right)v_{i}+\mathcal{O}(\left|\vec{v}\right|^{5})\,,
\end{equation}

\medskip{}

\noindent where again $\epsilon^{\prime},\, P\,\ll\rho$ has been
used. The result reads: 
\begin{equation}
T_{0i}^{\prime}=-\rho v_{i}\,.
\end{equation}

\noindent In the non-relativistic case momentum density $T_{i0}$
and energy flux $T_{0i}$ are not the same. To obtain $T_{0i}$, we
need to subtract the term $\rho v_{i}$ coming from the rest energy
first and obtain 
\begin{equation}
T_{i0}^{\prime}=-\left(\epsilon^{\prime}+P+\frac{\rho}{2}\vec{v}^{2}\right)v_{i}\,.
\end{equation}

\noindent Analogously we obtain the spatial components:
\begin{equation}
T_{ij}^{\prime}=\delta_{ij}P+\rho v_{i}v_{j}\,.
\end{equation}

\noindent Identical results can be obtained by starting right away
from a non-relativistic Lagrangian and repeating the analysis of section
\ref{sub:Zero-temperature-hydrodynamics:-Landau}. The $\varphi^{4}$
theory in the non-relativistic limit is discussed for example in \cite{Phi4nonrel}. 

\newpage{}

\section{The 1PI and 2PI effective action \label{sec:Preface-1PI2PI}}

~

\noindent In this section, we review the one-particle irreducible
(1PI) and two-particle irreducible (2PI) formalism which we shall
apply in the finite temperature calculations. Both formalisms are
non-perturbative in the sense that the coupling constant is present
to all orders and a certain class of diagrams is resummed. We have
discussed in section \ref{sec:Lagrangian} that such a non-perturbative
and self-consistent formulation is necessary to extend our calculations
to finite temperatures. We can restrict the discussion in this chapter
to the zero temperature case. In the imaginary time formalism, the
introduction of finite temperature amounts to a ``straightforward''
modification of the results which we will obtain in the following.
As a preparation, it is necessary to discuss the difference between
full, connected and 1PI diagrams. 

~

\subsection{Full, connected and one-particle irreducible Green`s functions.\label{sub:Full,-connected-and-1PI}}

\noindent ~

\noindent In what follows, bare propagators are represented by $S_{ij}^{0}$,
interaction vertices by $\gamma_{ijk...}$ and sources (sinks) by
$J_{i}$. The abstract index $i$ represents continuous as well as
discrete parameters such as space-time $x^{\mu}$ or the internal
$2\times2$ internal space spanned by the fields $\varphi_{1}$ and
$\varphi_{2}$. We consider only the bosonic case where the ordering
of $i,j,k,..$ is irrelevant. For the purpose of introducing the 1PI
and 2 PI formalism, we shall denote a bosonic scalar \textit{quantum
field} by $\varphi$ and a corresponding \textit{classical field }by
$\phi^{c}$. Later, when we perform actual calculations, we will denote
the classical field simply by $\phi$ (we have already done so in
section \ref{sub:Spontaneous-symmetry-breaking}). 

\begin{figure}
\fbox{\begin{minipage}[t]{1\columnwidth}%
\includegraphics[scale=0.7]{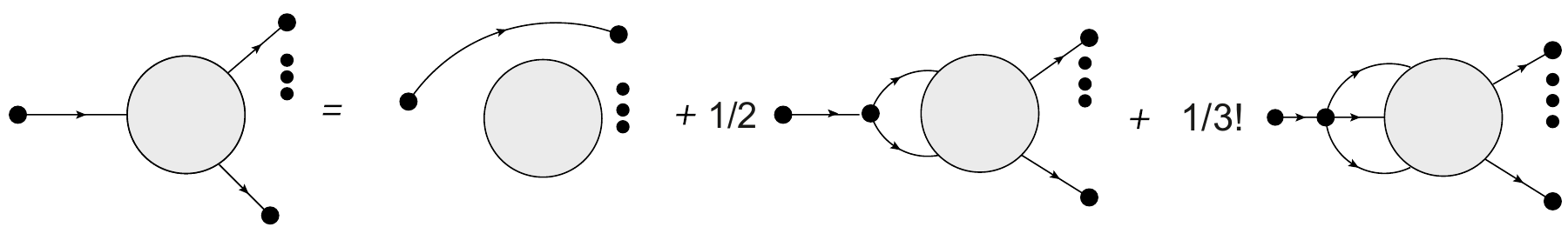}%
\end{minipage}}\protect\caption{Dyson-Schwinger equation (DSE) of a full N-point Green`s function
with 1 particle in the initial and (N-1) particles in the final state.
Propagators are represented by black lines with black dots at the
endpoints. The endpoint of a propagator can also correspond to a vertex.
The initial particle can either not interact at all or interact at
least once, twice,..etc. The structure of these interaction is given
by the Feynman rules of the theory. The Dyson-Schwinger equation is
the sum of all possible processes weighted with the probability for
the occurrence of each process. The combinatorical prefactors are
chosen such that an iteration of the DSE results in the perturbative
expansion. It is important to realize that the full Green`s function
of the left hand side appears again on the right hand side. The DSE
is therefore self-consistent. \label{fig:Schwinger-Dyson-equation-of}}
\end{figure}

\noindent Physical processes, algebraically expressed in terms of
Green`s functions, can be graphically illustrated by Feynman diagrams.
The full n-point Green`s function defined as the time-ordered vacuum
expectation value of Heisenberg field operators $\left\langle 0\left|T[\hat{\varphi}(x_{1})...\hat{\varphi}(x_{n})]\right|0\right\rangle $
corresponds to an infinite sum of Feynman diagrams which take into
account all possible interactions as we discuss in figure \ref{fig:Schwinger-Dyson-equation-of}.
The Dyson-Schwinger equation (DSE) represents a systematic way to
generate all these diagrams. 

\noindent 
One can see right away that the DSE is self-consistent and it is exact
only if all diagrams can be resummed (which is practically never the
case). In principle, the DSE can be used to reproduce the perturbative
expansion of a given theory to arbitrary order by iteration (see for
example \cite{Cvitanovic}). We define the functional $Z[J]$ from
which we can generate all vacuum Green\textasciiacute s functions
for a process with $m$ different sources. (Since $Z$ depends on
continuous as well as discrete parameters, it is a functional.) Obviously,
we can obtain the full N point Green`s functions by acting on $Z$
with $N$ (functional) derivatives:

\bigskip{}

\noindent 
\begin{equation}
Z[J]=\sum_{m=0}^{\infty}\frac{1}{m!}G_{i_{1}i_{2}...i_{m}}J_{i_{1}}J_{i_{2}}...J_{i_{m}},\,\,\,\,\,\, G_{ijk..}=\frac{\delta}{\delta J_{i}}\frac{\delta}{\delta J_{j}}\frac{\delta}{\delta J_{k}}(...)Z[J]\left|_{J=0}\right.\,.
\end{equation}

\medskip{}

\noindent This allows us to express the DSE as:

\bigskip{}

\noindent 
\begin{equation}
\frac{\delta}{\delta J_{i}}Z[J]=S_{ij}^{0}\left[J_{j}+\frac{1}{2}\gamma_{jkl}\frac{\delta}{\delta J_{k}}\frac{\delta}{\delta J_{l}}+\frac{1}{3!}\gamma_{jklm}\frac{\delta}{\delta J_{k}}\frac{\delta}{\delta J_{l}}\frac{\delta}{\delta J_{m}}+...\right]Z[J]\,.
\end{equation}

\medskip{}

\noindent Collecting bare (inverse) propagators and vertices in the
definition of the action of the field $\varphi$

\bigskip{}

\noindent 
\begin{equation}
S[\varphi]=-\frac{1}{2}\varphi_{i}\left(S_{ij}^{0}\right)^{-1}\varphi_{j}+S_{I},\,\,\,\,\, S_{I}[\varphi]=\sum_{m}\gamma_{ijk..l}\frac{\varphi_{i}\varphi_{j}...\varphi_{l}}{m!}\,,
\end{equation}

\medskip{}

\noindent and using:

\bigskip{}

\noindent 
\begin{equation}
\frac{\delta S}{\delta\varphi_{i}}\left[\frac{\delta}{\delta J}\right]:=\frac{\delta S[\varphi]}{\delta\varphi_{i}}\left|_{\varphi=\frac{\delta}{\delta J}}=-\left(S_{ij}^{0}\right)^{-1}\frac{\delta}{\delta J_{j}}+\frac{1}{2}\gamma_{ijk}\frac{\delta}{\delta J_{j}}\frac{\delta}{\delta J_{k}}+\frac{1}{3!}\gamma_{ijkl}\frac{\delta}{\delta J_{j}}\frac{\delta}{\delta J_{k}}\frac{\delta}{\delta J_{l}}+....\right.
\end{equation}

\medskip{}

\noindent we can elegantly write the DSE for any full N-point functions
as:

\bigskip{}

\noindent 
\begin{equation}
\left(\frac{\delta S}{\delta\varphi_{i}}\left[\frac{\delta}{\delta J}\right]+J_{i}\right)Z[J]=0\,.\label{eq:SDE1}
\end{equation}

\medskip{}

\newpage{}

\noindent This equation is the \textit{quantum version} of the classical
equation of motion which is given by

\medskip{}

\noindent 
\begin{equation}
\delta S[\varphi]/\delta\varphi_{i}\left|_{\varphi=\phi^{c}}\right.=-J_{i}\,.
\end{equation}

\begin{figure}
\fbox{\begin{minipage}[t]{1\columnwidth}%
\includegraphics[scale=0.7]{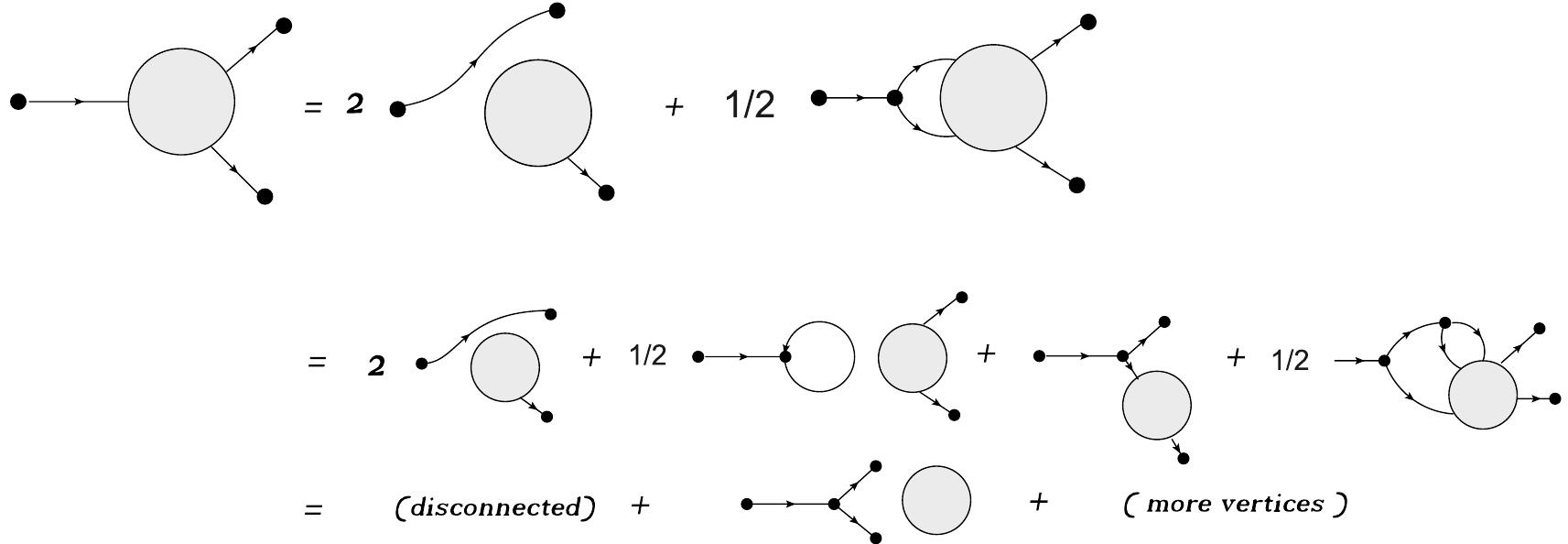}%
\end{minipage}}\protect\caption{First iteration of the DSE for a full three-point function.\label{fig:First-iteration-of} }
\end{figure}

\noindent As an example, we perform one iteration of the DSE for a
full three-point function in figure \ref{fig:First-iteration-of}.
As we can see, the result is a factorization into\textit{ mutually
connected and disconnected diagrams} which represents a new way of
decomposing all Feynman diagrams. 

\noindent 
\begin{wrapfigure}{o}{0.5\columnwidth}%
\includegraphics{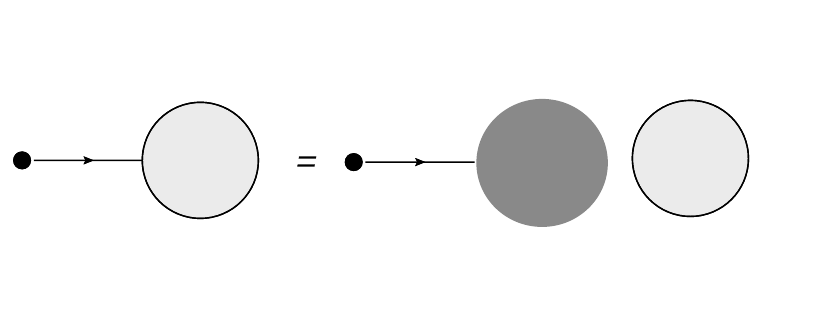}\protect\caption{Any full Green`s function factorizes in a connected Green`s function
(dark gray ``blob'') and a full disconnected Green`s function. \label{fig:factorize}}
\end{wrapfigure}%
This factorization is graphically illustrated in figure \ref{fig:factorize}.
Formally, one can write this factorization for an arbitrary N-point
function as:

\bigskip{}

\noindent 
\begin{equation}
\frac{\delta}{\delta J_{i}}Z[J]=\frac{\delta W[J]}{\delta J_{i}}\cdot Z[J]\,,\label{eq:connDconn}
\end{equation}

\medskip{}

\noindent where $W[J]$, the generating functional of all connected
Green`s functions $G^{(c)}$, is defined in exactly the same way as
$Z[J]$ (with the index $m$ starting from 1 as there is at least
once source attached to a connected Green`s function):

\bigskip{}

\noindent \begin{flushleft}
\begin{equation}
W[J]=\sum_{m=1}^{\infty}\frac{1}{m!}G_{i_{1}i_{2}...i_{m}}^{(c)}J_{i_{1}}J_{i_{2}}...J_{i_{m}}\,.
\end{equation}

\par\end{flushleft}

\medskip{}

\newpage{}

\noindent The differential equation (\ref{eq:connDconn}) relating
$Z$ and $W$ can easily be solved resulting in

\bigskip{}

\noindent 
\begin{equation}
Z[J]=e^{W[J]}\,.
\end{equation}

\medskip{}

\noindent Using this relation, one can easily construct the DSE in
terms of connected diagrams from (\ref{eq:SDE1}). Taking into account
the chain rule

\bigskip{}

\noindent 
\[
\frac{1}{Z}\frac{\delta}{\delta J}(\frac{\delta}{\delta J}Z[J]\cdot f[J])=\left(\frac{\delta}{\delta J}W[J]+\frac{\delta}{\delta J}\right)f[J]\,,
\]

\medskip{}

\noindent we find the following elegant formulation of the DSE in
terms of connected Green\textasciiacute s functions

\bigskip{}

\noindent 
\begin{equation}
\left(\frac{\delta S}{\delta\varphi_{i}}\left[\frac{\delta}{\delta J}W[J]+\frac{\delta}{\delta J}\right]+J_{i}\right)=0\,.\label{eq:SDE2}
\end{equation}

\medskip{}

\begin{figure}
\fbox{\begin{minipage}[t]{1\columnwidth}%
\includegraphics[scale=0.7]{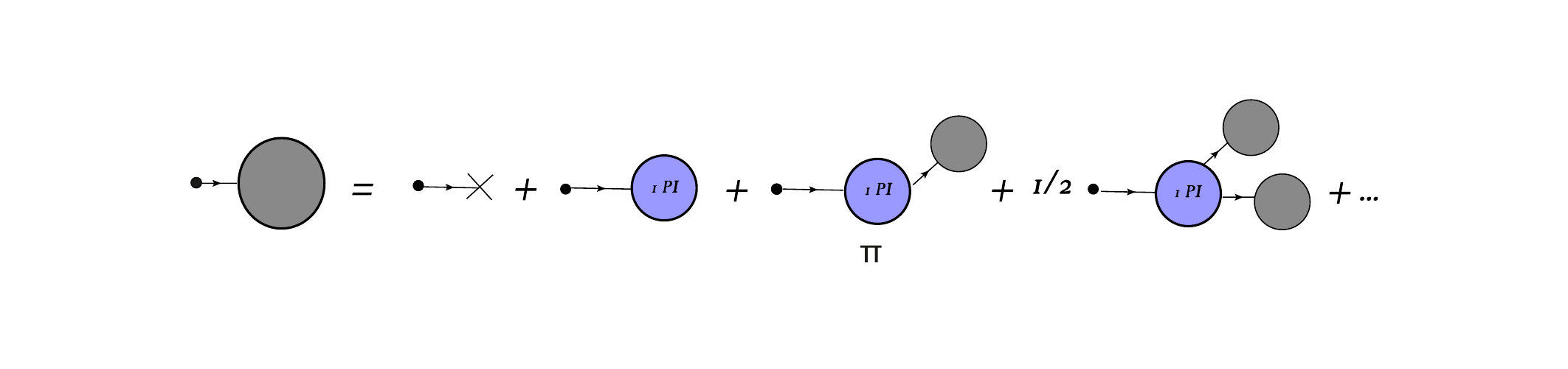} %
\end{minipage}}\protect\caption{Decomposition of a connected Green`s function into 1PI Green`s functions.
1PI diagrams connecting two external legs are called proper self energy,
here denoted by $\Pi_{ij}$. \label{fig: 1PI} }
\end{figure}

\noindent Finally, we can further decompose connected Green`s functions
in terms of \textit{one-particle irreducible} (1PI) ones. By definition,
1PI Green`s function remains connected even if one internal line is
cut. If one follows an external leg into a connected Green`s function
one inevitably ends up on a 1PI diagram which again has zero, one
or more external legs which lead to another connected part and whose
cutting would disconnect the diagram (at tree level, one does not
end up at a 1PI diagram at all). Iteration leads to yet another decomposition
in terms of 1PI diagrams. This is summarized in figure \ref{fig: 1PI}. 

\noindent 
The 1PI Green`s functions can be viewed as the fundamental building
blocks of any Feynman diagram. The coefficients in this expansion
are called\textit{ proper self energy} $\Pi_{ij}$ (connecting two
external legs) and\textit{ proper vertices $\Gamma_{ijk..}$ }(connecting
three or more external legs). We introduce the field $\phi_{i}^{c}=\frac{\delta}{\delta J_{i}}W$
, a connected Green`s function with one external leg attached. This
object is precisely the classical field

~

\noindent defined as the vacuum expectation value of the quantum field
$\varphi$ in the presence of an external source 

\medskip{}

\noindent 
\begin{equation}
\phi^{c}=\frac{\left\langle 0\left|\varphi\right|0\right\rangle \left|_{J}\right.}{\left\langle 0\left|0\right.\right\rangle \left|_{J}\right.}\,.
\end{equation}
Algebraically, we can write the decomposition displayed in figure
\ref{fig: 1PI} as 
\begin{equation}
\phi_{i}^{c}=S_{ij}^{0}\left(J_{j}+\Gamma_{j}+\Pi_{jk}\phi_{k}^{c}+\frac{1}{2}\Gamma_{jkl}\phi_{k}^{c}\phi_{l}^{c}+....\right)\,.\label{eq:2PI1}
\end{equation}

\noindent Here, we have defined the proper self energy as $\Pi_{ij}=\left(S_{ij}^{0}\right)^{-1}+\Gamma_{ij}$
. The reason for this definition is that we want to reformulate equation
(\ref{eq:2PI1}) such that the left side is zero:

\bigskip{}

\noindent 
\begin{equation}
0=J_{i}+\Gamma_{i}+(-\left(S_{ij}^{0}\right)^{-1}+\Pi_{ij})\phi_{j}^{c}+\frac{1}{2}\Gamma_{ijk}\phi_{j}^{c}\phi_{k}^{c}+...=J_{i}+\Gamma_{i}+\Gamma_{ij}\phi_{j}^{c}+\frac{1}{2}\Gamma_{ijk}\phi_{j}^{c}\phi_{k}^{c}+...\,\,.\label{eq:2PI2}
\end{equation}

\medskip{}

\noindent Again we introduce the corresponding generating functional:

\bigskip{}

\noindent 
\begin{equation}
\Gamma[\phi^{c}]=\sum_{m=1}^{\infty}\frac{1}{m!}\Gamma_{i,j...m}\phi_{i}^{c}\phi_{j}^{c}...\phi_{m}^{c}\,,
\end{equation}

\medskip{}

\noindent where the argument of $\Gamma$ denotes a vector of the
classical fields $\phi^{c}=\left[\left\{ \phi_{i}^{c}\right\} \right]$.
Acting with $\delta/\delta\phi_{i}^{c}\,\delta/\delta\phi_{j}^{c}...$
on $\Gamma[\phi^{c}]$ generates vertices of 1PI diagrams - unlike
full or connected Green`s functions, 1PI Green`s functions do not
have propagators attached on their external legs. With these preparations,
we can reformulate equation (\ref{eq:2PI2}) in the following way

\bigskip{}

\noindent 
\begin{equation}
0=J_{i}+\frac{\delta\Gamma[\phi^{c}]}{\delta\phi_{i}^{c}}\,,\,\,\,\,\,\phi_{i}^{c}=\frac{\delta W[J]}{\delta J_{i}}\,\,.\label{eq:effPot}
\end{equation}

\medskip{}

\noindent Equations (\ref{eq:effPot}) reveal the true nature of the
relationship between connected and 1PI green functions: $W[J]$ and
$\Gamma[\phi^{c}]$ are connected by a Legendre transform 

\bigskip{}

\noindent 
\begin{equation}
W[J]=\Gamma[\phi^{c}]+\phi_{i}^{c}J_{i}\,.\label{eq:effPot-1}
\end{equation}

\medskip{}

\noindent This equation summarizes that $W$ is independent of $\phi^{c}$
and $\Gamma$ is independent of $J$

\bigskip{}

\noindent 
\begin{equation}
\frac{\delta W}{\delta\phi_{i}^{c}}=J_{i}+\frac{\delta\Gamma[\phi^{c}]}{\delta\phi_{i}^{c}}=0\,,\,\,\,\,\,\,\,\,\frac{\delta\Gamma}{\delta J_{i}}=\phi_{i}^{c}-\frac{\delta W[J]}{\delta J_{i}}=\phi_{i}^{c}-\phi_{i}^{c}=0\,.
\end{equation}

\medskip{}

\noindent Once again, we can construct the DSE in terms of 1PI diagrams.
To do so, we eliminate $J$ derivatives

~

~

\noindent in favor derivatives of $\phi^{c}$:

\bigskip{}

\noindent 
\begin{equation}
\frac{\delta}{\delta J_{i}}=\frac{\delta\phi_{j}^{c}}{\delta J_{i}}\frac{\delta}{\delta\phi_{j}^{c}}=\frac{\delta^{2}W[J]}{\delta J_{i}\delta J_{j}}\frac{\delta}{\delta\phi_{j}^{c}}\,,
\end{equation}

\medskip{}

\noindent and further use $J_{i}=-\delta\Gamma[\phi^{c}]/\delta\phi_{i}^{c}$
and $\phi_{i}^{c}=\delta W[J]/\delta J_{i}$ to exchange $\left\{ J,\delta J\right\} $with
$\left\{ \phi^{c},\delta\phi^{c}\right\} $ in the DSE (\ref{eq:SDE1}):

\medskip{}

\noindent 
\begin{equation}
\frac{\delta\Gamma[\phi^{c}]}{\delta\phi_{i}^{c}}=\frac{\delta S}{\delta\phi_{i}^{c}}\left[\phi^{c}+W^{\prime\prime}[J]\frac{\delta}{\delta\phi^{c}}\right]\,.\label{eq:SDE3}
\end{equation}

\medskip{}

\noindent The 1PI DSE equation (\ref{eq:effPot}) still plays the
role of a quantum analog to the classical equation of motion. Motivated
by equation (\ref{eq:SDE3}), $\Gamma$ is called \textit{effective
action}. To obtain the classical equation of motion from the 1PI quantum
DSE (\ref{eq:SDE3}), we can simply neglect the contribution coming
from $W^{\prime\prime}$. The role of $W^{\prime\prime}$ is to create
loops. %
\footnote{\noindent The classical DSE can be obtained from (\ref{eq:SDE3})
by neglecting the contribution coming from $W^{\prime\prime}$. It
is a tree expansion of the form

\noindent 
\[
\phi_{i}^{c}=S_{ij}^{0}\left(J_{j}+\frac{\delta S[\phi^{c}]}{d\phi_{j}^{c}}\right)=S_{ij}^{0}\left(J_{j}+\frac{1}{2}\gamma_{jkl}\phi_{k}^{c}\phi_{l}^{c}+\frac{1}{6}\gamma_{jklm}\phi_{k}^{c}\phi_{l}^{c}\phi_{m}^{c}+...\right)\,.
\]

\noindent We can interpret the right hand side diagramatically:

\noindent \includegraphics[scale=0.5]{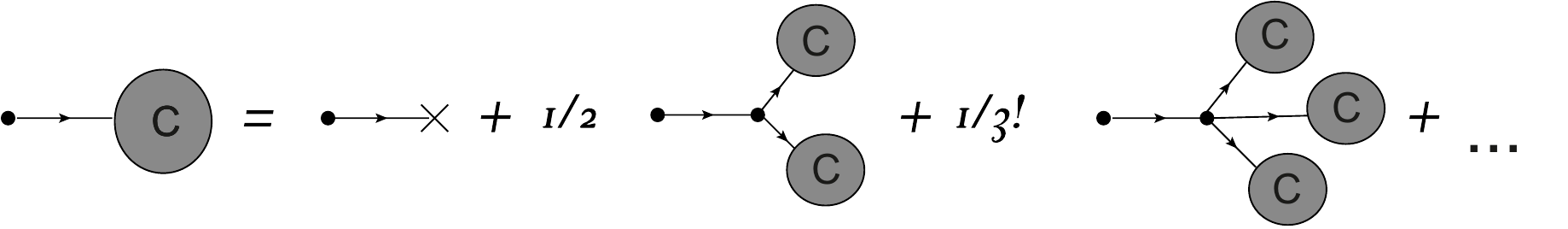}

\noindent Comparing this to the 1PI expansion displayed in figure
\ref{fig: 1PI}, we can see that the vertices $\gamma_{ijkl..}$ of
the tree expansion have been replaced by 1PI diagrams. In other words
in the 1PI formalism, the quantum loop corrections to the classical
expansion are embodied in the proper vertices $\Gamma_{ijkl..}$ ,
i.e. they sit in the spot of the classical vertices. %
}

\noindent In summary, we can take the generating functional $Z$ of
full Green`s functions, replace $S[\varphi]$ with $\Gamma[\varphi]$
and and instead of evaluating the full path integral substitute the
classical approximation $\varphi\rightarrow\phi^{c}$ in the integrand
and obtain an \textit{exact }result.

\noindent Spontaneous symmetry breaking manifests itself in the 1PI
formalism by a nonzero vacuum expectation value of $\varphi$ , even
though the source J is set to zero

~

~

\noindent 
\begin{equation}
\left.\frac{\delta\Gamma}{\delta\varphi}\right|_{\varphi=\phi^{c}}=0,\,\,\,\,\,\,\,\phi^{c}\neq0\,.
\end{equation}

\noindent ~

~

~

\subsection{\noindent 1PI effective action in the one loop approximation\label{sub:1PI-effective-action-1Loop}}

\noindent ~

\noindent In order to obtain the one loop approximation for the effective
action for a complex scalar field $\varphi$ we introduce the path
integral representation for the generating functional of connected
Green`s functions 
\begin{equation}
Z[J]=e^{W[J]}=\frac{\int D\varphi e^{S[\varphi]+\int J\varphi}}{\int D\varphi e^{iS[\varphi]}}\,.
\end{equation}

\noindent Next, we perform the Legendre transform to obtain the 1PI
effective action \bigskip{}

\noindent 
\begin{equation}
e^{\Gamma[\phi^{C}]}=e^{i(W[J]-\int\,\phi^{C}J)}=\frac{\int D\varphi e^{\left(S[\varphi]+\int J(\varphi-\phi^{c})\right)}}{\int D\varphi e^{iS[\varphi]}}\,,\label{eq:NR1}
\end{equation}

\noindent and expand the action $S$ around the classical solution
which is the main contribution to the path integral

\bigskip{}

\noindent 
\begin{equation}
S[\varphi]=\int\mathcal{L}(\varphi)=\int\mathcal{L}(\phi^{c})+\int\underbrace{\left(\varphi-\phi^{c}\right)}_{\delta\varphi}\left.\frac{\delta S}{\delta\varphi}\right|_{\varphi=\phi^{c}}+\underbrace{\frac{1}{2}\int\delta\varphi_{i}\left.\frac{\delta\mathcal{L}}{\delta\varphi_{i}\delta\varphi_{j}}\right|_{\varphi=\phi^{c}}\delta\varphi_{j}}_{one\,\, loop\,\, correction}+...\,.\label{eq:epand}
\end{equation}

\noindent This amounts to \bigskip{}

\noindent 
\begin{equation}
\Gamma[\phi^{C}]\sim\int\mathcal{L}[\phi^{c}]+\int\delta\varphi\underbrace{\left(J+\left.\frac{\delta S}{\delta\varphi}\right|_{\varphi=\phi^{C}}\right)}_{=0}+\frac{1}{2}\int\delta\varphi_{i}\left.\frac{\delta\mathcal{L}}{\delta\varphi_{i}\delta\varphi_{j}}\right|_{\varphi=\phi^{c}}\delta\varphi_{j}\,.
\end{equation}

\noindent Inserting the result in the exponent of (\ref{eq:NR1})
and performing the Gaussian integral%
\footnote{The linear shift in $\varphi\rightarrow\varphi-\phi^{c}$ doesn\textquoteright t
affect the integration measure.

~%
} yields

\bigskip{}

\noindent 
\[
e^{\Gamma[\phi^{C}]}\sim e^{\int\mathcal{L}[\phi^{c}]}\cdot\left.\textrm{det}\left(\frac{\delta^{2}\mathcal{L}}{\delta\varphi_{i}\delta\varphi{}_{j}}\right)^{-1/2}\right|_{\varphi=\phi^{c}}\,,
\]

\noindent and therefore 

\bigskip{}

\noindent 
\begin{equation}
\Gamma[\phi^{c}]\thickapprox\int\mathcal{L}[\phi^{c}]+\left.\frac{1}{2}\textrm{Tr\,\ ln}\left(\frac{\delta^{2}\mathcal{L}}{\delta\varphi_{i}\delta\varphi{}_{j}}\right)\right|_{\varphi=\phi^{c}}=\int\mathcal{L}[\phi^{c}]+\left.\frac{1}{2}\textrm{Tr\,\ ln}\, S_{0}^{-1}\right|_{\varphi=\phi^{c}}\,,\label{eq:1PIresummed}
\end{equation}

\noindent where $\textrm{det\,\ M=exp[Tr\,\ ln\,\ M]}$ has been used.
This is the one loop effective action which we shall use (within certain
approximations) in section \ref{sub:The-finite-temperature-setup}.
The term proportional to the quadratic fluctuations $\delta\varphi$
is the propagator, see for example (\ref{eq:TestProp}). In summary,
the Legendre transform has introduced new vertices to our theory.
To check that the diagrammatic expansion of $\Gamma[\phi^{c}]$ is
indeed given by all one loop diagrams of these vertices, we separate
contributions from the condensate which are resummed to all orders
in $\lambda$ in (\ref{eq:1PIresummed}), from the free propagator
in absence of a condensate $G_{0}$, $S_{0}^{-1}=G_{0}^{-1}+\lambda\phi_{c}^{2}\xi$
where $\xi$ is matrix of the dimension of the internal space%
\footnote{Comparing to our case of a complex scalar filed, $G_{0}$ and $\xi$
are given by 
\[
G_{0}^{-1}=\left(\begin{array}{cc}
-k^{2}+m^{2}-\mu^{2} & 2i\, k_{0}\mu\\
-2i\, k_{0}\mu & -k^{2}+m^{2}-\mu^{2}
\end{array}\right)\,,\,\,\,\,\,\,\xi=\left(\begin{array}{cc}
3 & 0\\
0 & 1
\end{array}\right)\,.
\]

The situation will be slightly more complicated once we treat modulus
and phase of the condensate separately (see also (\ref{eq:fluct2}))
as $S_{0}^{-1}$ becomes anisotropic.%
}. 

\noindent Then the logarithmic term is decomposed as follows

\noindent 
\[
\textrm{ln}\, S_{0}^{-1}=\textrm{ln}\, G_{0}^{-1}+\textrm{ln}\,\left(1+\lambda G_{0}\phi_{c}^{2}\,\xi\right)\,.
\]

\noindent We can now expand the second term in a Taylor series and
obtain 

\medskip{}

\noindent 
\[
\textrm{Tr}\,\textrm{ln}\,\left(1+\lambda G_{0}\left(\phi^{c}\right)^{2}\,\xi\right)=-\sum_{n}\left(-\lambda\left(\phi^{c}\right)^{2}\,\right)^{n}\textrm{Tr}\left[G_{0}\,\xi\right]^{n}\,.
\]

\noindent Diagrammatically, this series corresponds to a one loop
diagram with 2, 4, 6.. classical fields (condensates) attached:

\noindent \includegraphics{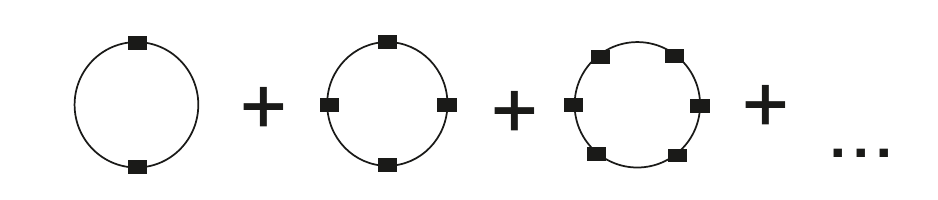}

\noindent As expected, no external legs are connected to these 1PI
diagrams. A more physical way to say this, is that external legs would
introduce a momentum dependence of $\Gamma[\phi^{c}]$. We will later
identify this functional with the pressure, which, being a thermodynamic
quantity, must not depend on external momenta. 

~

\subsection{\noindent Two-particle irreducible formalism\label{sub:Two-particle-irreducible-formalism}}

~

\noindent The construction of the 2PI effective action is a straightforward
generalization of the 1PI case. As a starting point we use the generating
functional of connected Green`s function and introduce two source
terms, one for the quantum field $\varphi$ and one for the quantum
propagator $S_{0}$:

\bigskip{}

\noindent 
\begin{equation}
Z=Z[J,K]=\textrm{exp}(iW[J,K])=\int\mathcal{D}\varphi\textrm{exp}\left(S[\varphi]+J_{i}\varphi_{i}+\frac{1}{2}K_{ij}\varphi_{i}\varphi_{j}\right)\,.\label{eq:doubleSource}
\end{equation}

\noindent The\textit{ classical field and two point function (propagator)}
are obtained in the usual way:

\bigskip{}

\noindent 
\begin{equation}
\frac{\delta W[J,K]}{\delta J_{i}}=\phi_{i}^{c}\,,\,\,\,\,\,\,\,\,\,\,\frac{\delta W[J,K]}{\delta J_{i}\delta J_{j}}=S_{ij}^{c}\,,\label{eq:classSoldouble}
\end{equation}

\noindent and therefore variation with respect to the source $K_{ij}$
results in

\bigskip{}

\noindent 
\[
\frac{\delta W[J,K]}{\delta K_{ij}}=\frac{1}{2}\left(\phi_{i}^{c}\phi_{j}^{c}+S_{ij}^{c}\right)\,.
\]

\medskip{}

\noindent Before we turn to the construction of the 2PI effective
action, we remark that we can easily construct the \textit{1PI effective
action} of the functional $Z[J,K]$ (\ref{eq:doubleSource}). The
quadratic term including $K$ can formally be treated as an additional
``mass'' term and the (single) Legendre transform can be directly
applied:

\bigskip{}

\noindent 
\begin{equation}
\Gamma_{K}[\phi^{c}]=W[J,K]-\phi_{i}^{c}J_{i}\,,\,\,\,\,\,\,\textrm{with\,\,\,\,}\frac{\delta W[J,K]}{\delta J_{i}}=\phi_{i}^{c}\,.\label{eq:1PI2PI}
\end{equation}

\noindent In analogy to the 1PI effective action, we define now the
2PI effective action by a double Legendre Transform with respect to
$\phi_{i}^{c}$ and $S_{ij}^{c}$

\bigskip{}

\noindent 
\begin{equation}
\Gamma[\phi^{c},S^{c}]=W[J,K]-\phi_{i}^{c}J_{i}-\frac{1}{2}\phi_{i}^{c}K_{ij}\phi_{j}^{c}-\frac{1}{2}S_{ij}^{c}K_{ij}\,.\label{eq:"PIfunctional}
\end{equation}

\noindent Furthermore, the equations of motion for the fields $\phi^{c}$
and $S^{c}$ are given by

\noindent 
\begin{eqnarray}
\frac{\delta\Gamma}{\delta\phi_{i}^{c}} & = & -J_{i}-K_{ij}J_{j}\,,\\
\frac{\delta\Gamma}{\delta S_{ij}^{c}} & = & -\frac{1}{2}K_{ij}\,.\label{eq:DSE2PI}
\end{eqnarray}

\noindent Using equation (\ref{eq:classSoldouble}) and the 2PI effective
action (\ref{eq:"PIfunctional}), it is again easy to show that $\delta\Gamma[\phi^{c},S^{c}]/\delta J_{i}=0$
and $\delta\Gamma[\phi^{c},S^{c}]/\delta K_{ij}=0$. The physically
interesting case is again given by vanishing sources in which case
the equations of motions turn into 

\bigskip{}

\noindent 
\begin{equation}
\left.\frac{\delta\Gamma}{\delta\varphi_{i}}\right|_{\varphi_{i}=\phi_{i}^{c}}=0,\,\,\,\,\,\,\,\left.\frac{\delta\Gamma}{\delta S_{ij}^{0}}\right|_{S_{ij}^{0}=S_{ij}^{c}}=0\,,\label{eq:stat2PI}
\end{equation}

\noindent with non-vanishing expectation values $\phi_{i}^{c}\neq0$
and $S_{ij}^{c}\neq0$. What is remarkable about the stationary equations
(\ref{eq:stat2PI}), is that they already include the DSE for the
propagator on the classical level. We shall exploit this fact extensively
in section \ref{sec:The-two-fluid-model-at-arbitrary-T}. In particular
cases such as BCS theory, spontaneous symmetry breaking does not result
in a vacuum expectation value of the field, but in a vacuum expectation
value of the two point function. The 2PI formalism is therefore particularly
well suited to describe such systems. In the path-integral representation,
the 2PI generating functional is given by 

\bigskip{}

\noindent 
\[
e^{\Gamma[\phi^{c},S^{c}]}=\frac{\int D\varphi\,\textrm{exp}\left[S[\varphi]+\int J(\varphi_{i}-\phi_{i}^{c})+\frac{1}{2}\int\int K_{ij}(\varphi_{i}\varphi_{j}-\phi_{i}^{c}\phi_{j}^{c}-S_{ij}^{c})\right]}{\int D\varphi\,\textrm{exp}\left[S[\varphi]\right]}\,.
\]

\noindent We can again study the loop expansion systematically. To
leading order, we can use the \textit{one loop }approximation derived
in the last section to evaluate the \textit{1PI effective action}
defined by (\ref{eq:1PI2PI}) (i.e. we use the generating functional
$W[J,K]$ from equation (\ref{eq:doubleSource}) but perform only
a single Legendre transform with respect to $J$). Remember that we
can absorb the term proportional to $\varphi_{i}\varphi_{j}$ as an
additional mass term into the propagator of the action $S[\varphi]$.
Now all we have to do is to replace $S[\phi^{c}]$ by $S[\phi_{c}]+K_{ij}\phi_{i}^{c}\phi_{j}^{c}$
and the free propagator in the background of the classical field $S_{0}^{^{-1}}$
by $S_{0}^{-1}+K$ in the expansion (\ref{eq:epand}). The result
reads:

\bigskip{}

\noindent 
\[
\Gamma_{K}[\phi^{c}]=S[\phi^{c}]+\frac{1}{2}\textrm{Tr}\,\textrm{ln}\,\left[S_{0}^{^{-}1}+K\right]+\frac{1}{2}K_{ij}\phi_{i}^{c}\phi_{j}^{c}\,.
\]

\noindent From the dual Legendre transform (\ref{eq:"PIfunctional})
we can obtain the one loop approximation of the 2PI functional:

\bigskip{}

\noindent 
\[
\Gamma[\phi^{c},S^{c}]=S[\phi^{c}]+\frac{1}{2}\textrm{Tr}\,\textrm{ln}\,\left[S_{0}^{^{-}1}+K\right]-\frac{1}{2}\textrm{Tr}[KS^{c}]\,.
\]

\noindent To reformulate this equation in a more useful way, we need
to eliminate the remaining source $K$ in favor of the field $S^{c}$.
This can be achieved with the aid of $\delta\Gamma/\delta K=0$ from
which we obtain (at lowest loop order) $S^{c{}^{-1}}=S_{0}^{-1}+K$
and therefore 
\[
\Gamma[\phi^{c},S^{c}]=S[\phi^{c}]+\frac{1}{2}\textrm{Tr}\,\textrm{ln}\,\left(S^{c}\right)^{-1}+\frac{1}{2}\textrm{Tr}\,\left[\left(S^{c^{-1}}-S_{0}^{-1}\right)S^{c}\right]\,.
\]
To go beyond the one loop approximation, we introduce the potential
$V_{2}$ which includes 2PI diagrams with \textit{at least } two loops
built from the propagator $S^{c}$: 

\bigskip{}

\noindent 
\begin{equation}
\Gamma[\phi^{c},S^{c}]=S[\phi^{c}]+\frac{1}{2}\textrm{Tr}\,\textrm{ln}\,\left(S^{c}\right)^{-1}+\frac{1}{2}\textrm{Tr}\,\left[\left(S^{c^{-1}}-S_{0}^{-1}\right)S^{c}\right]+V_{2}\,.\label{eq:2PIFunc}
\end{equation}

\noindent We will use the two-loop Hartree approximation of this object
as will be explained in section \ref{sec:The-two-fluid-model-at-arbitrary-T}.

\newpage{}

\section{Two-fluid formalism in the low-temperature approximation\label{sec:Finite-temperature:-Two-fluid}}

~

\subsection{Effective action and dispersion relations. \label{sub:The-finite-temperature-setup}}

~

\noindent We the aid of the non-perturbative 1PI and 2PI formalisms
which we have reviewed in the last section, we can now continue our
derivation of the two-fluid model at finite temperature. The nonzero-temperature
calculation is complicated because we are interested in the situation
of a nonzero (homogeneous) superflow; this renders the system and
in particular the dispersions of the Goldstone mode anisotropic. In
this part, we are mostly interested in analytical results to discuss
the translation to hydrodynamics thoroughly. Since we have argued
in section \ref{sec:Lagrangian} that a calculation for arbitrary
temperatures requires the more complicated 2PI formalism in which
results are obtained purely numerically, we shall for the moment restrict
ourselves to small temperatures. In particular this means that we
shall not determine the condensate self-consistently by solving $\delta\Gamma/\delta\rho=0$,
and rather work with its zero-temperature value obtained from $\partial U/\partial\rho=0$.
Starting from the full self-consistent formalism, one can show that
the temperature-dependence of the condensate includes an additional
power of the coupling constant compared to the terms we are keeping.
We are thus working in the low-temperature, weak-coupling approximation.
Note however that the melting of the condensate, as well as other
contributions from the full action may induce terms proportional to
$\lambda\mu^{2}T^{2}$ in the pressure, while we shall only keep terms
proportional to $T^{4}$ and higher order in $T$. This approximation
is only consistent if $T^{2}\gg\lambda\mu^{2}$, i.e., strictly speaking,
our approximation leaves a ``gap'' between $T=0$ and the small temperatures
we are discussing, although this gap can be made arbitrarily small
by choosing the coupling constant $\lambda$ sufficiently small. 

\noindent Within this approximation, we can work with the simple 1PI
effective action in the one-loop approximation

\begin{equation}
\Gamma=-\frac{V}{T}\, U(\rho,\sigma)-\frac{1}{2}\sum_{k}{\textstyle Tr}\ln\frac{S_{0}^{-1}(k)}{T^{2}}\,,\label{eq:effAction}
\end{equation}
with the tree-level potential $U$ from equation (\ref{eq:PotentialMin})
and the inverse tree-level propagator in momentum space $S_{0}^{-1}(k)$
which can be read off from the terms quadratic in the fluctuations
in equation (\ref{eq:fluct2}). The trace is taken over the $2\times2$
space of the two real degrees of freedom of the complex scalar field,
and the sum is taken over four-momenta $k_{\mu}=(k_{0},\vec{k})$,
$k_{0}=-i\omega_{n}$ with the bosonic Matsubara frequencies $\omega_{n}$.
The sum over three-momenta, here written for a finite volume $V$
and thus over discrete momenta, becomes an integral in the thermodynamic
limit. At the stationary point, i.e., with equation (\ref{eq:rhomin}),
we have

\begin{equation}
S_{0}^{-1}(k)=\left(\begin{array}{cc}
-k^{2}+2(\sigma^{2}-m^{2}) & 2ik\cdot\partial\psi\\[2ex]
-2ik\cdot\partial\psi & -k^{2}
\end{array}\right)\,.\label{eq:Propagator}
\end{equation}

~

\noindent Here, $k^{2}\equiv k_{\mu}k^{\mu}$, $k\cdot\partial\psi=k_{\mu}\partial^{\mu}\psi$
and $\sigma=\partial_{\mu}\psi\partial^{\mu}\psi$ . In all our microscopic
calculations we evaluate thermal fluctuations in the background of
a \textit{uniform} superflow, so we take $\partial_{\mu}\psi$ (and
hence $\sigma$) to be space-time independent%
\footnote{Observe that the relation between $S_{0}^{-1}$ with and without superflow
is given by a simple boost. To see this, set $\vec{\nabla}\psi=\vec{0}$
in (\ref{eq:Propagator}) and then boost every wave vector $k^{\mu}$
by an arbitrary velocity $\vec{v}$:

\[
k^{\mu\prime}\rightarrow\text{\ensuremath{\Lambda}}_{\text{\ensuremath{\nu}}}^{\mu}(v)k^{\nu}\,.
\]

This results in 

\[
S_{0}^{-1\prime}\rightarrow\left(\begin{array}{cc}
-k^{2}+2(\sigma^{2}-m^{2}) & -2i\sigma\frac{k_{0}-\vec{k}\cdot\vec{v}}{\sqrt{1-v^{2}}}\\
2i\sigma\frac{k_{0}-\vec{k}\cdot\vec{v}}{\sqrt{1-v^{2}}} & -k^{2}
\end{array}\right)
\]

Inserting the microscopic expressions $\vec{v}=-\vec{\nabla}\psi/\partial_{0}\psi$
and $\sigma=\sqrt{\partial_{\mu}\psi\partial^{\mu}\psi}$ we arrive
precisely at (\ref{eq:Propagator}). %
}.

\noindent Since we neglect the melting of the condensate, we have
inserted the zero-temperature solution into the propagator. In order
to compute the effective action, one may employ partial integration
with respect to the $|\vec{k}|$ integral. The effective action density
then becomes

\begin{eqnarray}
\frac{T}{V}\Gamma & = & \frac{(\sigma^{2}-m^{2})^{2}}{4\lambda}+\frac{1}{6}\frac{T}{V}\sum_{k}|\vec{k}|\,\textrm{Tr}\left[S_{0}(k)\frac{\partial S_{0}^{-1}(k)}{\partial|\vec{k}|}\right]\nonumber \\
 & = & \frac{(\sigma^{2}-m^{2})^{2}}{4\lambda}-\frac{2}{3}\frac{T}{V}\sum_{k}\frac{\vec{k}^{2}(k^{2}-\sigma^{2}+m^{2})+2k\cdot\partial\psi\,\vec{k}\cdot\vec{\nabla}\psi}{{\rm det}\, S_{0}^{-1}(k)}\,,\label{eq:effActionExplicit}
\end{eqnarray}
where we have used the explicit form of the tree-level propagator
and have performed the $2\times2$ trace. We have also inserted the
zero temperature solution for $\rho$ into the potential $U$. The
evaluation of the Matsubara sum and the simplification of the result
are explained in detail in appendix \ref{sec:Matsubara-sum-with-anisotropy}.
Neglecting the thermal contribution of the massive mode, we can write
the result as 

\begin{equation}
\frac{T}{V}\Gamma\simeq\frac{(\sigma^{2}-m^{2})^{2}}{4\lambda}-\int\frac{d^{3}\vec{k}}{(2\pi)^{3}}\frac{F(\epsilon_{1,\vec{k}},\vec{k})}{(\epsilon_{1,\vec{k}}+\epsilon_{1,-\vec{k}})(\epsilon_{1,\vec{k}}+\epsilon_{2,-\vec{k}})(\epsilon_{1,\vec{k}}-\epsilon_{2,\vec{k}})}\,\coth\frac{\epsilon_{1,\vec{k}}}{2T}\,,\label{eq:effectiveAction2}
\end{equation}
 where $F(k_{0},\vec{k})$ denotes the numerator in the momentum sum
of equation (\ref{eq:effActionExplicit})

\begin{equation}
F(k_{0},\vec{k})\equiv-\frac{2}{3}\left[\vec{k}^{2}(k^{2}-\sigma^{2}+m^{2})+2k\cdot\partial\psi\,\vec{k}\cdot\vec{\nabla}\psi\right]\,,\label{eq:Fk}
\end{equation}
and where $\epsilon_{1/2,\vec{k}}$ are complicated excitation energies
whose small-momentum approximations are

\begin{eqnarray}
\epsilon_{1,\vec{k}} & = & \sqrt{\frac{\sigma^{2}-m^{2}}{3\sigma^{2}-m^{2}}}\;\zeta(\hat{k})\,|\vec{k}|+{\cal O}(|\vec{k}|^{3})\,,\label{eq:DispGold}\\
\epsilon_{2,\vec{k}} & = & \sqrt{2}\sqrt{3\sigma^{2}-m^{2}+2(\vec{\nabla}\psi)^{2}}+{\cal O}(|\vec{k}|)\,.\label{eq:DispMass}
\end{eqnarray}
Here we have abbreviated

\begin{equation}
\zeta(\hat{k})\equiv\left[\sqrt{1+2\frac{(\vec{\nabla}\psi)^{2}-(\vec{\nabla}\psi\cdot\hat{k})^{2}}{3\sigma^{2}-m^{2}}}-\frac{2\partial_{0}\psi\vec{\nabla}\psi\cdot\hat{k}}{\sqrt{\sigma^{2}-m^{2}}\sqrt{3\sigma^{2}-m^{2}}}\right]\left[1+\frac{2(\vec{\nabla}\psi)^{2}}{3\sigma^{2}-m^{2}}\right]^{-1}\,.
\end{equation}
with $\theta$ being the angle between $\vec{k}$ and $\vec{\nabla}\psi$.
The physically relevant low-energy excitation is $\epsilon_{1,\vec{k}}$.
This is the Goldstone mode which, as equation (\ref{eq:DispGold})
confirms, is massless and linear in the momentum for small momenta,
as it should be. The coefficient in front of the linear term is related
to the speed of (first) sound as we explain in part III. For the case
without superflow we have $\zeta(\hat{k})=1$, and thus, in the limit
$m=0$, we recover the well-known value of $1/\sqrt{3}$. The superflow
introduces an angular dependence $\zeta(\hat{k})$ into the Goldstone
dispersion and thus also into the sound velocity. This angular-dependent
function also shows that, in contrast to the zero-temperature case,
we cannot write the result in a covariant way since the temporal and
spatial components of $\partial_{\mu}\psi$ appear separately. Another
observation is the complicated factor in the integrand of equation
(\ref{eq:effectiveAction2}). This factor indicates a mixing between
the original modes of the complex field (between particles and antiparticles,
essentially) due to condensation. Such a factor appears also, although
considerably simpler, for the case without superflow. It is analogous
to a Bogoliubov coefficient in the case of Cooper pairing which, in
that case, accounts for the mixing between fermions and fermion-holes.
We shall now calculate the field theoretic representation of the stress-energy
tensor and the Noether current.

\newpage{}

\subsection{Stress-energy tensor and current\label{sub:Stress-energy-tensor-and-current}}

~

\noindent As we have emphasized, stress-energy tensor and charge-current
are particularly important in our derivation of the two-fluid model
as we posses a precise recipe how to construct them from field theory:

\begin{equation}
T^{\mu\nu}=\left\langle \frac{2}{\sqrt{-g}}\frac{\delta(\sqrt{-g}\,{\cal L})}{\delta g_{\mu\nu}}\right\rangle \,,\qquad j^{\mu}=\left\langle \frac{\partial{\cal L}}{\partial(\partial_{\mu}\psi)}\right\rangle \,.\label{eq:TensorFunc}
\end{equation}
The angular brackets denote the expectation value of an operator in
the finite-temperature ensemble of the microscopic theory,

\begin{equation}
\langle A\rangle\equiv\frac{1}{Z}\int{\cal D}\varphi_{1}'{\cal D}\varphi_{2}'\, A\,\exp\left(\int d^{4}x\,\sqrt{-g}\,{\cal L}\right)\,,\label{eq:Integrals}
\end{equation}
 with the partition function $Z$ defined so that $\langle1\rangle=1$.
The calculation of the functional integral is lengthy and carried
out in appendix \ref{sec:Path-integrals-over-complexF}. The results
reads 

\begin{eqnarray}
T^{\mu\nu} & = & -\left(2\frac{\partial U}{\partial g_{\mu\nu}}-g^{\mu\nu}U\right)-\frac{T}{V}\sum_{k}\textrm{Tr}\left[S_{0}(k)\frac{\partial S_{0}^{-1}(k)}{\partial g_{\mu\nu}}-\frac{g^{\mu\nu}}{2}\right]\,,\label{eq:TPath}\\
j^{\mu} & = & \partial^{\mu}\psi\,\frac{\sigma^{2}-m^{2}}{\lambda}-\frac{1}{2}\frac{T}{V}\sum_{k}\textrm{Tr}\left[S_{0}(k)\frac{\partial S_{0}^{-1}(k)}{\partial(\partial_{\mu}\psi)}\right]\,.\label{eq:JPath}
\end{eqnarray}
In the case of the stress-energy tensor, one can check explicitly
that the sum over the Matsubara frequencies leads to an infinite result.
A renormalization is thus required. As a renormalization condition,
we require for the case without superflow the obvious interpretation
of the diagonal components $T^{\mu\nu}$ in terms of the energy density
$\epsilon$ and the pressure $P$, 

\begin{equation}
T^{00}=\epsilon\,,\qquad T^{ij}=\delta^{ij}P\,.
\end{equation}
Then, switching on a nonzero superflow, does not yield any additional
divergences. These conditions can be implemented on a very general
level, without explicit evaluation of the stress-energy tensor. We
do so in detail in appendix \ref{sec:Renormalization-and-useful-identities}.
The calculation in this appendix also leads to a very useful formulation
of the effective action, the stress-energy tensor, and the current,
which will later facilitate the interpretation in terms of the hydrodynamic
two-fluid picture. We define 

\begin{equation}
\Psi_{k}\equiv\Psi_{k}(\sigma^{2},k\cdot\partial\psi,k^{2})\equiv-\frac{1}{2}\textrm{Tr}\ln\frac{S_{0}^{-1}(k)}{T^{2}}\,,\label{eq:PsiK}
\end{equation}
and 

\vspace{-0.5cm}

\begin{equation}
A_{k}\,\equiv\frac{\partial\Psi_{k}}{\partial(k\cdot\partial\psi)}=4\frac{k\cdot\partial\psi}{{\rm det}\, S_{0}^{-1}},\,\,\, B_{k}\,\equiv2\frac{\partial\Psi_{k}}{\partial\sigma^{2}}=\frac{2k^{2}}{{\rm det}\, S_{0}^{-1}},\,\,\,\, C_{k}\equiv2\frac{\partial\Psi_{k}}{\partial k^{2}}=-\frac{2(k^{2}-\sigma^{2}+m^{2})}{{\rm det}\, S_{0}^{-1}}\,.\label{eq:AkBkCk}
\end{equation}

\noindent In the zero-temperature discussion, we have related the
tree-level potential $U$ to the generalized pressure $\Psi$. The
notation of equation (\ref{eq:PsiK}) anticipates that we can identify
the effective action $\Gamma$ with the generalized pressure at finite
temperature. In section \ref{sub:The-two-fluid-formalism-from-field-theory},
we will see that this assumption is justified. The notations $A_{k}$,
$B_{k}$, $C_{k}$ are chosen in analogy to $\overline{{\cal A}}$,
$\overline{{\cal B}}$, $\overline{{\cal C}}$, which we introduced
in section \ref{sub:Relativistic-thermodynamics-and-entrain} (we
already know the zero temperature contribution to the coefficient
$\overline{{\cal B}}$ from the single-fluid treatment of section
\ref{sub:Zero-temperature-hydrodynamics:-generalized}). With the
help of these quantities we can rewrite the effective action as (see
appendix \ref{sec:Renormalization-and-useful-identities})

\begin{equation}
\frac{T}{V}\Gamma=-U-\frac{1}{3}(g^{\mu\nu}-u^{\mu}u^{\nu})\frac{T}{V}\sum_{k}\left(C_{k}k_{\mu}k_{\nu}+A_{k}k_{\mu}\partial_{\nu}\psi\right)\,,\label{eq:EffActionk}
\end{equation}
 where we have abbreviated $u^{\mu}=(1,0,0,0)$, while the renormalized
stress-energy tensor and the current become 

\begin{eqnarray}
T^{\mu\nu} & = & -\left(2\frac{\partial U}{\partial g_{\mu\nu}}-g^{\mu\nu}U\right)+\frac{T}{V}\sum_{k}\left[C_{k}k^{\mu}k^{\nu}+B_{k}\partial^{\mu}\psi\partial^{\nu}\psi+A_{k}(k^{\mu}\partial^{\nu}\psi+k^{\nu}\partial^{\mu}\psi)+2u^{\mu}u^{\nu}\right]\,,\nonumber \\
\label{eq:TFk}\\
j^{\mu} & = & \partial^{\mu}\psi\,\frac{\sigma^{2}-m^{2}}{\lambda}+\frac{T}{V}\sum_{k}\left(B_{k}\partial^{\mu}\psi+A_{k}k^{\mu}\right)\,.\label{eq:JFk}
\end{eqnarray}
These expressions are useful because, firstly, they are written in
a covariant way and, secondly, they anticipate the two-fluid formulation
which as we know is constructed from two basic four-vectors. We have
already identified the conjugate momentum to $j^{\mu}$ with $\partial^{\mu}\psi$
; the other conjugate momentum can of course not be identified with
$k^{\mu}$, which is a purely microscopic quantity, but we shall see
that this formulation is close enough to the two-fluid formulation
that it can easily be cast in a more ``macroscopic'' form. At the
moment, the introduction of the four-vector $u^{\mu}$ is for notational
purposes only, introduced to write the above equations in terms of
four-vectors. We will see however that it really corresponds to the
four-velocity of the normal fluid introduced in section \ref{sub:A-relativistic-version-of-Landau}. 

\newpage{}

\noindent It is instructive to compute the trace of the stress-energy
tensor. With the help of equation (\ref{eq:auxTrace}) we immediately
find

\begin{equation}
T_{\;\;\mu}^{\mu}=m^{2}\left[\rho^{2}+2\frac{T}{V}\sum_{k}\frac{k^{2}}{{\rm det}\, S_{0}^{-1}(k)}\right]\,,\label{eq:TraceT}
\end{equation}
i.e., the trace vanishes when we set $m=0$, as expected. 

~

\subsection{The two-fluid formalism from field theory\label{sub:The-two-fluid-formalism-from-field-theory}}

~

\noindent So far we have related the zero-temperature limit of the
microscopic theory to hydrodynamics where we identified $\partial^{\mu}\psi$
as the conjugate momentum to $j^{\mu}$. Furthermore, we have carefully
prepared field theoretic results at finite temperature, in particular
the conserved current $j^{\mu}$ and the stress energy tensor $T^{\mu\nu}$.
With these results we can continue to apply the machinery of the two-fluid
formalism to successively identify all remaining hydrodynamic and
thermodynamic quantities in terms of field theoretic variables. Our
covariant formulation from section \textit{\ref{sub:Stress-energy-tensor-and-current}
before }performing the Matsubara sum will turn out to be very useful
for this purpose. 

\noindent Before we can proceed with this program, we have to clarify
the issue of frame dependence of the microscopic calculations at finite
temperature once and for all. Motivated by the discussion of section
\ref{sub:Temperature-and-chemical-pot}, it seems reasonable to assume
that the microscopic calculation is performed in the normal-fluid
rest frame (or the rest frame of the heat bath) defined by $u^{\mu}=(1,0,0,0)$.
We shall now explicitly demonstrate that this is indeed consistent
with the results we obtained in the last section and identify the
thermodynamic parameters $\mu$, $T$ and $\vec{v}_{s}$ in this frame.
As a first step, we may use Landau`s definition of the charge current
from equation (\ref{eq:RelLandauCurrent}) 

\bigskip{}

\noindent  
\[
j^{\mu}=n_{n}u^{\mu}+n_{s}\frac{\partial^{\mu}\psi}{\sigma}\,,
\]

\noindent and evaluate it the normal-fluid rest frame defined by $u^{\mu}=(1,0,0,0)$.
From the temporal components of $j^{\mu}$ we know that in the normal-fluid
rest frame the total number density is given by $n=n_{n}+\partial\psi_{0}\, n_{s}/\sigma$
where $\partial_{0}\psi/\sigma=1/\sqrt{1-\vec{v}_{s}^{2}}$ plays
the role of a Lorentz factor (see also the discussion of table 1).
While $\sigma$ is the invariant expression of the chemical potential,
$\partial_{0}\psi$ clearly plays the role of the chemical potential
in the normal-fluid rest frame. The superfluid velocity measured in
the normal-fluid rest frame is given by $\vec{v}_{s}=-\vec{\nabla}\psi/\partial_{0}\psi$.
We could have introduced the normal-fluid rest frame already in the
discussion of the zero temperature results in section \ref{sub:Zero-temperature-hydrodynamics:-Landau}.
To avoid confusion, we have simply distinguished between results obtained
in the superfluid rest frame ($\vec{v}_{s}=\vec{0}$) and outside
the superfluid rest frame ($\vec{v}_{s}\neq\vec{0}$). From the finite
temperature point of view, measurements from outside the superfluid
rest frame correspond to measurements in the normal-fluid rest frame.
In case of $\vec{v}_{s}=\vec{0}$ , both fluids share a common rest
frame and we have $\mu_{s}=\sigma=\partial_{0}\psi=\mu_{n}$. We shall
from now on denote the chemical potential in the normal-fluid rest
frame by $\mu$ rather than $\mu_{n}$ (we have used the latter notation
in (\ref{eq:Framedef}) for the purpose of introducing both rest frames).

\noindent As a side remark, observe that we can also evaluate the
spatial components of $j^{\mu}$ in the normal-fluid rest frame, contract
the equation with $\vec{\nabla}\psi$ and solve for $n_{s}$. This
yields

\bigskip{}

\noindent 
\begin{equation}
n_{s}=-\sigma\frac{\vec{\nabla}\psi\cdot\vec{j}}{(\vec{\nabla}\psi)^{2}}\,.\label{eq:NSMicro}
\end{equation}
The right-hand side is now purely defined in terms of the field-theoretic
quantities (the spatial components of the Noether current and the
gradients of the phase $\psi$). We will however use a more systematic
way to obtain the fluid densities from ${\cal A}$, ${\cal B}$ and
${\cal C}$ in section \ref{sub:Entrainment-and-superfluid-from-field-theory}. 

~

\subsubsection{Generalized thermodynamics from field theory\label{sub:Generalized-thermodynamics-from-field-theory}}

~

\noindent Now that we have obtained chemical potential $\mu=\partial_{0}\psi$
and the velocity of the superflow $\vec{v}_{s}=-\vec{\nabla}\psi/\mu$
in the normal-fluid rest frame, we continue with the discussion of
the generalized pressure $\Psi$ and the generalized thermodynamic
relation which in addition to $\partial^{\mu}\psi$ involves the conjugate
momentum $\Theta^{\mu}$. With the stress-energy tensor (\ref{eq:RelGenT})
and using $u^{\mu}=s^{\mu}/s$ we can write

\begin{equation}
\Psi=\frac{1}{3}(g^{\mu\nu}-u^{\mu}u^{\nu})(j_{\mu}\partial_{\nu}\psi-T_{\mu\nu})\,.\label{eq:PsiProjectU}
\end{equation}
Let us now compute the right-hand side with the microscopic expression
for $T^{\mu\nu}$. With $T^{\mu\nu}$ and $j^{\mu}$ from equations
(\ref{eq:microT}), (\ref{eq:microCurrent}) we compute 

\begin{equation}
\frac{1}{3}(g^{\mu\nu}-u^{\mu}u^{\nu})(j_{\mu}\partial_{\nu}\psi-T_{\mu\nu})=-U-\frac{1}{3}(g^{\mu\nu}-u^{\mu}u^{\nu})\frac{T}{V}\sum_{k}\left(C_{k}k_{\mu}k_{\nu}+A_{k}k_{\mu}\partial_{\nu}\psi\right)\,.\label{eq:IdentifyPsi}
\end{equation}
Here we have used $u^{\mu}=(1,0,0,0)$, while equation (\ref{eq:PsiProjectU})
is a general relation for arbitrary four-velocities $u^{\mu}$. The
tensor $(g^{\mu\nu}-u^{\mu}u^{\nu})$ appearing in equation (\ref{eq:PsiProjectU})
can be interpreted as a projector onto the 4-dimensional hypersurface
orthogonal to $u^{\mu}$. It is important to keep in mind that we
cannot simply promote $u^{\mu}$ to an arbitrary four-velocity in
the microscopic calculation. It would occur additionally in different
places in the calculation, which we cannot identify in our present
treatment because the spatial components of $u^{\mu}$ have been projected
out in our finite temperature calculation. In a manner of speaking,
we are ``stuck'' in the normal-fluid rest frame. But, of course,
since $\Psi$ is a Lorentz scalar, the normal-fluid rest frame is
as good as any other frame to compute $\Psi$. By comparing with equation
(\ref{eq:EffActionk}) we see that the right-hand side of equation
(\ref{eq:IdentifyPsi}) is exactly ($T/V$ times) the effective action.
Therefore, we have obtained the important result,

\begin{equation}
\Psi=\frac{T}{V}\Gamma\,.\label{eq:PsiIsGamma}
\end{equation}
This relation is somewhat expected since we already know that, at
zero temperature (and at tree-level), $\Psi$ corresponds to the Lagrangian,
which in this case gives the microscopic pressure. At nonzero temperature
(without superflow), the effective action gives the (isotropic) pressure.
Therefore, the relation (\ref{eq:PsiIsGamma}) is a natural generalization
to the anisotropic case with a nonzero superflow. This motivates our
choice of notation in equation (\ref{eq:PsiK}) because now we have 

\begin{equation}
\Psi=-U+\frac{T}{V}\sum_{k}\Psi_{k}\,.
\end{equation}
 Next, let us discuss the generalized thermodynamic relation $\Lambda=-\Psi+j\cdot\partial\psi+s\cdot\Theta$.
In the normal-fluid rest frame, $s^{\mu}=(s^{0},0,0,0)$ and thus
$s\cdot\Theta=s^{0}\Theta^{0}$, which is the product of entropy and
temperature, measured in this particular frame. To confirm this microscopically,
we use the thermodynamical definition of the entropy density, 

\begin{equation}
s=\frac{\partial\Psi}{\partial T}=\frac{1}{V}\sum_{k}\left(\Psi_{k}+2+C_{k}k_{0}^{2}+A_{k}k_{0}\partial_{0}\psi\right)\,,\label{eq:Sk}
\end{equation}
 with $A_{k}$ and $C_{k}$ defined in equation (\ref{eq:AkBkCk}).
Note that the first two terms ($\Psi_{k}$ and 2) come from the explicit
$T$-dependence in the prefactor $T/V$ and in the $1/T^{2}$ within
the logarithm. On the other hand, we can compute $s\cdot\Theta$ via
the generalized thermodynamic relation. We find

\begin{eqnarray}
s\cdot\Theta & = & \Lambda+\Psi-j\cdot\partial\psi\nonumber \\
 & = & \frac{T}{V}\sum_{k}\left[\Psi_{k}+C_{k}k_{0}^{2}+A_{k}k_{0}\partial_{0}\psi-(C_{k}k^{2}+B_{k}\sigma^{2}+2A_{k}k\cdot\partial\psi)+\frac{2m^{2}k^{2}}{{\rm det}\, S_{0}^{-1}}\right]\,,
\end{eqnarray}
 where we have used $\Lambda=T_{\;\;\mu}^{\mu}+3\Psi$ , the trace
of the stress-energy tensor (\ref{eq:TraceT}), the effective action
(which is $\Psi$) (\ref{eq:EffActionk}), and the current (\ref{eq:JFk}).
With the help of the identity (\ref{eq:auxTrace}) we see that this
is indeed the same as $T$ times the entropy from equation (\ref{eq:Sk}).
In other words, the microscopic temperature $T$ can be identified
with generalized quantity $\Theta^{0}$ in the normal-fluid rest frame.
In this frame, we have now identified all terms in the generalized
thermodynamic relation. By now, we have obtained the means to correctly
interpret microscopic results in the frame of the two-fluid model.
We shall analyze the parameters of this model now in a low-temperature
approximation.

\subsection{Explicit results in the low-temperature approximation\label{sub:Explicit-results-in-lowT}}

~

\noindent We want to compute the effective action, the components
of the stress-energy tensor, and the current for small temperatures
explicitly. To this end, we have to approximate momentum sums of the
form

\begin{equation}
\frac{T}{V}\sum_{k}\frac{F(k)}{{\rm det}\, S_{0}^{-1}(k)}\,,
\end{equation}
where, for the case of the effective action, $F(k)$ is given by equation
(\ref{eq:Fk}), and for the stress-energy tensor and the current we
need to replace $F(k)$ by

\begin{eqnarray}
F^{\mu\nu}(k) & \equiv & 2\left[-(k^{2}-\sigma^{2}+m^{2})k^{\mu}k^{\nu}+k^{2}\partial^{\mu}\psi\partial^{\nu}\psi\right]\label{eq:FTmunu}\\
 & + & 2\left[2(k\cdot\partial\psi)(k^{\mu}\partial^{\nu}\psi+k^{\nu}\partial^{\mu}\psi)+u^{\mu}u^{\nu}{\rm det}\, S_{0}^{-1}(k)\right]\,,\nonumber \\
\nonumber \\
F^{\mu}(k) & \equiv & 2(k^{2}\partial^{\mu}\psi+2k\cdot\partial\psi\, k^{\mu})\,,\label{eq:FJmu}
\end{eqnarray}
which can be read off from equations (\ref{eq:TFk}), (\ref{eq:JFk}).
For all three cases, we can write the result of the Matsubara sum
in the form (\ref{eq:effectiveAction2}). Then, we write $\coth[\epsilon_{1,\vec{k}}/(2T)]=1+2f(\epsilon_{1,\vec{k}})$
with the Bose distribution function $f(x)=1/(e^{x/T}-1)$. The integral
over the first term is, in the present approximation, temperature
independent. It is divergent and has to be renormalized by subtracting
the vacuum contribution. After doing so, a finite term remains which
however is suppressed by one power of $\lambda$ compared to the zero-temperature
term $\propto\lambda^{-1}$ we have already computed, for instance
in equation (\ref{eq:effectiveAction2}). We shall thus neglect this
contribution and only keep the thermal contribution, i.e., the integral
over the second term that contains the Bose function. This contribution
is finite and unaffected by the renormalization.%
\footnote{The subleading zero-temperature contribution we are neglecting here
gives rise to the difference between the\textit{ superfluid density}
and the\textit{ condensate density}. At zero temperature, the charge
density is always identical to the superfluid density, $n(T=0)=n_{s}$,
while the condensate density, defined as $\mu\rho^{2}$ with the modulus
of the condensate $\rho$, might be smaller. In our leading-order
approximation, condensate and superfluid densities are identical.%
}We explain the small-temperature expansion in detail in appendix \ref{sec:Small-temperature-expansion}.
Before we come to the main results, let us discuss the simpler example
without superflow, $\vec{\nabla}\psi=\vec{0}$ and use $\partial_{0}\psi=\mu$.
In the absence of a superflow, the dispersions reduce to $\epsilon_{1,\vec{k}}=\epsilon_{\vec{k}}^{+}$,$\epsilon_{2,\vec{k}}=\epsilon_{\vec{k}}^{-}$
with 

\begin{equation}
\epsilon_{\vec{k}}^{\pm}=\sqrt{\vec{k}^{2}+3\mu^{2}-m^{2}\mp\sqrt{4\mu^{2}\vec{k}^{2}+(3\mu^{2}-m^{2})^{2}}}\,,\label{eq:DispNoSF}
\end{equation}
which is equivalent to (\ref{eq:DispPrev}) after the condensate from
equation (\ref{eq:CondPrev}) has been inserted. As usual, $\epsilon_{\vec{k}}^{+}$
is the Goldstone mode and $\epsilon_{\vec{k}}^{-}$ the massive mode.
This yields the small-temperature result for the pressure to order
$T^{6}$ (see also appendix \ref{sub:Sound-velocities-at-arbitraryM}
for the calculation with finite mass parameter $m$)

\begin{equation}
P=\frac{T}{V}\Gamma(|\vec{\nabla}\psi|=0)\simeq\frac{(\mu^{2}-m^{2})^{2}}{4\lambda}+\frac{(3\mu^{2}-m^{2})^{3/2}}{(\mu^{2}-m^{2})^{3/2}}\frac{\pi^{2}T^{4}}{90}-\frac{\mu^{6}(3\mu^{2}-m^{2})^{1/2}}{(\mu^{2}-m^{2})^{7/2}}\frac{4\pi^{4}T^{6}}{63\mu^{2}}\,.\label{eq:EffActionLowT}
\end{equation}
 \begin{table*}[t] \begin{tabular}{|c||c|c|c|} \hline \rule[-1.5ex]{0em}{6ex} & $\;\;\displaystyle{\frac{\mu^4}{4\lambda}}(1-{\bf v}_s^2)\;\;$ & $\;\;\displaystyle{\frac{\pi^2T^4}{10\sqrt{3}}\,\frac{1-{\bf v}_s^2}{(1-3{\bf v}_s^2)^3}}\;\;$ & $\;\;\displaystyle{\frac{4\pi^4T^6}{105\sqrt{3}\,\mu^2}\,\frac{1-{\bf v}_s^2}{(1-3{\bf v}_s^2)^6}}\;\;$ \\[2ex] \hline\hline \rule[-1.5ex]{0em}{6ex} 
$\;\;\displaystyle{\frac{T}{V}\Gamma}\;\;$ & $\;\;\displaystyle{1-{\bf v}_s^2}\;\;$ & $\displaystyle{\;\;(1-{\bf v}_s^2)(1-3{\bf v}_s^2)\;\;}$ & $\displaystyle{\;\;-(1-{\bf v}_s^2)(1-3{\bf v}_s^2)(5+30{\bf v}_s^2+9{\bf v}_s^4)\;\;}$ \\[2ex] \hline\hline \rule[-1.5ex]{0em}{6ex} 
$\;\;\displaystyle{T^{00}}\;\;$ & $\;\;\displaystyle{3+{\bf v}_s^2}\;\;$ & $\displaystyle{\;\;3-20{\bf v}_s^2+9{\bf v}_s^4\;\;}$ & $\displaystyle{\;\;- (15-160{\bf v}_s^2-774{\bf v}_s^4+432{\bf v}_s^6+135{\bf v}_s^8)\;\;}$ \\[2ex] \hline \rule[-1.5ex]{0em}{6ex} 
$\;\;\displaystyle{T^{0i}}\;\;$ & $\;\;\displaystyle{4{\bf v}_{si}}\;\;$ & $\displaystyle{\;\;-8{\bf v}_{si}\;\;}$ & $\displaystyle{\;\;2{\bf v}_{si} (95+243{\bf v}_s^2-135{\bf v}_s^4-27{\bf v}_s^6)\;\;}$ \\[2ex] \hline \rule[-1.5ex]{0em}{6ex} 
$\;\;\displaystyle{T_{\perp}}\;\;$ & $\;\;\displaystyle{1-{\bf v}_s^2}\;\;$ & $\displaystyle{\;\;(1-{\bf v}_s^2)(1-3{\bf v}_s^2)\;\;}$ & $\displaystyle{\;\;-(1-{\bf v}_s^2)(1-3{\bf v}_s^2)(5+30{\bf v}_s^2+9{\bf v}_s^4)\;\;}$ \\[2ex] \hline \rule[-1.5ex]{0em}{6ex} 
$\;\;\displaystyle{T_{||}}\;\;$ & $\;\;\displaystyle{1+3{\bf v}_s^2}\;\;$ & $\displaystyle{\;\;1-12{\bf v}_s^2+3{\bf v}_s^4\;\;}$ & $\displaystyle{\;\;- (5-180{\bf v}_s^2-582{\bf v}_s^4+324{\bf v}_s^6+81{\bf v}_s^8)\;\;}$ \\[2ex] \hline\hline \rule[-1.5ex]{0em}{6ex} 
$\;\;\displaystyle{\mu\, j^0}\;\;$ & $\;\;\displaystyle{4}\;\;$ & $\displaystyle{\;\;-8{\bf v}_s^2\;\;}$ & $\displaystyle{\;\;2 (5+105{\bf v}_s^2+147{\bf v}_s^4-81{\bf v}_s^6)\;\;}$ \\[2ex] \hline \rule[-1.5ex]{0em}{6ex} 
$\;\;\displaystyle{\mu\, {\bf j}}\;\;$ & $\;\;\displaystyle{4{\bf v}_s}\;\;$ & $\displaystyle{\;\;-8{\bf v}_s\;\;}$ & $\displaystyle{\;\;2 {\bf v}_s (95+243{\bf v}_s^2-135{\bf v}_s^4-27{\bf v}_s^6)\;\;}$ \\[2ex] \hline
\end{tabular} \caption{Microscopic results for the effective action $\Gamma$, the stress-energy tensor $T^{\mu\nu}$, and the conserved charge current $j^\mu$ up to order $T^6$ for $m=0$. Each row is the result for the quantity given in the left column; this quantity is a sum of the $\mu^4$, $T^4$, and $T^6/\mu^2$ terms given in the top row, each multiplied by the specific entry of the table. These results are obtained without making any assumptions about the magnitude of the superfluid velocity ${\vec{v}}_s$. We see that there is a divergence in all nonzero temperature results for ${\vec{v}}_s^2\to 1/3$, indicating an instability of the superflow, in accordance with Landau's critical velocity for superfluidity. } \end{table*}The $T^{4}$ term has two interesting properties related to the resulting
charge density, which is obtained by taking the derivative with respect
to $\mu$. Firstly, if we set $m=0$, the $\mu$-dependence drops
out, such that there is no $T^{4}/\mu$ contribution to the charge
density in this case. Secondly, in the presence of a finite $m$,
one finds that the $T^{4}/\mu$ contribution to the charge density
is negative, i.e., for small temperatures the density\textit{ decreases}
with temperature. (The second derivative $\partial^{2}P/\partial\mu^{2}$
is positive, i.e., the system is thermodynamically stable.) The $T^{6}$
term has neither of these properties, it contributes to the charge
density even for $m=0$ and gives rise to a positive $T^{6}/\mu^{3}$
term in the density. 

\noindent The case with superflow is of course more complicated. In
particular, the momentum integration now involves a nontrivial angular
integral over the angle between the momentum and $\vec{\nabla}\psi$.
Nevertheless, it turns out that this angular integral can be performed
analytically for all cases we consider. For brevity, we set $m=0$
in the following. Then, we obtain for the effective action density
to order $T^{6}$

\begin{equation}
\frac{T}{V}\Gamma\simeq\frac{\mu^{4}}{4\lambda}(1-\vec{v}_{s}^{2})^{2}+\frac{\pi^{2}T^{4}}{10\sqrt{3}}\,\frac{(1-\vec{v}_{s}^{2})^{2}}{(1-3\vec{v}_{s}^{2})^{2}}-\frac{4\pi^{4}T^{6}}{105\sqrt{3}\,\mu^{2}}\,\frac{(1-\vec{v}_{s}^{2})^{2}}{(1-3\vec{v}_{s}^{2})^{5}}(5+30\vec{v}_{s}^{2}+9\vec{v}_{s}^{4})\,.\label{eq:PressureLowT}
\end{equation}

\noindent Obviously, for $\vec{v}_{s}=0$ we recover the $m=0$ limit
of equation (\ref{eq:EffActionLowT}). Note that the result is valid
to all orders in the superfluid velocity. We have not applied any
expansion in $|\vec{v}_{s}|$. Analogously, we compute the components
of the stress-energy tensor and the current. For the spatial components
of the stress-energy tensor we use

\begin{eqnarray}
T_{\perp} & \equiv & \frac{1}{2}\left[\delta_{ij}-\frac{\partial_{i}\psi\partial_{j}\psi}{(\nabla\psi)^{2}}\right]T_{ij}\,,\\
\nonumber \\
T_{||} & \equiv & \frac{\partial_{i}\psi\partial_{j}\psi}{(\nabla\psi)^{2}}T_{ij}\,.
\end{eqnarray}
The results are collected in table 2. All expressions in this table
are written in terms of quantities measured in the normal-fluid rest
frame. At zero temperature it is very natural to write them in terms
of the Lorentz scalar $\sigma=\mu\sqrt{1-\vec{v}_{s}^{2}}$, for instance
$T_{\perp}=\sigma^{4}/(4\lambda)$. As expected, such a formulation
is less obvious for the nonzero temperature terms. Since the microscopic
calculation has a preferred rest frame where the thermodynamic variables
$\mu$, $T$, and $\vec{v}_{s}$ are measured, the velocity-dependence
shown here is a mixture of trivial Lorentz boosts and complicated
effects of the superflow on the collective excitations (which in turn
can be interpreted as boosts of the \textit{microscopic }vector $k^{\mu}$,
see footnote below equation (\ref{eq:Propagator}) ).

\noindent We may compare the results presented in table 2 with the
frame dependent quantities listed in table 1. One can for instance
check from the explicit results that the difference between longitudinal
and transverse pressure is indeed $-\vec{j}\cdot\nabla\psi=\mu\,\vec{j}\cdot\vec{v}_{s}$,
and that the momentum density $T^{i0}$ is indeed $\partial^{0}\psi=\mu$
times the current $j^{i}$. More importantly, we see that $T_{\perp}=\frac{T}{V}\Gamma$,
which, since $T_{\perp}=\Psi$, confirms that the generalized pressure
can be identified with the effective action $\Psi=\frac{T}{V}\Gamma$. 

\noindent With the results of \noun{$T^{\mu\nu}$} in hand, we may
also ask for the superfluid density $\rho_{s}$ . This \textit{energy}
density is defined in the low-velocity limit, in generalization of
the mass densities in the non-relativistic framework (and in contrast
to $n_{s}$, $n_{n}$ which we calculate in the next section and which
are \textit{number} densities). With $T^{0i}=j^{i}\partial^{0}\psi$,
$\vec{j}=-n_{s}\vec{\nabla}\psi/\sigma$, and $\vec{v}_{s}=-\vec{\nabla}\psi/\partial^{0}\psi$
, we obtain

\[
T^{0i}=(\partial^{0}\psi)^{2}\frac{n_{s}}{\sigma}v_{si}=\rho_{s}v_{si}+{\cal O}(|\vec{v}_{s}|^{3})\,,
\]
(Alternatively, we can write $\rho_{s}=\sigma^{2}/{\cal B}$.) This
expression for the superfluid density is obviously in agreement with
the zero-temperature result (\ref{eq:microRho}). The superfluid density
appears also in the energy density as part of the kinetic energy,
$T^{00}=\Lambda+\rho_{s}\vec{v}_{s}^{2}+{\cal O}(|\vec{v}_{s}|^{4})$.

~

\subsubsection{Entrainment and superfluid density from field theory\label{sub:Entrainment-and-superfluid-from-field-theory}}

~

\noindent The independent degrees of freedom of our microscopic calculation
are the chemical potential $\mu=\partial^{0}\psi$, the temperature
$T=\Theta^{0}$, and the superfluid three-velocity $\vec{v}_{s}=-\vec{\nabla}\psi/\mu$,
all measured in the normal-fluid rest frame, where the entropy current
vanishes by definition, $s^{i}=0$. We have thus given 8 independent
components out of the 16 components of the 4 four-vectors $j^{\mu}$,
$s^{\mu}$, $\Theta^{\mu}$, $\partial^{\mu}\psi$ of the two-fluid
formalism. The other 8 components are $j^{\mu}$, $s^{0}$, and $\Theta^{i}$.
For the Noether current $j^{\mu}$ and the entropy $s^{0}$ we have
field-theoretic and thermodynamic definitions respectively. It remains
to compute the spatial components of the thermal four-vector $\Theta^{\mu}$.
Additionally, we have to compute the coefficients $\overline{{\cal A}}$,
$\overline{{\cal B}}$, $\overline{{\cal C}}$. They are defined as
the derivatives of $\Psi$ with respect to the Lorentz scalars $\sigma^{2}$,$\Theta^{2}$,
and $\Theta\cdot\partial\psi$. However, this is not the form in which
our $\Psi$ is given. Therefore, we need to find a different way to
compute these coefficients. This can be done with the help of equations
(\ref{eq:Entrainment3}) and (\ref{eq:Entrainment4}). First, we solve
the spatial part of equation (\ref{eq:Entrainment3}) for $\Theta^{i}$
and insert the result into the spatial part of equation (\ref{eq:Entrainment4}).
Together with the temporal components, this yields three equations
for the three variables $\overline{{\cal A}}$, $\overline{{\cal B}}$,
$\overline{{\cal C}}$, whose solutions are listed in table 3, where,
for completeness, we also give the coefficients ${\cal A}$, ${\cal B}$,
${\cal C}$, which are obtained from the inverse of equation (\ref{eq:AAbar}).
With these results we immediately find

\noindent \begin{flushleft}
\begin{table*}[t] \begin{tabular}{|c|c||c|c|}  \hline \rule[-1.5ex]{0em}{6ex}  $\;\;\overline{\cal A}\;\;$ & $\;\; \displaystyle{\frac{s^0}{\partial^0\psi}\frac{\eta}{\eta+{\bf v}_s^2s^0\Theta^0}}\;\;$ & $\;\;{\cal A}\;\;$ &  $\displaystyle{\frac{\partial^0\psi}{s^0{\bf j}\cdot\nabla\psi}\eta}$ \\[2ex] \hline \rule[-1.5ex]{0em}{6ex}  $\overline{\cal B}$ & $\;\; \displaystyle{\frac{1}{(\partial^0\psi)^2}\frac{j^0\partial^0\psi\eta -{\bf j}\cdot\nabla\psi s^0\Theta^0}{\eta+{\bf v}_s^2s^0\Theta^0}}\;\;$& ${\cal B}$ &  $\displaystyle{-\frac{(\nabla\psi)^2}{{\bf j}\cdot\nabla\psi}}$ \\[2ex] \hline \rule[-1.5ex]{0em}{6ex}  $\overline{\cal C}$ & $\;\; \displaystyle{\frac{{\bf v}_s^2(s^0)^2}{\eta +{\bf v}_s^2s^0\Theta^0}}\;\;$& ${\cal C}$ &  $\;\;\displaystyle{-\frac{j^0\partial^0\psi\eta-{\bf j}\cdot\nabla\psi s^0\Theta^0}{(s^0)^2\,{\bf j}\cdot\nabla\psi}}\;\;$ \\[2ex] \hline \end{tabular} \caption{Coefficients that relate the currents $j^\mu$, $s^\mu$ with the conjugate momenta $\partial^\mu\psi$, $\Theta^\mu$, given  in terms of ``microscopic'' quantities: the Noether current $j^\mu$, the space-time derivative of the phase of the condensate  $\partial^\mu\psi$, the temperature $\Theta^0$, and the entropy density $s^0$, all measured in the normal-fluid rest frame. For simplicity, we have abbreviated $\eta=\vec{v}_{s}^{2}j^{0}\partial^{0}\psi+\vec{j}\cdot\vec{\nabla}\psi$. The low-temperature approximations for $\overline{\cal A}$,  $\overline{\cal B}$, $\overline{\cal C}$ are given in equation (10.36)-(10.38). } \label{table2} \end{table*}\bigskip{}

\par\end{flushleft}

\noindent 
\begin{equation}
\vec{\Theta}=-\frac{{\cal A}}{{\cal B}}\vec{\nabla}\psi=\frac{\vec{\nabla}\psi}{s^{0}}\left[j^{0}+\partial^{0}\psi\frac{\vec{j}\cdot\vec{\nabla}\psi}{(\vec{\nabla}\psi)^{2}}\right]\,.
\end{equation}
 Since all results are expressed in terms of quantities accessible
from our microscopic calculation, we can for instance compute (for
$m=0$ and in the low-temperature limit) \bigskip{}

\noindent 
\begin{eqnarray}
\overline{{\cal A}} & \simeq & \frac{4\pi^{2}T^{3}}{15\sqrt{3}\,\mu}\,\frac{1-\vec{v}_{s}^{2}}{(1-3\vec{v}_{s}^{2})^{2}}-\frac{16\pi^{4}T^{5}}{315\sqrt{3}\,\mu^{3}}\,\frac{1-\vec{v}_{s}^{2}}{(1-3\vec{v}_{s}^{2})^{5}}(25+78\vec{v}_{s}^{2}-27\vec{v}_{s}^{4})\,,\label{eq:AbarlowT}\\
\nonumber \\
\overline{{\cal B}} & \simeq & \frac{\mu^{2}}{\lambda}(1-\vec{v}_{s}^{2})-\frac{4\pi^{2}T^{4}}{15\sqrt{3}\,\mu^{2}}\,\frac{1-\vec{v}_{s}^{2}}{(1-3\vec{v}_{s}^{2})^{3}}+\frac{8\pi^{4}T^{6}}{315\sqrt{3}\,\mu^{4}}\frac{65+256\vec{v}_{s}^{2}-402\vec{v}_{s}^{4}+81\vec{v}_{s}^{8}}{(1-3\vec{v}_{s}^{2})^{6}}\,,\label{eq:BbarlowT}\\
\nonumber \\
\overline{{\cal C}} & \simeq & \frac{2\pi^{2}T^{2}}{15\sqrt{3}}\frac{1-\vec{v}_{s}^{2}}{1-3\vec{v}_{s}^{2}}+\frac{8\pi^{4}T^{4}}{315\sqrt{3}\,\mu^{2}}\frac{5-59\vec{v}_{s}^{2}+27\vec{v}_{s}^{4}+27\vec{v}_{s}^{6}}{(1-3\vec{v}_{s}^{2})^{4}}\,.\label{eq:CbarlowT}
\end{eqnarray}

~

\noindent We see that the coefficients $\overline{{\cal A}}$ and
$\overline{{\cal C}}$ vanish at $T=0$. This is in accordance to
our zero-temperature discussion, where only $\overline{{\cal B}}$
was nonzero. The connection between the coefficients ${\cal A}$,
${\cal B}$ and the number densities $n_{s}$, $n_{n}$ is given in
equation (\ref{eq:LandauABC}). We can insert ${\cal A}$ and ${\cal B}$
as functions of $n_{s}$, $n_{n}$ into the temporal component of
equation (\ref{eq:Entrainment2}) to get also ${\cal C}$ as a function
of $n_{s}$, $n_{n}$. The result is the useful translation

\noindent 
\begin{equation}
{\cal A}=-\frac{\sigma n_{n}}{sn_{s}}\,,\qquad{\cal B}=\frac{\sigma}{n_{s}}\,,\qquad{\cal C}=\frac{\sigma n_{n}^{2}}{s^{2}n_{s}}+\frac{\mu n_{n}+sT}{s^{2}}\,.
\end{equation}
(To avoid confusion: $s$, $T$, $\mu$, $n_{n}$ are quantities measured
in the normal frame, while $n_{s}$ is the superfluid density measured
in the superfluid frame. Remember that the superfluid density measured
in the normal frame is $n_{s}\mu/\sigma$.) As a check, we see that
${\cal B}$ given in table 3 is indeed the same as ${\cal B}=\sigma/n_{s}$
with $n_{s}$ calculated from equation (\ref{eq:NSMicro}). It is
now of course straightforward to also express $\overline{{\cal A}}$,
$\overline{{\cal B}}$, $\overline{{\cal C}}$ in terms of $n_{s}$
and $n_{n}$. In the small-temperature approximation, the superfluid
and normal number densities, measured in the normal-fluid rest frame,
become\bigskip{}

\noindent 
\begin{eqnarray}
n_{s}\frac{\mu}{\sigma} & \simeq & \frac{\mu^{3}}{\lambda}(1-\vec{v}_{s}^{2})-\frac{4\pi^{2}T^{4}}{5\sqrt{3}\,\mu}\,\frac{1-\vec{v}_{s}^{2}}{(1-3\vec{v}_{s}^{2})^{3}}+\frac{8\pi^{4}T^{6}}{105\sqrt{3}\,\mu^{3}}\,\frac{1-\vec{v}_{s}^{2}}{(1-3\vec{v}_{s}^{2})^{6}}(95+243\vec{v}_{s}^{2}-135\vec{v}_{s}^{4}-27\vec{v}_{s}^{6})\,,\nonumber \\
\label{eq:nsLowT}\\
n_{n} & \simeq & \frac{4\pi^{2}T^{4}}{5\sqrt{3}\,\mu}\,\frac{(1-\vec{v}_{s}^{2})^{2}}{(1-3\vec{v}_{s}^{2})^{3}}-\frac{16\pi^{4}T^{6}}{35\sqrt{3}\,\mu^{3}}\,\frac{(1-\vec{v}_{s}^{2})^{2}}{(1-3\vec{v}_{s}^{2})^{6}}(15+38\vec{v}_{s}^{2}-9\vec{v}_{s}^{4})\,.\label{eq:nnlowT}
\end{eqnarray}

~

\noindent One can check that the sum of both densities gives the total
charge density $j^{0}$ from table 2. As expected, the normal density
vanishes for $T=0$ and begins to increase with increasing temperature,
while the superfluid density decreases. In the more complete treatment
of section \ref{sec:The-two-fluid-model-at-arbitrary-T}, we expect
the superfluid density to vanish at the critical temperature because
the condensate melts. Remember that this melting is, in our one-loop
effective action, a higher-order effect in the coupling constant,
which we have neglected. The decrease of $n_{s}$ is therefore only
due to the interaction between the two fluids. As an aside, note that
for $\vec{v}_{s}=\vec{0}$ the $T^{4}$ contributions in superfluid
and normal densities cancel each other exactly. We have made this
observation already below equation (\ref{eq:EffActionLowT}) where
we have seen that in the $m=0$ limit there is no $T^{4}$ contribution
to the density. 

\noindent Finally, we may express the generalized pressure $\Psi$
in terms of Lorentz scalars. This reformulation is instructive because
it gives $\Psi$ in the form that is usually assumed in the two-fluid
formalism.

\noindent \newpage{}Our quantities in the normal-fluid rest frame
$T$, $\mu$, $\vec{v}_{s}$ are translated into the relevant Lorentz
scalars $\sigma^{2}$, $\Theta^{2}$, $\partial\psi\cdot\Theta$ via 

\bigskip{}

\noindent 
\begin{eqnarray}
\sigma^{2} & = & \,\,\mu^{2}-(\vec{\nabla}\psi)^{2}\,\,\,\,\,\,=\mu^{2}(1-\vec{v}_{s}^{2})\,,\label{eq:InvSig}\\
\nonumber \\
\Theta^{2} & = & T^{2}-\frac{{\cal A}^{2}}{{\cal B}^{2}}(\vec{\nabla}\psi)^{2}=\frac{(1-\vec{v}_{s}^{2})(1-9\vec{v}_{s}^{2})}{(1-3\vec{v}_{s}^{2})^{2}}\, T^{2}+{\cal O}(T^{4})\,,\label{eq:InvT}\\
\nonumber \\
\partial\psi\cdot\Theta & = & \,\,\mu T-\frac{{\cal A}}{{\cal B}}(\vec{\nabla}\psi)^{2}\,=\frac{1-\vec{v}_{s}^{2}}{1-3\vec{v}_{s}^{2}}\,\mu T+{\cal O}(T^{3})\,.\label{eq:InvMix}
\end{eqnarray}

~

\noindent We solve these equations for $T$, $\mu$, and $\vec{v}_{s}$
and insert the result into the effective action. Then, up to fourth
order in the temperature we can write the generalized pressure as
\bigskip{}

\noindent 
\begin{equation}
\Psi(\sigma^{2},\Theta^{2},\Theta\cdot\partial\psi)\simeq\frac{\sigma^{4}}{4\lambda}+\frac{\pi^{2}}{90\sqrt{3}}\left[\Theta^{2}+2\frac{(\partial\psi\cdot\Theta)^{2}}{\sigma^{2}}\right]^{2}\,.\label{eq:InvarPressure}
\end{equation}

~

\noindent The term in the square brackets can be written as ${\cal G}^{\mu\nu}\Theta_{\mu}\Theta_{\nu}$
with the ``sonic metric'' ${\cal G}^{\mu\nu}\equiv g^{\mu\nu}+2v^{\mu}v^{\nu}$,
see equation (8.9) of reference \cite{CarterLanglois}. In other words,
the Lorentz invariant $T^{4}$ term of the pressure in the presence
of a superflow is obtained by replacing $T^{2}\to{\cal G}^{\mu\nu}\Theta_{\mu}\Theta_{\nu}$
in the $T^{4}$ term of the pressure in the absence of a superflow.
In principle, we can use the higher order terms in equation (\ref{eq:InvSig}-\ref{eq:InvMix})
to write the $T^{6}$ contribution in terms of Lorentz scalars. However,
we have not found a compact way of writing this contribution. But,
we have checked that it is not simply given by the same replacement
as for the $T^{4}$ term. This is no surprise since the sonic metric
is constructed solely for systems for a Goldstone mode with linear
dispersion. The $T^{6}$ term however, knows about the cubic term
in the dispersion. 

~

\subsection{Conclusion \label{sub:Conclusion1}}

~

\noindent We have discussed the dissipationless hydrodynamics of a
relativistic superfluid, starting from a complex scalar field. Our
main goal has been to relate the field theory with the covariant two-fluid
framework of superfluidity. To achieve these goals analytically, we
have so far restricted our calculations to the limit of small temperatures.
The results from these studies and the approximations necessary to
obtain them shall now be summarized. 
\begin{enumerate}
\item \textit{Microscopic calculation.} We have started from Bose-Einstein
condensation in a $\varphi^{4}$ theory. The condensate has been assumed
to be static and homogeneous (which corresponds to the simple hydrodynamic
scenario of a static, homogeneous superflow). A crucial role is played
by the phase of the condensate. While small oscillations of the phase
correspond to the excitations of the Goldstone mode, rotations of
the phase around the full $U(1)$ circle give rise to a chemical potential
(speed of the rotation) and a superfluid three-velocity (number of
rotations per unit length). In the presence of a non-zero superflow,
the excitations of the Goldstone mode become anisotropic, and we have
computed the resulting components of the conserved current and stress-energy
tensor for nonzero temperatures. We have restricted ourselves to a
weak-coupling, low-temperature approximation, including terms up to
sixth order in the temperature. This has allowed us to present the
microscopic results in an analytical form. 
\item \textit{The two-fluid formalism.} In the relativistic two-fluid formalism,
the basic variables are the charge current and the entropy current.
In the dissipationless case, both are conserved. For both currents,
we can define conjugate momenta. Now, if the two fluids are interacting
with each other, neither of the currents is (four-)parallel to its
own momentum, but also receives a contribution from the other momentum.
This contribution is called entrainment, and it must be computed from
an underlying microscopic theory. This formalism was explained in
detail in sections \ref{sec:Relativistic-thermodynamics-and-hydro}
and \ref{sub:A-relativistic-version-of-Landau} where different formulations
based on ``mixed'' and ``pure'' variable sets where described
and related to each other. For instance, the entrainment coefficient
(and related coefficients) can be expressed in terms of the superfluid
and normal-fluid charge densities $n_{n}$ and $n_{s}$ (whose corresponding
superfluid and normal-fluid four-currents are, even in the dissipationless
case, not separately conserved). 
\item \textit{Relationship between them}. There are several concepts and
quantities in the two-fluid formalism that are usually not used in
field theory, such as the generalized pressure that depends on Lorentz
scalars. Therefore, one important aspect of this work has been a translation
of these quantities into field-theoretic language. For instance, we
have proven that, once we assume that the microscopic calculation
is performed in the rest frame of the normal fluid, it follows that
the generalized pressure is given by the effective action. We have
also expressed the coefficients that relate the currents with the
conjugate momenta in terms of quantities that are well defined in
field theory, namely the space-time derivative of the phase of the
condensate, the components of the Noether current, and the entropy.
As a result, we have been able to compute these coefficients explicitly
as a function of temperature, chemical potential, and superfluid velocity.
Certain combinations of these coefficients yield $n_{n}$ and $n_{s}$.
Our calculation shows for instance that $n_{s}$ is not simply given
by the condensate density: even though we have neglected the temperature
dependence of the condensate, $n_{s}$ depends on temperature. It
also confirms that $n_{n}$ is not identical to the phonon number,
as one might naively expect; while the phonon number goes like $T^{3}$
for small temperatures (see for instance \cite{MannarelliManuel}),
$n_{n}\propto T^{4}/\mu$. 
\end{enumerate}
\newpage{}

\section{The two fluid model at arbitrary temperatures\label{sec:The-two-fluid-model-at-arbitrary-T}}

~

\noindent We will now go beyond the low-temperature approximation
(i.e. extend calculations up to the critical temperature), and go
beyond the weak-coupling limit by resumming certain contributions
to all orders in the coupling constant. We still neglect dissipation
and keep the superflow uniform in time and space. In the extension
to high temperatures we have to make use of the more elaborate 2-particle
irreducible (2PI) formalism \cite{BaymCJT,Luttinger,CJT} (also called
Cornwall-Jackiw-Tomboulis (CJT) formalism or $\Phi$-derivable approximation
scheme) which we have introduced in section \ref{sub:Two-particle-irreducible-formalism}.
We will use this formalism in the Hartree approximation at two-loop
level. We have seen that this formalism is particularly well suited
to systems with spontaneously broken symmetries. It has been used
previously, among many other applications, to describe meson condensation
in the CFL phase \cite{Andersen2007,CritTKaon,AndersenLeganger},
but without including the effects of a superflow. This is the first
time that a superflow has been implemented in this formalism. The
extension to all temperatures below the critical temperature for superfluidity
is relevant in the context of compact stars because temperatures in
the star may well be of the order of the critical temperature or higher,
in particular in its early evolutionary stages. 

\noindent ~

\subsection{The effective action in CJT\label{sub:The-effective-action-in-CJT}}

~

\noindent Based on the results of section \ref{sub:Generalized-thermodynamics-from-field-theory},
we relate the 2PI effective action $\Gamma[\rho,S]$ which is a functional
of the (modulus of the) condensate and the full propagator $S$, to
the generalized pressure $\Psi$. Remember that the index ``c''
for classical fields and propagators will be suppressed. In the two
loop approximation, the effective action density ($T/V$ times the
effective action) is then given by (see equation (\ref{eq:2PIFunc}))

\noindent \medskip{}

\noindent 
\begin{equation}
\Psi[\rho,S]=-U(\rho)-\frac{1}{2}\frac{T}{V}\sum_{k}\textrm{Tr}\ln\frac{S^{-1}}{T^{2}}-\frac{1}{2}\frac{T}{V}\sum_{k}\textrm{Tr}[S_{0}^{-1}(\rho)S-1]-V_{2}[\rho,S]\,,\label{eq:Action2PI}
\end{equation}
 where the trace is taken over the internal $2\times2$ space and
$V$ is the three-volume. The the potential $V_{2}[\rho,S]$ includes
all two-loop, two-particle-irreducible, diagrams. These diagrams are
shown in figure \ref{fig:Two-particle-irreducible}. Here, we have
used the definitions of the tree level potential 

\noindent \medskip{}

\noindent 
\begin{equation}
U(\rho)=-\frac{\rho^{2}}{2}(\sigma^{2}-m^{2})+\frac{\lambda}{4}\rho^{4}\,,
\end{equation}

\noindent and the \textit{free propagator} in the presence of a condensate

\noindent \medskip{}
\begin{equation}
S_{0}^{-1}(k)=\left(\begin{array}{cc}
-k^{2}-\sigma^{2}+m^{2}+3\lambda\rho^{2} & 2ik_{\mu}\partial^{\mu}\psi\\[2ex]
-2ik_{\mu}\partial^{\mu}\psi & -k^{2}-\sigma^{2}+m^{2}+\lambda\rho^{2}
\end{array}\right)\,\,.
\end{equation}

\noindent For convenience, we first discuss the unrenormalized effective
action and include a counterterm $\delta\Psi$ later, see section
\ref{sub:Renormalized-stationarity-equations-and-pressure} and appendix
\ref{sec:Renormalization-in-2PI}. The Hartree approximation implies
that we only consider self interaction terms coming from the ``double
bubble diagram'' (left diagram in figure \ref{fig:Two-particle-irreducible}).
The exchange contributions to $V_{2}$ coming from the ``sunset diagram''
(right diagram in figure \ref{fig:Two-particle-irreducible}) are
neglected. We are thus left with a single two-loop diagram which is
generated by the quartic interactions and whose algebraic expression
is 

\noindent \medskip{}

\noindent 
\begin{equation}
V_{2}[S]\simeq\frac{\lambda}{4}\left(\frac{T}{V}\right)^{2}\sum_{K,Q}\Big\{\textrm{Tr}[S(K)]\,\textrm{Tr}[S(Q)]+\textrm{Tr}[S(K)S(Q)]+\textrm{Tr}[S(K)S(Q)^{T}]\Big\}\,.\label{eq:twoPiPotential}
\end{equation}

\noindent Had we included the ``sunset diagram'' originating from
the cubic interactions, $V_{2}$ would also depend explicitly on $\rho$.
Moreover, the self-energy would depend on momentum. Therefore, neglecting
the contribution from the cubic interaction is a tremendous simplification,
even though only an explicit calculation can show whether its contribution
is indeed small. Naively, the additional factor of the condensate
at the cubic vertex suggests that for chemical potentials only slightly
above the mass $m$ our simplification is a good approximation. However,
it was that shown the contribution we neglect is important to obtain
a second order phase transition, i.e., the Hartree approximation shows,
unphysically, a first order phase transition \cite{Baacke,Marko1,Marko2}.
We shall come back to this issue when we present our results in section
\ref{sub:Critical-temperature,-condensate-critV-forallT}.

\noindent In our approximation the self-energy does not depend on
momentum and is given by

\noindent \medskip{}

\noindent 
\begin{equation}
\Sigma\equiv2\frac{\delta V_{2}}{\delta S}\simeq\lambda\frac{T}{V}\sum_{k}\textrm{Tr}[S(k)]+\lambda\frac{T}{V}\sum_{k}[S(k)+S(k)^{T}]\,,
\end{equation}
where the first term is proportional to the unit matrix. One can now
easily confirm the useful relation

\noindent \medskip{}

\noindent 
\begin{equation}
V_{2}[S]=\frac{1}{4}\frac{T}{V}\sum_{k}\textrm{Tr}[\Sigma\, S(k)]\,.\label{eq:2PiPot}
\end{equation}
 To determine the ground state of the system, we need to find the
stationary points of the effective action. To this end, we take the
(functional) derivatives of the effective action with respect to $\rho$
and 

\noindent \begin{flushleft}
\newpage{}$S$ and set these to zero,
\par\end{flushleft}

\noindent 
\begin{eqnarray}
0 & = & \frac{\partial U}{\partial\rho}+\frac{1}{2}\frac{T}{V}\sum_{k}\textrm{Tr}\left[\frac{\partial S_{0}^{-1}}{\partial\rho}S\right]\,,\label{eq:StatRho}\\
S^{-1} & = & S_{0}^{-1}+\Sigma\,.\label{eq:DSE}
\end{eqnarray}
 With an ansatz for the full propagator we can bring these equations
into a more explicit form. Within the present approximation, the most
general form of the propagator is \cite{CritTKaon} 

\noindent \medskip{}

\noindent 
\begin{equation}
S^{-1}(k)=\left(\begin{array}{cc}
-k^{2}-\sigma^{2}+M^{2}+\delta M^{2} & 2ik_{\mu}\partial^{\mu}\psi\\[2ex]
-2ik_{\mu}\partial^{\mu}\psi & -k^{2}-\sigma^{2}+M^{2}-\delta M^{2}
\end{array}\right)\,,\label{eq:FullProp}
\end{equation}

~

\noindent with two mass parameters $M$, $\delta M$, that have to
be determined self-consistently. With this ansatz, the off-diagonal
components of the Dyson-Schwinger equation (\ref{eq:DSE}) are automatically
fulfilled. We are left with the scalar equation (\ref{eq:StatRho})
and the two diagonal components of equation (\ref{eq:DSE}). Inserting
the first of the diagonal components into equation (\ref{eq:StatRho}),
and adding and subtracting the two diagonal components to/from each
other, yields the following (yet unrenormalized) three equations for
the three variables $\rho$, $M$, and $\delta M$,

\noindent 
\begin{eqnarray}
M^{2}+\delta M^{2}-\sigma^{2} & = & 2\lambda\rho^{2}\,,\label{eq:stat1}\\
M^{2} & = & m^{2}+2\lambda\rho^{2}+2\lambda\frac{T}{V}\sum_{k}[S_{11}(k)+S_{22}(k)]\,,\label{eq:stat2}\\
\delta M^{2} & = & \lambda\rho^{2}+\lambda\frac{T}{V}\sum_{k}[S_{11}(k)-S_{22}(k)]\,,\label{eq:stat3}
\end{eqnarray}
where $S_{11}(k)$ and $S_{22}(k)$ are the diagonal elements of the
full propagator $S$, and where we have already assumed that the condensate
$\rho$ is nonzero (there is also the trivial solution $\rho=0$ which
we briefly discuss in the context of renormalization, see appendix
\ref{sec:Renormalization-in-2PI}). With the help of equation (\ref{eq:2PiPot})
and (\ref{eq:DSE}), the pressure at the stationary point can be written
as

\noindent \medskip{}

\noindent 
\begin{equation}
\Psi_{{\rm stat}}=-U-\frac{1}{2}\frac{T}{V}\sum_{k}\textrm{Tr}\ln\frac{S^{-1}}{T^{2}}-\frac{1}{4}\frac{T}{V}\sum_{k}\textrm{Tr}[S_{0}^{-1}S-1]\,.
\end{equation}

\newpage{}

\subsection{Renormalized stationarity equations and pressure\label{sub:Renormalized-stationarity-equations-and-pressure}}

~

\noindent Renormalization in the 2PI formalism has been discussed
in numerous works in the literature, for instance in references \cite{Andersen2007,AndersenLeganger,Marko2,AmelinoReno,Berges2Reno,Blaizot1Reno,Blaizot2Reno,FejoesReno,GrinsteinReno,HeesReno,IvanovReno,JackiwReno,PilaftsisReno,RischkeReno,SeelReno}.
For our purposes, the methods developed and used in references \cite{Andersen2007,AndersenLeganger,FejoesReno}
are most useful. While reference \cite{FejoesReno} introduces counterterms
``directly'' in the effective action, \cite{Andersen2007,AndersenLeganger}
use an iterative method, based on \cite{Blaizot1Reno,Blaizot2Reno},
where the counterterms are introduced order by order in the coupling.
Both methods are equivalent. We shall follow the ``direct'' approach
of \cite{FejoesReno}. All details of the renormalization are discussed
in appendix \ref{sec:Renormalization-in-2PI}. Here we simply summarize
the main steps and give the results. 

\noindent The renormalization requires to add appropriate counterterms
to the effective action (\ref{eq:Action2PI}), proportional to the
(infinite) parameters $\delta m^{2}$, $\delta\lambda_{1}$, $\delta\lambda_{2}$.
In the condensed phase, two different parameters $\delta\lambda_{1}$,
$\delta\lambda_{2}$ for the renormalization of the coupling are necessary.
Then one can show, after regularizing the ultraviolet divergent integrals
in the action and the stationarity equations that the parameters $\delta m^{2}$,
$\delta\lambda_{1}$, $\delta\lambda_{2}$ can be expressed in terms
of the (finite) renormalized parameters $m^{2}$, $\lambda$, an ultraviolet
cutoff $\Lambda$, and a renormalization scale $\ell$. And, importantly,
these parameters do not depend on the medium, i.e., on $\mu$, $T$,
and $\vec{\nabla}\psi$. The relation between the cutoff dependent
quantities and the renormalized ones becomes a bit more compact if
we introduce (infinite) bare parameters via $m_{{\rm bare}}^{2}=m^{2}+\delta m^{2}$,
$\lambda_{1/2,{\rm bare}}=\lambda+\delta\lambda_{1/2}$. Then, we
can write the renormalized parameters as

\noindent \bigskip{}

\noindent 
\begin{equation}
\frac{1}{\lambda}=\frac{1}{\lambda_{1,{\rm bare}}}+\frac{1}{4\pi^{2}}\ln\frac{\Lambda^{2}}{\ell^{2}}=\frac{1}{\lambda_{2,{\rm bare}}}+\frac{1}{8\pi^{2}}\ln\frac{\Lambda^{2}}{\ell^{2}}\,,\qquad\frac{m^{2}}{\lambda}=\frac{m_{{\rm bare}}^{2}}{\lambda_{1,{\rm bare}}}+\frac{\Lambda^{2}}{4\pi^{2}}\,.
\end{equation}
For the regularization of the divergent momentum integrals we use
Schwinger's proper time regularization \cite{Schwinger1951}, where
the cutoff $\Lambda$ is introduced by setting the lower boundary
of the proper time integral to $1/\Lambda^{2}$. More precisely, we
separate a ``vacuum'' contribution from each of the divergent integrals
such that a finite integral remains and the ``vacuum'' term can be
regularized. This term is not exactly a vacuum term because the ultraviolet
divergences depend on the self-consistent mass $M$ (and thus implicitly
on $\mu$, $T$, and $\vec{\nabla}\psi$), and therefore the subtraction
term must be (implicitly) medium dependent. In the presence of a superflow,
we even find that the ultraviolet divergences depend explicitly on
$\vec{\nabla}\psi$, see discussion in section \ref{sub:Condensed-phase-with-SF}.

\noindent To write down the result of the renormalization procedure
we first introduce the following abbreviations for the momentum sums, 

\noindent \medskip{}

\noindent 
\begin{equation}
I^{\pm}\equiv\frac{T}{V}\sum_{k}[S_{11}(k)\pm S_{22}(k)]\,,\qquad J\equiv-\frac{1}{2}\frac{T}{V}\sum_{k}\textrm{Tr}\ln\frac{S^{-1}}{T^{2}}\,.
\end{equation}
 Then, the renormalized stationarity equations (\ref{eq:stat1}-\ref{eq:stat3})
are 

\noindent 
\begin{eqnarray}
M^{2}+\delta M^{2}-\sigma^{2} & = & 2\lambda\rho^{2}\,,\label{eq:stat1reno}\\
M^{2} & = & m^{2}+2\lambda\rho^{2}+2\lambda I_{{\rm finite}}^{+}\,,\label{eq:stat2reno}\\
\delta M^{2} & = & \lambda\rho^{2}+\lambda I_{{\rm finite}}^{-}\,,\label{eq:stat3reno}
\end{eqnarray}
 where the finite parts of the momentum sums are
\begin{eqnarray}
I_{{\rm finite}}^{+} & = & \frac{M^{2}}{8\pi^{2}}(\gamma-1)+\frac{M^{2}+\delta M^{2}}{16\pi^{2}}\ln\frac{M^{2}+\delta M^{2}}{\ell^{2}}+\frac{M^{2}-\delta M^{2}}{16\pi^{2}}\ln\frac{M^{2}-\delta M^{2}}{\ell^{2}}\label{eq:IPlus}\\
 & + & \sum_{e=\pm}\int\frac{d^{3}\vec{k}}{(2\pi)^{3}}\left\{ \frac{2[(\epsilon_{\vec{k}}^{e})^{2}-\vec{k}^{2}-M^{2}+\sigma^{2}][1+2f(\epsilon_{\vec{k}}^{e})]}{(\epsilon_{\vec{k}}^{e}+\epsilon_{-\vec{k}}^{e})(\epsilon_{\vec{k}}^{e}+\epsilon_{-\vec{k}}^{-e})(\epsilon_{\vec{k}}^{e}-\epsilon_{\vec{k}}^{-e})}-\frac{1}{2\omega_{\vec{k}}^{e}}\right\} \,,\nonumber \\
\nonumber \\
I_{{\rm finite}}^{-} & = & \frac{\delta M^{2}}{8\pi^{2}}(\gamma-1)+\frac{M^{2}+\delta M^{2}}{16\pi^{2}}\ln\frac{M^{2}+\delta M^{2}}{\ell^{2}}-\frac{M^{2}-\delta M^{2}}{16\pi^{2}}\ln\frac{M^{2}-\delta M^{2}}{\ell^{2}}\label{eq:IMinus}\\
 & + & \sum_{e=\pm}\int\frac{d^{3}\vec{k}}{(2\pi)^{3}}\left\{ \frac{2\delta M^{2}[1+2f(\epsilon_{\vec{k}}^{e})]}{(\epsilon_{\vec{k}}^{e}+\epsilon_{-\vec{k}}^{e})(\epsilon_{\vec{k}}^{e}+\epsilon_{-\vec{k}}^{-e})(\epsilon_{\vec{k}}^{e}-\epsilon_{\vec{k}}^{-e})}-\frac{e}{2\omega_{\vec{k}}^{e}}\right\} \,,\nonumber 
\end{eqnarray}
\\

\noindent with the Euler-Mascheroni constant $\gamma\simeq0.5772$,
the Bose distribution function $f(x)=1/(e^{x/T}-1)$, and the quasiparticle
excitations $\epsilon_{\vec{k}}^{e}$ that are given by the positive
solutions of ${\rm det}\, S^{-1}=0$. The energies

\noindent 
\[
\omega_{\vec{k}}^{e}=\sqrt{(\vec{k}+e\nabla\psi)^{2}+M^{2}+e\delta M^{2}}
\]
 appear in the ``vacuum'' subtractions whose regularized versions
give rise to the (medium dependent) finite terms in the first lines
of equation (\ref{eq:IPlus}) and (\ref{eq:IMinus}) and to (medium
independent) infinite terms which are absorbed in the renormalized
coupling constant and the renormalized mass. 

\noindent The renormalized version of the pressure at the stationary
point is 

\noindent \medskip{}

\noindent 
\begin{equation}
\Psi_{{\rm stat}}=\frac{\rho^{2}}{2}(\mu^{2}-m^{2})-\frac{\lambda}{4}\rho^{4}+J_{{\rm finite}}+\frac{(M^{2}-m^{2}-2\lambda\rho^{2})^{2}}{8\lambda}+\frac{(\delta M^{2}-\lambda\rho^{2})^{2}}{4\lambda}\,,\label{eq:PsiStat}
\end{equation}
 with $M$, $\delta M$, and $\rho$ being solutions of the stationarity
conditions (\ref{eq:stat1reno}-\ref{eq:stat3reno}), and the finite
part of

\newpage{}

\noindent the momentum sum 

\noindent 
\begin{eqnarray}
J_{{\rm finite}} & = & \frac{M^{4}+\delta M^{4}}{64\pi^{2}}(3-2\gamma)-\frac{(M^{2}+\delta M^{2})^{2}}{64\pi^{2}}\ln\frac{M^{2}+\delta M^{2}}{\ell^{2}}-\frac{(M^{2}-\delta M^{2})^{2}}{64\pi^{2}}\ln\frac{M^{2}-\delta M^{2}}{\ell^{2}}\nonumber \\
 & - & \frac{1}{2}\sum_{e=\pm}\int\frac{d^{3}\vec{k}}{(2\pi)^{3}}\left[\epsilon_{\vec{k}}^{e}-\omega_{\vec{k}}^{e}+2T\ln\left(1-e^{-\epsilon_{\vec{k}}^{e}/T}\right)\right]\,.\label{eq:J}
\end{eqnarray}

~

\subsection{\noindent Goldstone mode\label{sub:Goldstone-mode}}

~

\begin{figure}[t]
~~~~~~~~~~~~\includegraphics[scale=0.4]{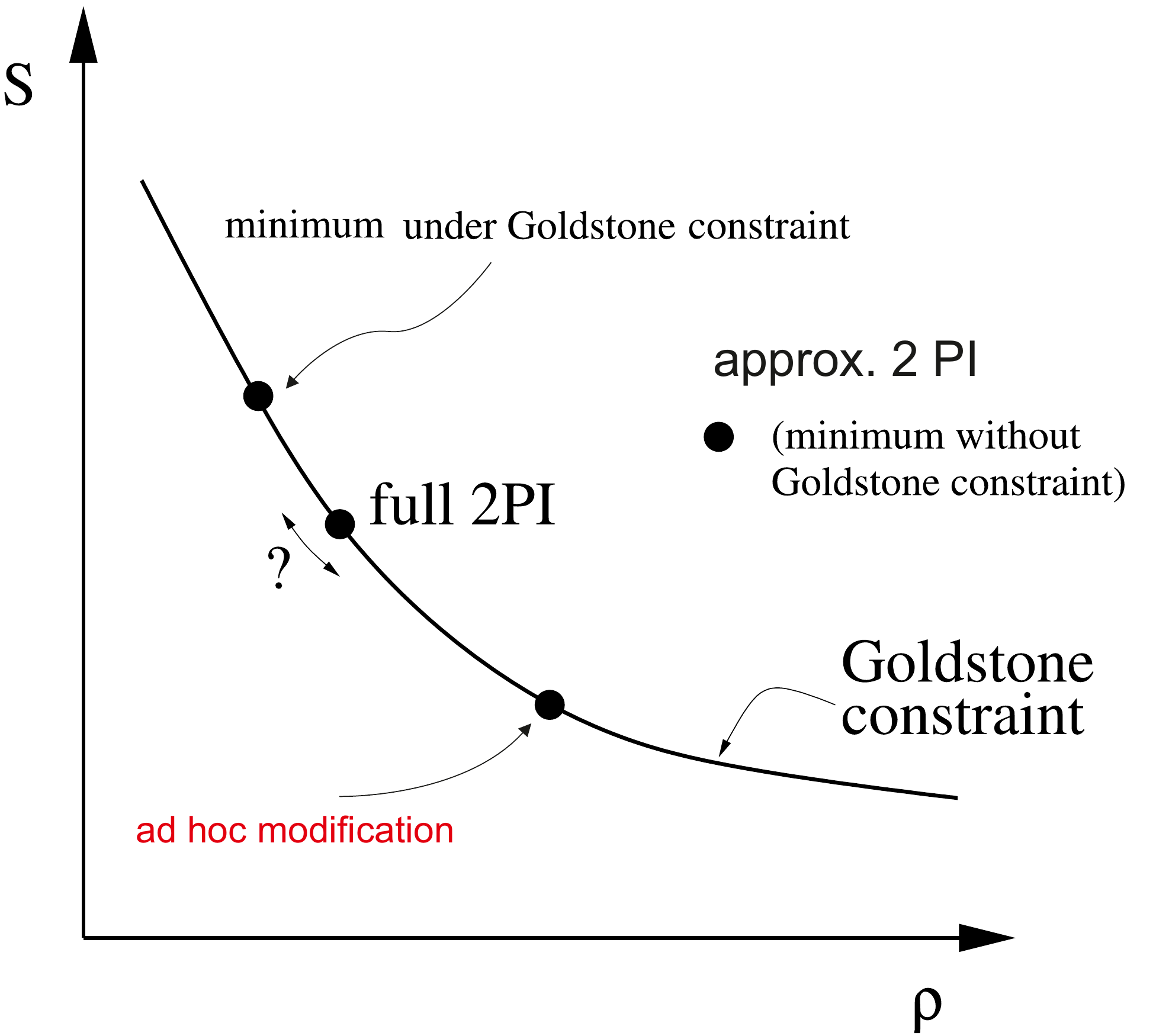}\protect\caption{Minima of the 2PI effective action. It can be shown \cite{PilaftsisReno}
that the Goldstone theorem is fulfilled on a curve in the space spanned
by $\rho$ and $S$. The full solution to the complete 2PI functional
lies somewhere on that curve. The two-loop Hartree approximation leads
to a violation of the Goldstone theorem and thus the minimum is not
located on that curve. In our approach, the Goldstone theorem is nevertheless
enforced by an ad hoc modification. A similar approach is followed
in reference \cite{PilaftsisReno}, where the stationary point in
the constrained subspace given by the condition of an exact Goldstone
mode is determined. It is unknown, which of the latter two approaches
is closer to the real minimum obtained from the full 2PI effective
action.\label{fig:Statpoin}}
\end{figure}

\noindent The quasiparticle dispersion relations $\epsilon_{\vec{k}}^{e}$
are determined by the zeros of ${\rm det}\, S^{-1}$. In analogy to
the zeros of ${\rm det}\, S_{0}^{-1}$ which lead to the dispersions
(\ref{eq:DispGold}), (\ref{eq:DispMass}) we expect one massive mode
and one Goldstone mode where the Goldstone mode fulfills $k_{0}=0$
at $\vec{k}=0$. Setting $\vec{k}=0$ in the inverse propagator (\ref{eq:FullProp}),
we see that $k_{0}=0$ is only a zero of ${\rm det}\, S^{-1}$ if
$M^{2}-\sigma^{2}-\delta M^{2}=0$ (or if $M^{2}-\sigma^{2}+\delta M^{2}=0$).
However, this condition for the existence of a Goldstone mode is in
contradiction to equations (\ref{eq:stat1reno}) and (\ref{eq:stat3reno})
which imply $M^{2}-\sigma^{2}-\delta M^{2}=-2\lambda I_{{\rm finite}}^{-}$,
where $I_{{\rm finite}}^{-}$ might be small but does not vanish.
Consequently, the Goldstone theorem is violated in our approach \cite{CritTKaon,Andersen2007,AndersenLeganger,AmelinoReno,Berges2Reno,FejoesReno,GrinsteinReno,IvanovReno,PilaftsisReno,RischkeReno,SeelReno}.
For our discussion of the superfluid properties it is crucial to work
with an exact Goldstone mode. Therefore, we shall ignore the contribution
from the momentum sum in equation (\ref{eq:stat3reno}), thereby giving
up the exact self-consistency of our approach \cite{CritTKaon,GrinsteinReno}.
This is an ad hoc modification of the stationarity equations, i.e.,
we do not consider the true minimum of the full self-consistency equations,
but a point away from this minimum. The benefit of this modification
is that the Goldstone theorem is built into our calculation. Of course,
our choice of enforcing the Goldstone theorem is not unique, and there
are infinitely many ``Goldstone points'' once the exact self-consistency
is sacrificed. This situation is illustrated in figure \ref{fig:Statpoin}.
A similar, but not identical, procedure is followed in reference \cite{PilaftsisReno},
where the stationary point in the constrained subspace given by the
condition of an exact Goldstone mode is determined. 

\noindent Our modification results in a particularly simple set of
equations, because the three stationarity equations now reduce to
two trivial ones and only one that still contains a momentum integral,

\noindent 
\begin{eqnarray}
\delta M^{2} & = & \lambda\rho^{2}=M^{2}-\sigma^{2}\,,\label{eq:dMequ}\\
2\sigma^{2}-m^{2} & = & M^{2}+2\lambda I_{{\rm finite}}^{+}\,.\label{eq:Mequ}
\end{eqnarray}

\newpage{}

\noindent The two dispersion relations $\epsilon_{\vec{k}}^{\pm}$
are then determined from 

\noindent \medskip{}

\noindent 
\begin{equation}
0={\rm det}\, S^{-1}=k^{2}[k^{2}-2(M^{2}-\sigma^{2})]-4(k_{\mu}\partial^{\mu}\psi)^{2}\,,\label{eq:DispEq}
\end{equation}
where equation (\ref{eq:dMequ}) has been used to eliminate $\delta M$.
Since we are interested in arbitrary temperatures below $T_{c}$,
we shall need the full dispersion of both modes. Their explicit form
is too lengthy to write down, but we can at least demonstrate the
linear part of the dispersion relation of the Goldstone mode as a
function of the mass parameter $M$ (this result is identical to (\ref{eq:DispGold})
if the zero temperature expression for $M$ is inserted)

\noindent \medskip{}

\noindent 
\begin{equation}
\epsilon_{\vec{k}}^{+}=\frac{\sqrt{(M^{2}-\sigma^{2})(M^{2}+\sigma^{2}+2[(\vec{\nabla}\psi)^{2}-(\hat{k}\cdot\vec{\nabla}\psi)^{2}])}-2\partial_{0}\psi\,\hat{k}\cdot\vec{\nabla}\psi}{M^{2}+\sigma^{2}+2(\vec{\nabla}\psi)^{2}}\,|\vec{k}|+\ldots\,\,.\label{eq:Goldstone2PI}
\end{equation}

\newpage{}

\noindent For vanishing superflow we have $\sigma=\mu$ and obtain 

\noindent \medskip{}

\noindent 
\begin{equation}
\epsilon_{\vec{k}}^{+}(\vec{\nabla}\psi=0)=\sqrt{\frac{M^{2}-\mu^{2}}{M^{2}+\mu^{2}}}\,|\vec{k}|+\ldots\,.\label{eq:Goldstone2PIv0}
\end{equation}

\noindent If we work at a point that is not exactly the stationary
point, we cannot use the pressure (\ref{eq:PsiStat}). We rather have
to evaluate the effective action density at the ''Goldstone point''.
For the renormalization it is crucial that we only modify the finite
part of the stationarity equations. Therefore, all infinities cancel
in the same way as above, see appendix \ref{sub:Renormalization-with-Goldstone}
for a more detailed discussion; for the pressure at the `''Goldstone
point'' we then find

\noindent \medskip{}

\noindent 
\begin{equation}
\Psi_{{\rm Gold}}=\frac{(M^{2}-\sigma^{2})(3\sigma^{2}-M^{2}-2m^{2})}{4\lambda}+J_{{\rm finite}}+\frac{(M^{2}-2\sigma^{2}+m^{2})^{2}}{8\lambda}-\frac{\lambda}{4}(I_{{\rm finite}}^{-})^{2}\,.\label{eq:Pfinite}
\end{equation}
 Here we have already eliminated $\rho$ and $\delta M$ with the
help of equation (\ref{eq:Mequ}). The stationarity equation (\ref{eq:Mequ})
and the pressure (\ref{eq:Pfinite}) are the starting point for our
calculations. We shall solve the stationarity equation numerically
for the self-consistent mass $M$, which in turn gives the condensate
via $\lambda\rho^{2}=M^{2}-\sigma^{2}$ as well as the dispersion
relations of the Goldstone mode and the massive mode. We need the
pressure for various thermodynamic derivatives which are needed to
compute for example the sound velocities. 

~

\subsection{Critical temperature, condensate and critical velocity for all temperatures\label{sub:Critical-temperature,-condensate-critV-forallT}}

~

\begin{figure}

\fbox{\begin{minipage}[t]{1\columnwidth}%
\includegraphics[scale=0.7]{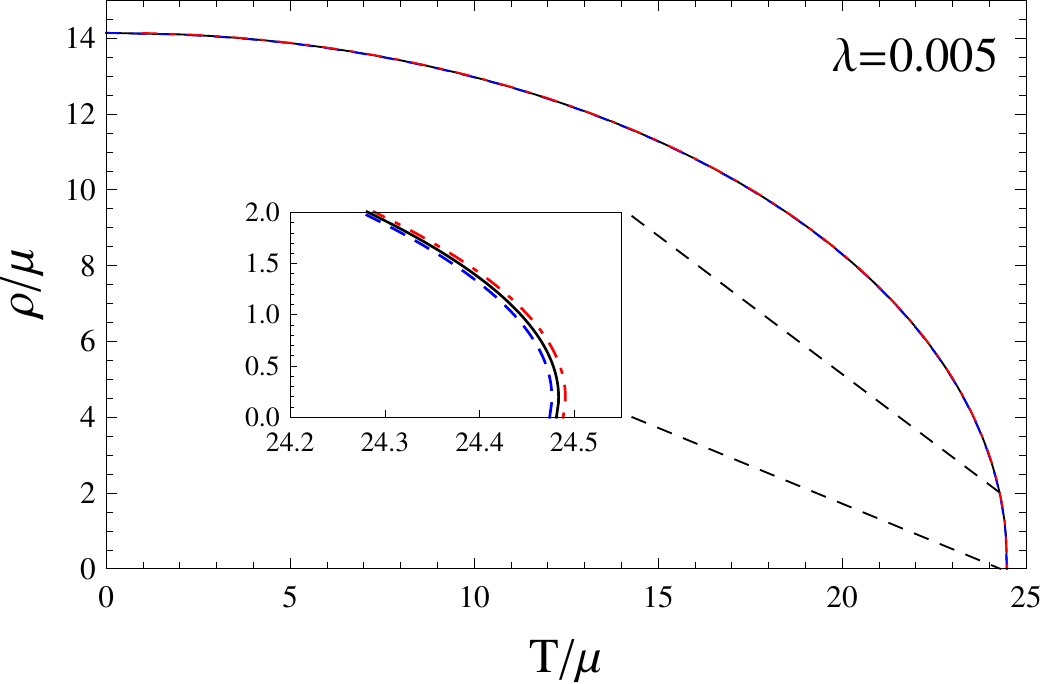}~~~~~~~~\includegraphics[scale=0.7]{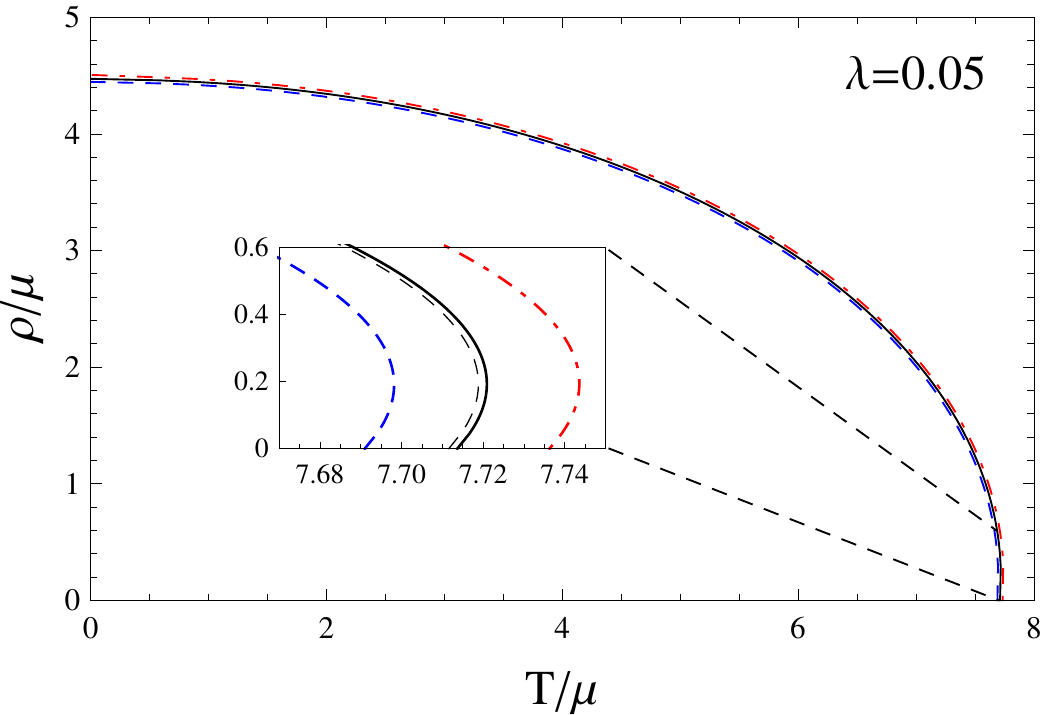}%
\end{minipage}}\protect\caption{Condensate $\rho$ as a function of temperature for $m=\left|\nabla\psi\right|=0$
and coupling constants $\lambda=0.005$ (left panel) and $\lambda=0.05$
(right panel). Even though barely visible on the large scale, the
phase transition is first order, and the results depend on the renormalization
scale $\ell$. For the small value of the coupling, even large variations
of the renormalization scale are barely visible, while for the larger
coupling, the result is more, but still only mildly, sensitive to
variations in $\ell$. In both panels, $\ell=0.1\mu$, $\mu$, $10\mu$
for the dashed (blue), solid (black), and dashed-dotted (red) lines,
respectively. The thin (black) dashed line in the inset of the right
panel is obtained with the approximation (\ref{eq:noscale1})-(\ref{eq:noscale3}),
where the dependence of the renormalization scale drops out. \label{fig:CondensatePlot}}

\end{figure}

\noindent To get started, we need to determine the critical temperature
$T_{c}$ by solving the stationarity equation (\ref{eq:Mequ}) for
$T$ at the point $M^{2}=\sigma^{2}-2(\nabla\psi)^{2}$ - we shall
explain that this is precisely the critical value of $M$, see discussion
around equation (\ref{eq:vcrit}). To do so, one has to choose specific
values for the parameters  $\mu$ and $|\vec{\nabla}\psi|$ as well
as $\lambda$, $m$, and the renormalization scale $\ell$. Then we
can obtain $M$ at any temperature between $T=0$ and $T=T_{c}$ by
a numerical evaluation of the stationary equation (\ref{eq:Mequ}).
Once we have determined $M$ from the stationarity equation we obtain
the condensate $\rho$ through $\lambda\rho^{2}=M^{2}-\sigma^{2}$.
In figure \ref{fig:CondensatePlot} we show the condensate as a function
of temperature for the simple case without superflow and for two different
coupling strengths. We have set $m=0$, but the conclusions we draw
from this figure are valid for all values of $m$. As mentioned below
equation (\ref{eq:twoPiPotential}), the phase transition to the non-superfluid
phase turns out to be of first order, although this is barely visible
if we plot the condensate for all temperatures. Moreover, there is
a dependence on the renormalization scale $\ell$ through the logarithmic
terms discussed in section \ref{sub:Renormalized-stationarity-equations-and-pressure}.
This dependence and the first order transition are very weak because
of the smallness of the coupling constants chosen here. As the figure
shows, the stronger the coupling, the stronger the dependence on the
renormalization scale and the stronger the first order transition.

\noindent Since we know that the first-order nature of the phase transition
is an artifact of the Hartree approximation, we shall restrict ourselves
to sufficiently weak coupling constants. We shall work with the two
couplings chosen in figure \ref{fig:CondensatePlot}. In this case
we find that we can, to a very good approximation, work with a simplified
stationarity equation and a simplified pressure, using%
\footnote{In the notation of appendix \ref{sec:Renormalization-in-2PI}, this
means that we approximate $I_{{\rm finite}}^{\pm}(T,\mu,\ell)=I_{{\rm vac,finite}}^{\pm}(\ell)+I_{\mu}^{\pm}(0)+I_{T}^{\pm}(\mu)\simeq I_{T}^{\pm}(\mu)$
and $J_{{\rm finite}}(T,\mu,\ell)\simeq J_{T}(\mu)$.%
} 

\noindent 
\begin{eqnarray}
I_{{\rm finite}}^{+} & \simeq & 4\sum_{e=\pm}\int\frac{d^{3}\vec{k}}{(2\pi)^{3}}\frac{(\epsilon_{\vec{k}}^{e})^{2}-\vec{k}^{2}-M^{2}+\sigma^{2}}{(\epsilon_{\vec{k}}^{e}+\epsilon_{-\vec{k}}^{e})(\epsilon_{\vec{k}}^{e}+\epsilon_{-\vec{k}}^{-e})(\epsilon_{\vec{k}}^{e}-\epsilon_{\vec{k}}^{-e})}\, f(\epsilon_{\vec{k}}^{e})\,,\label{eq:noscale1}\\
I_{{\rm finite}}^{-} & \simeq & 4\sum_{e=\pm}\int\frac{d^{3}\vec{k}}{(2\pi)^{3}}\frac{\delta M^{2}}{(\epsilon_{\vec{k}}^{e}+\epsilon_{-\vec{k}}^{e})(\epsilon_{\vec{k}}^{e}+\epsilon_{-\vec{k}}^{-e})(\epsilon_{\vec{k}}^{e}-\epsilon_{\vec{k}}^{-e})}\, f(\epsilon_{\vec{k}}^{e})\,,\label{eq:noscale2}
\end{eqnarray}

\newpage{}

\noindent and

\noindent 
\begin{equation}
J_{{\rm finite}}\simeq-T\sum_{e=\pm}\int\frac{d^{3}\vec{k}}{(2\pi)^{3}}\ln\left(1-e^{-\epsilon_{\vec{{\bf k}}}^{e}/T}\right)\,.\label{eq:noscale3}
\end{equation}

~

\noindent In this approximation, the contributions from loop diagrams
that do not depend on temperature explicitly are neglected. As a consequence,
the zero-temperature results are identical to the tree-level results.
All dependence on the renormalization scale $\ell$ is gone and thus
we do not have to specify $\ell$. We have checked numerically that
the subleading terms that we have dropped do not visibly change any
of the curves we show, see also right panel of figure \ref{fig:CondensatePlot},
where the approximation is compared to the full result.

\begin{figure}[t]
\fbox{\begin{minipage}[t]{1\columnwidth}%
\includegraphics[scale=0.7]{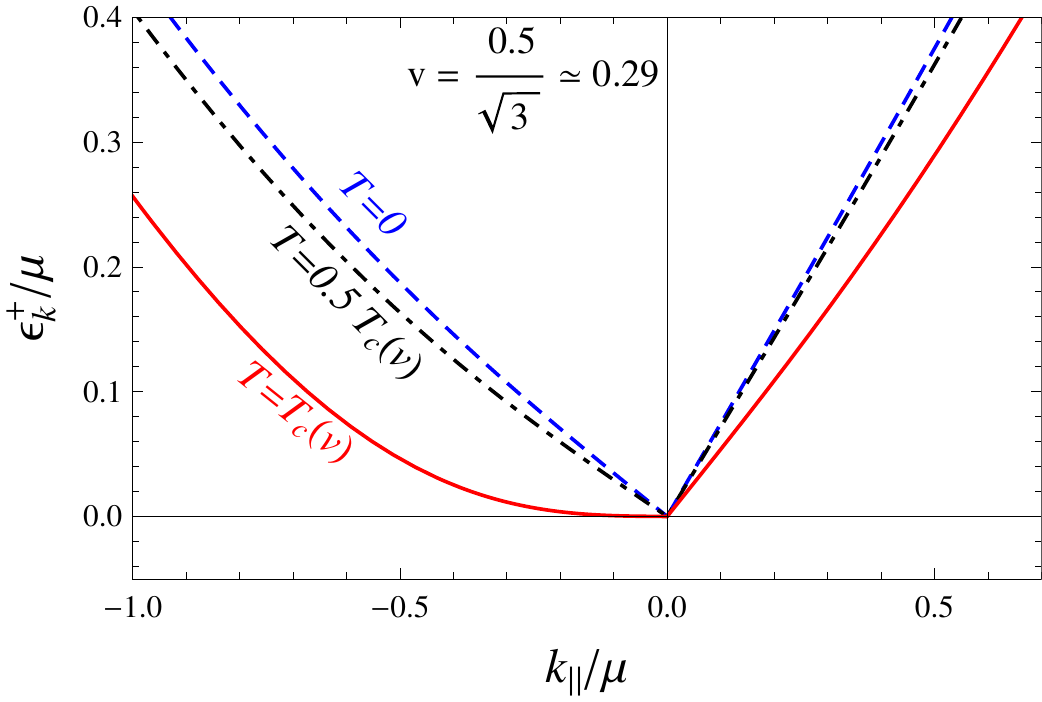}~~~~~~~~~\includegraphics[scale=0.7]{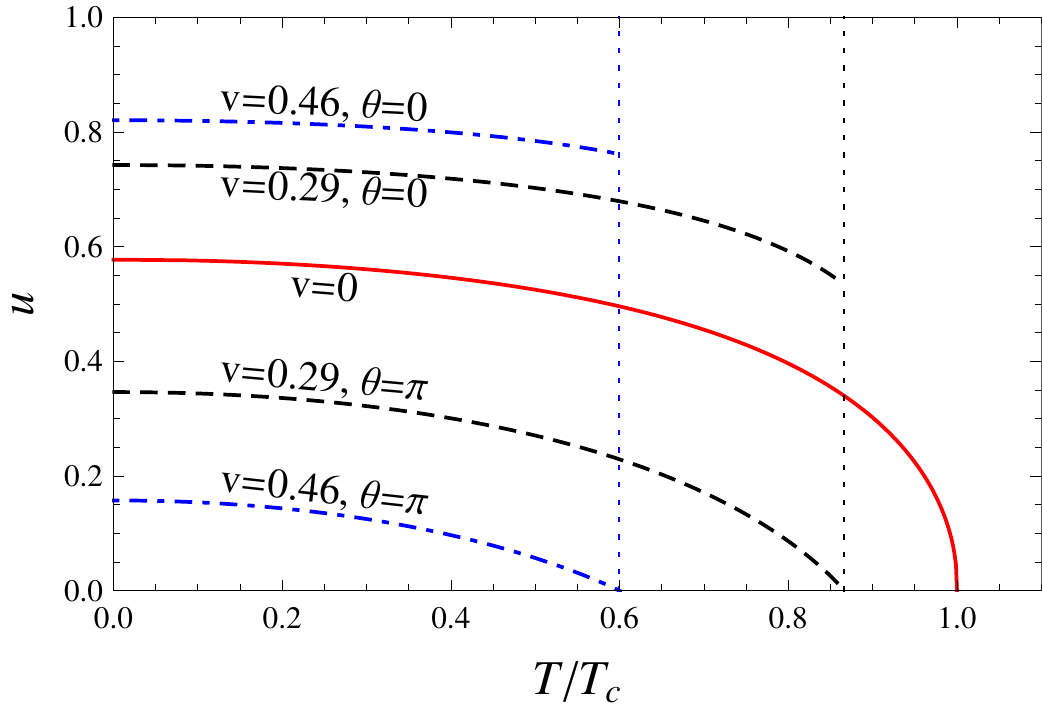}%
\end{minipage}}\protect\caption{Dispersion of the Goldstone mode parallel ($k_{\parallel}>0$) and
anti-parallel ($k_{\parallel}<0$) to the superflow for a superfluid
velocity $v=0.5/\sqrt{3}$ and three different temperatures with $T_{c}(v)$
being the temperature beyond which $\epsilon_{\vec{k}}^{+}$ would
become negative for small momenta. Right panel: slope $u(\theta)$
of the low-energy dispersion of the Goldstone mode, $\epsilon_{\vec{k}}^{+}\simeq u(\theta)|\vec{k}|$,
in the directions parallel, $\theta=0$, and antiparallel, $\theta=\pi$,
to the superflow as a function of temperature for three different
values of the superfluid velocity. The vertical dotted lines indicate
the critical temperatures beyond which there is a negative excitation
energy. Only for the case of vanishing superflow is the critical temperature
the point where the condensate has completely melted away. This temperature
is denoted by $T_{c}\equiv T_{c}(v=0)$. The case of the intermediate
superfluid velocity $v\simeq0.29$ corresponds to the left panel,
i.e., $T\simeq0.86\, T_{c}$ in the right panel is identical to $T_{c}(v)$
in the left panel. We have set $\lambda=0.005$ and $m=0$. \label{fig:dispersionInst}}
\end{figure}

\noindent The effects of a nonzero superflow on the solution are shown
in figure \ref{fig:dispersionInst}, for the parameters $\lambda=0.005$
and $m=0$. In the left panel of this figure we show the Goldstone
mode dispersion relation $\epsilon_{\vec{k}}^{+}$, evaluated using
the value of $M$ that solves the stationarity equation. We see that
for a given superfluid velocity, there is a temperature at which the
dispersion becomes flat in the direction opposite to the superflow.
For higher temperatures, there would be negative energies, indicating
an instability of the system. Therefore, this particular temperature
is a critical temperature, even though the condensate has not yet
melted completely. Only at vanishing superflow is the critical temperature
the same as the point where the condensate has become zero (if our
approach gave an exact second-order phase transition). The plot shows
that the low-energy part of the dispersion, where $\epsilon_{\vec{k}}^{+}$
is linear in $k$, is sufficient to locate the instability. Therefore,
in the right panel of the figure, we show the slope of the linear
part, see equation (\ref{eq:Goldstone2PI}), as a function of temperature
for three different values of the superflow. The superfluid state
breaks down when the slope in the anti-parallel direction vanishes.
This defines a critical temperature for any given velocity, or a critical
velocity for any given temperature. With the help of the low-energy
dispersion (\ref{eq:Goldstone2PI}) we can derive a semi-analytical
result for the critical velocity. We find that the linear part of
the dispersion is positive for all angles between the momentum $\vec{k}$
and the superflow $\vec{\nabla}\psi$ only if $M^{2}-\sigma^{2}>2(\vec{\nabla}\psi)^{2}$.
This shows explicitly that the condensate $\rho^{2}=(M^{2}-\sigma^{2})/\lambda$
cannot become arbitrarily small for nonzero superflow. Using $\sigma^{2}=\mu^{2}-(\vec{\nabla}\psi)^{2}$
and $\vec{v}=-\vec{\nabla}\psi/\mu$, we can rewrite the condition
for the positivity of the excitation energy in the equivalent, but
more instructive, form

\begin{figure}[t]
\fbox{\begin{minipage}[t]{1\columnwidth}%
\noindent \begin{center}
\includegraphics[scale=0.8]{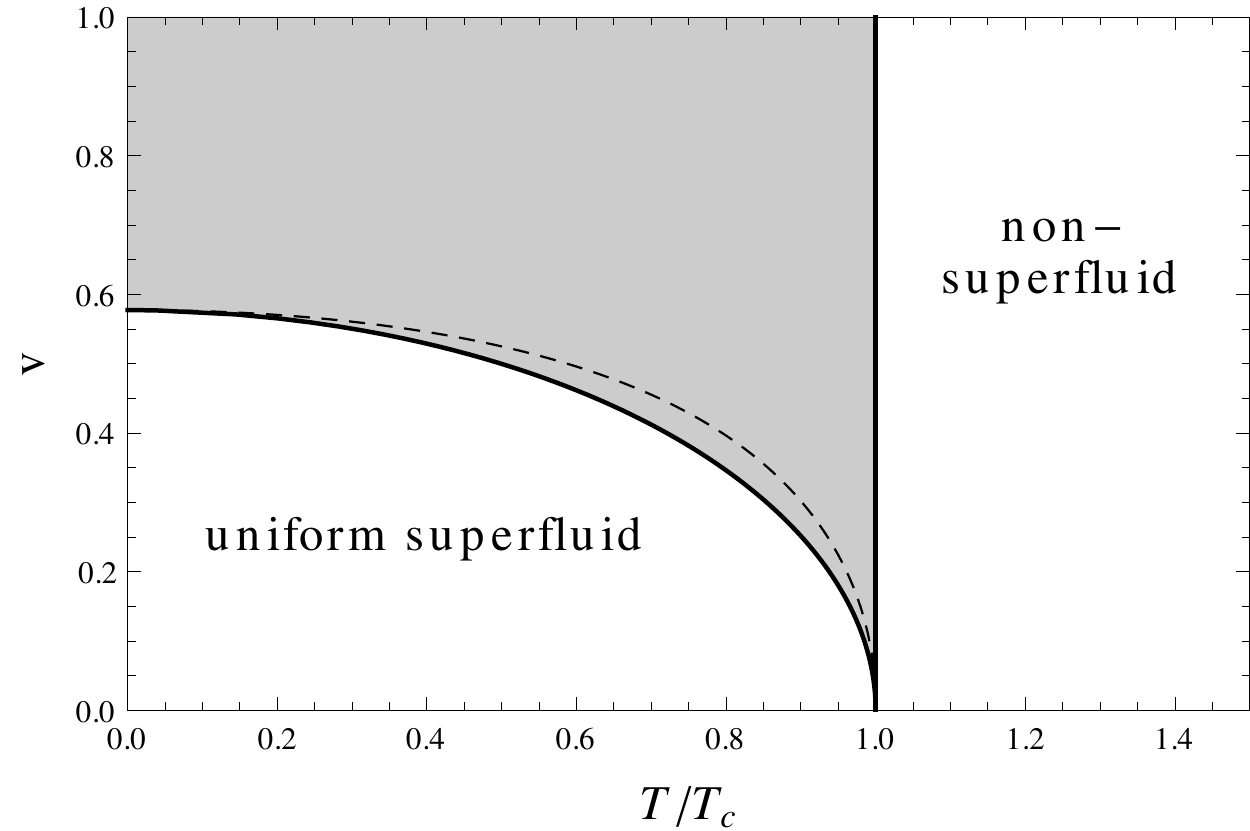}
\par\end{center}%
\end{minipage}}\protect\caption{Phase diagram resulting from the instability shown in figure \ref{fig:dispersionInst},
in the plane of superfluid velocity $v$ and temperature. Within our
ansatz, which only allows for spatially homogeneous condensates, there
is no stable phase in the shaded region. The dashed line is the slope
of the Goldstone dispersion at $v=0$ and thus shows - for comparison
- the would-be critical velocity if the superflow only acted as a
Lorentz transformation on the dispersion, and not also on the self-consistently
determined condensate.\label{fig:PhaseDiagCrit}}
\end{figure}

\noindent \medskip{}

\noindent 
\begin{equation}
v<\sqrt{\frac{M^{2}-\sigma^{2}}{M^{2}+\sigma^{2}}}\,.\label{eq:vcrit}
\end{equation}
 This is an implicit condition for allowed values of the superfluid
velocity $v$. (Remember that $M$ is a complicated function of this
velocity.) We plot the critical line, given implicitly by equation
(\ref{eq:vcrit}), in the plane of superfluid velocity and temperature
in figure \ref{fig:PhaseDiagCrit}.

\noindent The right-hand side of the inequality (\ref{eq:vcrit}),
evaluated at $v=0$, is simply the slope of the Goldstone mode at
$v=0$, see equation (\ref{eq:Goldstone2PIv0}). This is related to
Landau's original argument for the critical velocity (see also section
\ref{sub:The-critical-velocity}): based on a Lorentz transformation
of the excitation energy (in the original non-relativistic context
a Galilei transformation), the critical velocity is determined by
the slope of the Goldstone mode at $v=0$ (unless there are non-trivial
features such as rotons in superfluid helium, which is not the case
here). One might thus think that we would just have to do the $v=0$
calculation to determine the critical line in the phase diagram. However,
switching on a superflow is not equivalent to a Lorentz transformation
of the excitation energy; it is a Lorentz transformation \textit{plus}
a change in the self-consistently determined condensate, which in
turn back-reacts on the excitation energies. This additional effect
is contained in the $v$ dependence of $M$ in equation (\ref{eq:vcrit}).
For comparison, we have plotted the (incorrect) critical curve obtained
from the $v=0$ dispersion in the phase diagram as a dashed line.
We see that the full result is smaller for nonzero temperatures. In
the weak-coupling case considered here, the effect of the superflow
(in addition to being a Lorentz boost) on the Goldstone dispersion
becomes negligibly small for low temperatures. Therefore, at $T=0$
the critical velocity is $\frac{1}{\sqrt{3}}$, in exact agreement
with the slope of the low-energy dispersion at $v=0$. For a similar
recent discussion in a holographic approach see reference \cite{LandsteinerSuperfl}
and in particular the phase diagram in figure 6 of this reference.

\noindent The critical line seems to suggest a (strong) first-order
phase transition to the non-superfluid phase at the critical velocity.
However, at temperatures below $T_{c}$ ($T_{c}$ being the critical
temperature in the absence of a superflow), the system still ``wants''
to condense, even for velocities beyond the critical value. In other
words, the uncondensed phase also turns out to be unstable. In our
calculation, this is seen as follows. First we note that the stationarity
equation for $M$ in the case $\rho=0$, see equation (\ref{eq:RenoM}),
does not depend on $\vec{\nabla}\psi$. This is clear since $\psi$
is the phase of the condensate, so the uncondensed phase must be independent
of $\vec{\nabla}\psi$. For supercritical temperatures the solution
to the $\rho=0$ stationarity equation gives a value of $M$ that
is greater than $\mu$, but at subcritical temperatures $M$ is less
than $\mu$. The excitation energies are simply given by $\epsilon_{\vec{k}}^{e}=\sqrt{\vec{k}^{2}+M^{2}}-e\mu$,
so $\epsilon_{k}^{+}$ becomes negative for certain momenta if $M<\mu$.
Therefore, the non-superfluid phase is unstable below $T_{c}$. As
a consequence, within our non-dissipative, uniform ansatz, we cannot
construct a stable phase for sufficiently large superfluid velocities
and low temperatures (shaded area in figure \ref{fig:PhaseDiagCrit}).
Beyond the critical velocity there may be no stable phase, if dissipative
effects such as vortex creation arise in that regime. There is some
evidence for this in liquid helium \cite{Gorter1949},\cite{Allum1977}
and ultra-cold bosonic \cite{Raman1999} and fermionic \cite{Miller2007}
gases. Possibly, a more complicated, dissipative and/or inhomogeneous
phase already replaces the homogeneous superfluid for superfluid velocities
below our critical line. In this sense we have only determined an
upper limit for the critical velocity as a function of temperature
below which the homogeneous superfluid is stable. This limit can for
instance be reduced by the onset of unstable sound modes due to the
two-stream instability \cite{SchmittTwostream}.

\subsection{Two-fluid properties for all temperatures\label{sub:Two-fluid-properties-for-allT}}

\subsubsection{Algorithm\label{sub:Algorithm}}

Besides solving the stationarity equation we need to compute derivatives
of the pressure with respect to $T$, $\mu$, and $|\vec{\nabla}\psi|$
to obtain the thermodynamic and hydrodynamic parameters of the two-fluid
model. As we will later also be interested in sound excitations, we
need to compute all first and second derivatives. In this section,
we explain the algorithm with which these results are obtained. The
most direct way to do so is via brute force numerical evaluation,
for instance with the method of finite differences. In order to obtain
results less prone to numerical uncertainties, we compute the derivatives
in the following semi-analytical way. First we note that the pressure
$\Psi$ at the point that respects the Goldstone theorem (as well
as at the point where the stationarity equations are exactly fulfilled)
depends explicitly as well as implicitly via $M$ on the relevant
variables, 

\begin{equation}
\Psi=\Psi[M(T,\mu,|\vec{\nabla}\psi|),T,\mu,|\vec{\nabla}\psi|]\,,
\end{equation}
and each thermodynamic derivative we are interested in also sees the
implicit dependence in $M$. (Had we only been interested in first
derivatives \textit{and} had we considered the exact solution of the
stationarity equations - sacrificing the Goldstone theorem - we could
have restricted ourselves to the explicit dependence, since then the
derivative of the pressure with respect to the self-consistently determined
mass would have vanished by construction at the stationary point.)
Denoting the variables by $x,y\in\{T,\mu,|\nabla\psi|\}$, we can
thus write 

\begin{eqnarray}
\frac{d\Psi}{dx} & = & \frac{\partial M}{\partial x}\frac{\partial\Psi}{\partial M}+\frac{\partial\Psi}{\partial x}\,,\label{eq:deriv1}\\
\frac{d^{2}\Psi}{dxdy} & = & \frac{\partial^{2}M}{\partial x\partial y}\frac{\partial\Psi}{\partial M}+\frac{\partial M}{\partial x}\frac{\partial M}{\partial y}\frac{\partial^{2}\Psi}{\partial M^{2}}+\frac{\partial M}{\partial x}\frac{\partial^{2}\Psi}{\partial y\partial M}+\frac{\partial M}{\partial y}\frac{\partial^{2}}{\partial x\partial M}+\frac{\partial^{2}\Psi}{\partial x\partial y}\,.\label{eq:deriv2}
\end{eqnarray}
 The derivatives of $M$ can be obtained from taking the first and
second derivatives of the stationarity equation (\ref{eq:Mequ}).
Writing this equation as $0=g(M,T,\mu,|\nabla\psi|)$, we find 

\begin{eqnarray}
\frac{\partial M}{\partial x} & = & -\frac{\partial g}{\partial x}\left(\frac{\partial g}{\partial M}\right)^{-1}\,,\label{eq:derivM1}\\
\frac{\partial^{2}M}{\partial x\partial y} & = & -\left(\frac{\partial g}{\partial M}\right)^{-1}\left[\frac{\partial^{2}g}{\partial M\partial y}\frac{\partial M}{\partial x}+\frac{\partial^{2}g}{\partial M\partial x}\frac{\partial M}{\partial y}+\frac{\partial^{2}g}{\partial M^{2}}\frac{\partial M}{\partial x}\frac{\partial M}{\partial y}+\frac{\partial^{2}g}{\partial x\partial y}\right]\,.\label{eq:derivM2}
\end{eqnarray}
 We can thus use the following algorithm to compute the properties
of the superfluid: 
\begin{itemize}
\item Choose values for the thermodynamic parameters  $\mu$ and $|\vec{\nabla}\psi|$
as well as the parameters $\lambda$, $m$, and the renormalization
scale $\ell$. 
\item Determine the critical temperature $T_{c}$ by solving the stationarity
equation (\ref{eq:Mequ}) for $T$ at the point $M^{2}=\sigma^{2}-2(\vec{\nabla}\psi)^{2}$
. 
\item Find the solution for $M$ of the stationarity equation (\ref{eq:Mequ})
for various values of the temperature $0<T<T_{c}$. 
\item Compute the first and second derivatives of the integrands of $I_{{\rm finite}}^{\pm}$
and $J_{{\rm finite}}$ with respect to $M$, $T$, $\mu$, and $|\vec{\nabla}\psi|$;
for the case without superflow one may use a simplification for this
step as we explain below. This is done algebraically, i.e, before
choosing numerical values. Nevertheless, it is useful to do all this
with a computer because the results get very complicated. 
\item Perform the three-momentum integrals numerically over all expressions
obtained in the previous step. Since there are three integrands ($I_{{\rm finite}}^{+}$,
$I_{{\rm finite}}^{-}$, $J_{{\rm finite}}$) and four variables ($M$,
$T$, $\mu$, $|\vec{\nabla}\psi|$), we have to perform $3\times4=12$
integrals for the first derivatives and $3\times10=30$ integrals
for the second derivatives at each temperature. In the presence of
a superflow, each of the integrals contains a non-trivial integration
over the polar angle; without superflow, only the integrals needed
for the superfluid density contain such an angular integral.
\item Use equations (\ref{eq:deriv1}-\ref{eq:derivM2}), the results of
the previous step, and some trivial derivatives of terms outside the
momentum integrals to put together the first and second derivatives
of $\Psi$ with respect to $T$, $\mu$, and $|\vec{\nabla}\psi|$.
There are many terms to handle but this is a trivial task for a computer
since the non-trivial numerical calculation has already been done
in the step before. We have checked that the derivatives thus obtained
are much cleaner in terms of numerical errors compared to a brute
force numerical calculation using finite differences. 
\item Insert the obtained derivatives into the definitions of the physical
quantities such as $n_{s}$,$n_{n}$, $\overline{{\cal A}}$, $\overline{{\cal B}}$,
$\overline{{\cal C}}$ .
\end{itemize}

\subsubsection{Limit of vanishing superflow \label{sub:Limit-of-vanishing-sF}}

Even though we are also interested in the general case of a non-vanishing
superflow, let us briefly discuss how the calculation simplifies in
the limit $\left|\nabla\psi\right|\to0$. In that case, when we compute
derivatives with respect to $T$ and $\mu$ we can set $\vec{\nabla}\psi=0$
straightforwardly. But, when we compute $n_{s}$ and $n_{n}$ we have
to be more careful. Consider the microscopic definition of $n_{s}$
according to (\ref{eq:NSMicro}). These quantities describe the response
of the system to a superflow, i.e., even if we are eventually interested
in the case $\vec{\nabla}\psi\to0$, we need to work initially with
a nonzero superflow. We can write the superfluid density for $\vec{\nabla}\psi\to0$
as

\begin{equation}
n_{s}\Big|_{\vec{\nabla}\psi=0}=-\mu\left(\frac{\partial^{2}\Psi}{\partial|\vec{\nabla}\psi|^{2}}\right)_{\vec{\nabla}\psi=0}\,,\label{eq:NsSFaprox}
\end{equation}
where we have expanded $\Psi$ in a Taylor series for small $|\vec{\nabla}\psi|$.
In this series we have dropped the linear term because the first derivative
(i.e., the current $\vec{j}$) vanishes for $|\vec{\nabla}\psi|=0$,
which is obvious physically and can also be checked explicitly. It
seems that the derivatives with respect to $|\vec{\nabla}\psi|$ are
very complicated to compute because they involve the derivatives of
the dispersion relations $\epsilon_{\vec{k}}^{e}$ which are contained
in the momentum integrals in $I_{{\rm finite}}^{\pm}$ and $J_{{\rm finite}}$,
see equations (\ref{eq:IPlus}), (\ref{eq:IMinus}) and (\ref{eq:J}).
However, since we know that $\epsilon_{\vec{k}}^{e}$ are the solutions
to the quartic equation (\ref{eq:DispEq}), we can simplify the calculation
significantly by taking the first and second derivatives of 

\begin{eqnarray}
\left.\frac{\partial\epsilon_{\vec{k}}^{e}}{\partial|\vec{\nabla}\psi|}\right|_{\vec{\nabla}\psi=0} & = & \frac{2\mu k_{\parallel}}{(\epsilon_{\vec{k}}^{e})^{2}-\vec{k}^{2}-M^{2}-\mu^{2}}\,\label{eq:derEps2}\\
\left.\frac{\partial^{2}\epsilon_{\vec{k}}^{e}}{\partial|\vec{\nabla}\psi|^{2}}\right|_{\vec{\nabla}\psi=0} & = & \frac{(\epsilon_{\vec{k}}^{e})^{2}+2k_{\parallel}^{2}-\vec{k}^{2}}{\epsilon_{\vec{k}}^{e}[(\epsilon_{\vec{k}}^{e})^{2}-\vec{k}^{2}-M^{2}-\mu^{2}]}+\frac{8\mu^{2}k_{\parallel}^{2}}{\epsilon_{k}^{e}[(\epsilon_{k}^{e})^{2}-\vec{k}^{2}-M^{2}-\mu^{2}]^{2}}\,\label{eq:derEps1}\\
 & - & \frac{4\mu^{2}k_{\parallel}^{2}[3(\epsilon_{\vec{k}}^{e})^{2}-\vec{k}^{2}-M^{2}-\mu^{2}]}{\epsilon_{k}^{e}[(\epsilon_{k}^{e})^{2}-\vec{k}^{2}-M^{2}-\mu^{2}]^{3}}\,,\nonumber 
\end{eqnarray}
where $k_{\parallel}=|\vec{k}|\cos\theta$ with $\theta$ being the
angle between $\vec{\nabla}\psi$ and $\vec{k}$, and

\[
\epsilon_{\vec{k}}^{e}\equiv\epsilon_{\vec{k}}^{e}(\vec{\nabla}\psi=0)=\sqrt{\vec{k}^{2}+M^{2}+\mu^{2}-e\sqrt{4\vec{k}^{2}\mu^{2}+(M^{2}+\mu^{2})^{2}}}
\]
are the excitation energies at vanishing superflow which are equivalent
to (\ref{eq:DispNoSF}) if $M(T=0,\,|\vec{\nabla}\psi|=0)=2\mu^{2}-m^{2}$
is inserted. Equations (\ref{eq:derEps2},\ref{eq:derEps1}) are very
useful for the explicit calculations, especially the tree-level calculation
of the sound velocities in appendix \ref{sub:General-structure-of-solutions}.

~

\subsubsection{Superfluid density and entrainment for all temperatures\label{sub:Superfluid-density-and-entrainment-for-allT}}

~

\noindent 
\begin{figure}[t]
\fbox{\begin{minipage}[t]{1\columnwidth}%
\begin{center}
\includegraphics[clip,scale=0.73]{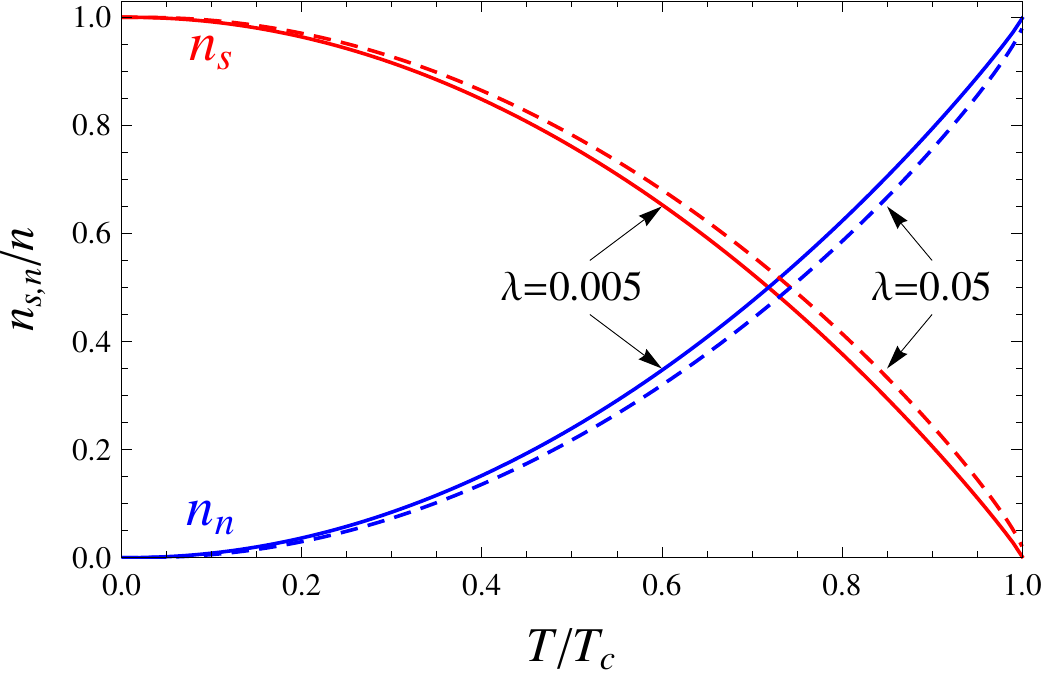}~~~~~\includegraphics[bb=0bp 0bp 360bp 195bp,scale=0.6]{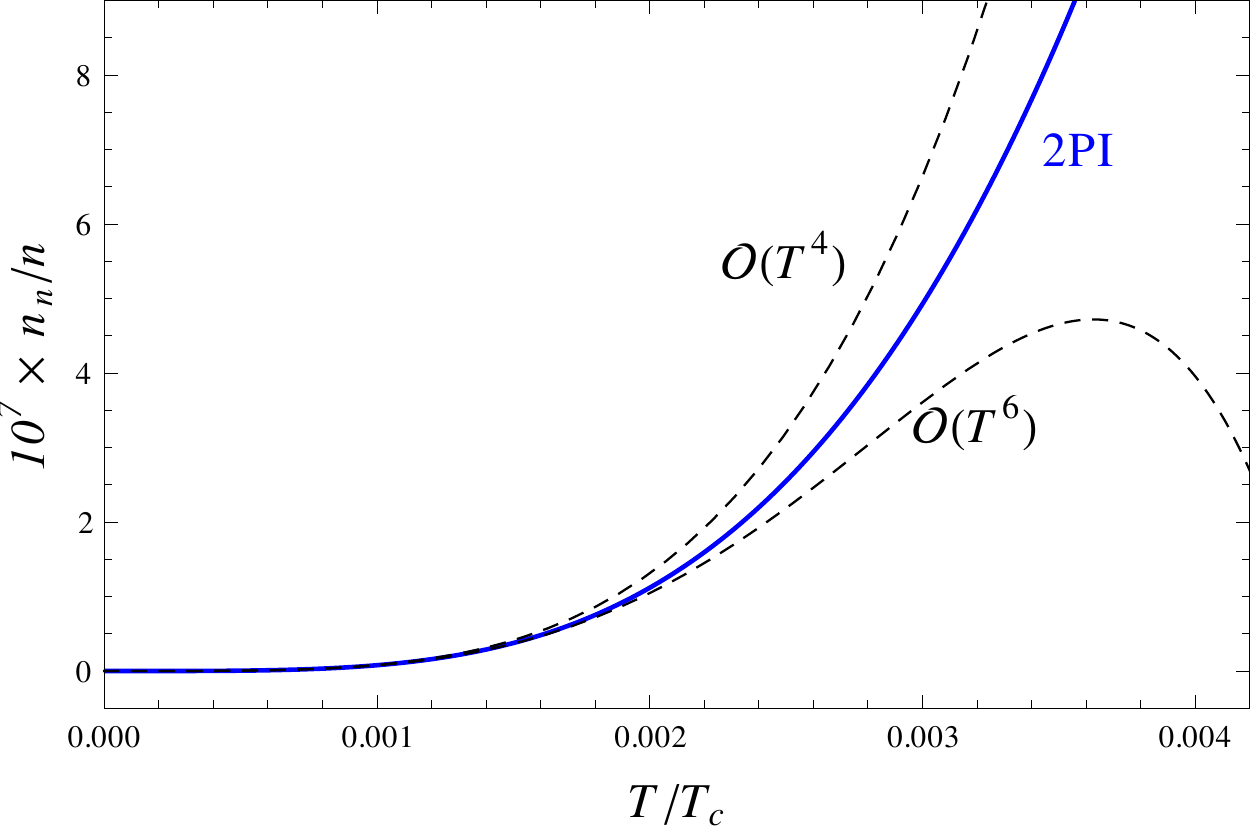}
\par\end{center}%
\end{minipage}}\protect\caption{Left panel: superfluid and normal-fluid charge densities, normalized
by the total charge density $n$, as a function of temperature for
all temperatures up to the critical temperature and two different
couplings $\lambda=0.005$ (solid lines) and $\lambda=0.05$ (dashed
lines). Since different couplings lead to different critical temperatures,
a given point on the horizontal axis $T/T_{c}$ corresponds to different
\textit{absolute} temperatures $T$ for solid and dashed lines. Right
panel: comparison of the full 2PI calculation with the analytical
low-temperature approximations from equations (\ref{eq:nsLowT}),
(\ref{eq:nnlowT}) for $\lambda=0.005$. We have set the superflow
and the mass parameter to zero, $|\vec{\nabla}\psi|=m=0$.\label{fig:SuperflAllT}}
\end{figure}

\noindent We shall now apply the algorithm introduced in the last
section to calculate the superfluid density as well as the coefficients
$\overline{{\cal A}}$, $\overline{{\cal B}}$, $\overline{{\cal C}}$
. The superfluid and normal-fluid densities for all temperatures up
to the critical temperature are shown in figure \ref{fig:SuperflAllT}.
Here we consider the case without superflow. As expected, the superfluid
density is identical to the total density at $T=0$ and decreases
monotonically with the temperature until it goes to zero continuously
at the critical temperature. The plot shows the densities for two
different values of the coupling constant. Different coupling strengths
lead to different critical temperatures. In the given plot, $T_{c}\simeq24.5\,\mu$
for the weaker of the two chosen couplings, $\lambda=0.005$, while
$T_{c}\simeq7.71\,\mu$ for the stronger coupling, $\lambda=0.05$
(the stronger the coupling, the stronger the repulsive force between
the bosons and hence the lower the critical temperature). Therefore,
the \textit{absolute} value of the temperature is different for the
two curves at a given point on the horizontal axis. This has to be
kept in mind for all following plots. We see that the stronger coupling
tends to favor the superfluid component, i.e., for a given \textit{relative}
temperature with respect to $T_{c}$ an increase of the coupling leads
to a (small) increase of the superfluid density fraction. 

\noindent In the right panel of the figure we compare the full 2PI
result with the tree-level approximation for low temperatures from
(\ref{eq:nsLowT}), (\ref{eq:nnlowT}). We have plotted the curves
where the expansion is truncated at order $T^{4}$ and where it is
truncated at order $T^{6}$. It is already clear from the comparison
of these two truncations that the series in $T$ converges very slowly.
Both truncations are only good approximations to the full result for
very low temperatures compared to the critical temperature, in this
case for $T\lesssim0.002\, T_{c}$. 

\noindent 
\begin{figure}[t]
\fbox{\begin{minipage}[t][0.15\paperheight]{1\columnwidth}%
\begin{center}
\includegraphics[bb=0bp 0bp 300bp 238bp,clip,scale=0.73]{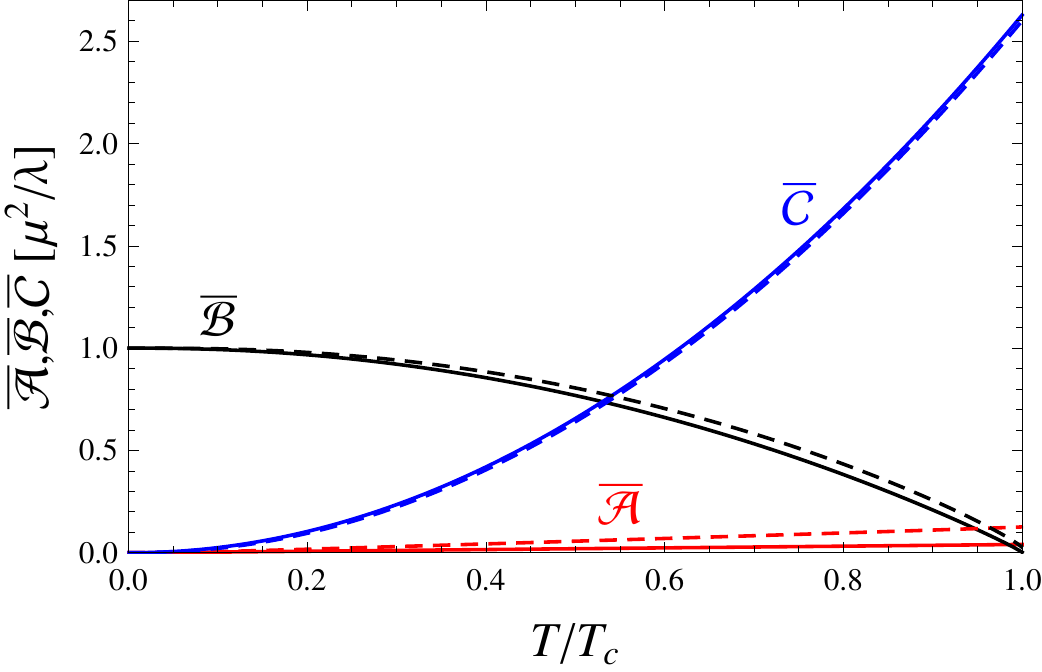}~~~~~\includegraphics[scale=0.59]{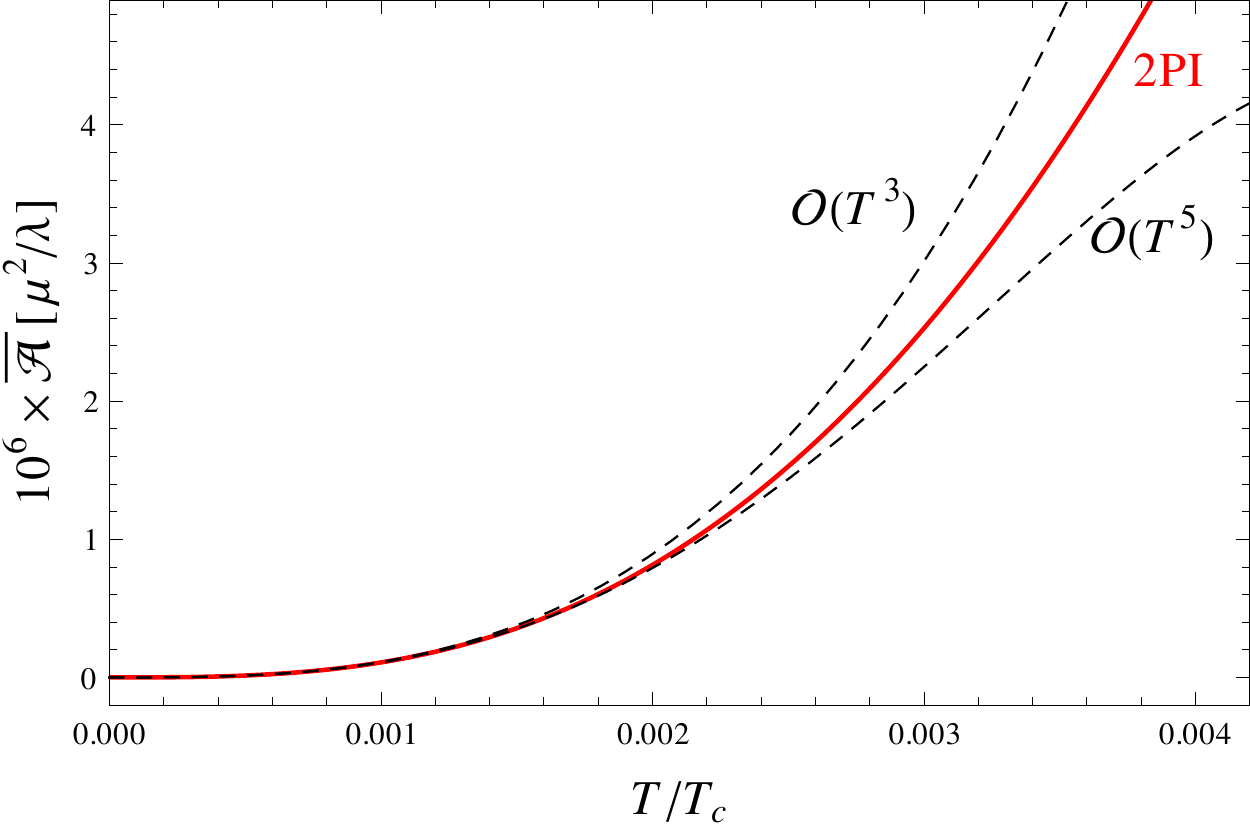}
\par\end{center}%
\end{minipage}}

\protect\caption{Left panel: coefficients $\overline{{\cal A}}$, $\overline{{\cal B}}$,
$\overline{{\cal C}}$ of the two-fluid formalism in units of $\mu^{2}/\lambda$
as a function of temperature for the same two couplings as in figure
\ref{fig:SuperflAllT}, $\lambda=0.005$ (solid lines) and $\lambda=0.05$
(dashed lines). Right panel: entrainment coefficient $\overline{{\cal A}}$
for low temperatures and comparison with the analytical results from
equation (\ref{eq:AbarlowT}) for $\lambda=0.005$. As in figure \ref{fig:SuperflAllT},
$|\vec{\nabla}\psi|=m=0$. \label{fig:ABC}}
\end{figure}

\noindent Next we compute the coefficients $\overline{{\cal A}}$,
$\overline{{\cal B}}$, $\overline{{\cal C}}$ and plot them in figure
\ref{fig:ABC}. Again, we have chosen the same two coupling strengths
as in figure \ref{fig:SuperflAllT}. We have normalized the coefficients
not only by dividing by $\mu^{2}$ (such that they become dimensionless),
but also by multiplying with a factor $\lambda$ such that the normalized
$\overline{{\cal B}}$ is 1 at zero temperature for all couplings,
which makes it easier to compare different couplings in a single plot.
At zero temperature, $\overline{{\cal A}}=\overline{{\cal C}}=0$,
which implies that there is no entropy current, $s^{\mu}=0$, as expected,
and we have a single-fluid system. At finite temperature, both currents
become nonzero and we have a two-fluid system. 

\noindent The dependence on the coupling seems to be relatively weak
for $\overline{{\cal B}}$, $\overline{{\cal C}}$, while the entrainment
coefficient $\overline{{\cal A}}$ increases significantly with the
coupling. We have checked that, for the case of the weaker coupling
$\lambda=0.005$, $\overline{{\cal A}}$ behaves linearly in the temperature
for all temperatures $T\gtrsim0.5\, T_{c}$. The low-temperature results
are given in equations (\ref{eq:AbarlowT}), (\ref{eq:BbarlowT})
and (\ref{eq:CbarlowT}). In the right panel of figure \ref{fig:ABC}
we compare the analytical low-temperature approximation for $\overline{{\cal A}}$
with the full result. As for the superfluid and normal-fluid densities
we see that we have to zoom in to very low temperatures compared to
$T_{c}$ in order to find agreement between the approximation and
the full result. 

~

\subsection{Conclusion\label{sub:Conclusion2}}

~

\noindent In this part, we have extended our study of the properties
of a bosonic relativistic superfluid to all temperatures below the
critical temperature within the 2PI formalism. These studies build
on the connection between field theory and the two-fluid picture that
we have developed in sections \ref{sec:Lagrangian}, \ref{sec:Zero-temperature:-Single-fluid}
and \ref{sec:Finite-temperature:-Two-fluid}. We shall now summarize
formal aspects of the 2PI approach which we had to deal with such
as renormalization as well as the physical results obtained within
that approach. 

~
\begin{itemize}
\item \textit{Formalism. }Even though the 2PI formalism is well suited to
the treatment of systems with spontaneous symmetry breaking, in practice
it has several difficulties, and we now describe how we have addressed
them. Firstly, the renormalization of the theory is nontrivial because
there are ultraviolet divergences in the action and stationarity equations
which implicitly depend on the medium through the self-consistent
masses. Presumably such unwanted dependences would be absent in a
more complete treatment that takes into account the momentum dependence
of the order parameter. We follow the approach adopted in the existing
literature, introducing counterterms on the level of the effective
action to achieve renormalizability. We have pointed out an additional
ultraviolet divergence in this approach, arising from nonzero superflow.\\
Secondly, the two-loop truncation of the 2PI effective action violates
the Goldstone theorem by giving a small mass to the Goldstone mode.
In the physics of a superfluid, however, the masslessness of the Goldstone
mode is crucial since it determines the low-energy properties of the
system. We have therefore built the Goldstone theorem into our calculation
by hand, using a modification of the stationarity equations. This
means that we do not work at the minimum of the potential, but at
a point slightly away from that minimum. In particular, we have evaluated
the effective action at that ``Goldstone point''.\\
Thirdly, we have employed the Hartree approximation, meaning that
we have neglected the exchange contribution to the effective action
from the cubic interactions that are induced by the condensate. This
approximation is particularly simple since the self-energy is then
momentum-independent. The price one has to pay, however, is that the
phase transition to the non-superfluid phase becomes first order,
while a complete treatment predicts a second order phase transition.
We control this problem by restricting our calculation to weak coupling,
in which case the unphysical discontinuity of the order parameter
at the critical point is small, as is the sensitivity of our final
results to the arbitrary renormalization scale.
\item \textit{Physical results.} We have analyzed Landau`s critical velocity
for superfluidity within the 2PI formalism. The critical velocity
manifests itself through the onset of an instability (negative energy)
in the dispersion relation of the Goldstone mode. We have computed
the critical velocity for all temperatures. At low temperatures, our
critical velocity is in agreement with the original version of Landau's
argument, which is based on a Lorentz (or Galilei) transformation
of the dispersions at vanishing superflow. In general, however, the
Goldstone dispersion at finite superflow is not just obtained by a
Lorentz transformation. A superflow also affects the condensate which
in turn influences the dispersion relation. This effect is taken into
account in our self-consistent formalism and turns out to decrease
the critical velocity sizably at intermediate temperatures. As a result
of this calculation, we have presented a phase diagram in the plane
of temperature and superfluid velocity. This phase diagram is incomplete
in the sense that we have restricted ourselves to homogeneous phases.
In particular, we have not constructed a superfluid phase for velocities
beyond the critical one.\\
We have computed the parameters of Landau`s two fluid formalism (i.e.
superfluid and normal-fluid densities) as well as of the generalized
hydrodynamical approach based on the conserved charge and entropy
current. Most notably is the entrainment coefficient, which expresses
the degree to which each current responds to the conjugate momentum
originally associated with the other current. We have seen that the
entrainment between the currents becomes larger with temperature and
is also increased significantly by increasing the microscopic coupling
$\lambda$. 
\end{itemize}
\newpage{}

\part{Sound modes in relativistic superfluids\label{part:Sound-modes-in-relSF}}

We shall now investigate one of the most striking consequences of
Landau`s two-fluid model: the existence of additional sound excitations
in a superfluid. As discussed in the introduction, the two-fluid equations
not only predict an ``ordinary'' sound wave which is basically an
oscillation in density (or chemical potential), but also oscillations
in temperature (or entropy) which have been termed second sound. Starting
from the conservation equations of the two-fluid model, we shall first
explain how to derive the wave equations which determine the speeds
of first and second sound. 

\noindent In a second step, we will use the low-temperature results
of the hydrodynamic parameters to describe the (very) low-temperature
behavior of the sound velocities and obtain explicit results which
depend on the temperature $T$, chemical potential $\mu$ and the
background superflow with velocity $\vec{v}_{s}$. Related calculations
can be found in the recent literature in the non-relativistic context
of superfluid atomic gases \cite{Taylor2009,Arahata2009,Hu2010,Bertaina2010},
where the experimental observation of both sound modes is in principle
possible, although challenging \cite{Meppelink2009,Arahata2011,Leonid2013}.
Our results will give the sound velocities in the presence of an arbitrary
superflow, i.e., an arbitrary relative velocity between the superfluid
and the normal fluid (limited by a critical velocity, as our results
will show). In particular, they will depend on the angle between the
direction of the sound wave and the direction of the superflow. A
similar calculation in the non-relativistic context of superfluid
helium has been performed in \cite{Adamenko2009} where, in contrast
to our calculation, the sound velocities are computed in the superfluid
rest frame and without temperature corrections. We will find temperature
corrections to the velocity of second sound, which, as we shall see,
arise from the cubic terms in momentum of the dispersion relation
of the Goldstone mode. We will also be able to confirm Landau`s prediction
that the speed of second sound $u_{2}$ approaches the limit of $u_{2}=u_{1}/\sqrt{3}$
at zero temperature. 

\noindent Finally we will make use of the numerical results obtained
from the 2PI formalism to calculate sound velocities for any temperature
up to $T_{c}$. In particular we will use the algorithm described
in section \ref{sub:Algorithm} and the limit of vanishing superflow.
In the frame of these calculations we will encounter a rather surprising
result: with increasing temperature, a pure density wave can transform
into a pure temperature wave and the other way around. We will term
this phenomenon \textit{role reversal }and discuss it in detail in
section \ref{sec:Sound-modes-for-allT}. 

\newpage{}

\section{Derivation of the wave equations\label{sec:Derivation-of-the-wavEQ}}

~

\noindent The sound wave equations are derived from the hydrodynamic
equations. We start from equations (\ref{eq:ConCurrents}) and (\ref{eq:RelEuler}),
that is, from the current and entropy conservation and the vorticity
equation. {[}Of course, equivalently, one can start from the current
conservation plus energy-momentum conservation.{]} In addition, we
need the expression for $d\Psi$ from (\ref{eq:VariationalMaster}),
which will allow us to rewrite derivatives of thermodynamic quantities
in terms of derivatives of the independent variables. These are the
chemical potential $\mu=\partial^{0}\psi$, the superfluid three-velocity
$\vec{v}_{s}$ (more precisely, $\vec{\nabla}\psi=-\mu\vec{v}_{s}$),
the temperature $T=\Theta^{0}$, and the normal-fluid three-velocity
$\vec{v}_{n}$. All these variables are now allowed to exhibit small
oscillations in space and time about their equilibrium values, $T\to T+\delta T(\vec{x},t)$,
$\mu\to\mu+\delta\mu(\vec{x},t)$. We perform the calculation in the
normal-fluid rest frame from the previous sections, i.e., the superfluid
velocity has a (large) static and homogeneous equilibrium value on
top of which the sound wave oscillations occur, $\vec{v}_{s}\to\vec{v}_{s}+\delta\vec{v}_{s}(\vec{x},t)$,
while the static and homogeneous part of the normal velocity can be
set to zero, $\vec{v}_{n}\to\delta\vec{v}_{n}(\vec{x},t)$. Of course,
we need to keep the oscillations of the normal-fluid velocity $\delta\vec{v}_{n}(\vec{x},t)$
because there is no global rest frame in which they vanish. 

\noindent We employ the linear approximation in the oscillations.
In this case, the temporal component of the vorticity equation is
trivially fulfilled. From the remaining equations one can eliminate
the normal velocity, such that one is left with two equations where
the sound wave oscillations are solely expressed in terms of oscillations
in $T$ and $\mu$ (oscillations of the superfluid velocity $\vec{\nabla}\psi$
can be expressed in terms of oscillations of $\mu$ by applying a
time derivative to the whole equation and using $\partial_{0}\vec{\nabla}\psi=\vec{\nabla}\mu$).
The derivation of the wave equations is quite lengthy, and we explain
the details in appendix \ref{sub:Derivation-of-the-waveEQ}. One obtains
the following system of two equations,

\noindent ~

\noindent 
\begin{eqnarray}
0 & \simeq & \frac{w}{s}\left(\frac{\partial n}{\partial T}\partial_{0}^{2}\mu+\frac{\partial s}{\partial T}\partial_{0}^{2}T\right)-n_{n}\Delta\mu-s\Delta T\label{eq:soundEq1}\\
 & + & \left[\frac{n_{s}}{\sigma}-\frac{w}{s}\frac{\partial(n_{s}/\sigma)}{\partial T}+\frac{n_{n}}{s}\frac{\partial n}{\partial T}-\frac{\partial n}{\partial\mu}-2\mu\frac{\partial n}{\partial(\vec{\nabla}\psi)^{2}}\right]\vec{\nabla}\psi\cdot\vec{\nabla}\partial_{0}\mu\nonumber \\
 & + & \left[\frac{n_{n}}{s}\frac{\partial s}{\partial T}-\frac{\partial s}{\partial\mu}-2\mu\frac{\partial s}{\partial(\vec{\nabla}\psi)^{2}}\right]\vec{\nabla}\psi\cdot\vec{\nabla}\partial_{0}T\\
 & - & \left[\frac{n_{n}}{s}\frac{\partial(n_{s}/\sigma)}{\partial T}-\frac{\partial(n_{s}/\sigma)}{\partial\mu}-2\mu\frac{\partial(n_{s}/\sigma)}{\partial(\vec{\nabla}\psi)^{2}}\right](\vec{\nabla}\psi\cdot\vec{\nabla})^{2}\mu\,,\nonumber 
\end{eqnarray}

\noindent ~
\begin{eqnarray}
0 & \simeq & \left(\mu\frac{\partial n}{\partial\mu}+T\frac{\partial n}{\partial T}\right)\partial_{0}^{2}\mu+\left(\mu\frac{\partial s}{\partial\mu}+T\frac{\partial s}{\partial T}\right)\partial_{0}^{2}T-n\Delta\mu-s\Delta T\label{eq:soundEq2}\\
 & + & \left[\frac{n_{s}}{\sigma}-\mu\frac{\partial(n_{s}/\sigma)}{\partial\mu}-T\frac{\partial(n_{s}/\sigma)}{\partial T}+\frac{n_{n}}{s}\frac{\partial n}{\partial T}-\frac{\partial n}{\partial\mu}\right]\vec{\nabla}\psi\cdot\vec{\nabla}\partial_{0}\mu+\left(\frac{n_{n}}{s}\frac{\partial s}{\partial T}-\frac{\partial s}{\partial\mu}\right)\vec{\nabla}\psi\cdot\vec{\nabla}\partial_{0}T\nonumber \\
 & - & \left[\frac{n_{n}}{s}\frac{\partial(n_{s}/\sigma)}{\partial T}-\frac{\partial(n_{s}/\sigma)}{\partial\mu}\right](\vec{\nabla}\psi\cdot\vec{\nabla})^{2}\mu\,,\nonumber 
\end{eqnarray}

\noindent where $w\equiv\mu n_{n}+sT$ is the enthalpy density of
the normal fluid. Each term is a product of a space-time derivative
- in which we can replace $T$ by $\delta T(\vec{x},t)$ and $\mu$
by $\delta\mu(\vec{x},t)$ - and a prefactor that only contains the
equilibrium values $T$, $\mu$, and $\vec{\nabla}\psi=-\mu\vec{v}_{s}$.
Before discussing results, let us write down the wave equations in
two limit cases. Firstly, let us set $T=0$. In this case, the normal
number density vanishes, $n_{n}=0$, and thus $n=n_{s}\mu/\sigma$.
With this relation and the zero-temperature expression $n_{s}=\sigma^{3}/\lambda$
from section \ref{sub:Zero-temperature-hydrodynamics:-Landau} (we
set $m=0$ for simplicity in this subsection) one finds that all terms
on the right-hand side of (\ref{eq:soundEq1}) vanish, and (\ref{eq:soundEq2})
can be compactly written as 

\bigskip{}

\noindent 
\begin{equation}
0\simeq(g^{\mu\nu}+2v^{\mu}v^{\nu})\partial_{\mu}\partial_{\nu}\mu\,.
\end{equation}
Again, we recover the sonic metric ${\cal G}^{\mu\nu}=g^{\mu\nu}+2v^{\mu}v^{\nu}$,
see remark below the generalized pressure (\ref{eq:InvarPressure}).
With $\delta\mu=\delta\mu_{0}e^{ik\cdot x}$ we obtain ${\cal G}^{\mu\nu}k_{\mu}k_{\nu}=0$,
which is equation (4.12) of \cite{CarterLanglois} (see also equation
(29) of reference \cite{MannarelliManuel}). This wave equation has
one physical solution $\omega=u_{1}|\vec{k}|$, with the velocity
of first sound $u_{1}$. The explicit solutions on the low-temperature
limit are given in (\ref{eq:solLOWT1}), (\ref{eq:solLOWT2}) (as
we shall see below, this solution is unaltered by temperature effects
up to the order we are working).

\noindent Secondly, we discuss the limit case without superflow, $\vec{\nabla}\psi=\vec{0}$.
In this case, only the first lines of equations (\ref{eq:soundEq1})
and (\ref{eq:soundEq2}) are nonvanishing. Now, with $\delta\mu=\delta\mu_{0}e^{i(\omega t-\vec{k}\cdot\vec{x})}$
and $\delta T=\delta T_{0}e^{i(\omega t-\vec{k}\cdot\vec{x})}$ we
obtain two equations for the two amplitudes $\delta\mu_{0}$, $\delta T_{0}$.
Since we are interested in nontrivial solutions, we need to require
the determinant of the coefficient matrix to vanish. After a bit of
algebra, using $n=n_{s}+n_{n}$ (which is true for $\vec{\nabla}\psi=\vec{0}$)
and $\frac{\partial n}{\partial T}=\frac{\partial s}{\partial\mu}$,
the resulting equation can be written as

\bigskip{}

\noindent 
\begin{equation}
0=\mu wT\left(\frac{\partial s}{\partial\mu}\frac{\partial n}{\partial T}-\frac{\partial n}{\partial\mu}\frac{\partial s}{\partial T}\right)\omega^{4}-n_{s}s^{2}T\,|\vec{k}|^{4}+\left[s^{2}\mu\frac{\partial n}{\partial\mu}+(\mu n_{n}^{2}+wn_{s})\frac{\partial s}{\partial T}-2\mu sn_{n}\frac{\partial s}{\partial\mu}\right]T\omega^{2}|\vec{k}|^{2}\,.\label{eq:SoundeqLowT}
\end{equation}
 This result is in exact agreement with the one given in equations
(19) - (22) of reference \cite{HerzigKovtun}. Now there are two physical
solutions, $\omega=u_{1,2}|\vec{k}|$, with the two sound velocities
$u_{1}$, $u_{2}$. The reason for the appearance of the second mode
is that the presence of the second fluid component allows for \textit{relative}
oscillations between the two fluids. As a check, one can confirm that
the solutions of equation (\ref{eq:SoundeqLowT}) are the $\vec{v}_{s}=0$
limit of the full (low-temperature) results (\ref{eq:solLOWT1}) and
(\ref{eq:solLOWT2}).

~

\subsection{General structure of the solutions\label{sub:General-structure-of-solutions}}

~

\noindent Before we explicitly calculate the speeds of sound for low
temperatures one can make interesting observations about the general
structure of the solutions which are obtained as follows: after replacing
$\mu$ and $T$ with their corresponding fluctuations $\delta\mu=\delta\mu_{0}e^{i(\omega t-\vec{k}\cdot\vec{x})}$,
$\delta T=\delta T_{0}e^{i(\omega t-\vec{k}\cdot\vec{x})}$ and introducing
the speed of sound $u=\omega/|\vec{k}|$, one can write equations
(\ref{eq:soundEq1}) and (\ref{eq:soundEq2}) compactly in the following
form

\noindent 
\begin{eqnarray}
0 & = & \Big[a_{1}u^{2}+(a_{2}+a_{4}|\vec{\nabla}\psi|^{2}\cos^{2}\theta)+a_{3}|\vec{\nabla}\psi|u\cos\theta\Big]\delta\mu_{0}+\Big(b_{1}u^{2}+b_{2}+b_{3}|\vec{\nabla}\psi|u\cos\theta\Big)\,\delta T_{0}\,,\nonumber \\
\label{eq:sound11}\\
0 & = & \Big[A_{1}u^{2}+(A_{2}+A_{4}|\vec{\nabla}\psi|^{2}\cos^{2}\theta)+A_{3}|\vec{\nabla}\psi|u\cos\theta\Big]\delta\mu_{0}+\Big(B_{1}u^{2}+B_{2}+B_{3}|\vec{\nabla}\psi|u\cos\theta\Big)\,\delta T_{0}\,,\nonumber \\
\label{eq:sound12}
\end{eqnarray}
The coefficients of this system of equations are complicated combinations
of first and second derivatives of the pressure (compare to equations
(\ref{eq:soundEq1}), (\ref{eq:soundEq2}) ) and are defined in appendix
\ref{sub:Solution-of-waveEQ}. Requiring the equations (\ref{eq:sound11})
and (\ref{eq:sound12}) to have nontrivial solutions for $\delta\mu_{0}$,
$\delta T_{0}$ yields a quartic equation for $u$ with four solutions,
two of which are physical (i.e. real an positive), the velocities
of first and second sound $u_{1}$ and $u_{2}$. Nevertheless, there
are more possible sound waves in a superfluid. They can be found by
starting from a certain subset of the conservation equations. For
instance the so-called fourth sound \cite{Atkins1959} can be excited
by fixing the normal fluid by an external force. It is thus calculated
after dropping momentum conservation \cite{Khalatnikov}\cite{HerzogYaron}.
We shall not be concerned with these solutions in the following.

\noindent We can now discuss on a general level whether first and
second sound are basically a propagation of a heat or a density wave.
To do so, we calculate the ratio of the amplitudes 

\bigskip{}

\noindent 
\begin{equation}
\frac{\delta T_{0}}{\delta\mu_{0}}=-\frac{a_{1}u^{2}+(a_{2}+a_{4}|\vec{\nabla}\psi|^{2}\cos^{2}\theta)+a_{3}|\vec{\nabla}\psi|u\cos\theta}{b_{1}u^{2}+b_{2}+b_{3}|\vec{\nabla}\psi|u\cos\theta}\,.
\end{equation}
For each sound mode, the one-dimensional space of solutions of equations
(\ref{eq:sound11}), (\ref{eq:sound12}) is a straight line through
the origin in the $\delta\mu_{0}$-$\delta T_{0}$ plane. It is convenient
to define the angle of that line with the $\delta\mu_{0}$ axis,

\noindent 
\begin{equation}
\alpha\equiv\arctan\frac{\delta T_{0}}{\delta\mu_{0}}\,.\label{eq:MixAngel}
\end{equation}
The sign of this angle tells us whether chemical potential and temperature
oscillate in phase ($\alpha>0$) or out of phase ($\alpha<0$). The
magnitude of $\alpha$ characterizes the mixture of oscillations in
temperature and chemical potential with $\alpha=0$ corresponding
to a pure oscillation in chemical potential and $|\alpha|=\pi/2$
to a pure oscillation in temperature. We shall investigate this ratio
in section \ref{sec:Sound-modes-for-allT} for all temperatures with
rather surprising results. We can also translate this into the amplitudes
in density and entropy. With the help of the thermodynamic relation
for the pressure $P$

\bigskip{}

\noindent 
\begin{equation}
dP=nd\mu+sdT-\frac{n_{s}}{\sigma}\vec{\nabla}\psi\cdot d\vec{\nabla}\psi\,,
\end{equation}
 we can derive (see also appendix \ref{sub:Solution-of-waveEQ}) a
similar expression for the amplitude ratio of $\delta n_{0}/\delta s_{0}$

\bigskip{}

\noindent 
\begin{equation}
\frac{\delta n_{0}}{\delta s_{0}}=\left[\frac{\partial n}{\partial\mu}+\frac{|\vec{\nabla}\psi|\cos\theta}{u}\frac{\partial(n_{s}/\sigma)}{\partial\mu}+\frac{\partial s}{\partial\mu}\frac{\delta T_{0}}{\delta\mu_{0}}\right]\left[\frac{\partial n}{\partial T}+\frac{|\vec{\nabla}\psi|\cos\theta}{u}\frac{\partial(n_{s}/\sigma)}{\partial T}+\frac{\partial s}{\partial T}\frac{\delta T_{0}}{\delta\mu_{0}}\right]^{-1}\,.
\end{equation}
In general, the sound modes and the corresponding amplitudes are very
complicated. Let us therefore begin with a discussion the case of
vanishing superflow, $|\vec{\nabla}\psi|\to0$. In this case, the
coefficients $a_{3}$, $a_{4}$, $b_{3}$, $A_{3}$, $A_{4}$, $B_{3}$
become irrelevant (for the explicit form of the wave equation in the
limit $|\vec{\nabla}\psi|\to0$ see (\ref{eq:SoundeqLowT})), and 

\noindent 
\begin{eqnarray}
a_{1} & = & \frac{w}{s}\frac{\partial n}{\partial T}\,,\qquad a_{2}=-n_{n}\,,\qquad b_{1}=\frac{w}{s}\frac{\partial s}{\partial T}\,,\qquad b_{2}=-s\,,\\
A_{1} & = & \mu\frac{\partial n}{\partial\mu}+T\frac{\partial n}{\partial T}\,,\qquad A_{2}=-n\,,\qquad B_{1}=\mu\frac{\partial s}{\partial\mu}+T\frac{\partial s}{\partial T}\,,\qquad B_{2}=-s\,.\nonumber 
\end{eqnarray}
We thus have the following simple quadratic equation for $u^{2}$, 

\bigskip{}

\noindent 
\begin{equation}
0=(a_{1}u^{2}+a_{2})(B_{1}u^{2}+B_{2})-(A_{1}u^{2}+A_{2})(b_{1}u^{2}+b_{2})\,.\label{eq:noSLsol}
\end{equation}
 It is instructive to solve this equation in the limit where there
are no other energy scales than $\mu$ and $T$. In our context, this
will be the case when we set the supercurrent and the mass parameter
to zero, $\vec{\nabla}\psi=m=0$. Then, we can write the pressure
as $\Psi=T^{4}h(T/\mu)$ with a dimensionless function $h$, and the
sound velocities assume a simple form \cite{HerzigKovtun}. The reason
is that now there are simple relations between first and second derivatives
of the pressure, for instance we find $A_{1}=3n$, $B_{1}=3s$. Then,
one computes the following two solutions of equation (\ref{eq:noSLsol})
for $u^{2}$,

\bigskip{}

\noindent 
\begin{equation}
\qquad u_{1}^{2}=\frac{1}{3}\,,\qquad u_{2}^{2}=\frac{n_{s}s^{2}}{w}\left(n\frac{\partial s}{\partial T}-s\frac{\partial n}{\partial T}\right)^{-1}\,.\label{eq:soundConf}
\end{equation}
We see that one solution is constant while the other depends on the
thermodynamic details of the 

\noindent ~

\noindent system. The ratios of the amplitudes become particularly
simple in this limit. We find

\bigskip{}

\noindent 
\begin{equation}
\left.\frac{\delta T_{0}}{\delta\mu_{0}}\right|_{u_{1}}=-\left.\frac{\delta n_{0}}{\delta s_{0}}\right|_{u_{2}}=\frac{T}{\mu}\,,\qquad\left.\frac{\delta n_{0}}{\delta s_{0}}\right|_{u_{1}}=-\left.\frac{\delta T_{0}}{\delta\mu_{0}}\right|_{u_{2}}=\frac{n}{s}\,.\label{eq:ConfRat}
\end{equation}
This result shows that, for a given pair of amplitudes, $\delta T_{0}$
and $\delta\mu_{0}$ or $\delta n_{0}$ and $\delta s_{0}$, first
sound is always an in-phase oscillation while second sound is always
an out-of-phase oscillation. Moreover, we can make an interesting
observation regarding the magnitude of the amplitudes. At $T=0$,
where also $s=0$, first sound is a pure chemical potential (and pure
density) wave, while second sound is a pure temperature (and pure
entropy) wave. This is no longer true for nonzero temperatures. If
at the critical temperature $T\gg\mu$ and $s\gg n$, the roles of
first and second sound completely reverse upon heating the superfluid
from $T=0$ to $T=T_{c}$. We shall discuss this role reversal in
more detail when we present our numerical results in section \ref{sec:Sound-modes-for-allT}. 

~

\section{The low-temperature approximation\label{sec:The-low-temperature-approx}}

~

\noindent In general, the full wave equations (\ref{eq:sound11})
and (\ref{eq:sound12}) yield very complicated results for the sound
velocities. However, in our approximation for low temperatures up
to order $T^{6}$ in the pressure, one can show that the resulting
quartic equation for $\omega$ factorizes into two quadratic equations.
This is explained in detail in appendix \ref{sub:Solution-of-waveEQ}.
In this appendix we also explain that our truncation of the low-temperature
series does not allow us to compute temperature corrections to the
sound velocities of order $T^{4}$ and higher. The $T^{2}$ corrections,
however, \textit{can} be reliably determined. As one can see from
equations (\ref{eq:Quadsol1}), (\ref{eq:quadsol2}) this is possible
because of the $T^{6}$ terms in the pressure which originate from
the $|\vec{k}|^{3}$ term in the dispersion of the Goldstone mode.
It turns out that there is a $T^{2}$ correction only to the second
sound $u_{2}$. The explicit results are 

\begin{eqnarray}
u_{1} & = & \frac{\sqrt{3-\vec{v}{}_{s}^{2}(1+2\cos^{2}\theta)}\sqrt{1-\vec{v}_{s}^{2}}+2|\vec{v}_{s}|\cos\theta}{3-\vec{v}_{s}^{2}}+{\cal O}(T^{4})\,,\label{eq:solLOWT1}\\
u_{2} & = & \frac{\sqrt{9(1-\vec{v}_{s}^{2})(1-3\vec{v}_{s}^{2})+\vec{v}_{s}^{2}\cos^{2}\theta}+|\vec{v}_{s}|\cos\theta}{9(1-\vec{v}_{s})}\label{eq:solLOWT2}\\
 & + & \frac{4}{63}\left(\frac{\pi T}{\mu}\right)^{2}\left[\frac{9(5-4\vec{v}_{s}^{2}-46\vec{v}_{s}^{4}+36\vec{v}_{s}^{6}+9\vec{v}_{s}^{8})-4(5-2\vec{v}_{s}^{2}-15\vec{v}_{s}^{4})\vec{v}_{s}^{2}\cos^{2}\theta}{(1-\vec{v}_{s}^{2})(1-3\vec{v}_{s}^{2})^{3}\sqrt{9(1-\vec{v}_{s}^{2})(1-3\vec{v}_{s}^{2})+\vec{v}_{s}^{2}\cos^{2}\theta}}\right.\nonumber \\
 &  & \,\,\,\,\,\,\,\,\,\,\,\,\,\,\,\,\,\,\,\,\,\,\,\,\,\,\,\,\,\,\,\,\left.-\frac{4(5-2\vec{v}_{s}^{2}-15\vec{v}_{s}^{2})|\vec{v}_{s}|\cos\theta}{(1-\vec{v}_{s}^{2})(1-3\vec{v}_{s}^{2})^{3}}\right]+{\cal O}(T^{4})\,,\nonumber 
\end{eqnarray}
 where $\theta$ is the angle between $\vec{v}_{s}$ and the direction
of the sound wave given by the wave vector $\vec{k}$. As a consistency
check, we confirm that $u_{1}$ is the ($m=0$ limit of the) linear
part of the dispersion of the Goldstone mode from (\ref{eq:DispGold}),
which was computed as one of the poles of the propagator. In the low-temperature
approximation, this dispersion does not depend on temperature since
the melting of the condensate has been taken into account (see section
\ref{sub:The-finite-temperature-setup}). 

\noindent The velocity of second sound $u_{2}$ becomes complex for
certain angles $\theta$ as soon as $|\vec{v}_{s}|>1/\sqrt{3}$. This
value corresponds to Landau`s critical velocity at zero temperature,
see also figure \ref{fig:PhaseDiagCrit}. Moreover, the $T^{2}$ term
of $u_{2}$ is divergent as $|\vec{v}_{s}|$ approaches $1/\sqrt{3}$.
We have seen in table 2 that all components of the stress-energy tensor
and the current exhibit this divergence too. The expressions in that
table show that due to this divergence the $T^{6}$ term (say, in
the energy density $T^{00}$) becomes comparable to or even larger
than the $T^{4}$ term for superfluid velocities close to (and below)
the zero temperature value of the critical velocity $1/\sqrt{3}$,
even if $T$ is very small. This suggests that a calculation to all
orders in $T$ must be performed to predict reliably the behavior
in this close-to-critical regime. In analogy, for the speed of second
sound close to the critical velocity and at nonzero temperatures we
also need the resummed result, and we cannot trust the truncated expression. 

\begin{figure}
\fbox{\begin{minipage}[t]{1.03\columnwidth}%
\begin{center}
\includegraphics[scale=0.582]{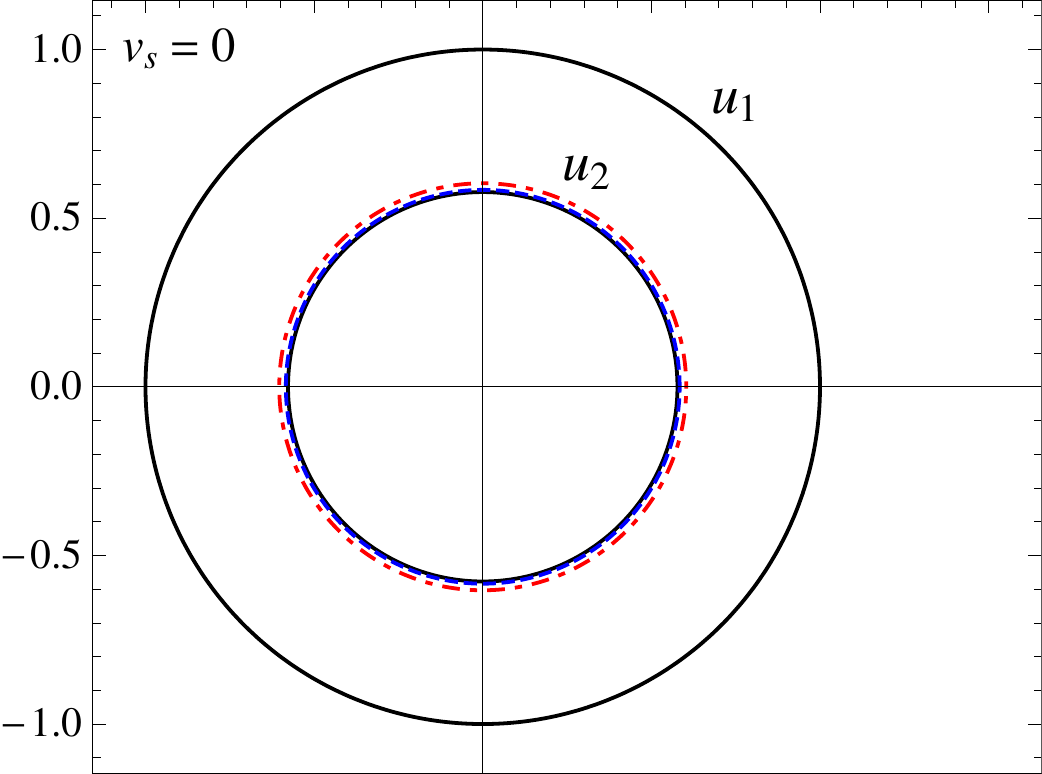}\includegraphics[scale=0.53]{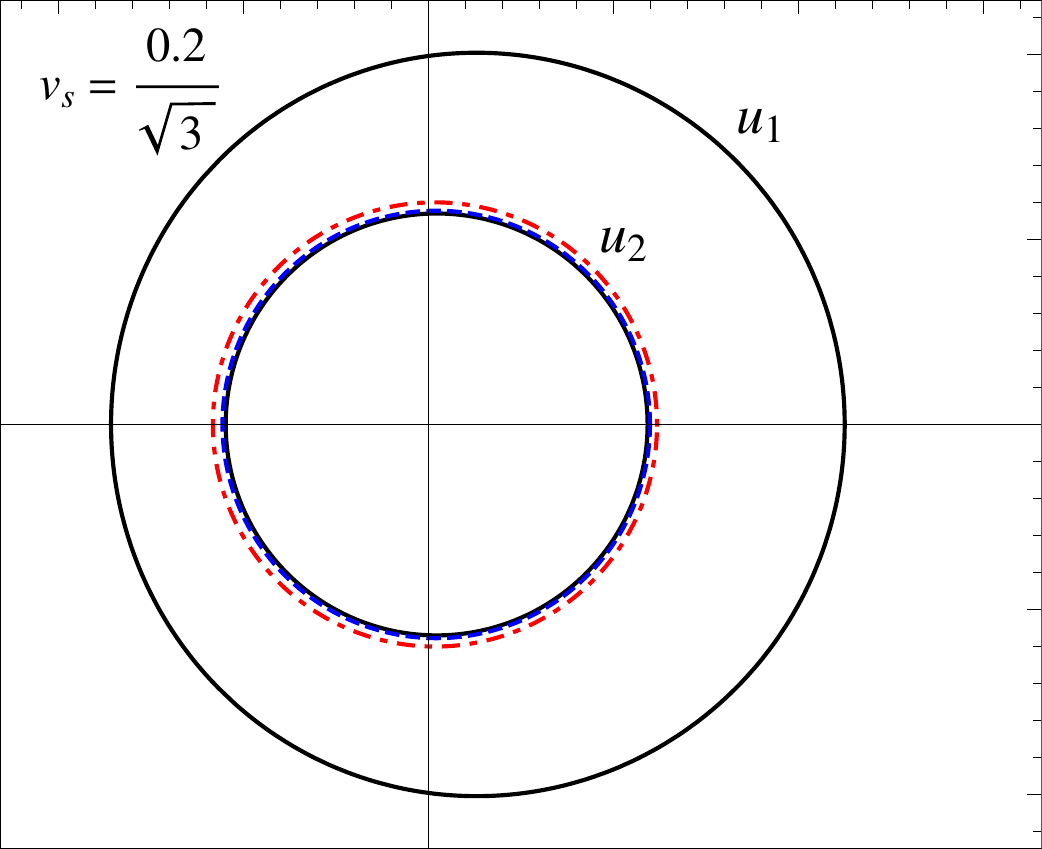}\includegraphics[scale=0.53]{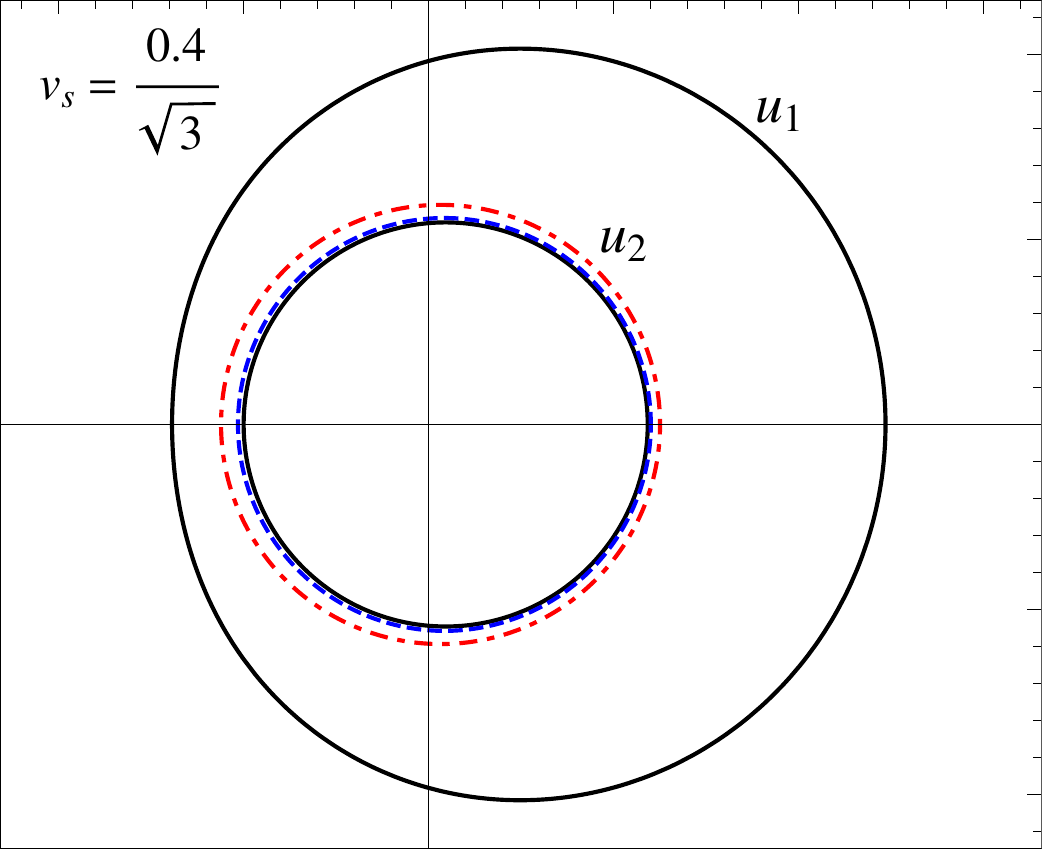}\\
\includegraphics[scale=0.581]{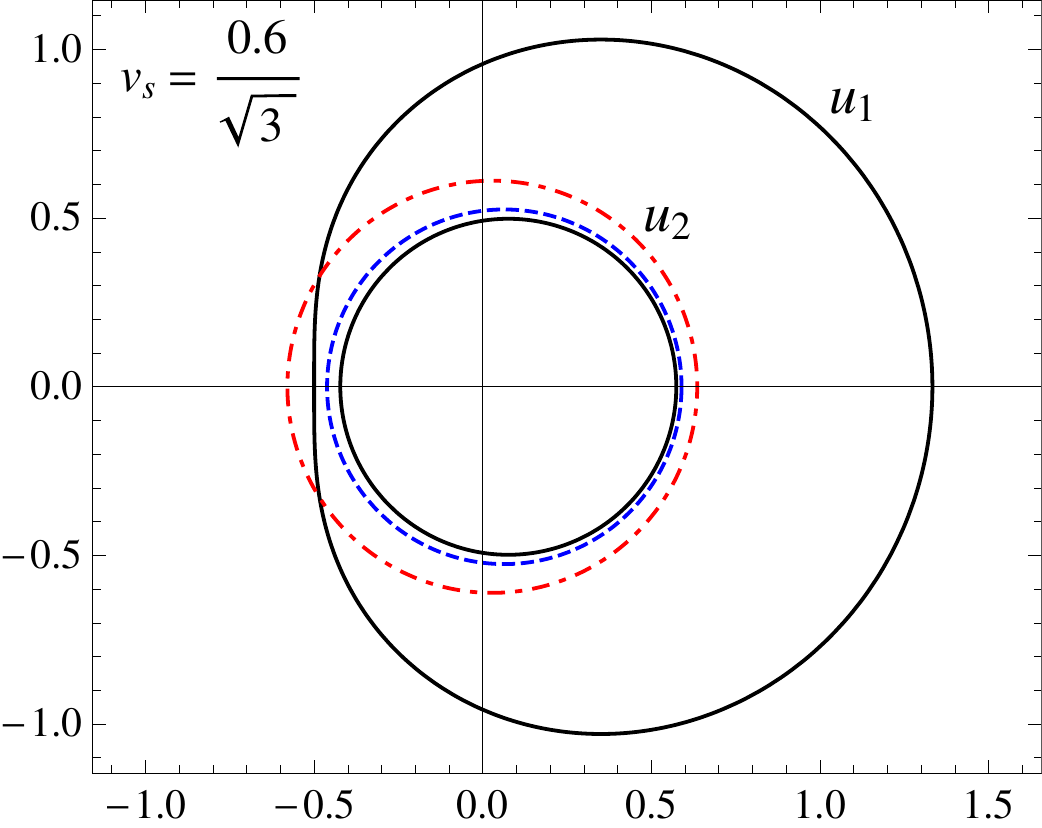}\includegraphics[scale=0.53]{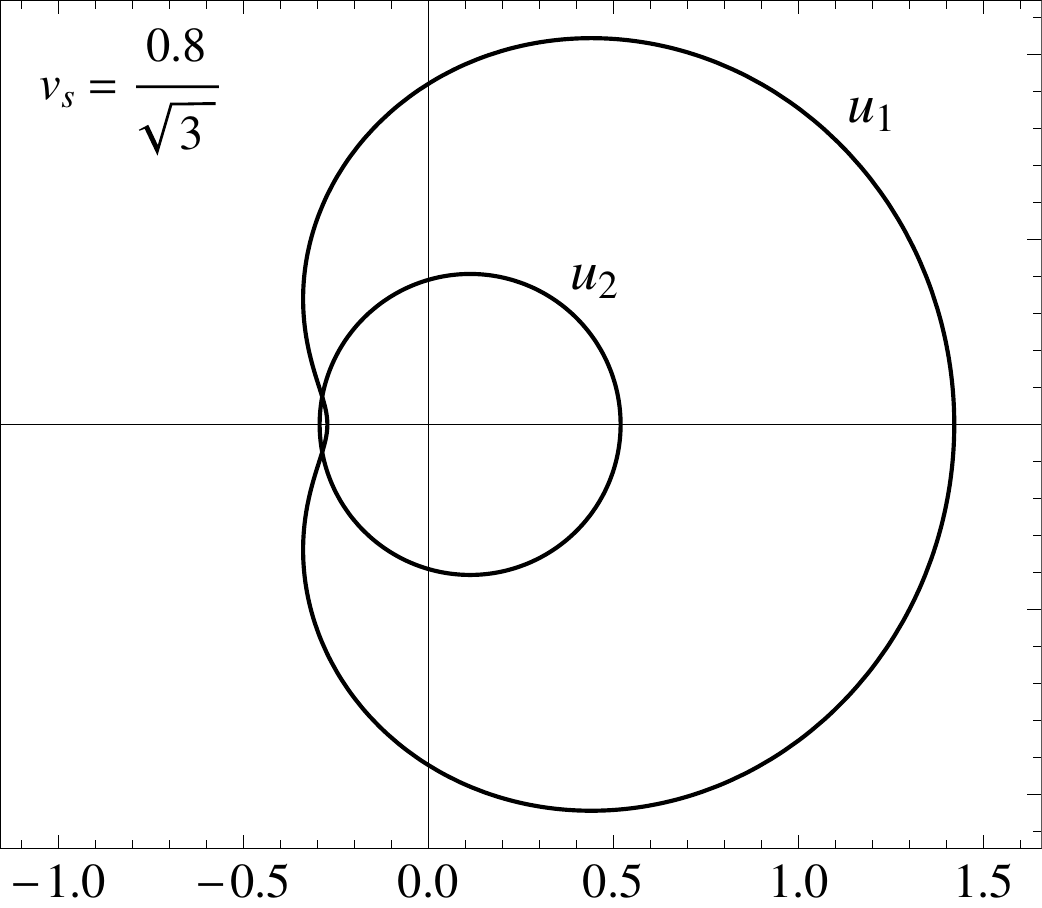}\includegraphics[scale=0.53]{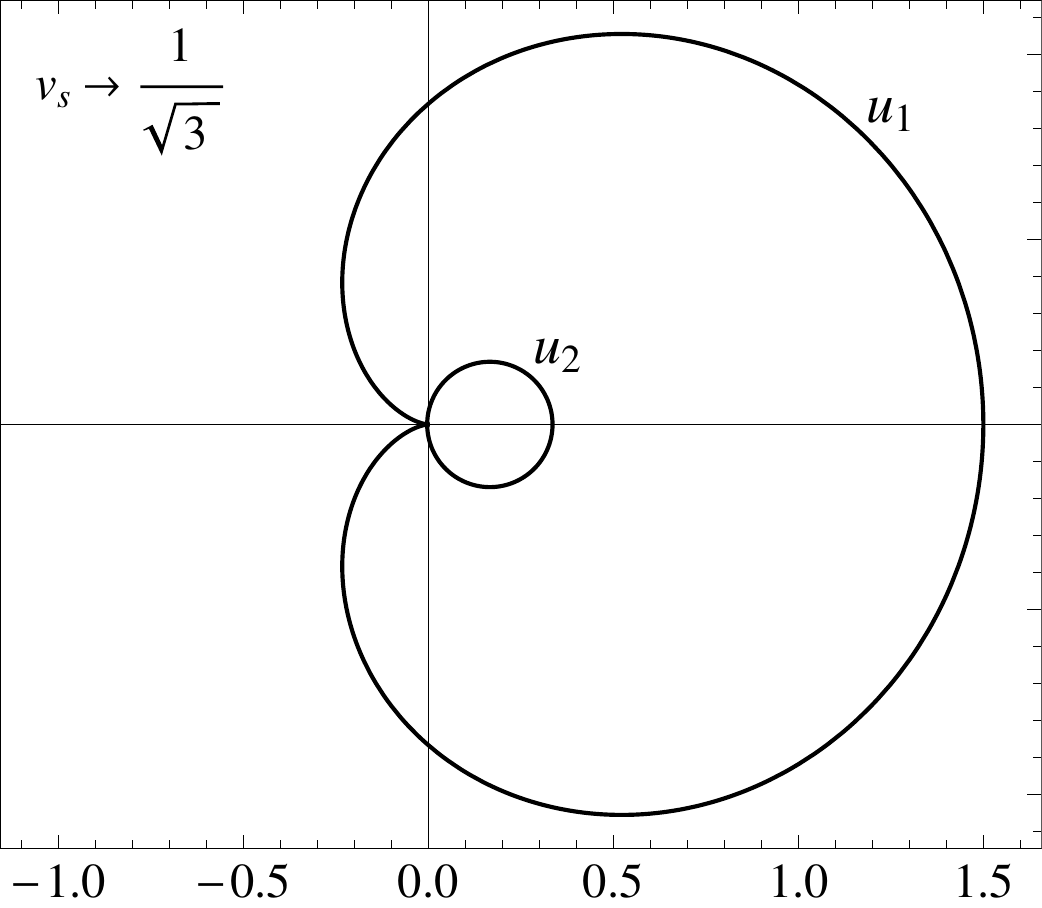}
\par\end{center}%
\end{minipage}}\protect\caption{Velocities of first and second sound $u_{1}$, $u_{2}$ from equations
(\ref{eq:solLOWT1}), (\ref{eq:solLOWT2}) for six different values
of the superfluid velocity $|\vec{v}_{s}|$ between 0 and $1/\sqrt{3}$.
All velocities are measured in the normal-fluid rest frame. In these
polar plots, the sound velocities for a given angle between the direction
of the wave vector and the superflow are given by the radial distance
of the curve to the origin; the direction of the superflow is parallel
to the horizontal axis and points to the right; the scale is normalized
to the velocity of first sound in the absence of a superflow, as one
can see in the upper left panel. The speed of first sound does not
depend on temperature within our approximation. The speed of second
sound is shown for three different temperatures: $T=0$ {[}(black)
solid{]}, $T/\mu=0.02$ {[}(blue) dashed{]}, $T/\mu=0.04$ {[}(red)
dashed-dotted{]}. For large superfluid velocities the temperature
expansion breaks down, and we have only shown the results for $T=0$.
\label{fig:Velocities-of-firstSoundLowT}}
\end{figure}

\noindent We plot the two sound velocities for all angles and for
various superfluid velocities in figure \ref{fig:Velocities-of-firstSoundLowT}.
Because of the breakdown of the temperature expansion we have just
explained, the results for nonzero temperature are only shown up to
a superfluid velocity where the $T^{2}$ correction is still smaller
than the $T=0$ term. We see that both sound velocities are increased
when they propagate parallel to the superflow and decreased when they
propagate in the opposite direction. At $T=0$, where the result can
be taken seriously for all $|\vec{v}_{s}|<1/\sqrt{3}$, the speed
of second sound decreases significantly when the critical velocity
is approached, and goes to zero for all ``backward'' angles $\pi/2<\theta<3\pi/2$
(while the velocity of first sound only goes to zero for propagation
exactly antiparallel to the superflow, $\theta=\pi$). Interestingly,
for a given superfluid velocity, the temperature effect always \textit{increases}
the speed of second sound for all angles. We know that for larger
temperatures it must decrease again, because it has to vanish at the
critical temperature where there is only one fluid in the system.
Within our low-temperature approximation we cannot see this decrease. 

\noindent Finally, let us discuss the low-temperature results at non
vanishing mass $m$. A finite mass is of importance, since its effect
as an additional energy scale will turn out to be interesting, and
we can use large values of $m$ to approach the non-relativistic limit.
Since the expressions become very complicated if both $m$ and $v$
are nonzero, we present the results for $\vec{v}_{s}=\vec{0}$. We
defer all details of the calculation to appendix \ref{sub:Sound-velocities-at-arbitraryM}.
The final result for the two sound velocities up to quadratic corrections

\noindent ~

~

\noindent in the temperature is 

\bigskip{}

\noindent 
\begin{equation}
u_{1}=\sqrt{\frac{\mu^{2}-m^{2}}{3\mu^{2}-m^{2}}}+{\cal O}(T^{4})\,\,\,\,,u_{2}=\frac{1}{\sqrt{3}}\sqrt{\frac{\mu^{2}-m^{2}}{3\mu^{2}-m^{2}}}+\left(\frac{\pi T}{\mu}\right)^{2}\frac{20\sqrt{3}\mu^{6}}{7(3\mu^{2}-m^{2})^{3/2}(\mu^{2}-m^{2})^{3/2}}+{\cal O}(T^{4})\,.\label{eq:uLowTwithM}
\end{equation}
As we can see, the speed of first sound is indeed identical to the
slope of the low-energy dispersion of the Goldstone mode (\ref{eq:DispGold})
- at least at very low temperatures. The speed of second sound at
zero temperature is simply $1/\sqrt{3}$ times the speed of first
sound - we recover Landau`s result in the presence of a mass $m$.
Also here we can see that the temperature corrections are positive,
even though we expect the speed of second sound to decrease eventually
and vanish at the critical temperature. When we perform a self-consistent
calculation for all temperatures up to $T_{c}$ in the next section,
we will see that this is indeed the case.

\section{Sound modes for all temperatures\label{sec:Sound-modes-for-allT}}

~

\begin{figure}
\fbox{\begin{minipage}[t]{1\columnwidth}%
~~\includegraphics[scale=0.66]{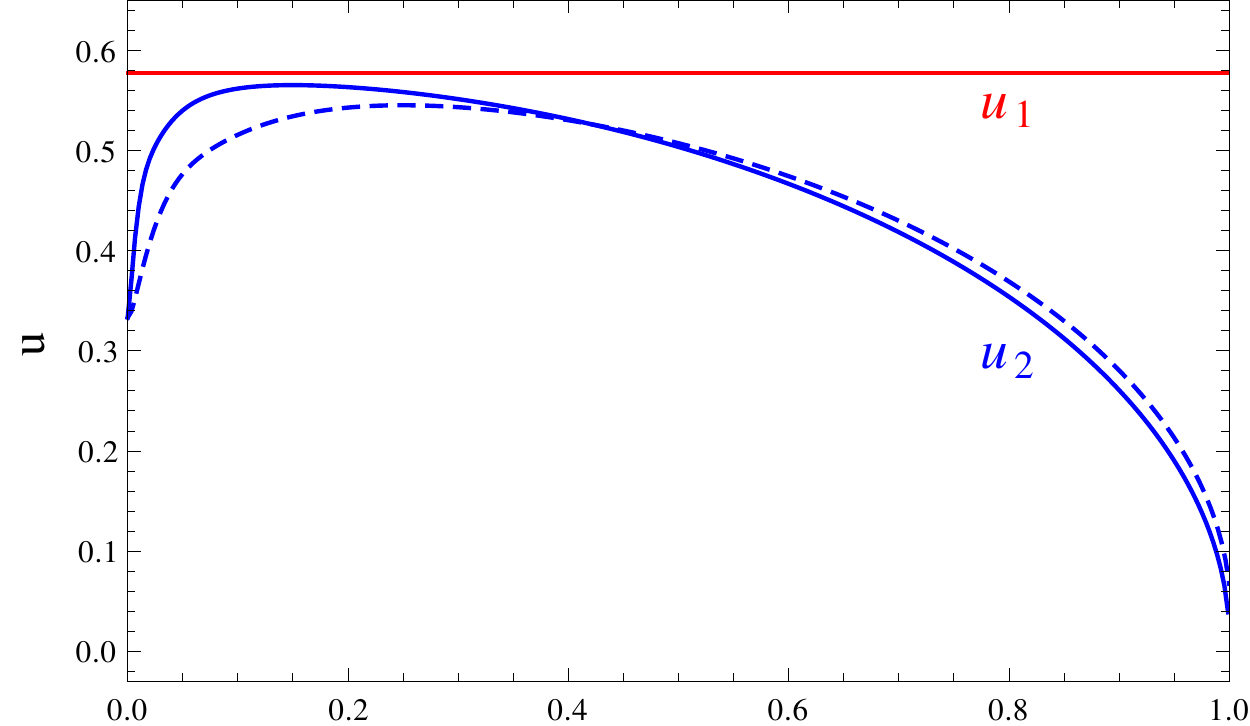}~~\includegraphics[scale=0.62]{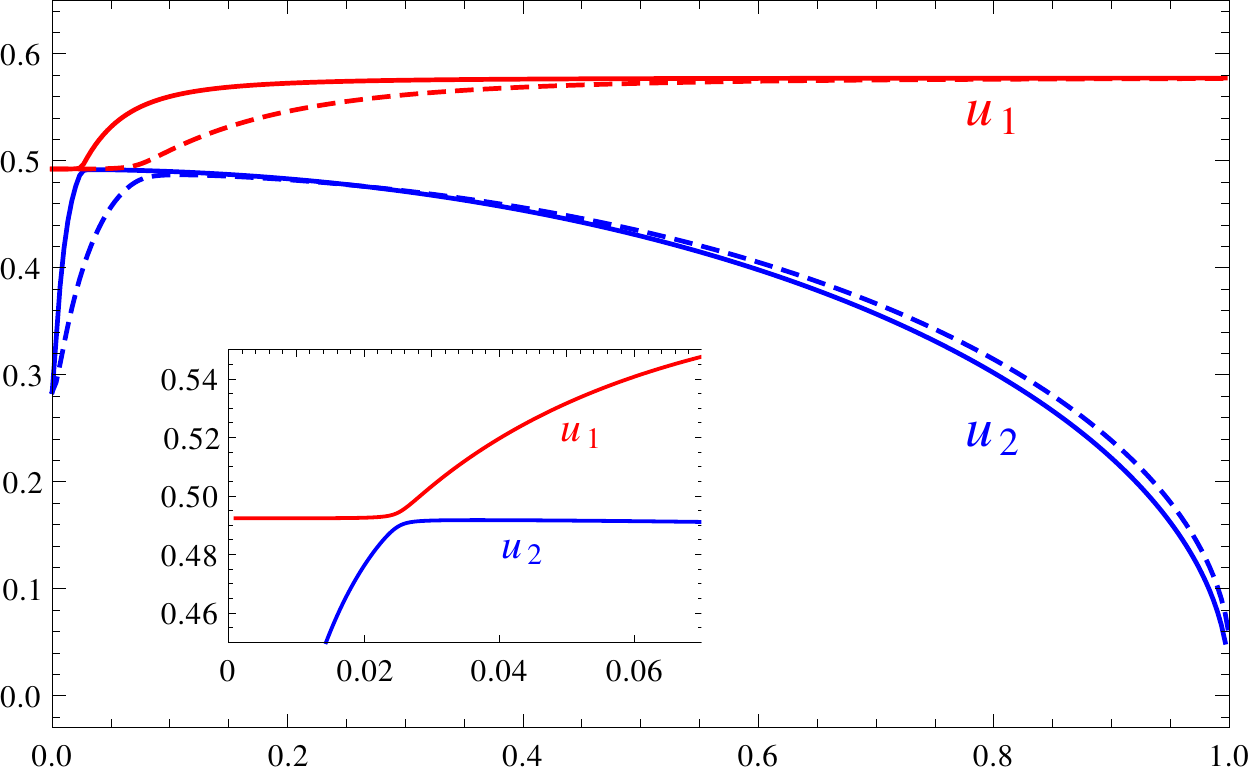}\\
\includegraphics[scale=0.82]{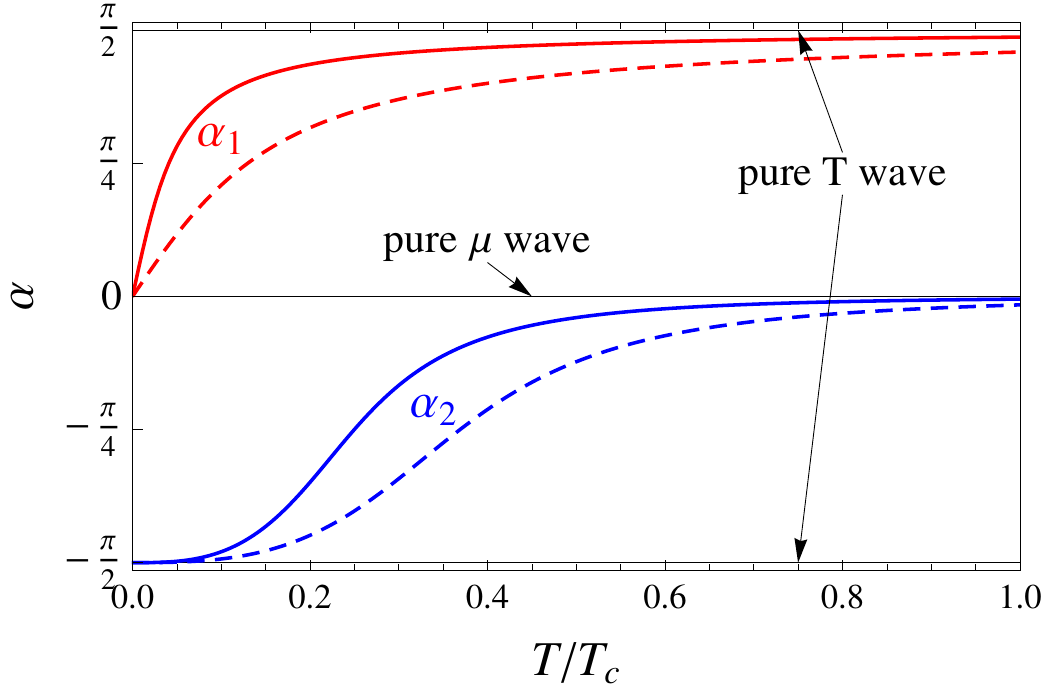}\includegraphics[scale=0.77]{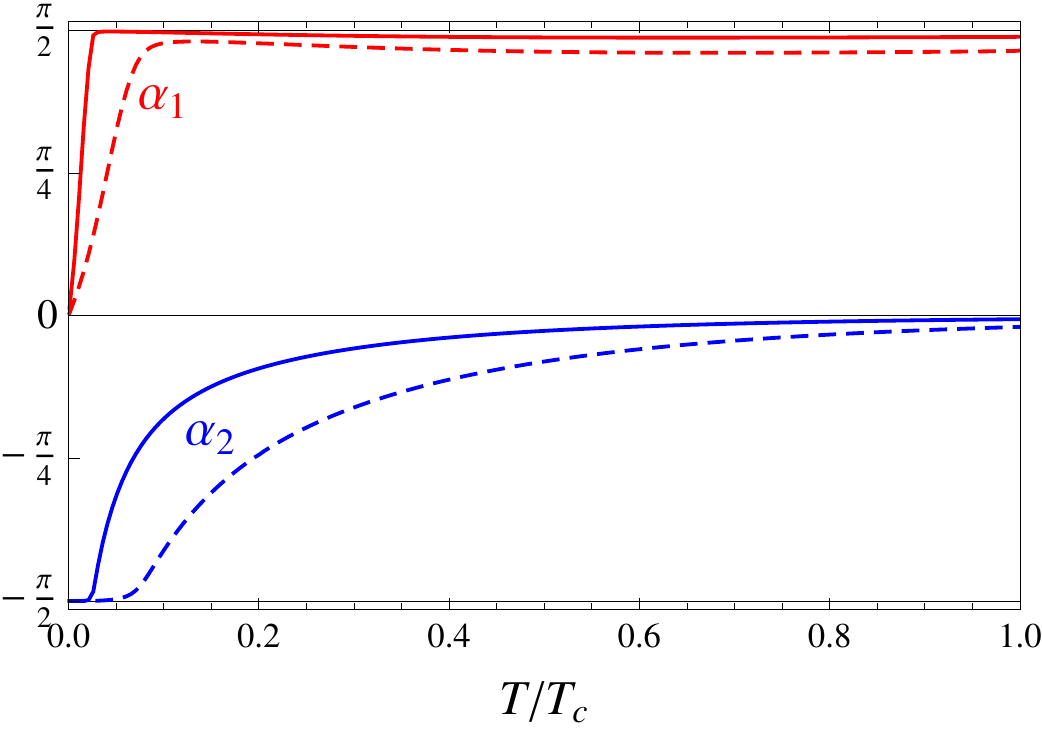}%
\end{minipage}}\protect\caption{Upper panels: speed of first and second sound in the absence of a
superflow, $\vec{\nabla}\psi=\vec{0}$, as a function of temperature
for the ultra-relativistic limit $m=0$ (left panel) and (approaching)
the non-relativistic limit $m=0.6\,\mu$ (right panel), as well as
for two different coupling constants, $\lambda=0.005$ (solid lines)
and $\lambda=0.05$ (dashed lines). The inset in the upper right panel
magnifies the region of an avoided crossing between first and second
sound for the lower coupling constant. Lower panels: mixing angle
$\alpha$ for the amplitudes in temperature and chemical potential
(see equation (\ref{eq:MixAngel})) associated to each sound wave,
for the same values of $\lambda$ and $m$. Positive (negative) values
of $\alpha$ correspond to in-phase (out-of-phase) oscillations, while
$|\alpha|=\pi/2$ ($\alpha=0$) corresponds to a pure temperature
(chemical potential) wave. \label{fig:SoundAllTZeroFlow}}
\end{figure}

\begin{figure}
\fbox{\begin{minipage}[t]{1\columnwidth}%
\includegraphics[scale=0.7]{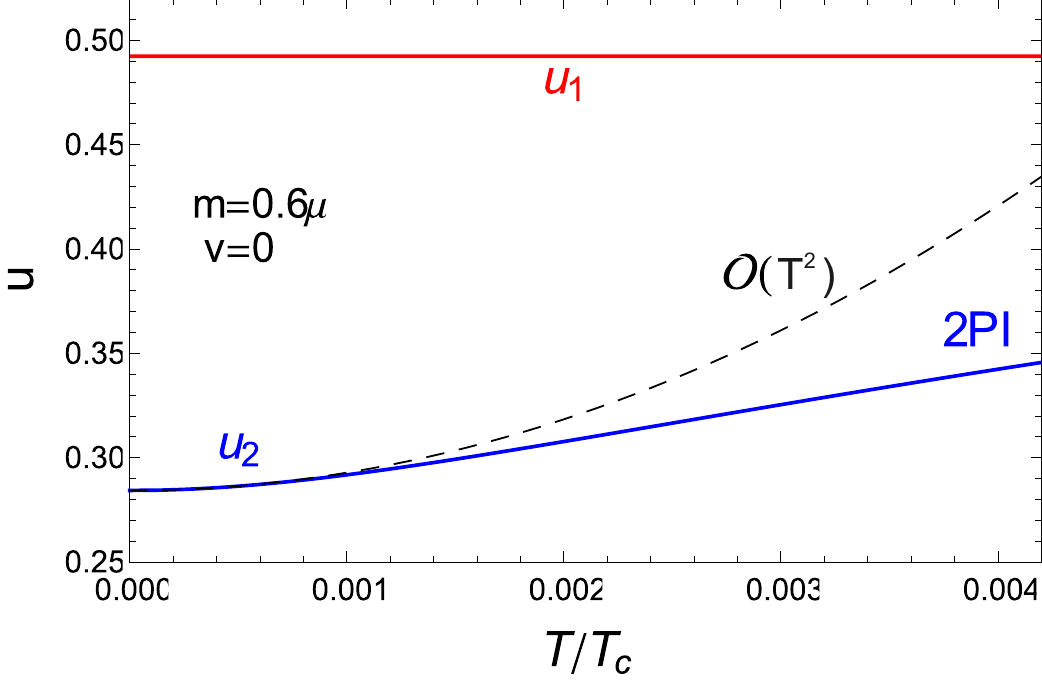}~~~~~~~~\includegraphics[scale=0.72]{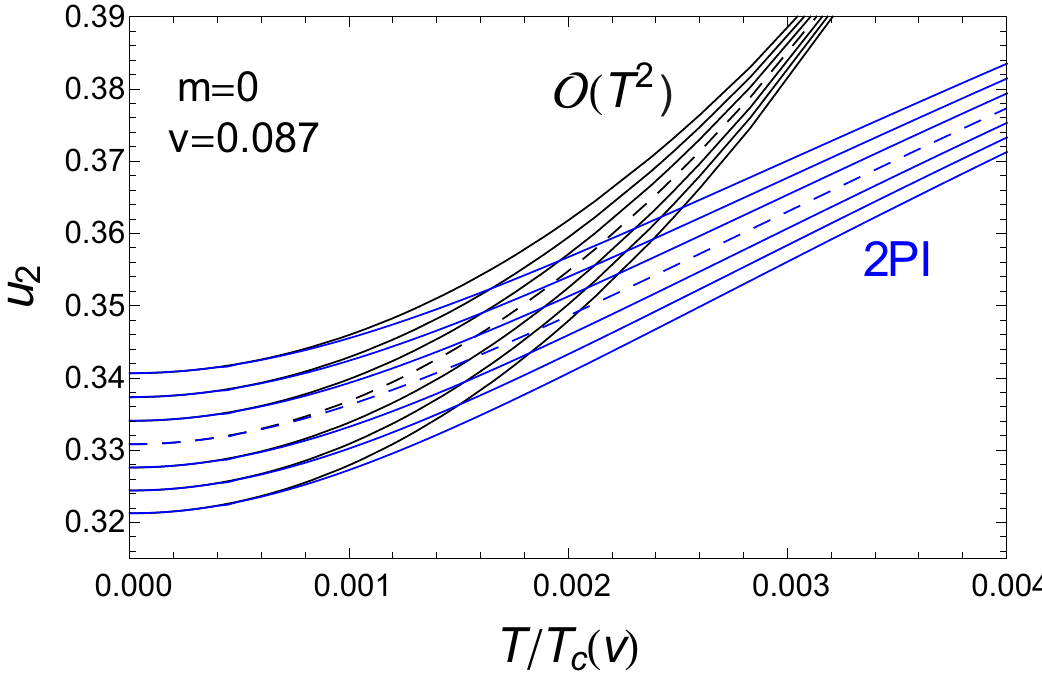}%
\end{minipage}}\protect\caption{Sound velocities for low temperatures and comparison with the analytical
low-temperature approximations for vanishing superflow and $m=0.6\mu$
(left panel) as well as for vanishing mass and $v=\frac{0.15}{\sqrt{3}}\simeq0.087$
(right panel). In both panels, $\lambda=0.005$. The approximations
are given in equations (\ref{eq:solLOWT1}) and (\ref{eq:solLOWT2})
(for the right panel) and in equations (\ref{eq:uLowTwithM}) (for
the left panel). The various curves in the right panel correspond
to different angles between the superflow and the sound wave, from
parallel (uppermost curve) to anti-parallel (lowermost curve) with
the middle (dashed) line corresponding to the perpendicular case.
\label{fig:Sound-velocities-forSoundApproxLowT}}
\end{figure}

\begin{figure}
\fbox{\begin{minipage}[t]{1\columnwidth}%
\begin{center}
~\includegraphics[scale=0.75]{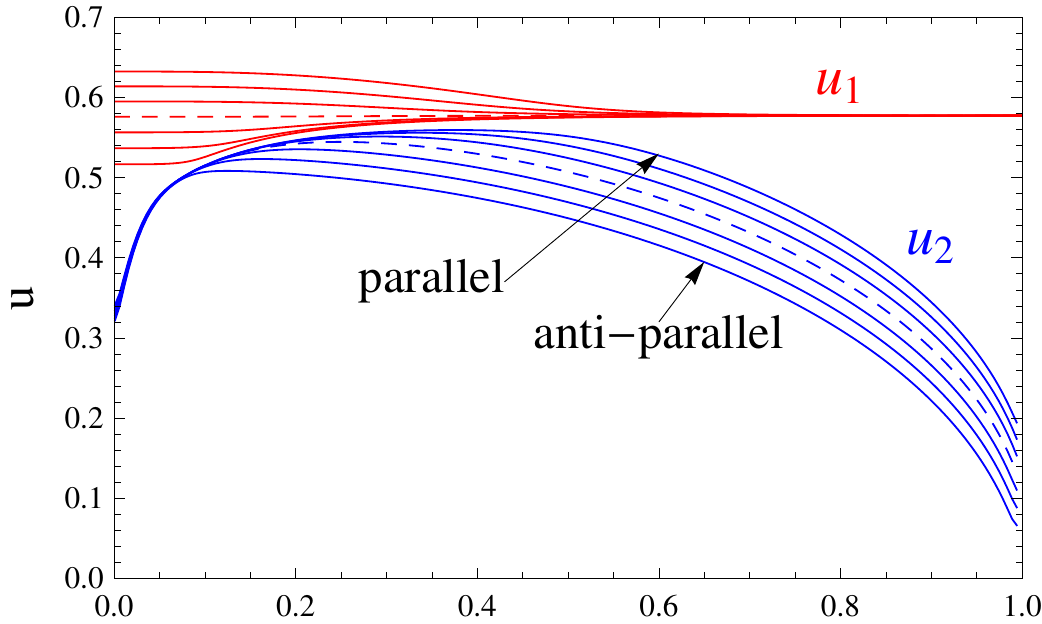}~~~\includegraphics[scale=0.7]{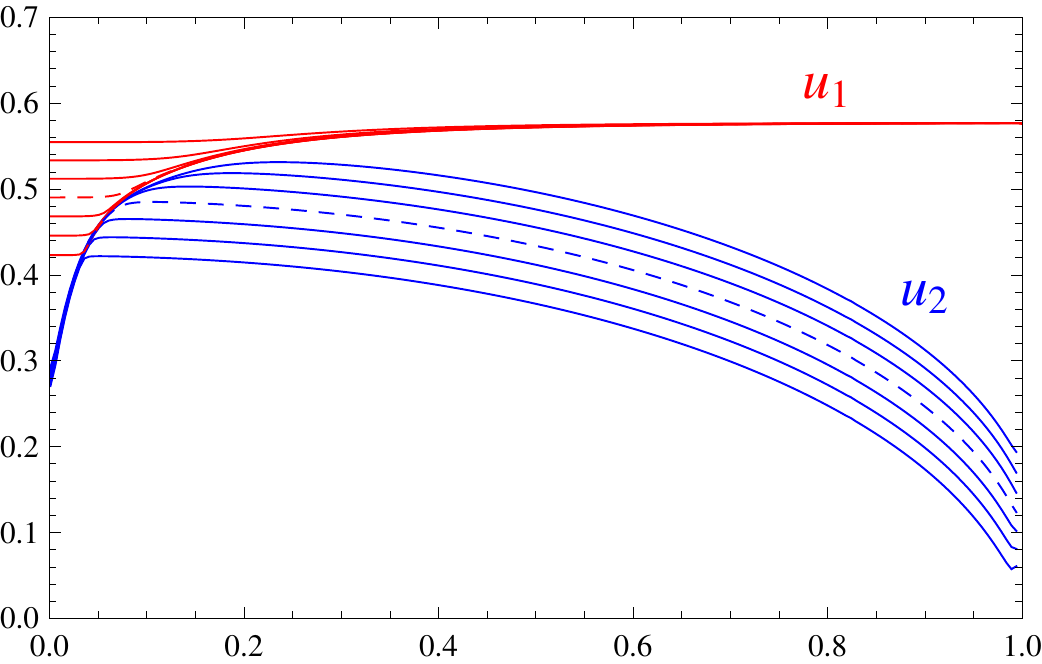}
\par\end{center}

\begin{center}
\includegraphics[scale=0.77]{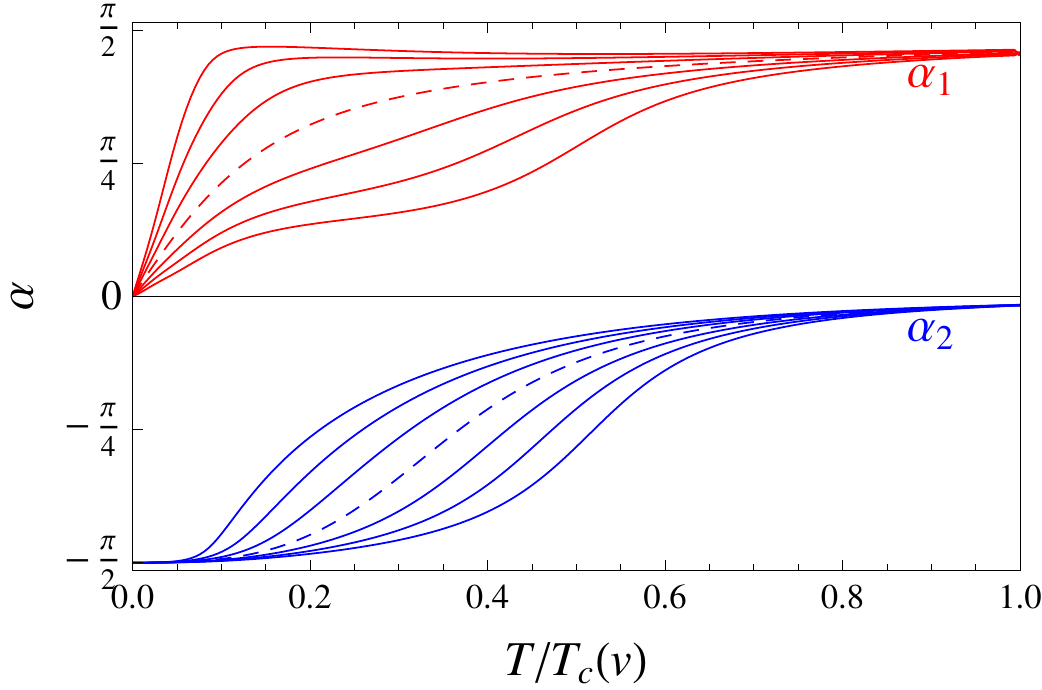}~~\includegraphics[scale=0.72]{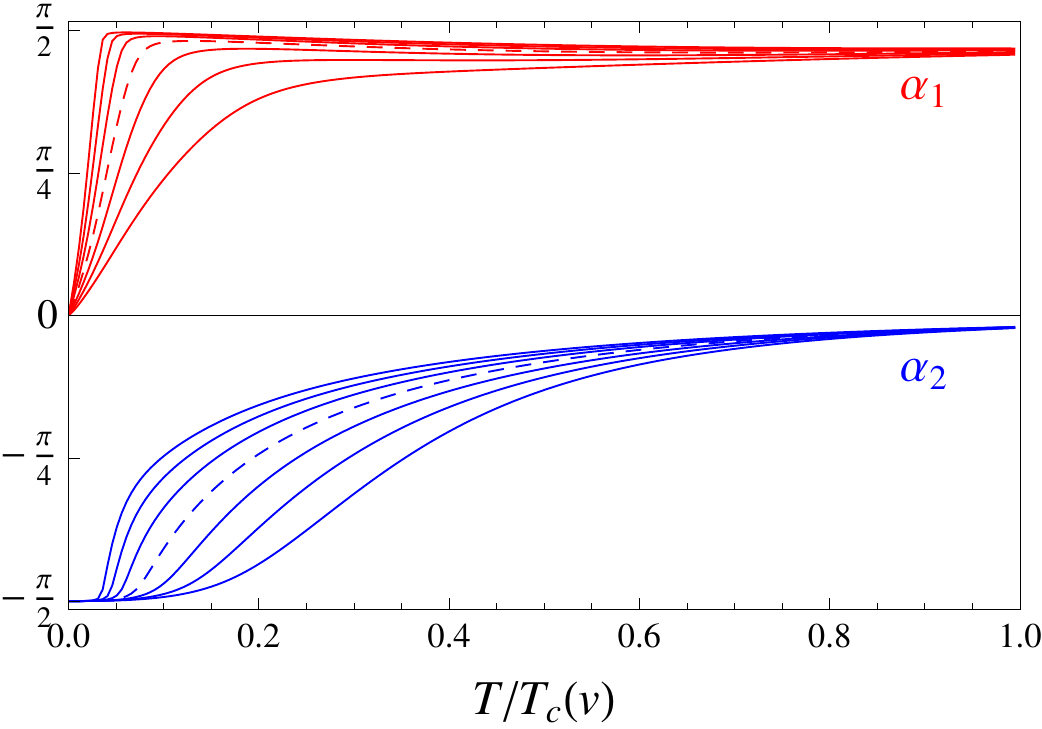}
\par\end{center}%
\end{minipage}}\protect\caption{Same as figure \ref{fig:SoundAllTZeroFlow}, but with a nonzero superfluid
velocity, chosen to be $v=\frac{0.15}{\sqrt{3}}$, i.e., 15 \% of
the critical velocity at $m=T=0$. The left panel shows results for
$m=0$ and the right panel for $m=0.6\,\mu$. Each plot shows the
results for seven different angles between the propagation of the
sound wave and the superflow, from parallel (uppermost curves) to
anti-parallel (lowermost curves) in equidistant steps of $\pi/6$
with the dashed lines corresponding to $\pi/2$. The coupling is chosen
to be $\lambda=0.05$. The dashed lines, where the effect of the superflow
is expected to be weakest, are comparable (however not exactly identical)
to the dashed lines of figure \ref{fig:SoundAllTZeroFlow}. \label{fig:SoundAllTWithFlow}}
\end{figure}

\noindent We compute the velocities of first and second sound $u_{1}$
and $u_{2}$ numerically for all temperatures by applying the algorithm
explained in section \ref{sub:Algorithm}. The results are shown in
figure \ref{fig:SoundAllTZeroFlow} (sound velocities and amplitudes
for zero superflow), figure \ref{fig:Sound-velocities-forSoundApproxLowT}
(sound velocities at very low temperatures and comparison with the
analytical results from last section) and figure \ref{fig:SoundAllTWithFlow}
(sound velocities and amplitudes for nonzero superflow). We now discuss
various aspects of the results separately.

~

\noindent \textit{Speed of first sound and scale-invariant limit.}
In the simplest case, with vanishing mass parameter and superflow,
the speed of first sound is $u_{1}=\frac{1}{\sqrt{3}}$ for all temperatures,
see analysis in section \ref{sub:General-structure-of-solutions}.
This is shown in the upper left panel of figure \ref{fig:SoundAllTZeroFlow}
and is in agreement with the analytical result (\ref{eq:soundConf}).
For low temperatures, this sound speed is identical to the slope of
the Goldstone dispersion. For higher temperatures, however, the slope
deviates from the speed of first sound and approaches zero at the
critical point, just like the speed of \textit{second} sound. In other
words, the Goldstone mode is, in general, not a solution to the wave
equations derived from the hydrodynamic conservation equations. Only
in certain temperature limits do these waves coincide with the Goldstone
mode. 

\noindent The upper right panel of figure shows that for a nonzero
mass parameter $m$, the speed of first sound deviates from the scale-invariant
value at low temperatures, but approaches this value for high temperatures
$T\gg m$. Notice that we have chosen the same mass parameter in units
of $\mu$ for both coupling strengths. As a consequence, the sound
velocities for the two coupling strengths coincide at zero temperature,
but the mass is different in units of $T_{c}$: for the smaller coupling
(solid lines) we have $m\simeq0.03\, T_{c}$, while for the larger
coupling (dashed lines) $m\simeq0.1\, T_{c}$. This is the reason
why $u_{1}$ appears to approach the scale-invariant value more slowly
for the case of the larger coupling.

\noindent \textit{Speed of second sound.} In all cases we consider,
the speed of second sound increases strongly at low temperatures.
We can see this increase in the low-temperature approximation, see
discussion in section \ref{sec:The-low-temperature-approx}. Even
though figure \ref{fig:Sound-velocities-forSoundApproxLowT} shows
that the analytic approximation is only valid for very low temperatures,
we see that the strong increase continues beyond the validity of the
analytical approximation (although it becomes less strong than the
approximation suggests). One can see from equations (\ref{eq:solLOWT1}),
(\ref{eq:solLOWT2}), (\ref{eq:uLowTwithM}) that the $T^{2}$ contribution
does not, to leading order, depend on the coupling constant. Therefore,
since smaller coupling strengths correspond to higher critical temperatures,
the increase of $u_{2}$ can be made arbitrarily sharp (on the relative
temperature scale $T/T_{c}$) by decreasing the coupling. This tendency
is borne out in figure \ref{fig:SoundAllTZeroFlow}. 

\noindent In the upper panels of figure \ref{fig:SoundAllTZeroFlow},
the velocity of second sound does not go to zero at the critical point.
This is an artifact of our Hartree approximation: as we have discussed
in section \ref{sub:Critical-temperature,-condensate-critV-forallT},
in our approach the phase transition is strictly speaking first order.
Therefore, the condensate is not exactly zero at our critical point.
The speed of second sound turns out to be sensitive to this effect,
and therefore $u_{2}$ does not approach zero at $T_{c}$. The superfluid
density appears to be less sensitive to this effect since it approaches
zero to a very good accuracy, see figure \ref{fig:SuperflAllT}.

\newpage{}

\noindent \textit{Role reversal of the sound modes.} To discuss the
physical nature of the sound waves, we first notice that the speeds
of sound show a feature that is reminiscent of an ``avoided level
crossing'' in quantum mechanics. This feature is most pronounced for
small coupling and nonzero mass parameter $m$, see upper right panel
of figure \ref{fig:SoundAllTZeroFlow} and the zoomed inset in this
panel. It suggests that there is a physical property that neither
first nor second sound possesses for all temperatures, but that is
rather ``handed over'' from first to second sound in the temperature
region where the curves almost touch. We find this property by computing
the amplitudes of the oscillations associated to the sound modes,
as discussed in section \ref{sub:General-structure-of-solutions}.
In particular, we are interested in the mixing angle $\alpha$ defined
in equation (\ref{eq:MixAngel}) that indicates whether a given sound
mode is predominantly an oscillation in chemical potential or in temperature
or something in between. Our results show that $u_{1}$ always corresponds
to $\alpha>0$ while $u_{2}$ always corresponds to $\alpha<0$. Therefore,
the first sound is always an in-phase oscillation, while the second
sound is always an out-of-phase oscillation. However, whether first
or second sound is a density wave or an entropy wave is a temperature
dependent statement, as already discussed in the scale-invariant limit
where there are simple expressions for the amplitudes, see equation
(\ref{eq:ConfRat}). In all cases we consider, $u_{1}$ transforms
from a pure density wave at $T=0$ to a pure entropy wave at $T=T_{c}$
and vice versa for $u_{2}$. This role reversal becomes sharper for
larger $m$ and/or smaller $\lambda$, i.e., it is smoothest in the
ultra-relativistic regime at strong coupling, see lower left panel
of figure \ref{fig:SoundAllTZeroFlow}. (Remember that we compare
two relatively weak coupling strengths, the ``strong coupling'' is
$\lambda=0.05$.)

~

\begin{figure}
\fbox{\begin{minipage}[t]{1\columnwidth}%
\begin{center}
~\includegraphics[scale=0.75]{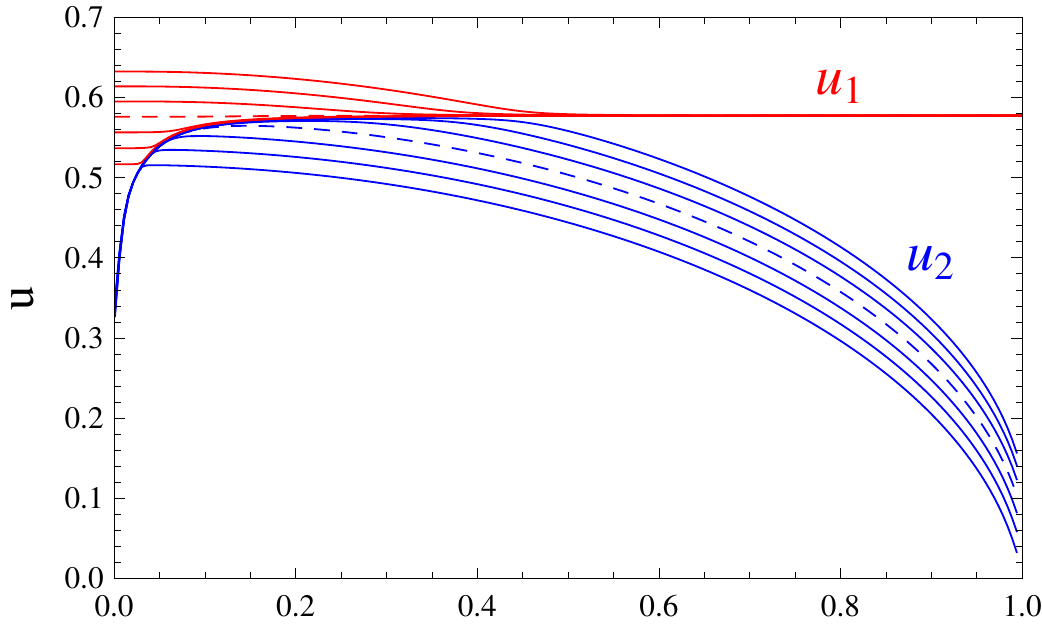}~~~\includegraphics[scale=0.7]{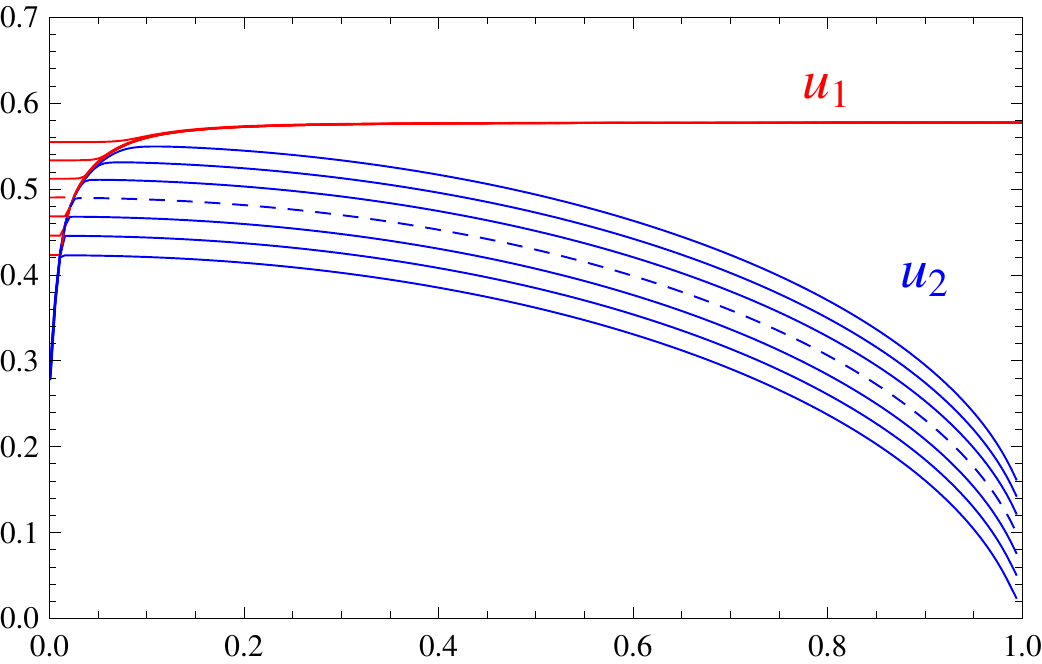}
\par\end{center}

\begin{center}
\includegraphics[scale=0.77]{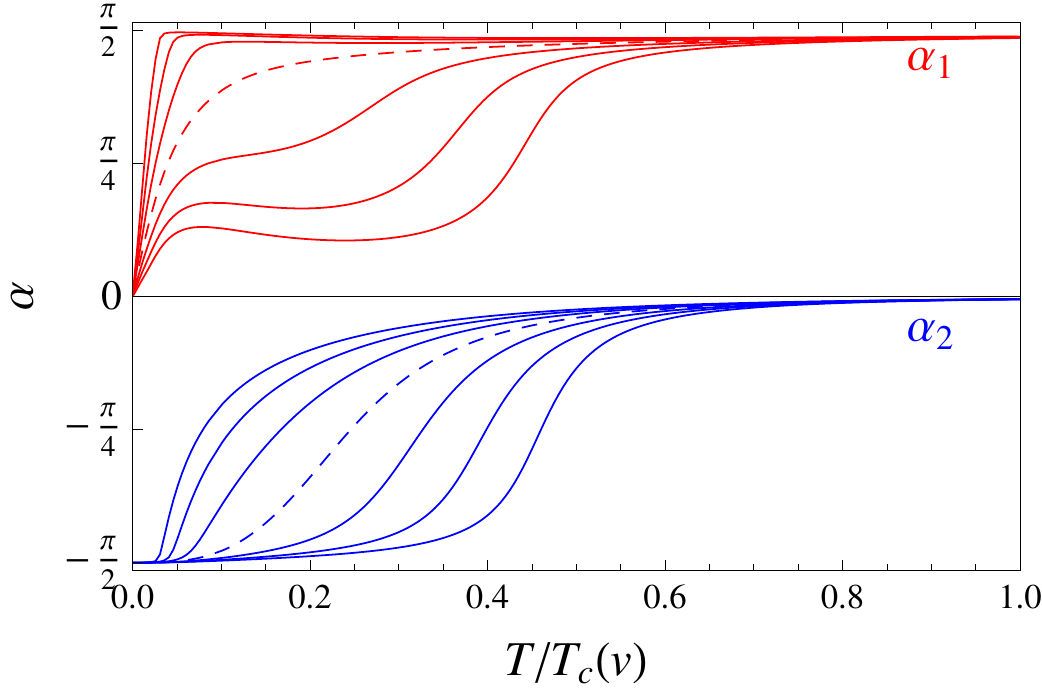}~~\includegraphics[scale=0.72]{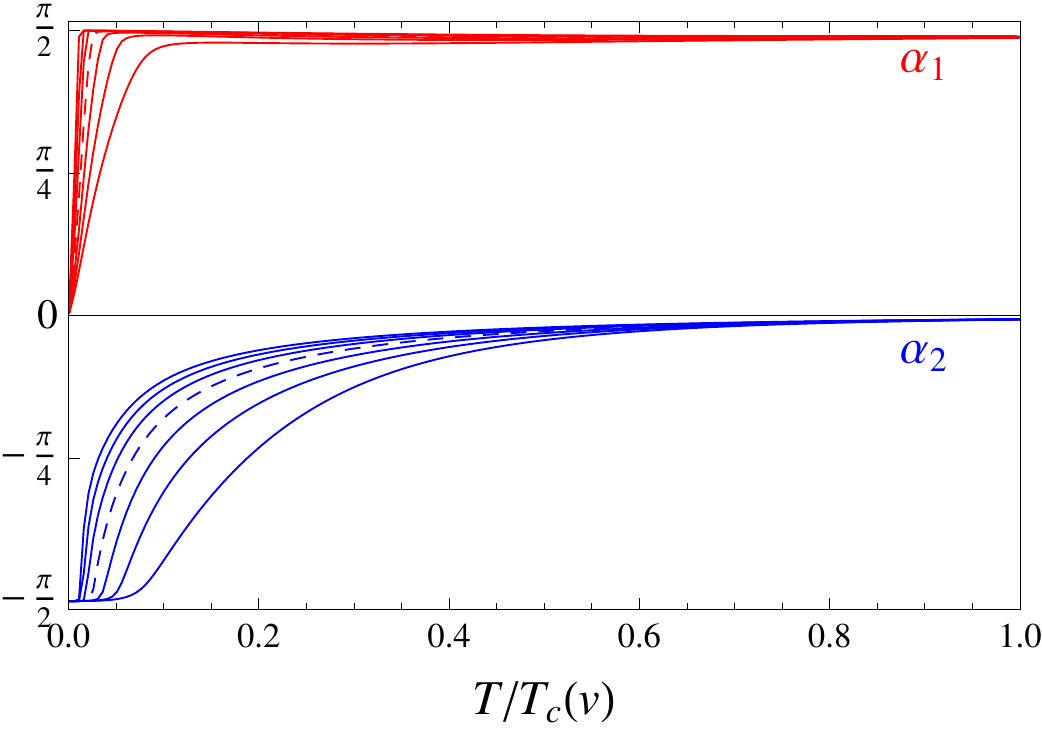}
\par\end{center}%
\end{minipage}}\protect\caption{Same as figure \ref{fig:SoundAllTWithFlow}, but at weaker coupling,
$\lambda=0.005$. The left panel shows results for $m=0$ and the
right panel for $m=0.6\,\mu$.\label{fig:SoundLargerCoupling}}
\end{figure}

~

\noindent \textit{Comparison to non-relativistic systems.} We can
view $m$ as a parameter with which we can go continuously from the
ultra-relativistic limit $m=0$ to the non-relativistic limit of large
$m$ (always keeping $m$ smaller than $\mu$ in order to allow for
condensation). Therefore, the right panels of figure \ref{fig:SoundAllTZeroFlow}
are comparable to the results in non-relativistic calculations. Of
course, $m=0.6\,\mu$, as chosen in the plots, is not actually a non-relativistic
value; for instance, for this value of $m$ the speed of first sound
at low temperatures is still about 50 \% of the speed of light, while,
for comparison, the speed of first sound in superfluid helium is about
$240\,{\rm m}/{\rm s}$, i.e., about $10^{-8}$ times the speed of
light. Nevertheless, already for this moderate value of $m$ we find
qualitative agreement with the non-relativistic results of reference
\cite{Hu2010} see in particular Fig. 6 in this reference which also
exhibits the avoided crossing and the sharp role reversal at a low
temperature. As in this reference, we also find that a stronger coupling
smooths out both of these features. The results shown in figures \ref{fig:SoundAllTWithFlow}
and \ref{fig:SoundLargerCoupling} generalize the results of \cite{Hu2010}
to the relativistic regime and to the case of nonzero superflow (For
a zero-temperature calculation of the sound velocities in the presence
of a superflow in $^{4}$He see reference \cite{Adamenko2009}).

\noindent Our results (and those of reference \cite{Hu2010}) for
the sound modes differ from the calculations and measurements for
superfluid helium \cite{Donelli}, \cite{Khalatnikov} and a (unitary)
Fermi gas. For instance, in neither of these experimentally accessible
cases does the speed of second sound increase significantly at low
temperatures. Another difference is that in superfluid $^{4}$He,
second sound is predominantly a temperature wave for almost all temperatures,
except for a regime close to the critical temperature. This shows
that the behavior of the sound waves is very sensitive to the details
of the underlying theory, i.e., the details of the interaction. We
see from (\ref{eq:soundConf}) that even in the ultra-relativistic,
scale-invariant limit the speed of second sound depends on thermodynamic
functions that can be significantly different in different theories.
Another feature of the second sound in $^{4}$He is a rapid decrease
in a regime where rotons start to become important \cite{Donelli},
\cite{Khalatnikov}. Our model for a complex scalar field also gives
rise to a massive mode whose mass is $\epsilon_{k=0}^{-}=\sqrt{6}\,\mu$
(the difference from rotons being that the minimum of the dispersion
is at zero momentum). For instance for the case $m=v=0$ this means
that the mass in units of the critical temperature is $\epsilon_{k=0}^{-}=0.1\, T_{c}$
(for the weaker coupling $\lambda=0.005$) and $\epsilon_{k=0}^{-}=0.3\, T_{c}$
(for the stronger coupling $\lambda=0.005$). Therefore, our sound
velocities are dominated by the Goldstone mode only for temperatures
$T\ll0.1\, T_{c}$ while for higher temperatures the massive mode
plays an important role, even though there appears to be no characteristic
drop in $u_{2}$ at the onset of that mode. 

\noindent \textit{Nonzero superflow. }In the presence of a nonzero,
uniform superflow - a relative flow between superfluid and normal-fluid
components, measured in the normal-fluid rest frame - the sound velocities
obviously become anisotropic (see equations \ref{eq:solLOWT1} and
\ref{eq:solLOWT2} for the explicit anisotropic result in the low-temperature
approximation). In figures \ref{fig:SoundAllTWithFlow} and \ref{fig:SoundLargerCoupling}
($\lambda=0.05$ and $\lambda=0.005$, respectively) we plot the speeds
of sound and the corresponding mixing angles for the amplitudes for
seven different directions of the sound wave with respect to the superfluid
velocity $\vec{v}_{s}$, from downstream propagation (uppermost curves
in all panels) through perpendicular propagation (dashed curves in
all panels) to upstream propagation (lowermost curves in all panels).
We see that both sound speeds are faster in the forward direction,
as was already observed in the low-temperature results. Since the
superflow $\vec{\nabla}\psi$, like the mass parameter $m$, introduces
an additional energy scale, the speed of first sound $u_{1}$ deviates
from the scale-invariant value, at least at low temperatures. In the
ultra-relativistic limit, $u_{1}$ approaches the scale-invariant
value at high temperatures from above (from below) for a downstream
(upstream) sound wave. The value of the superflow used in the figures
corresponds to about 1 \% of the critical temperature, $|\vec{\nabla}\psi|\sim0.01\, T_{c}(\left|\vec{v}\right|)$
for the stronger coupling, figure \ref{fig:SoundAllTWithFlow}, and
to about 0.4 \% of the critical temperature for the weaker coupling,
figure \ref{fig:SoundLargerCoupling}. The critical temperatures are
$T_{c}(\left|\vec{v}\right|)\simeq7.62\,\mu\,[6.04\,\mu]$ for $\lambda=0.05$
and $m=0\,[0.6\,\mu]$ and $T_{c}(\left|\vec{v}\right|)\simeq24.2\,\mu\,[19.23\,\mu]$
for $\lambda=0.005$ and $m=0\,[0.6\,\mu]$.

\noindent To compare to the analytical results in the low-temperature
limit, we use equations (\ref{eq:solLOWT1}) and (\ref{eq:solLOWT2})
for the massless case and equations (\ref{eq:uLowTwithM}) for finite
mass $\textrm{m}$ , see figure \ref{fig:Sound-velocities-forSoundApproxLowT}.
Even though we only show the comparison for $u_{2}$, we have checked
that the numerical results agree with the low-temperature approximation
also for $u_{1}$. 

\noindent At the critical point, there is a sizable nonzero value
of the speed of second sound for all angles. In contrast to the case
without superflow, this is not only due to our use of the Hartree
approximation. Remember from the discussion in section \ref{sub:Critical-temperature,-condensate-critV-forallT}
that $T_{c}(v)$ is the point beyond which there is no stable uniform
superfluid, see in particular the phase diagram in figure \ref{fig:PhaseDiagCrit}.
At that critical point, the condensate is not zero (and is not expected
to be zero in a more complete treatment), and therefore we do not
expect $u_{2}$ to go to zero.

\noindent Comparing figures \ref{fig:SoundAllTWithFlow} and \ref{fig:SoundLargerCoupling}
we observe that a weaker coupling leads again to a more pronounced
avoided crossing effect. This is particularly obvious from the upper
right panel of figure (\ref{fig:SoundLargerCoupling}), where we observe
the avoided crossing effect now for each angle separately. Like for
vanishing superflow, a weaker coupling tends to shift the point of
the role reversal to lower temperatures, even though this statement
is not completely general. Namely, in the ultra-relativistic limit
we see that changing the coupling has a more complicated effect for
the sound waves that propagate in the backward direction, see curves
below the dashed one in the lower left panels of figures \ref{fig:SoundAllTWithFlow}
and \ref{fig:SoundLargerCoupling}. As a consequence, we find the
following interesting phenomenon: depending on the external parameters,
there can be a sizable temperature regime of intermediate temperatures
where a second sound wave, sent out in the forward direction, is almost
a pure chemical potential wave while sent out in the backward direction
it is almost a pure temperature wave (and vice versa for the first
sound). This effect is most pronounced for weak coupling and the ultra-relativistic
limit. We have checked that it gets further enhanced by a larger value
of the superflow. In other words, the role reversal in the sound modes
does not only occur by changing temperature (most pronounced in the
\textit{non-relativistic} case at weak coupling), but can also occur
by changing the direction of the sound wave (most pronounced in the
\textit{ultra-relativistic} case at weak coupling). 

\newpage{}

\section{\noindent Conclusion\label{sec:Conclusion}}

\noindent ~

\noindent We have derived the wave equations which described sound
excitations of a relativistic superfluid and calculated the corresponding
solutions which are the velocities of first and second sound. In a
first step, we calculated the sound excitations in a low-temperature
approximation in the presence of a nonzero superflow. (The calculation
of the sound modes always requires at least an infinitesimal superflow;
by nonzero superflow we mean larger than infinitesimal.)

\noindent To obtain non zero-temperature corrections it has turned
out to be crucial to go beyond the linear approximation of the dispersion
of the Goldstone mode. Cubic corrections in the dispersion give rise
to $T^{6}$ corrections in the pressure and $T^{2}$ corrections in
the velocity of second sound, while the velocity of first sound remains
temperature-independent. We have found that the velocity of second
sound increases with low temperatures.

\noindent The numerical calculation up to $T_{c}$ within the 2PI
formalism requires us to numerically evaluate first and second derivatives
of the pressure with respect to the temperature, chemical potential,
and superflow. To avoid numerical uncertainties we have computed these
derivatives in a semi-analytical way (see section \ref{sub:Algorithm}).
The low-temperature limit of the self consistent calculation agree
perfectly with the analytic results obtained in the low-temperature
calculation and show that this approximation is valid only for a very
low temperatures regime whose size depends on the value of the coupling.
Even for the smallest coupling we have used, $\lambda=0.005$, the
approximation deviates significantly from the full 2PI numerical result
for all temperatures higher than about 0.1 \% of the critical temperature.
The main reason seems to be the temperature dependence of the condensate
which is not included in the low-temperature calculation. This approximation
also neglects the massive mode that is present in our theory (and
which is not unlike the roton in superfluid helium); the massive mode
becomes important for temperatures higher than about 10 \% of the
critical temperature. 

\noindent We have investigated the dependence of the sound velocities
on the boson mass $m$ and the superflow $\vec{\nabla}\psi$. For
$m=|\vec{\nabla}\psi|=0$ the speed of first sound assumes the universal
value $\frac{1}{\sqrt{3}}$ for all temperatures, while an additional
scale, provided by $m$ and/or $|\vec{\nabla}\psi|$, leads to a deviation
from this result. For temperatures higher than that scale but still
lower than the critical temperature (if such a regime exists) the
speed of first sound again approaches $\frac{1}{\sqrt{3}}$. The speed
of second sound is more sensitive to details of the system and has
a non-universal behavior even for $m=|\vec{\nabla}\psi|=0$. In our
particular model we found a strong increase for low temperatures before
a decrease sets in, eventually leading to a vanishing speed of second
sound at the critical point, if the superflow is zero. 

\noindent By computing the amplitudes in chemical potential and temperature
of the sound waves for all temperatures, one finds that first sound
is always an in-phase oscillation of chemical potential and temperature
(and thus also of density and entropy) and second sound is always
an out-of-phase oscillation, which can thus be viewed as their defining
property. However, the degree to which a given sound wave is a density
or entropy wave depends on the temperature. We have shown that, with
respect to this property, first and second sound typically reverse
their roles as a function of temperature: the in-phase (out-of-phase)
mode is a pure density (entropy) wave at low temperatures and becomes
a pure entropy (density) wave at high temperatures. This observation
is in agreement with non-relativistic studies. While in the non-relativistic
case this role reversal occurs rather abruptly in the very low-temperature
regime, it is more continuous in the ultra-relativistic case. We have
also found that for certain parameters of the model and intermediate
temperatures, there can be a role reversal at a fixed temperature:
if there is a nonzero superflow, the first sound is (almost) a pure
entropy wave parallel to the superflow and (almost) a pure density
wave anti-parallel to the superflow, while the second sound behaves
exactly opposite. This interesting effect is most pronounced in the
ultra-relativistic limit at weak coupling.

\newpage{}

\part{A mixture of two superfluids\label{sec:A-mixture-of-two-SF}}

\section{Effective description\label{sub:Effective-description}}

~

\noindent In the final part of this thesis, we will investigate another
kind of two-fluid system which is qualitatively different from the
one we considered before: a mixture of two superfluids. First, we
will construct an effective theory for a such a mixture. We will do
so strictly at zero temperature (i.e. in the absence of a normal-fluid
component) and discuss the translation into a set of coupled hydrodynamic
equations. Such a study is useful in the context of compact stars
where usually more than one superfluid is present and also for example
for mixtures of superfluid $^{3}\textrm{He}$ and $^{4}\textrm{He}$
or cold atoms (see for instance \cite{Bao2003}, \cite{Watanabe2013}).
Hydrodynamics for such complicated systems have for example been discussed
in \cite{AndreevBashkin,mineev1975}, for a modern relativistic description
of multicomponent superfluids see reference \cite{LReviewsAnderssonComer}.
The key ingredient is once again entrainment - even at zero temperature
(see section \ref{sub:Relativistic-thermodynamics-and-entrain} for
a discussion and references). We have discussed that the microscopic
mechanisms leading to this zero temperature entrainment can be very
complicated. At this point, we shall not be concerned with these details
but rather introduce entrainment in an effective way, by requiring
that the current $j_{1}^{\mu}$ of particle species $1$ is proportional
to the conjugate momenta of both particles species , say $\partial_{\mu}\psi_{1}$
and $\partial_{\mu}\psi_{2}$. As we will see, this can be achieved
by introducing a gradient coupling in the effective Lagrangian. The
strength of this coupling will be tunable by the parameter $\lambda_{12}$.
Such a coupling makes a rigorous renormalization as we have discussed
it in the case of a single particles species impossible (which is
of no further concern to the zero temperature calculations of this
section). At finite temperature, one would need to introduce a cutoff
to execute thermal integrals. We work in the same approximations as
introduced in section \ref{sub:Zero-temperature-hydrodynamics} and
assume a uniform density and flow of the superfluid. Since there are
now two \textit{superfluid} velocities, there are two different rest
frames even at zero temperature. In other words, a temperature limit
in which the pressure is isotropic does no longer exist. We shall
therefore rely on the generalized two-fluid formalism of section \ref{sec:Relativistic-thermodynamics-and-hydro}
right from the beginning. 

\noindent The starting point is again a complex scalar field theory

\medskip{}

\noindent 
\begin{equation}
\mathcal{L}=\partial_{\mu}\varphi_{1}\partial^{\mu}\varphi_{1}^{*}-m_{1}^{2}\left|\varphi_{1}\right|^{2}-\lambda_{1}\left|\varphi_{1}\right|^{4}+\partial_{\mu}\varphi_{2}\partial^{\mu}\varphi_{2}^{*}-m_{2}^{2}\left|\varphi_{2}\right|^{2}-\lambda_{2}\left|\varphi_{2}\right|^{4}+\mathcal{L}_{int}\,,\label{eq:2SFL}
\end{equation}

\noindent where the fields $\varphi_{1}$ and $\varphi_{2}$ are coupled
by a gradient interaction

\noindent 
\begin{equation}
\mathcal{L}_{int}=-\frac{1}{2}\lambda_{12}\varphi_{1}\varphi_{2}^{*}\partial_{\mu}\varphi_{1}^{*}\partial^{\mu}\varphi_{2}+\, h.c.\,.
\end{equation}

\noindent The interaction term ins manifestly $U(1)_{1}\times U(1)_{2}$
invariant (the ``chemical index'' $i$ denotes the particle species)
and the complex conjugates are chosen such that the currents calculated
as $\partial\mathcal{L}/\partial\partial_{\mu}\varphi_{i}$ and $\partial\mathcal{L}/\partial\partial_{\mu}\varphi_{i}^{*}$
are complex conjugates of each other. To compensate for the extra
mass dimensions of the gradients, the coupling $\lambda_{1,2}$ cannot
be dimensionless but rather has $\textrm{dim}[\lambda_{12}]=m^{-2}$.
In complete analogy to section \ref{sec:Lagrangian} we introduce
the condensates to the fields $\varphi_{1}$ and $\varphi_{2}$ 

\noindent 
\begin{eqnarray}
\varphi_{1}(x) & \rightarrow & \frac{1}{\sqrt{2}}e^{i\psi_{1}(x)}\left(\rho_{1}(x)+\varphi_{11}(x)+i\varphi_{12}(x)\right)\,,\\
\varphi_{2}(x) & \rightarrow & \frac{1}{\sqrt{2}}e^{i\psi_{2}(x)}\left(\rho_{2}(x)+\varphi_{21}(x)+i\varphi_{22}(x)\right)\,.\nonumber 
\end{eqnarray}

\noindent For our tree-level analysis, we can neglect all fluctuations
$\varphi_{ij}$ and are only concerned with the potential 

\medskip{}

\noindent 
\begin{equation}
U(\rho_{1},\rho_{2})=-\mathcal{L}(\rho_{1},\rho_{2})=\frac{m_{1}^{2}-\sigma_{1}^{2}}{2}\rho_{1}^{2}+\frac{m_{2}^{2}-\sigma_{2}^{2}}{2}\rho_{2}^{2}+\frac{1}{4}\lambda_{1}\rho_{1}^{4}+\frac{1}{4}\lambda_{2}\rho_{2}^{4}+\frac{1}{2}\lambda_{12}\sigma_{12}\rho_{1}^{2}\rho_{2}^{2}\,.\label{eq:2SLpotential}
\end{equation}

\noindent Here we have introduced the usual definitions $\sigma_{1}^{2}=\partial_{\mu}\psi_{1}\partial^{\mu}\psi_{1}$,~
$\sigma_{2}^{2}=\partial_{\mu}\psi_{2}\partial^{\mu}\psi_{2}$ and
$\sigma_{12}=\partial_{\mu}\psi_{1}\partial^{\mu}\psi_{2}$. For dimensional
considerations, it is helpful to keep in mind that the combination
$\lambda_{12}\sigma_{12}$ is dimensionless. The equations of motions
for $\psi_{1}$ and $\psi_{2}$ are

\noindent 
\begin{eqnarray}
\partial_{\mu}\frac{\partial\mathcal{L}}{\partial\partial_{\mu}\psi_{1}} & = & \partial_{\mu}\left[\rho_{1}^{2}\left(\partial^{\mu}\psi_{1}-\frac{1}{2}\lambda_{12}\rho_{2}^{2}\partial^{\mu}\psi_{2}\right)\right]=0\,,\label{eq:EOMP11}\\
\partial_{\mu}\frac{\partial\mathcal{L}}{\partial\partial_{\mu}\psi_{2}} & = & \partial_{\mu}\left[\rho_{2}^{2}\left(\partial^{\mu}\psi_{2}-\frac{1}{2}\lambda_{12}\rho_{1}^{2}\partial^{\mu}\psi_{1}\right)\right]=0\,.\label{eq:EOMP12}
\end{eqnarray}

\noindent As expected, the gradient coupling leads to entrainment
in equations (\ref{eq:EOMP11}) and (\ref{eq:EOMP12}). The equations
of motion for $\rho_{1}$ and $\rho_{2}$ are

\noindent 
\begin{eqnarray}
\frac{\partial\mathcal{L}}{\partial\rho_{1}} & = & \rho_{1}\left(-m_{1}^{2}+\sigma_{1}^{2}-\lambda_{1}\rho_{1}^{\text{2}}-\lambda_{12}\sigma_{12}\rho_{2}^{2}\right)=0\,,\label{eq:EOMR11}\\
\frac{\partial\mathcal{L}}{\partial\rho_{2}} & = & \rho_{2}\left(-m_{2}^{2}+\sigma_{2}^{2}-\lambda_{2}\rho_{2}^{\text{2}}-\lambda_{12}\sigma_{12}\rho_{1}^{2}\right)=0\,.\label{eq:EOMR12}
\end{eqnarray}

\newpage{}

\section{\noindent Hydrodynamics\label{sub:Hydrodynamics}}

~

\noindent In the zero temperature case, the microscopic representation
of the coefficients $\bar{{\cal A}}$, $\bar{{\cal B}}$ and $\bar{{\cal C}}$
can be conveniently read off from equations (\ref{eq:EOMP11}) and
(\ref{eq:EOMP12}), 

\noindent 
\begin{eqnarray}
j_{1}^{\mu} & = & \bar{{\cal B}}\,\partial^{\mu}\psi_{1}+\bar{{\cal A}}\,\partial^{\mu}\psi_{2}\,,\label{eq:2FLcurrents}\\
j_{2}^{\mu} & = & \bar{{\cal A}}\,\partial^{\mu}\psi_{1}+\bar{{\cal C}}\,\partial^{\mu}\psi_{2}\,.\label{eq:2FLcurrents2}
\end{eqnarray}

\noindent where:

\noindent 
\[
\bar{{\cal B}}=\rho_{1}^{2},\,\,\,\,\,\,\,\,\,\,\bar{{\cal C}}=\rho_{2}^{2},\,\,\,\,\,\,\,\,\,\,\bar{{\cal A}}=-\frac{1}{2}\lambda_{12}\rho_{1}^{2}\rho_{2}^{2}\,.
\]

\noindent Here, we have implicitly assumed that the phase-gradients
$\partial_{\mu}\psi_{i}$ are indeed the conjugate momenta to $j_{\mu,i}$
as we derived for the single fluid case in section \ref{sub:Zero-temperature-hydrodynamics:-generalized}.
We shall see that this assumption is consistent with the generalized
two-fluid framework. The microscopic stress-energy tensor reads:

\noindent 
\begin{eqnarray}
T^{\mu\nu} & = & 2\frac{\partial\mathcal{L}}{\partial g^{\mu\nu}}-g^{\mu\nu}\mathcal{L}\nonumber \\
\nonumber \\
 & = & \rho_{1}^{2}\partial^{\mu}\psi_{1}\partial^{\nu}\psi_{1}+\rho_{2}^{2}\partial^{\mu}\psi_{2}\partial^{\nu}\psi_{2}+\lambda_{12}\rho_{1}^{2}\rho_{2}^{2}\partial^{\mu}\psi_{1}\partial^{\nu}\psi_{2}-g^{\mu\nu}\mathcal{L}\,,\label{eq:2SFT}
\end{eqnarray}

\noindent and the corresponding two-fluid expression is given by

\medskip{}

\noindent 
\begin{equation}
T^{\mu\nu}=-g^{\mu\nu}\Psi+j_{1}^{\mu}\partial^{\nu}\psi_{1}+j_{2}^{\mu}\partial^{\nu}\psi_{2}\,.\label{eq:TwoFLT0}
\end{equation}

\noindent After contraction with $g_{\mu\nu}$, equation (\ref{eq:TwoFLT0})
can be solved for the generalized pressure density,

\medskip{}

\noindent 
\begin{equation}
\Psi=-\frac{1}{4}\left(g_{\mu\nu}T^{\mu\nu}-j_{1}\cdot\partial\psi_{1}-j_{2}\cdot\partial\psi_{2}\right)\,.
\end{equation}

\noindent With the aid of equations (\ref{eq:2FLcurrents}) and (\ref{eq:2FLcurrents2}),
we can obtain the generalized pressure in its natural form expressed
solely in terms of the momenta $\partial_{\mu}\psi_{i}$ 

\medskip{}

\noindent 
\begin{equation}
\Psi=\Psi[\sigma_{1}^{2},\,\sigma_{2}^{2},\,\sigma_{12}]=-\frac{m_{1}^{2}-\sigma_{1}^{2}}{2}\rho_{1}^{2}-\frac{m_{2}^{2}-\sigma_{2}^{2}}{2}\rho_{2}^{2}-\frac{1}{4}\lambda_{1}\rho_{1}^{4}-\frac{1}{4}\lambda_{2}\rho_{2}^{4}-\frac{1}{2}\lambda_{12}\sigma_{12}\rho_{1}^{2}\rho_{2}^{2}\,.\label{eq:2SLPsi}
\end{equation}

\noindent This on the other hand is precisely the negative tree level
potential from equation (\ref{eq:2SLpotential}). We have now reestablished
the important relation between the effective pressure and the (tree
level) effective action:

\medskip{}

\noindent 
\begin{equation}
\Psi=-U=\mathcal{L}_{T=0}\,.
\end{equation}

\newpage{}

\noindent Starting from $\Psi$ one can obtain the coefficients $\bar{{\cal A}}$,
$\bar{{\cal B}}$ and $\bar{{\cal C}}$ from

\medskip{}

\noindent 
\begin{equation}
\bar{{\cal B}}=2\frac{\partial\Psi}{\partial\sigma_{1}^{2}},\,\,\,\,\,\,\,\,\,\,\bar{{\cal C}}=2\frac{\partial\Psi}{\partial\sigma_{2}^{2}},\,\,\,\,\,\,\,\,\,\,\bar{{\cal A}}=\frac{\partial\Psi}{\partial\sigma_{12}}\,,
\end{equation}

\noindent which can be used as a consistency check. The generalized
energy density $\Lambda$ can be obtained with the help of equations
(\ref{eq:2SFT}) and (\ref{eq:2SLPsi}) ,

\noindent 
\begin{eqnarray}
\Lambda & = & Tr[T]-3\Psi=\frac{3}{2}\left(j_{1}\cdot\partial\psi_{1}+j_{2}\cdot\partial\psi_{2}\right)-\frac{1}{2}T_{\,\,\text{\ensuremath{\mu}}}^{\mu}\nonumber \\
 & = & \frac{1}{2}\left[\rho_{1}^{2}\left(\sigma_{1}^{2}+m_{1}^{2}\right)+\rho_{2}^{2}\left(\sigma_{2}^{2}+m_{2}^{2}\right)+\frac{1}{2}\lambda_{1}\rho_{1}^{4}+\frac{1}{2}\lambda_{2}\rho_{2}^{4}-3\lambda_{12}\sigma_{12}\rho_{1}^{2}\rho_{2}^{2}\right]\,.
\end{eqnarray}

\noindent This expression is formulated in terms of momenta rather
than currents. A microscopic construction of $\Lambda=\Lambda\left[j_{1}^{2},\, j_{2}^{2},\, j_{1}\cdot j_{2}\right]$
is difficult even in the zero temperature case since the condensates
$\rho_{i}$ are complicated functions of the momenta determined by
equations (\ref{eq:EOMR11}) and (\ref{eq:EOMR12}) (see next section
for explicit results). One can check the limit case without entrainment
($\lambda_{12}=0$) where one obtains the generalized energy density
as the sum two of non interacting superfluid energy-densities (compare
to equation (\ref{eq:microEP}))

\noindent 
\begin{equation}
\Lambda[\lambda_{12}=\text{0}]=\sum_{i}\epsilon_{s,i}=\sum_{i}\frac{(3\sigma_{i}^{2}+m_{i}^{2})(\sigma_{i}^{2}-m_{i}^{2})}{4\lambda_{i}}\,.
\end{equation}

\noindent With the above results at hand, it is easy to check the
validity of the generalized thermodynamic relation

\noindent 
\begin{equation}
\Psi+\Lambda=j_{1}\cdot\partial\psi_{1}+j_{2}\cdot\partial\psi_{2}\,,
\end{equation}

\noindent which confirms the role of $\partial^{\mu}\psi_{i}$ as
the conjugate momenta to $j_{i}^{\mu}$.

~

\subsection{\noindent Frame dependent hydrodynamics\label{sub:Frame-dependent-hydrodynamics}}

~

\noindent Now that the invariant formalism is set up, it remains to
determine the physical meaning of the conjugate momenta $\partial_{\mu}\psi_{i}$.
From the discussion of sections \ref{sub:Temperature-and-chemical-pot}
and \ref{sub:Entrainment-and-superfluid-from-field-theory} it should
be clear that the field theoretic expressions are obtained in a global
rest frame, the rest frame of the heat bath (or equivalently of the
normal fluid, see also discussion in section \ref{sub:The-two-fluid-formalism-from-field-theory}).
If we set the coupling $\lambda_{12}$ to zero, we are left with a
hydrodynamic system of two independent superfluids and the results
of section \ref{sub:Zero-temperature-hydrodynamics:-generalized}
can be adopted. In particular, $\sigma_{i}$ plays the role of a chemical
potential in the respective rest frame of each superfluid while $\partial_{0}\psi_{i}$
and $\vec{v}_{i}=-\vec{\nabla}\psi_{i}/\partial_{0}\psi_{i}$ are
chemical potential and superfluid velocity in the rest frame of the
heat bath. At first glance, it might seem obvious that these identifications
still remain intact once the coupling $\lambda_{12}$ is turned on.
However, in particular the microscopic definition of the chemical
potential has to be carefully checked. As we know, the chemical potential
enters the Lagrangian similar to the zeroth component of a gauge field
(i.e. $\partial_{0}\varphi\rightarrow\mathcal{D}_{0}\varphi=\partial_{0}\varphi-i\mu$).
The situation at hand is more complicated as the derivatives of $\psi_{1}$
and $\psi_{2}$ appear coupled to each other in \ref{eq:2SFL}. In
appendix \ref{sec:Chemical-potential-in-2SF} we clarify that the
microscopic definition of the chemical potential remains intact (i.e.
even in the presence of a nonzero coupling $\lambda_{12}$ the chemical
potential is introduced to the Lagrangian by replacing \textit{all
}time derivatives by $\partial_{0}\varphi_{i}-i\mu_{i}$) and there
is no reason to doubt that we can identify $\partial_{0}\psi_{i}$
as the chemical potential $\mu_{i}$ in the normal-fluid rest frame.

\noindent Let`s now discuss the role of the spatial components of
$\partial_{\mu}\psi_{i}$. By setting $\vec{\nabla}\psi_{1}=\vec{0}$
we assume that superfluid 1 is at rest in the frame of the heat bath
and only superfluid 2 is in motion. To simplify notation, we define
the generalized pressure expressed in this frame as $\Psi_{1}=\Psi(\vec{\nabla}\psi_{1}=\vec{0})$.
The components of the pressure thus are $T_{1\perp}$ and $T_{1\parallel}$
with respect to $\vec{\nabla}\psi_{2}$. Then, using the projection
operators introduced in section \ref{sub:From-generalized-to-frames},
one can easily confirm that in this particular frame the following
relations hold (compare also with the left column of table 1):

\noindent 
\begin{eqnarray}
T_{1\perp} & = & \frac{1}{2}\left(\delta_{ij}-\frac{\partial^{i}\psi_{2}\partial^{j}\psi_{2}}{\vec{\nabla}\psi_{2}^{2}}\right)T_{1}^{ij}=\Psi_{1}\,,\\
\nonumber \\
T_{1\parallel} & = & \frac{\partial^{i}\psi_{2}\partial^{j}\psi_{2}}{\vec{\nabla}\psi_{2}^{2}}\, T_{1}^{ij}=\Psi_{1}-\,\vec{j}_{2,1}\cdot\vec{\nabla}\psi_{2}\,.
\end{eqnarray}

\noindent Furthermore, we can confirm for $\Lambda_{1}=\Lambda(\vec{\nabla}\psi_{1}=\vec{0})$
\begin{equation}
\Lambda_{1}=T_{1}^{00}+T_{1\perp}-T_{1\parallel}\,.
\end{equation}

\newpage{}

\section{\noindent Phase diagram at zero temperature\label{sub:Phase-diagram-at-zeroT}}

~

\subsection{Stability condition and global minima}

~

\noindent So far, we have ignored the possibility that the coexistence
of the condensates $\rho_{1}$ and $\rho_{2}$ might be forbidden
for some values of the parameter space spanned by $\mu_{1}$, $\mu_{2}$,
the couplings $\lambda_{1}$, $\lambda_{2}$, $\lambda_{12}$, the
velocities $\left|\vec{v}_{1}\right|$,$\left|\vec{v}_{2}\right|$
and the angle $\theta$ between $\vec{v}_{1}$ and $\vec{v}_{2}$.
Possible scenarios include no condensation, the existence of a single
condensate $\rho_{1}$ or $\rho_{2}$ and the coexistence of both.
Therefore we have to construct a phase diagram where we compare the
potentials $U(0,0)$, $U(\rho_{1},0)$, $U(0,\rho_{2})$ and $U(\rho_{1},\rho_{2})$. 

\noindent Before we do so, we have to check the stability of the potential
(\ref{eq:2SLpotential}). The potential must be bounded from below
in all directions in the space spanned by $\rho_{1}$ and $\rho_{2}$.
The asymptotic behavior of the tree-level potential is determined
by the interaction terms $\frac{1}{4}\lambda_{1}\rho_{1}^{4}+\frac{1}{4}\lambda_{2}\rho_{2}^{4}+\frac{1}{2}\lambda_{12}\sigma_{12}\rho_{1}^{2}\rho_{2}^{2}$.
We first observe that as long as we move along a trajectory of fixed
$\rho_{2}$, $\lambda_{1}$ has to be larger than zero and vice versa
which leads to the restrictions $\lambda_{1}>0$ and $\lambda_{2}\mathcal{>}0$.
It remains to find a similar condition for the coupling $\lambda_{12}$.
Obviously the case $\lambda_{12}\sigma_{12}>0$ leads to a stable
potential. To investigate the case of $\lambda_{12}\sigma_{12}<0$,
we define:

\noindent 
\begin{eqnarray}
P_{1} & = & \frac{1}{4}\lambda_{1}\rho_{1}^{4}+\frac{1}{4}\lambda_{2}\rho_{2}^{4}\,,\\
P_{2} & = & \frac{1}{2}\left|\lambda_{12}\sigma_{12}\right|\rho_{1}^{2}\rho_{2}^{2}\,.\nonumber 
\end{eqnarray}

\noindent Along the $\rho_{1}$ and $\rho_{2}$ axes the potential
$P_{2}$ is zero and lies below the potential $P_{1}$, which has
a parabolic shape and is only zero at the origin. An instability occurs,
if $\left|\lambda_{12}\sigma_{12}\right|$ becomes large enough compared
to $\lambda_{1}$ and $\lambda_{2}$ for the potentials $P_{1}$ and
$P_{2}$ to intersect. This leads to ($P_{1}=P_{2}$)

\medskip{}

\noindent 
\begin{equation}
\rho_{2}^{2}=\rho_{1}^{2}\left[\frac{\left|\lambda_{12}\sigma_{12}\right|}{\lambda_{1}}\pm\sqrt{\left(\frac{\lambda_{12}\sigma_{12}}{\lambda_{1}}\right)^{2}-\frac{\lambda_{2}}{\lambda_{1}}}\right]\,.
\end{equation}

\noindent In order for such an intersection not to occur, the square
root must become imaginary which happens for

\noindent 
\begin{equation}
\lambda_{1}\lambda_{2}>\lambda_{12}^{2}\sigma_{12}^{2}\,.\label{eq:MinStable}
\end{equation}

\noindent Observe that $\lambda_{12}\sigma_{12}=\lambda_{12}\mu_{1}\mu_{2}\left(1-\vec{v}_{1}\cdot\vec{v}_{2}\right)$
can change its sign if either $\lambda_{12}$ or the product $\mu_{1}\mu_{2}$
becomes negative (the factor $(1-\vec{v}_{1}\cdot\vec{v}_{2})$ is
always larger than or equal to zero). It is interesting to observe
that entrainment manifests itself in a coupling of the chemical potentials
$\mu_{1}$ and $\mu_{2}$ and is thus present even if there is no
relative motion between both superfluids (i.e. $\vec{v}_{1}=\vec{v}_{2}$). 

\noindent We can now continue with the discussion of the phase structure.
The axes of the phase diagrams will be given by $\mu_{1}$ and $\mu_{2}$
while the parameters $\lambda_{1}$, $\lambda_{2}$, $\lambda_{12}$,$\left|\vec{v}_{1}\right|$,$\left|\vec{v}_{2}\right|$
and $\theta$ are fixed in each diagram. Any phase must fulfill the
following set of conditions in order to represent the ground state: 
\begin{enumerate}
\item \noindent The (minimized) condensates in each phase must be real.
\item \noindent The phase under consideration must be a local minimum.
\item \noindent In addition, the phase under consideration must also be
a global minimum.
\end{enumerate}
\noindent Condition (2) is fulfilled, if the eigenvalues of the Hesse
matrix $H(\rho_{1},\rho_{2})$ are positive. In the most general case,
$H=H(\rho_{1},\rho_{2})$ is given by

\bigskip{}

\noindent 
\begin{equation}
H(\rho_{1},\rho_{2})=\left(\begin{array}{cc}
m_{1}^{2}-\sigma_{1}^{2}+3\lambda_{1}\rho_{1}^{2}+\lambda_{12}\sigma_{12}\rho_{2}^{2} & 2\lambda_{12}\sigma_{12}\rho_{1}\rho_{2}\\
2\lambda_{12}\sigma_{12}\rho_{1}\rho_{2} & m_{2}^{2}-\sigma_{2}^{2}+3\lambda_{2}\rho_{2}^{2}+\lambda_{12}\sigma_{12}\rho_{1}^{2}
\end{array}\right)\,.\label{eq:Hesse}
\end{equation}

~

\noindent \textit{No condensation. }We begin with the simplest case
of no condensation ($\rho_{1}=0$ and $\rho_{2}=0$). The Hesse matrix
(\ref{eq:Hesse}) takes the simple form 

\noindent 
\begin{equation}
H(0,0)=\left(\begin{array}{cc}
m_{1}^{2}-\sigma_{1}^{2} & 0\\
0 & m_{2}^{2}-\sigma_{2}^{2}
\end{array}\right)\,.
\end{equation}

\noindent The eigenvalues can be read off right away and the local
minimum requirements are

\noindent 
\begin{eqnarray}
\sigma_{1}^{2} & < & m_{1}^{2}\,,\label{eq:nocond}\\
\sigma_{2}^{2} & < & m_{2}^{2}\,.\label{eq:nocond2}
\end{eqnarray}

\noindent As long as the conditions (\ref{eq:nocond}), (\ref{eq:nocond2})
are met, $U(\rho_{1},0)$, $U(0,\rho_{2})$ and $U(\rho_{1},\rho_{2})$
are all larger than zero and $U(0,0)$ represents the global minimum. 

~

\noindent \textit{Single condensation. }Next we consider $\rho_{1}\neq0$
and $\rho_{2}=0$ . The case of $\rho_{2}\neq0$ and $\rho_{1}=0$
can be obtained from this case by exchanging the chemical index 1
with 2. The Hesse matrix reads

\bigskip{}

\noindent 
\begin{equation}
H(\rho_{1},0)=\left(\begin{array}{cc}
m_{1}^{2}-\sigma_{1}^{2}+3\lambda_{1}\rho_{1}^{2} & 0\\
0 & m_{2}^{2}-\sigma_{2}^{2}+\lambda_{12}\sigma_{12}\rho_{1}^{2}
\end{array}\right)\,.
\end{equation}

\noindent Observe that the coupling between both condensates stills
plays a role, even though $\rho_{2}$ is set to zero.

\noindent \begin{flushleft}
\newpage{}Condition (1) is given by 
\par\end{flushleft}

\bigskip{}

\noindent 
\begin{equation}
\frac{\sigma_{1}^{2}-m_{1}^{2}}{\lambda_{1}}=\rho_{1}^{2}\,,\label{eq:rho1min}
\end{equation}

\noindent which can be used to simplify $H(\rho_{1},0)$. The local
minimum conditions (2) then read:

\noindent 
\begin{eqnarray}
\sigma_{1}^{2} & > & m_{1}^{2}\,,\label{eq:OneCond}\\
\sigma_{2}^{2} & < & m_{2}^{2}+\frac{\lambda_{12}\sigma_{12}}{\lambda_{1}}\left(\sigma_{1}^{2}-m_{1}^{2}\right)\,,\,\,\,\,\,\textrm{for}\,\lambda_{12}\sigma_{12}>0\,,\label{eq:eqOneCond}\\
\sigma_{2}^{2} & < & m_{2}^{2}-\frac{\left|\lambda_{12}\sigma_{12}\right|}{\lambda_{1}}\left(\sigma_{1}^{2}-m_{1}^{2}\right)\,,\,\,\,\,\,\textrm{for}\,\lambda_{12}\sigma_{12}<0\,,\label{eq:OneCond3}
\end{eqnarray}

\noindent From equation (\ref{eq:rho1min}) and (\ref{eq:OneCond}),
we find that condition (1) is fulfilled in a region where $U(\rho_{1},0)$
represents a local minimum. The minimized potential reads

\bigskip{}

\noindent 
\begin{equation}
U(\rho_{1},0)=-\frac{1}{4}\frac{(\sigma_{1}^{2}-m_{1}^{2})^{2}}{\lambda_{1}}\,.\label{eq:singlePot}
\end{equation}

\noindent Since this expression is manifestly negative, it will be
the preferred ground state compared to $U(0,0)$ (as long as inequalities
(\ref{eq:eqOneCond}) and (\ref{eq:OneCond3}) are fulfilled). To
determine whether it represents the global minimum we still have to
compare it to the case of the coexistence of $\rho_{1}$ and $\rho_{2}$. 

~

\noindent \textit{Coexistence phase. }If both condensates are present,
the Hesse matrix is given by

\bigskip{}

\noindent 
\begin{equation}
H(\rho_{1},\rho_{2})=2\left(\begin{array}{cc}
\lambda_{1}\rho_{1}^{2} & \lambda_{12}\sigma_{12}\rho_{1}\rho_{2}\\
\lambda_{12}\sigma_{12}\rho_{1}\rho_{2} & \lambda_{2}\rho_{2}^{2}
\end{array}\right)\,.
\end{equation}

\noindent Here we have used the equations of motion (\ref{eq:EOMR11})
and (\ref{eq:EOMR12}) to simplify the Hesse matrix (\ref{eq:Hesse}).
The eigenvalues calculate to

\medskip{}

\noindent 
\begin{equation}
\omega_{1,2}=(\lambda_{1}\rho_{1}^{2}+\lambda_{2}\rho_{2}^{2})\pm\sqrt{(\lambda_{1}\rho_{1}^{2}+\lambda_{2}\rho_{2}^{2})^{2}-4\rho_{1}^{2}\rho_{2}^{2}(\lambda_{1}\lambda_{2}-\lambda_{12}^{2}\sigma_{12}^{2})}\,.\label{eq:LocMin}
\end{equation}

\noindent In case of $\lambda_{1}\lambda_{2}\leq\lambda_{12}^{2}\sigma_{12}^{2}$
we can see right away that one eigenvalue turns out to be negative,
while the other is positive (i.e. we found a saddle point rather than
a minimum). A \textit{local} minimum is therefore only given for $\lambda_{1}\lambda_{2}>\lambda_{12}^{2}\sigma_{12}^{2}$.
Observe that this is also precisely the stability condition from equation
(\ref{eq:MinStable}). Condition (1) can be checked by solving the
equations of motion (\ref{eq:EOMR11}) and (\ref{eq:EOMR12}) for
$\rho_{1}$ and

\noindent ~

\noindent ~

~

\noindent $\rho_{2}$ whereby we find the following expressions

\noindent 
\begin{eqnarray}
\rho_{1} & = & \frac{1}{N}\sqrt{\frac{\sigma_{1}^{2}-m_{1}^{2}}{\lambda_{1}}-\frac{\lambda_{12}\sigma_{12}}{\lambda_{1}\lambda_{2}}\left(\sigma_{2}^{2}-m_{2}^{2}\right)}\,,\label{eq:rho11min}\\
\rho_{2} & = & \frac{1}{N}\sqrt{\frac{\sigma_{2}^{2}-m_{2}^{2}}{\lambda_{2}}-\frac{\lambda_{12}\sigma_{12}}{\lambda_{1}\lambda_{2}}\left(\sigma_{1}^{2}-m_{1}^{2}\right)}\,,\label{eq:rho12min}\\
N & = & \sqrt{1-\frac{\lambda_{12}^{2}\sigma_{12}^{2}}{\lambda_{1}\lambda_{2}}}\,.
\end{eqnarray}

\noindent Using condition (\ref{eq:MinStable}), we can see that $N$
is real and positive and it is sufficient to check the expressions
in the numerators of $\rho_{1}$ and $\rho_{2}$:

\noindent 
\begin{eqnarray}
\sigma_{2}^{2} & < & m_{2}^{2}+\frac{\lambda_{2}}{\lambda_{12}\sigma_{12}}\left(\sigma_{1}^{2}-m_{1}^{2}\right)\,,\label{eq:sigma2min1}\\
\sigma_{2}^{2} & > & m_{2}^{2}+\frac{\lambda_{12}\sigma_{12}}{\lambda_{1}}\left(\sigma_{1}^{2}-m_{1}^{2}\right)\,,\label{eq:sigma2min2}\\
\lambda_{12} & > & 0\,.\nonumber 
\end{eqnarray}

\noindent While the combination $\lambda_{12}\sigma_{12}$ appears
quadratically in the stability condition (\ref{eq:MinStable}), it
appears linear in equations (\ref{eq:sigma2min1}) and (\ref{eq:sigma2min2}).
In case $\lambda_{12}\sigma_{12}$ is negative, condition (1) leads
to the following set of equations: 

\noindent 
\begin{eqnarray}
\sigma_{2}^{2} & > & m_{2}^{2}-\frac{\lambda_{2}}{\left|\lambda_{12}\sigma_{12}\right|}\left(\sigma_{1}^{2}-m_{1}^{2}\right)\,,\label{eq:sigmamin1neg}\\
\sigma_{2}^{2} & > & m_{2}^{2}-\frac{\left|\lambda_{12}\sigma_{12}\right|}{\lambda_{1}}\left(\sigma_{1}^{2}-m_{1}^{2}\right)\,,\label{eq:sigmamin2neg}\\
\lambda_{12} & < & 0\,.\nonumber 
\end{eqnarray}

\noindent Consider the plane spanned by positive and negative values
of $\mu_{1}$ and $\mu_{2}$. In quadrant I and III, equations (\ref{eq:sigma2min1})
and (\ref{eq:sigma2min2}) are valid for $\lambda_{12}>0$ while equations
(\ref{eq:sigmamin1neg}) and (\ref{eq:sigmamin2neg}) are valid for
$\lambda_{12}<0$. In quadrants II and IV it is the other way around. 

\noindent It remains to check whether the coexistence phase indeed
represents a global minimum.The full minimized potential reads:

\bigskip{}

\noindent 
\begin{equation}
U(\rho_{1},\rho_{2})=-\lambda_{2}\frac{(\sigma_{1}^{2}-m_{1}^{2})^{2}}{4(\lambda_{1}\lambda_{2}-\lambda_{12}^{2}\sigma_{12}^{2})}-\lambda_{1}\frac{(\sigma_{2}^{2}-m_{2}^{2})^{2}}{4(\lambda_{1}\lambda_{2}-\lambda_{12}^{2}\sigma_{12}^{2})}+\lambda_{12}\sigma_{12}\frac{(\sigma_{1}^{2}-m_{1}^{2})(\sigma_{2}^{2}-m_{2}^{2})}{2(\lambda_{1}\lambda_{2}-\lambda_{12}^{2}\sigma_{12}^{2})}\,.\label{eq:PotFullmin}
\end{equation}

~

~

\noindent For vanishing coupling $\lambda_{12}$, this reduces to
the potential of two non-interacting condensates

\bigskip{}

\noindent 
\begin{equation}
U(\rho_{1},\,\rho_{2},\,\lambda_{12}=0)=-\frac{(\sigma_{1}^{2}-m_{1}^{2})^{2}}{4\lambda_{1}}-\frac{(\sigma_{2}^{2}-m_{2}^{2})^{2}}{4\lambda_{2}}\,.\label{eq:PotL120}
\end{equation}
To compare (\ref{eq:PotFullmin}) to the case of no condensation and
the single condensate phases, we re-organize the numerator

\bigskip{}

\noindent 
\begin{eqnarray}
U(\rho_{1},\rho_{2}) & = & -\frac{(\sigma_{1}^{2}-m_{1}^{2})\left[\lambda_{2}(\sigma_{1}^{2}-m_{1}^{2})-\lambda_{12}\sigma_{12}(\sigma_{2}^{2}-m_{2}^{2})\right]}{4(\lambda_{1}\lambda_{2}-\lambda_{12}^{2}\sigma_{12}^{2})}\label{eq:DualPot}\\
 & - & \frac{(\sigma_{2}^{2}-m_{2}^{2})\left[\lambda_{1}(\sigma_{2}^{2}-m_{2}^{2})-\lambda_{12}\sigma_{12}(\sigma_{1}^{2}-m_{1}^{2})\right]}{4(\lambda_{1}\lambda_{2}-\lambda_{12}^{2}\sigma_{12}^{2})}\,.
\end{eqnarray}

\noindent The denominator is always positive. In the numerator, the
expressions in the square brackets are always positive due to inequalities
(\ref{eq:sigma2min1})-(\ref{eq:sigmamin2neg}). Thus, the potential
is always negative and preferred to case of no condensation. In order
to compare this to the single condensate case, we can ask for example
in which cases

\bigskip{}

\noindent 
\begin{equation}
U(\rho_{1},0)>U(\rho_{1},\rho_{2})
\end{equation}

\noindent is fulfilled. Inserting both potentials from equations (\ref{eq:singlePot})
and (\ref{eq:DualPot}), we find that this condition is equivalent
to

\bigskip{}

\noindent 
\begin{eqnarray*}
\lambda_{12}\sigma_{12}(\sigma_{1}^{2}-m_{1}^{2})\left[\lambda_{12}\sigma_{12}(\sigma_{1}^{2}-m_{1}^{2})-\lambda_{1}(\sigma_{2}^{2}-m_{2}^{2})\right] & < & \lambda_{1}(\sigma_{2}^{2}-m_{2}^{2})\left[\lambda_{12}\sigma_{12}(\sigma_{1}^{2}-m_{1}^{2})-\lambda_{1}(\sigma_{2}^{2}-m_{2}^{2})\right]\\
\rightarrow\lambda_{12}\sigma_{12}(\sigma_{1}^{2}-m_{1}^{2})-\lambda_{1}(\sigma_{2}^{2}-m_{2}^{2}) & < & 0\,.
\end{eqnarray*}

\noindent This is precisely the condition of $\sigma_{2}$ to be real.
The condition $U(\rho_{2},0)>U(\rho_{1},\rho_{2})$ leads to the second
reality condition $(\sigma_{2}^{2}-m_{2}^{2})\lambda_{12}\sigma_{12}-\lambda_{2}(\sigma_{1}^{2}-m_{1}^{2})>0$.
In other words, in regions of the parameter space, where the coupled
phase is a valid candidate according to condition (1) and (2) , it
indeed represents the ground state. We can also pose the inverse condition
$U(\rho_{1},0)<U(\rho_{1},\rho_{2})$ which is obviously equivalent
to

\bigskip{}

\noindent 
\[
\lambda_{1}(\sigma_{2}^{2}-m_{2}^{2})-\lambda_{12}\sigma_{12}(\sigma_{1}^{2}-m_{1}^{2})>0\,,
\]

\noindent the condition for the existence of a phase with condensate
of only one condensate (in this case $\rho_{1}$). Again, we could
repeat this analysis by comparing $U(0,\rho_{2})$ with $U(\rho_{1},\rho_{2})$. 

~

\newpage{}

\subsection{\noindent Phase diagrams\label{sub:Phase-diagrams}}

~

\begin{figure}[t]
\fbox{\begin{minipage}[t]{1\columnwidth}%
\begin{center}
\includegraphics[scale=0.6]{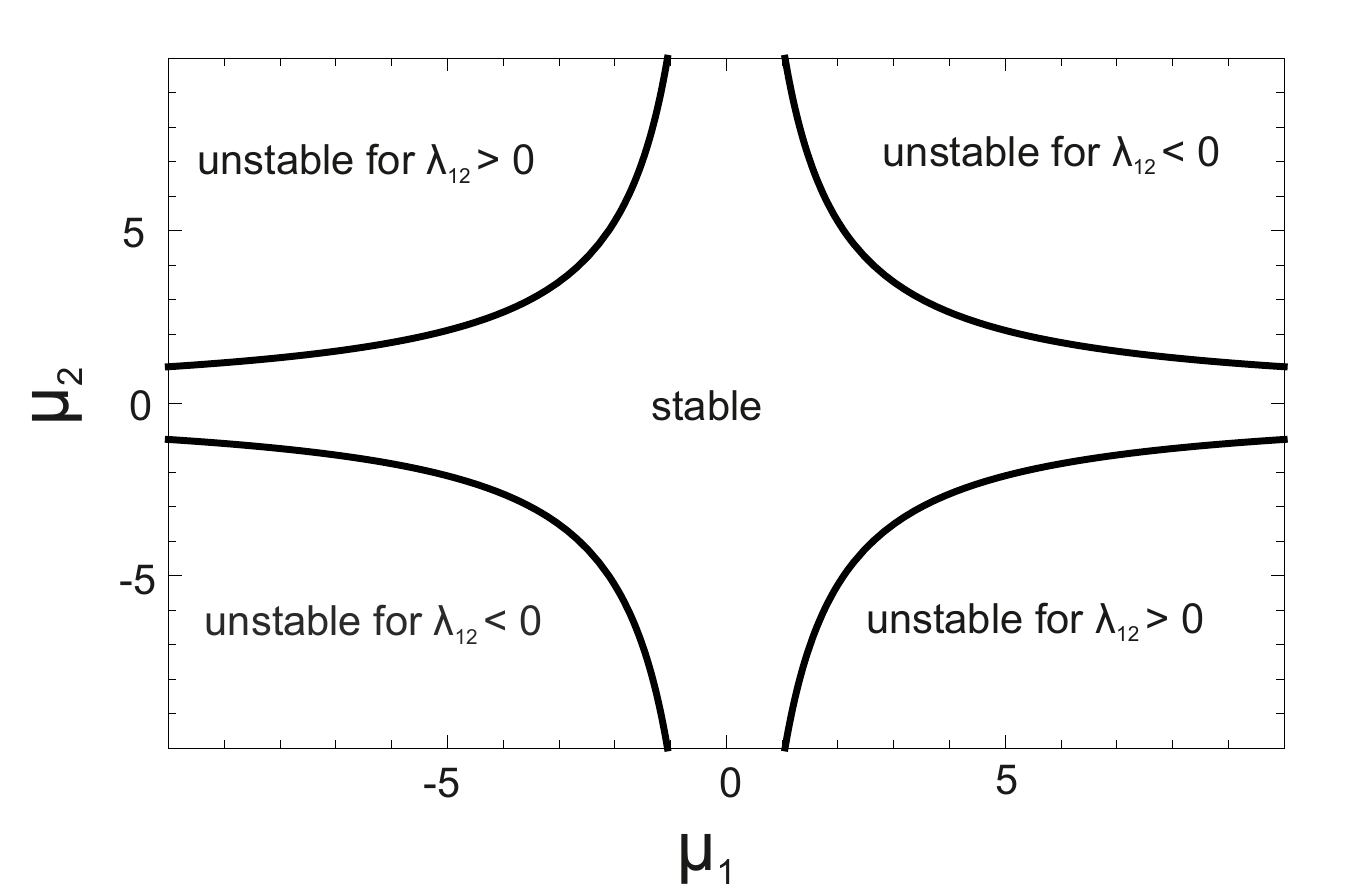}~~~~~\includegraphics[scale=0.6]{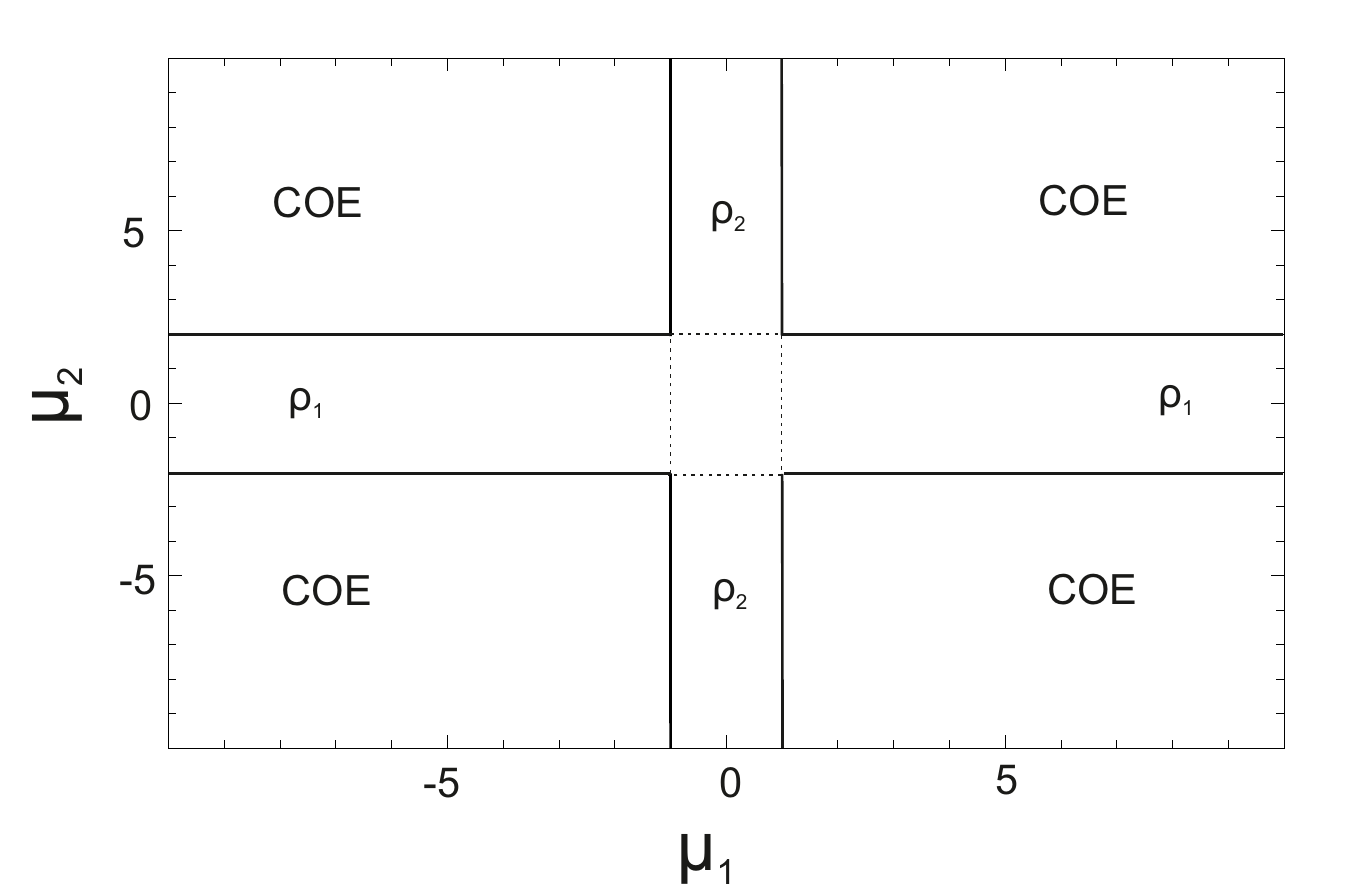}
\par\end{center}%
\end{minipage}}\protect\caption{Left panel: unstable regions in the space spanned by the chemical
potentials $\mu_{1}=\partial_{0}\psi_{1}$ and $\mu_{2}=\partial_{0}\psi_{2}$
for positive and negative couplings $\lambda_{12}$. The parameters
for this plot are chosen as follows: $m_{1}=1,\,\, m_{2}=2,\,\,\lambda_{1}=0.2,\,\,\lambda_{2}=0.5,\,\,\left|\lambda_{12}\right|=0.03,\,\,\left|\vec{v}_{1}\right|=\left|\vec{v}_{2}\right|=0$.
Right panel: outline of the phase diagram for a system of two decoupled
scalar fields (i.e. in the absence of entrainment, $\lambda_{12}=0$,
all other parameters remain unchanged). The boundaries for single
and dual condensation are given by $\mu_{i}=\pm m_{i}\gamma_{i}=\pm m_{i}/\sqrt{1-\vec{v}_{i}^{2}}$
(which reduces to $\mu_{i}=\pm m_{i}$ in the given parameter set).
In the area in the middle where $\mu_{1}<m_{1}$ \textit{and} $\mu_{2}<m_{2}$,
no condensation is possible. The coexistence phase (COE) exists in
regions where $\mu_{1}>m_{1}$ \textit{and} $\mu_{2}>m_{2}$. Single
condensate phases reside in the regions where only one chemical potential
is larger than its corresponding mass. Solid lines correspond to first-order
phase transitions, dashed lines correspond to second-order phase transitions.
\label{fig:PDiangNoEnt}}
\end{figure}

\noindent Conditions (\ref{eq:nocond}), (\ref{eq:nocond2}) as well
as (\ref{eq:OneCond})-(\ref{eq:OneCond3}) and (\ref{eq:sigma2min1})-(\ref{eq:sigmamin2neg})
together with the restrictions $\lambda_{1}>0,\,\,\lambda_{2}>0$
and $\lambda_{1}\lambda_{2}>\lambda_{12}^{2}\sigma_{12}^{2}$ contain
all necessary information to construct the zero temperature phase
diagrams. To determine the phase structure in a plane spanned by $\mu_{1}=\partial_{0}\psi_{1}$
and $\mu_{2}=\partial_{0}\psi_{2}$, we turn all inequalities into
equations which we solve for $\mu_{2}=\mu_{2}(\mu_{1})$. This leaves
the couplings, $\left|\vec{v}_{1}\right|$ ,$\left|\vec{v}_{2}\right|$
and the angle $\theta$ as tunable parameters. To obtain an analytic
condition for the boundary between the two single condensate phases,
we can solve $U(\rho_{1},0)=U(0,\rho_{2})$ for $\mu_{2}$. Since
some of these equations are cubic in $\mu_{2}$, an analytic construction
of the phase diagram is difficult. We shall rather compare $U(0,0)$,
$U(\rho_{1},0)$ , $U(0,\rho_{2})$ and $U(\rho_{1},\rho_{2})$ numerically
for each point in the phase diagram. This of course involves checking
the respective conditions (1)-(3) for each phase and each point in
the phase diagram. To check the numerical results, we can compare
the numerically calculated phase boundaries with the analytical solutions
of the set of equations discussed above. We shall not attempt to obtain
all possible topologies of the phase diagram but rather aim to identify
regions in the parameter space, in which the coexistence of both condensates
in the presence of entrainment is in principle allowed. The left panel
of figure \ref{fig:PDiangNoEnt} displays the phase space which is
``cut out'' by stability conditions for positive and negative entrainment
coupling. 

\noindent 
Let us begin with the very simple scenario of $\lambda_{12}=0$. The
potential is given by (\ref{eq:PotL120}) and single condensation
occurs if $\sigma_{1}^{2}>m_{1}^{2}$ or $\sigma_{2}^{2}>m_{2}^{2}$.
In regions where both conditions are satisfied, we find the coexistence
of both condensates. This scenario is illustrated in the figure \ref{fig:PDiangNoEnt}.
We do not specify, in which units the chemical potentials $\mu_{i}$
are measured in the phase diagrams we present here as this is a general
study. 

\begin{figure}[t]
\fbox{\begin{minipage}[t]{1\columnwidth}%
\begin{center}
\includegraphics[scale=0.65]{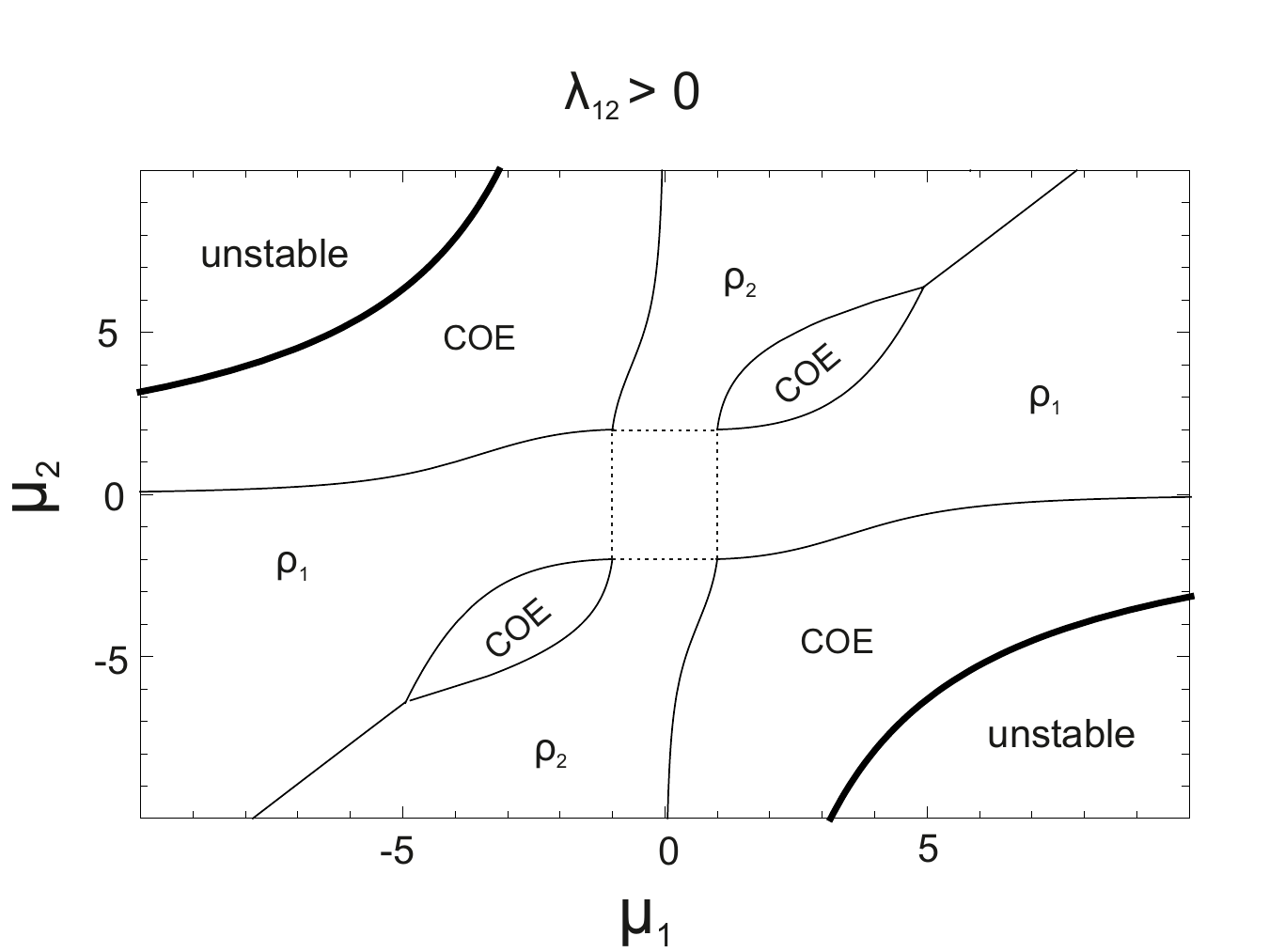}\includegraphics[scale=0.65]{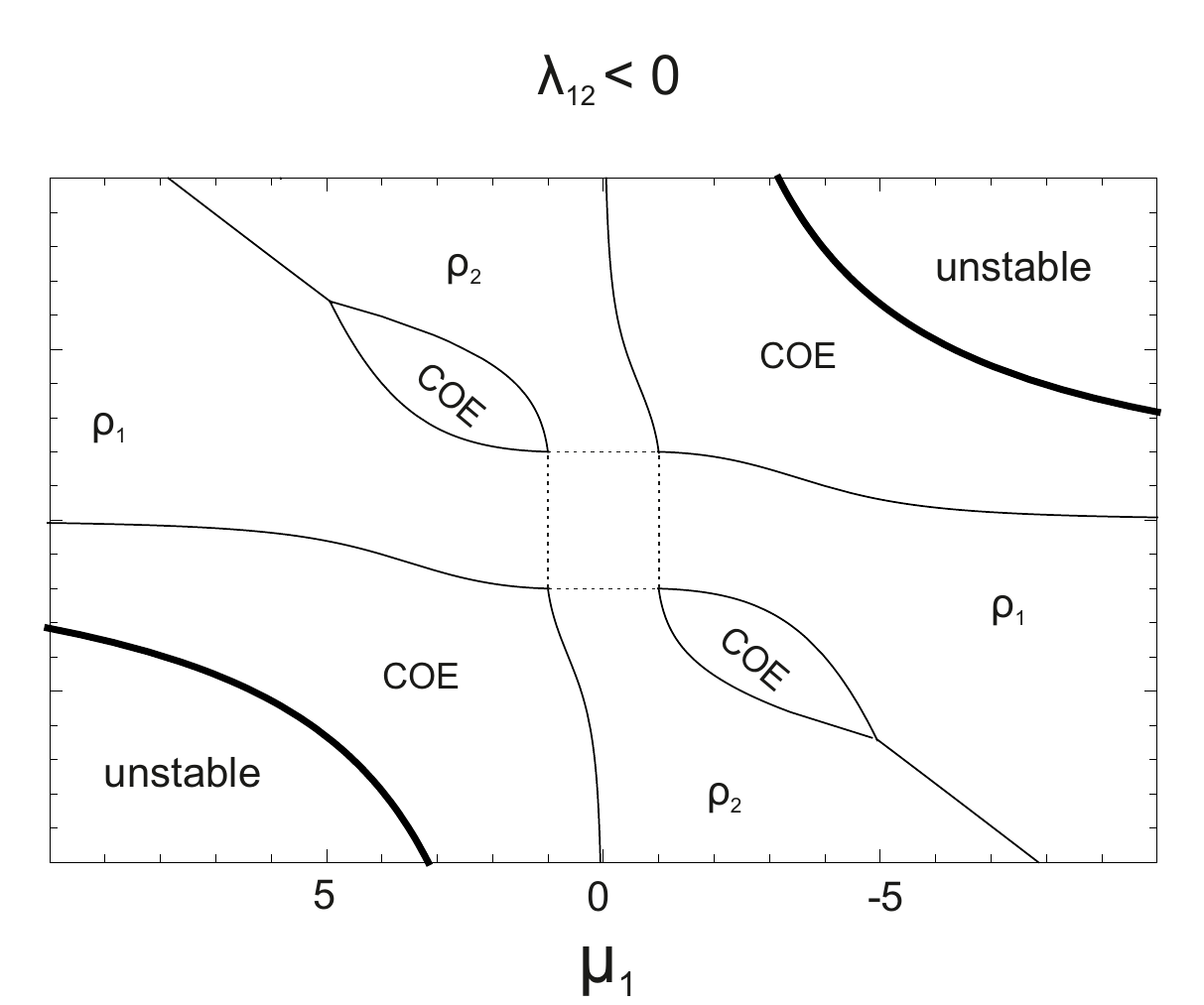}
\par\end{center}%
\end{minipage}}\protect\caption{Outline of the phase diagram for a system of two scalar fields coupled
by entrainment interactions. The parameter set is the same as in figure
\ref{fig:PDiangNoEnt} with the exception of $\lambda_{12}$ which
is now given by $\lambda_{12}=0.01$ (left panel) and $\lambda_{12}=-0.01$
(right panel). Thick black lines visualize the stability condition
(\ref{eq:MinStable}). Obviously, changing the of sign of $\lambda_{12}$
``rotates'' the phase diagram by 90\textdegree . \label{fig:PDiangwithEnt}}
\end{figure}

\noindent Once we switch on entrainment interactions, the phase diagram
changes in two ways: first of all, the straight lines of $\sigma_{i}=\pm m_{i}$
representing the phase boundaries in the $\lambda_{12}=0$ case become
deformed and are determined by (\ref{eq:OneCond}), (\ref{eq:eqOneCond})
and (\ref{eq:sigma2min1})-(\ref{eq:sigmamin2neg}) (with the exception
of the phase with no condensation which of course still resides in
the area defined by $\sigma_{i}<\pm m_{i}$). Secondly, condition
(\ref{eq:MinStable}) significantly reduces the available phase space,
see left panel of figure \ref{fig:PDiangNoEnt}. Different values
for the superfluid velocities lead to slight modifications of the
magnitude of $\sigma_{1}$, $\sigma_{2}$ and $\sigma_{12}$ which
will not induce a new topology of the phase structure. This was checked
numerically by varying the the superfluid velocities $\vec{v}_{i}$
between zero and $1/\sqrt{3}$, which is the upper bound for the critical
velocity - at least in the absence of entrainment. The impact on entrainment
on the critical velocity is an interesting subject for future studies.
Qualitative changes of the phase structure are driven by the magnitude
and sign of $\lambda_{12}$ (of course we need to guarantee $\lambda_{1}>0$
and $\lambda_{2}>0$!). To demonstrate the symmetry properties between
positive and negative values of $\mu_{i}$, we shall plot all four
quadrants of the phase diagram. The results for positive and negative
entrainment couplings $\lambda_{12}$ are displayed in figure \ref{fig:PDiangwithEnt}.
As expected from inequalities (\ref{eq:sigma2min1})-(\ref{eq:sigmamin2neg}),
changing the sign of $\lambda_{12}$ 
corresponds to a $90^{\text{\textdegree}}$ rotation of the phase
diagram. As we can see in the left panel of figure \ref{fig:PDiangwithEnt},
the coexistence phase occupies a larger region in the second and fourth
quadrant of the phase phase diagram which is constrained by the stability
condition (\ref{eq:MinStable}) of the tree-level potential as well
as a smaller region in the first and third quadrant constrained by
the phase boundary to the single condensate phases ($\lambda_{12}>0$).
In the second and fourth quadrant, we find regions which are generically
unstable. For $\lambda_{12}<0$, we see that these unstable regions
are ``rotated'' into the first and third quadrant. There is no choice
for $\lambda_{12}$ (except $\lambda_{12}=0$) at which the phase
diagram is generically stable for all chemical potentials. It is important
to realize that the instability can be directly related to the gradient
coupling. If we had coupled the fields directly by an interaction
term of the form 
\[
\mathcal{L}_{int}=g\left|\varphi_{1}\right|^{2}\left|\varphi_{2}\right|^{2}\,,
\]

\noindent with a dimensionless coupling $g$, we would have obtained
the stability condition $\lambda_{1}\lambda_{2}>g^{2}$. This condition
is independent of the chemical potentials $\mu_{i}$ and does not
induce any instabilities in the phase diagram.

\noindent Increasing $\lambda_{12}$ will result in an increase of
unstable regions. At a limit value of 

\medskip{}

\noindent 
\begin{equation}
\bar{\lambda}_{12}=\frac{\sqrt{\lambda_{1}\lambda_{2}}}{m_{1}m_{2}}\,,\label{eq:LambdaLimit}
\end{equation}

\begin{figure}[t]
\fbox{\begin{minipage}[t]{1\columnwidth}%
\begin{center}
\includegraphics[scale=0.7]{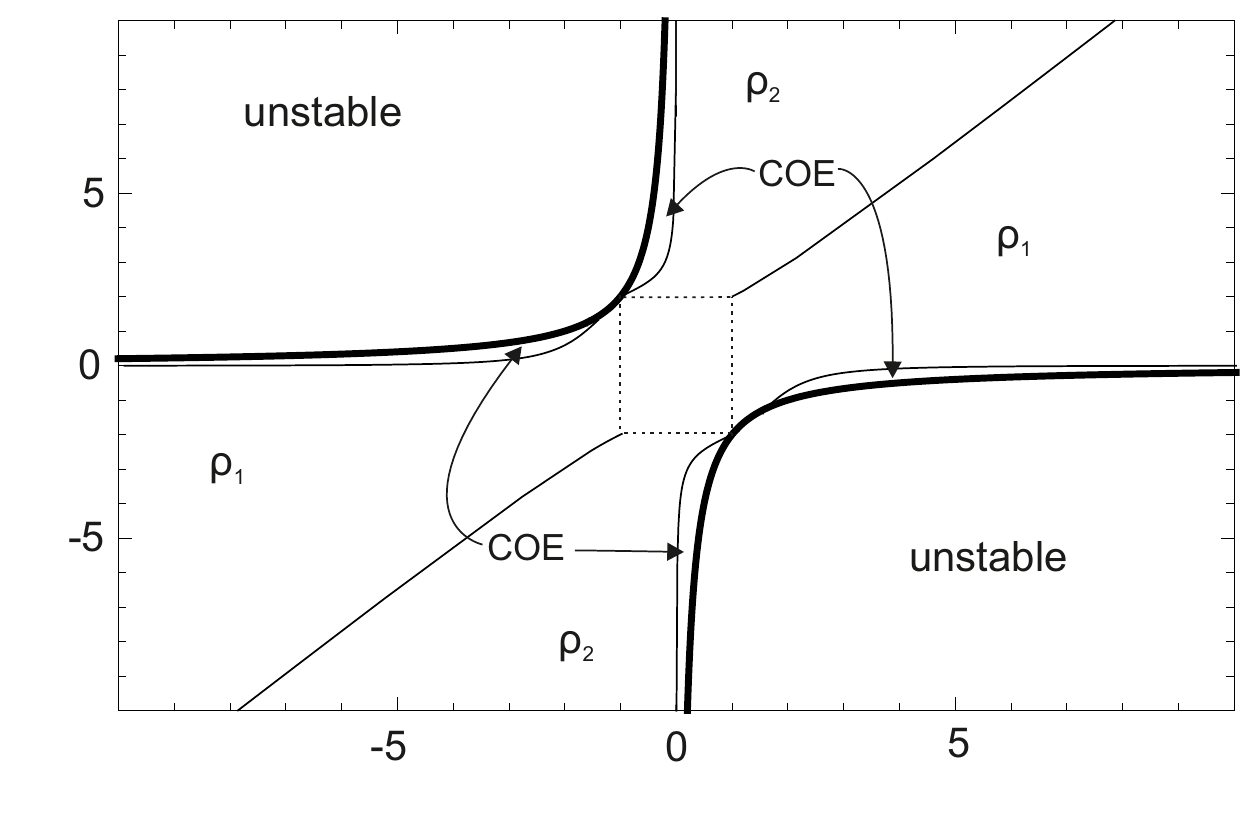}
\par\end{center}%
\end{minipage}}\protect\caption{Same parameter set as in figure \ref{fig:PDiangwithEnt} with the
exception of the entrainment coupling which is now set to $\bar{\lambda}_{12}$,
see equation (\ref{eq:LambdaLimit}). The coexistence phase disappears
completely in quadrants I and III. In quadrants II and IV on the other
hand, a small region of the coexistence phase persists even for values
of $\lambda_{12}$ of order one as one can check numerically. \label{fig:PDiangLargeEnt}}
\end{figure}

\noindent (here we have set $\vec{v}_{i}=\vec{0}$) the stability
condition intersects with the edges of the square defined by $\mu_{1}=\pm m_{1}$
and $\mu_{2}=\pm m_{2}$ and the ``bubble'' in which the coexistence
phase resides in quadrants I and III disappears ($\lambda_{12}>0$).
In quadrants II and IV one can check numerically that the coexistence
phase persists in a small area even in case of very large values of
$\lambda_{12}$ (see figure \ref{fig:PDiangLargeEnt}). 

\noindent 

\subsection{Conclusion}

~

\noindent We have studied the effect of two individual Bose-Einstein
condensates on a microscopic scalar field theory with gradient interaction.
As usual, we have decomposed the condensates $\phi_{i}$ in terms
of a modulus $\rho_{i}$ and a phase $\psi_{i}$. The overall Lagrangian
is invariant under two independent $U(1)_{1}\times U(1)_{2}$ transformations.
Hydrodynamically, this corresponds to a system of two coupled superfluids,
which can effectively be described by a generalized pressure of the
form $\Psi=\Psi[\sigma_{1}^{2},\,\sigma_{2}^{2},\,\sigma_{12}]$ which
is given by the (negative) tree-level potential. We have seen that
gradients of the phase $\psi_{i}$ constitute the conjugate momenta
to the currents $j_{i}^{\mu}$. Calculating the equations of motion,
one can see that the currents $j_{1}^{\mu}$ and $j_{2}^{\mu}$ exhibit
entrainment even at zero temperature. Even in a case of no relative
motion between both fluid components, entrainment plays a significant
role since it leads to a coupling of the chemical potentials. 

\noindent We have further studied the stability of the tree-level
potential. Depending on the sign of the product $\mu_{1}\mu_{2}$
the gradient interaction must either be positive or negative. There
is no nonzero value for $\lambda_{12}$ at which the phase diagram
is stable for any value of $\mu_{1}$ and $\mu_{2}$. The occurring
instabilities can be related to entrainment since the only appear
in the presence of a gradient coupling. Furthermore, we have identified
a critical value for $\lambda_{12}$ at which the coexistence phase
completely disappears in two out the four quadrants of the phase diagram. 

\newpage{}

\part{Outlook}

~

\noindent The results presented in this thesis are an important step
in connecting the microscopic foundations of superfluidity with the
corresponding macroscopic hydrodynamical approach. Hydrodynamically,
a superfluid can be described in terms of a two-fluid model. At zero
temperature only the superfluid is present and above the critical
temperature only the fluid component termed normal fluid persist.
The microscopic dynamics on the other hand are determined by a quantum
field theory. The necessary ingredients to study superfluidity on
a microscopic level are the spontaneous breaking of a continuous symmetry
by a Bose-Einstein condensate as well as the elementary excitations
of a given system. We have shown explicitly how the effective hydrodynamic
description of superfluidity evolves from the underlying microscopic
physics in part II of this thesis. Analytic results for all non-dissipative
hydrodynamic quantities were obtained in the low-temperature limit.
Based on these results, a numerical and self-consistent study was
carried out for all temperatures below the critical temperature. In
particular, superfluid and normal-fluid densities, entrainment effects
and the critical velocity were studied for all temperatures. A more
detailed summary of the obtained results and the methods which have
been applied can be found in the conclusions of sections 10 and 11. 

\noindent Based on this microscopic realization of the two-fluid model,
we have studied the sound excitations of a superfluid. Relative oscillations
between superfluid and normal-fluid densities allow for an additional
sound wave, which has been termed ``second sound''. We have studied
this sound modes for all temperatures with rather surprising results:
first and second sound undergo what we termed \textquotedblleft role
reversal\textquotedblright{} as a function of the temperature: while
the first sound behaves like a pure pressure (density) wave at low
temperatures and transforms into a pure temperature (entropy) wave
for high temperatures, the second sound mode behaves exactly the other
way around. In addition, we studied the near-nonrelativistic limit
of the sound modes and compared the results to other physical systems
such a superfluid helium and cold atomic gases. For a more detailed
summary of the results obtained for the sound modes, refer to the
conclusion in section 15. 

\noindent Finally we have investigated a two-fluid system consisting
of a mixture of two superfluids instead of one superfluid and one
normal fluid in section 16. 

\noindent The results of this thesis should prove extremely useful
in future applications. Among the many existing (equivalent) formulations
of non-viscous superfluid hydrodynamics, we put special emphasis on
the canonical formulation introduced in section \ref{sec:Relativistic-thermodynamics-and-hydro}
since this is the preferred framework to describe superfluidity in
modern astrophysics. With this important groundwork at hand, the consequent
next step is to approach applications in compact star physics. In
section \ref{sec:Compact-stars-from-a-modern-POV}, we have seen that
the rich and complex physics of dense matter in compact stars require
numerous extensions of the results presented in this thesis. We shall
now discuss some of these extensions. 
\begin{itemize}
\item \noindent We have seen in section \ref{sub:intermediate-densities}
that usually more than one superfluid component is present in the
interior of a compact star (e.g. neutron superfluids and proton superconductors,
dense quark or nuclear matter with meson condensation etc.). In part
\ref{sec:A-mixture-of-two-SF} we have discussed how such a coupled
system of superfluids can be described at least qualitatively at zero
temperature. At finite temperature an additional normal current appears.
The corresponding hydrodynamic equations to be derived in this case
are very complex since entrainment effects are expected to introduce
couplings between all three fluid components resulting in a generalized
pressure of the from\medskip{}
\[
\Psi=\Psi[\partial\psi_{1}^{2},\,\partial\psi_{2}^{2},\,\Theta^{2},\,\partial\psi_{1}\cdot\partial\psi_{2},\,\Theta\cdot\partial\psi_{1},\,\Theta\cdot\partial\psi_{2}]\,.
\]
In analogy to chapters \ref{sec:Finite-temperature:-Two-fluid} and
\ref{sec:The-two-fluid-model-at-arbitrary-T} all hydrodynamic parameters
can be calculated in terms of field-theoretic variables either analytically
in a low-temperature approximation or numerically for all temperatures. 
\item \noindent A local gauge symmetry has to be introduced to consider
charged particles (in particular to study proton-superconductivity).
While a \textit{global }U(1) symmetry led to a massless Goldstone
mode in the excitation spectrum, a \textit{local }will lead to a Meissner-mass
for the gauge boson. 
\item \noindent Actually most superfluids in a compact star are made of
fermions (be it neutrons, protons or even quarks). Thus, BCS theory
should be introduced to the existing microscopic calculations. From
this point of view, the bosonic model described above corresponds
to an effective low energy description when fermionic excitations
are gapped.  However, in light of recent measurements suggesting a
critical temperature for neutron superfluidity of about $T_{c}\sim5.5\cdot10^{8}$
K (see discussion in section \ref{sub:Phenomenology-of-compact-stars}),
fermionic excitations should definitely be considered. The temperatures
of compact stars, especially shortly after they a born in a supernova
explosion can be as high as $T_{star}\thicksim10^{11}K.$ 
\item \noindent The kaon condensate spontaneously breaks conservation of
strangeness which is explicitly broken due to weak interactions in
the first place. In which sense such a system still exhibits superfluid
behavior should be analyzed by adding an explicit symmetry breaking
term to the Lagrangian. The results can be implemented to refine the
derivation of hydrodynamics for $\textrm{CFL-\ensuremath{K^{0}}}$.
\item \noindent Probably the most significant complication is given by the
introduction of viscous effects. In hydrodynamic terms this corresponds
to a coupling of a superfluid to a viscous normal fluid described
by a Navier-Stokes equation. One of the peculiarities of the two-fluid
formalism is that more viscosity coefficients appear than in the description
of a normal fluid. In addition to the shear viscosity mode, three
bulk viscosities are present. If there are two superfluid components
in addition to the normal fluid, the number of coefficients is even
larger. For a discussion of nuclear matter see reference \cite{GusakovAndersson},
viscosity coefficients in CFL quark matter are discussed in \cite{Mannarelli2010}.
Bulk viscosity in kaon-condensed color-flavor locked quark matter
has been investigated before, see for example reference \cite{Alford2008}.
Equipped with a deeper understanding of the microscopic origin of
the two fluid model, it might be possible to improve on existing calculations
of the viscosity coefficients. In the context of compact stars, viscous
effects are essential since they are expected to affect the r-mode
instability.
\end{itemize}
\noindent In addition to these refinements in the derivation of the
hydrodynamic equations, there are plenty of interesting phenomenological
aspects which could be analyzed in this framework:
\begin{itemize}
\item \noindent As we have discussed, stable regions of superfluidity are
bounded by the critical temperature and velocity. In addition, another
boundary is given by regions where collective excitations (i.e. the
sound modes) become unstable. This phenomenon is also termed \textquotedblleft two-stream
instability\textquotedblright{} and can be related to observable phenomena
such as pulsar glitches discussed in section \ref{sub:Phenomenology-of-compact-stars}.
A study of the two-stream instability based on the hydrodynamics derived
in this thesis has already been carried out \cite{SchmittTwostream}
and is in agreement with existing literature \cite{Samuelsson2010}
in astrophysics. Taking into account two superfluid components, the
resulting three fluid system at finite temperature will show an even
richer spectrum of sound modes and additional instabilities are likely
to occur. The study of these instabilities might provide new insights
into the dynamics of pulsar glitches. 
\item \noindent Of special interest are properties of the crust of a compact
star which have a strong influence on most observables of a compact
star including X-ray bursts, magnetar flares or pulsar glitches. In
this context, the superfluid phase within the crust and especially
entrainment effects are important. These entrainment effects between
crust and superfluid are assumed to be of a different origin as those
discussed in section \ref{sub:Relativistic-thermodynamics-and-entrain}:
it is speculated that neutrons can be scattered by the crystalline
structure of the crust according to Bragg\textquoteright s law \cite{Chamel2012,LivingCrust}.
As a result, it might be necessary to take into account an effective
reduction of the flux of superfluid neutrons in the vicinity of the
crust. 
\item Another particularly interesting field of study are the heat transport
properties in superfluids. The most important channel for heat transport
in superfluid helium is convection. However, we have discussed in
section \ref{sub:Phenomenology-of-compact-stars} that convection
might be suppressed in the nuclear matter due presence of electrons
while in CFL it might indeed be dominant. A calculation of the thermal
conductivity (and probably also the specific heat) under careful consideration
of all available excitations in the various superfluid phases might
shed some light on this question. 
\end{itemize}
Many more projects could be added to this list. It should be emphasized
that the proposed studies are certainly not a \textquotedblleft one
way road\textquotedblright{} leading from fundamental field theory
to astrophysics. Much more, a fruitful symbiosis has evolved at the
interface of these two disciplines: a deeper understanding of dense
hadronic matter from first principles allows for a more accurate modeling
of compact stars whereas at the same time observations of compact
stars can be used as a testing ground for fundamental physics. In
particular, the outline of the QCD phase diagram at intermediate densities
is one of the most challenging open problems in modern particle physics:
as mentioned in the introduction, only in a region of asymptotically
high density, quark matter can be analyzed from first principles and
important information such as the magnitude of the superconducting
gap can be obtained. In intermediate density regions, a rich variety
of candidate ground states have been studied. Among them are the $\textrm{CFL-\ensuremath{K^{0}}}$
phase, two-color superconductors or crystalline phases. Even a continuous
transition from quark to hadronic matter is in principle possible.
These fundamental questions can most likely not be attacked by conventional
techniques: heavy ion collisions and lattice calculations are powerful
tools to probe the phase structure in a regime of high temperatures
and low densities, but their applicability at higher densities is
very limited. The study of compact stars as the only \textquotedblleft laboratory\textquotedblleft{}
where such intermediate densities are realized could prove to be invaluable
as the maximum value for the chemical potential inside a compact star
is estimated to be high enough for deconfined quark matter to be conceivable.

\noindent Finally it should be remarked that the results obtained
from the studies presented in this theses are not limited to relativistic
superfluids appearing in high energy physics. We have discussed how
the non-relativistic limit can be extracted from the relativistic
version of the two-fluid equations and the solutions for first and
second sound. In particular the effects of instabilities which are
induced by the critical velocity, the two-stream instability or the
instability of the (zero-temperature) thermodynamic pressure in a
system of two-coupled superfluids are subject of intense research
in the fields of cold atomic gases or liquid helium. The obtained
results and the outcome of possible future studies are therefore of
interest to a large community of scientist. 

\noindent \newpage{}

\part{Appendix\label{part:Appendix}}

\appendix

\section{\noindent Matsubara sum with anisotropic excitation energies and
the low-temperature limit\label{sec:Matsubara-sum-with-anisotropy}}

~

\noindent Here we derive the result (\ref{eq:effectiveAction2}) for
the effective action. The calculation shown here is formulated in
a general way, such that it is also applicable to the stress-energy
tensor and the current. In order to perform the Matsubara sum in (\ref{eq:effActionExplicit}),
we write the determinant of the inverse propagator in terms of its
zeros, 

\noindent \smallskip{}

\noindent 
\[
S_{0}^{-1}(k)=(k_{0}-\epsilon_{1,\text{\ensuremath{\vec{k}}}}^{+})(k_{0}-\epsilon_{1,\text{\ensuremath{\vec{k}}}}^{-})(k_{0}-\epsilon_{2,\text{\ensuremath{\vec{k}}}}^{+})(k_{0}-\epsilon_{2,\text{\ensuremath{\vec{k}}}}^{-})\,.
\]
In the presence of a superflow $\vec{\nabla}\psi$, the zeros are
very complicated. The reason is the linear term in $k_{0}$ in the
off-diagonal elements. With the help of Mathematica we obtain analytical,
but very lengthy expressions for $\epsilon_{i,\vec{k}}^{\pm}$. For
small momenta, we can write

\noindent 
\begin{eqnarray}
\epsilon_{1,\vec{k}}^{\pm} & = & \pm\sqrt{\frac{\sigma^{2}-m^{2}}{3\sigma^{2}-m^{2}}}\,\zeta_{\pm}(\hat{k})|\vec{k}|+{\cal O}(|\vec{k}|^{3})\,,\label{eq:epsSmallK1}\\
\epsilon_{2,\vec{k}}^{\pm} & = & \pm\sqrt{2}\sqrt{3\sigma^{2}-m^{2}+2(\vec{\nabla}\psi)^{2}}+{\cal O}(|\vec{k}|)\,,\label{eq:epsSmallK2}
\end{eqnarray}
where

\noindent \medskip{}

\noindent 
\[
\zeta^{\pm}(\hat{k})\equiv\left[\sqrt{1+2\frac{(\vec{\nabla}\psi)^{2}-(\vec{\nabla}\psi\cdot\hat{k})^{2}}{3\sigma^{2}-m^{2}}}\mp\frac{2\partial_{0}\psi\vec{\nabla}\psi\cdot\hat{k}}{\sqrt{\sigma^{2}-m^{2}}\sqrt{3\sigma^{2}-m^{2}}}\right]\left[1+\frac{2(\vec{\nabla}\psi)^{2}}{3\sigma^{2}-m^{2}}\right]^{-1}\,.
\]
 We now use the Matsubara sum 

\noindent \medskip{}

\noindent 
\[
T\sum_{k_{0}}\frac{F(k_{0},\vec{k})}{{\rm det}\, S^{-1}(k)}=-\frac{1}{2}\sum_{e=\pm}\sum_{i=1,2}\frac{F(\epsilon_{i,\vec{k}}^{e},\vec{k})}{(\epsilon_{i,\vec{k}}^{e}-\epsilon_{i,\vec{k}}^{-e})(\epsilon_{i,\vec{k}}^{e}-\epsilon_{j,\vec{k}}^{e})(\epsilon_{i,\vec{k}}^{e}-\epsilon_{j,\vec{k}}^{-e})}\coth\frac{\epsilon_{i,\vec{k}}^{e}}{2T}
\]
 with $j=2$ if $i=1$ and vice versa, and an arbitrary function $F(k_{0},\vec{k})$
(without poles in the complex $k_{0}$ plane). For the effective action,
$F(k_{0},\vec{k})$ is given by (\ref{eq:Fk}), for the stress-energy
tensor and the current see (\ref{eq:FTmunu}), (\ref{eq:FJmu}). In
the low-temperature approximation which we discuss in appendix \ref{sec:Small-temperature-expansion},
we may neglect the contribution from the massive mode, i.e., the two
of the four terms in the sum where $i=2$. More precisely: later,
after writing $\coth[\epsilon_{i,\vec{k}}^{e}/(2T)]=1+2f(\epsilon_{i,\vec{k}}^{e})$
with the Bose distribution $f$, we shall only keep the thermal contribution,
which, in the case of the massive mode, is suppressed for low temperatures.
For the non-thermal (divergent) contribution, all terms have to be
kept in principle. However, after renormalization, the contribution
is subleading since it contains an additional factor of the coupling
constant $\lambda$ and we shall neglect it. Therefore, after taking
the thermodynamic limit, we can write

\noindent 
\begin{eqnarray*}
\frac{T}{V}\sum_{k}\frac{F(k_{0},\vec{k})}{{\rm det}\, S^{-1}(k)} & \simeq & -\frac{1}{2}\int\frac{d^{3}\vec{k}}{(2\pi)^{3}}\frac{F(\epsilon_{1,\vec{k}}^{+},\vec{k})}{(\epsilon_{1,\vec{k}}^{+}-\epsilon_{1,\vec{k}}^{-})(\epsilon_{1,\vec{k}}^{+}-\epsilon_{2,\vec{k}}^{+})(\epsilon_{1,\vec{k}}^{+}-\epsilon_{2,\vec{k}}^{-})}\coth\frac{\epsilon_{1,\vec{k}}^{+}}{2T}\\
 &  & -\,\frac{1}{2}\int\frac{d^{3}\vec{k}}{(2\pi)^{3}}\frac{F(\epsilon_{1,\vec{k}}^{-},\vec{k})}{(\epsilon_{1,\vec{k}}^{-}-\epsilon_{1,\vec{k}}^{+})(\epsilon_{1,\vec{k}}^{-}-\epsilon_{2,\vec{k}}^{+})(\epsilon_{1,\vec{k}}^{-}-\epsilon_{2,\vec{k}}^{-})}\coth\frac{\epsilon_{1,\vec{k}}^{-}}{2T}\\
 & = & -\int\frac{d^{3}\vec{k}}{(2\pi)^{3}}\frac{F(\epsilon_{1,\vec{k}}^{+},\vec{k})}{(\epsilon_{1,\vec{k}}^{+}+\epsilon_{1,-\vec{k}}^{+})(\epsilon_{1,\vec{k}}^{+}+\epsilon_{2,-\vec{k}}^{+})(\epsilon_{1,\vec{k}}^{+}-\epsilon_{2,\vec{k}}^{+})}\,\coth\frac{\epsilon_{1,\vec{k}}^{+}}{2T}\,,
\end{eqnarray*}
where, in the last step, we have changed the integration variable
of the second integral $\vec{k}\to-\vec{k}$, and have used $F(k_{0},\vec{k})=F(-k_{0},-\vec{k})$
as well as

\noindent \medskip{}

\noindent 
\[
\epsilon_{i,-\vec{k}}^{+}=-\epsilon_{i,\vec{k}}^{-}\,.
\]
This relation is easily checked for the small-momentum expressions
(\ref{eq:epsSmallK1}), (\ref{eq:epsSmallK2}), and also holds for
the full results. Due to the symmetries with respect to reflection
of $\vec{k}$, the poles $\epsilon_{i,\vec{k}}^{-}$ have thus dropped
out of the result, and the only physical excitations are $\epsilon_{i,\vec{k}}^{+}$.
Therefore, in the main text, we have simply denoted $\epsilon_{i,\vec{k}}\equiv\epsilon_{i,\vec{k}}^{+}$
and $\zeta(\hat{k})\equiv\zeta^{+}(\hat{k})$.

~

\section{\noindent Path integrals over complex fields\label{sec:Path-integrals-over-complexF}}

\noindent In this section of the appendix, we evaluate path integrals
for expectation values of a microscopic (grand canonical) ensemble
of particles which are given by (see also equation (\ref{eq:Integrals}))

\noindent \bigskip{}

\noindent 
\begin{equation}
\langle A\rangle\equiv\frac{1}{Z}\int{\cal D}\varphi_{1}'{\cal D}\varphi_{2}'\,\hat{A}\,\exp\left(\int d^{4}x\,{\cal L}\right)\,.\label{eq:Operator}
\end{equation}

\noindent Obviously, this involves solving the path integral for the
partition function $Z$ which we shall discuss first. Later we turn
to the discussion of the slightly more complicated integral in the
numerator.

\subsection{\noindent The partition function in momentum space\label{sub:The-partition-function}}

\noindent We consider the Fourier transform of the partition function 

\noindent \bigskip{}

\noindent 
\begin{equation}
Z=\int D\varphi(k)D\varphi^{\dagger}(k)\, e^{-\frac{1}{2}\underset{k}{\sum}\varphi^{\dagger}(k)\, S(k)\,\varphi(k)}\,.\label{eq:IntegralComplex}
\end{equation}

\noindent By $k$ we denote the momentum four vector $k^{\mu}$, $\varphi(k)$
denotes a complex doublet field and $S(k)$ a hermitian $2\times2$
matrix. Since in position space $\varphi(x)$ is real, we seem to
have doubled dimensions by using the complex fields $\varphi(k)$
and $\varphi^{\dagger}(k)$. In order to account for this, we have
to restrict ourselves to positive values for $k$

\noindent \bigskip{}

\noindent 
\begin{equation}
\frac{1}{2}\sum_{k}\varphi^{\dagger}(k)\, S(k)\,\varphi(k)=\frac{1}{2}\left[\sum_{k>0}\varphi^{\dagger}(k)\, S(k)\,\varphi(k)+\sum_{k<0}\varphi^{\dagger}(k)\, S(k)\,\varphi(k)\right]\,.
\end{equation}

\noindent Rewriting the second term

\noindent \medskip{}
\begin{equation}
\sum_{k<0}\varphi^{\dagger}(k)\, S(k)\,\varphi(k)=\sum_{k>0}\varphi^{\dagger}(-k)\, S(-k)\,\varphi(-k)=\sum_{k>0}\varphi^{T}(k)\, S(-k)\,\varphi^{*}(k)\,,
\end{equation}

\noindent we find that both contributions add up to 
\begin{equation}
\frac{1}{2}\sum_{k}\varphi^{\dagger}(k)\, S(k)\,\varphi(k)=\sum_{k>0}\varphi^{\dagger}(k)\, S(k)\,\varphi(k)
\end{equation}

\noindent This identity holds for the case of vanishing as well as
for finite background superflow where $S$ is of the form

\noindent \medskip{}
\begin{equation}
S(k)=\left(\begin{array}{cc}
A(k^{2}) & -iC\, k\\
iC\, k & B(k^{2})
\end{array}\right)\,.
\end{equation}

\noindent For the integration measure, we have

\noindent \medskip{}

\noindent 
\begin{equation}
\prod_{k}d\varphi(k)=\prod_{k>0}d\varphi(k)\, d\varphi^{\dagger}(k)\,.
\end{equation}

\noindent We now diagonalize $S$ 

\noindent \medskip{}

\noindent 
\begin{equation}
S_{diag}=V\, S\, V^{-1}=diag(\lambda_{1},\lambda_{2})\,.
\end{equation}

\noindent The columns of the matrix $V$ are given by the eigenvectors
corresponding to the eigenvalues $\lambda_{1}$ , $\lambda_{2}$.
After reinserting $V^{-1}S_{diag}V$ into the exponent of (\ref{eq:IntegralComplex}),
we redefine the fields $\varphi(k)$ by using the matrix $V$: 
\begin{eqnarray*}
\varphi & \rightarrow & \bar{\varphi}=V\varphi\,,\\
\varphi^{\dagger} & \rightarrow & \bar{\varphi}^{\dagger}=\varphi^{\dagger}V^{-1}\,.
\end{eqnarray*}

\noindent The integration measure remains unaffected by this transformation.
We thus find:

\noindent \medskip{}
\begin{equation}
\prod_{k>0}\int d\varphi(k)d\varphi^{\dagger}(k)\, e^{-\varphi^{\dagger}(k)\, S(k)\,\varphi(k)}=\prod_{k>0}\int d\bar{\varphi}(k)d\bar{\varphi}^{\dagger}(k)\, e^{-\bar{\varphi}^{\dagger}(k)\, S_{diag}(k)\,\bar{\varphi}(k)}\,.
\end{equation}

\noindent After decomposing the complex fields into real and imaginary
part ($\varphi_{1}=u+iv$ , $\varphi_{2}=p+iq$) and calculating the
corresponding Jacobian

\noindent \medskip{}
\begin{equation}
J=det\left(\frac{\partial(\varphi_{1},\varphi_{2},\varphi_{1}^{\dagger},\varphi_{2}^{\dagger})}{\partial(u,v,p,q)}\right)=4\,,
\end{equation}

\noindent we can now independently integrate all real and imaginary
parts (which are now simple scalar Gauss functions):

\noindent \medskip{}
\begin{eqnarray}
4\prod_{k>0}\int du(k)dv(k)dp(k)dq(k)\, e^{-\lambda_{1}(u^{2}+v^{2})-\lambda_{2}(p^{2}+q^{2})} & = & \prod_{k>0}\frac{(2\pi)^{2}}{\lambda_{1}(k)\lambda_{2}(k)}\\
=\prod_{k>0}\frac{(2\pi)^{2}}{det(S_{diag}(k))} & = & \prod_{k>0}\frac{(2\pi)^{2}}{det(S(k))}\,.
\end{eqnarray}

\noindent In the last step, we used $\textrm{det}(S_{diag})=\textrm{det}(V\, S\, V^{-1})=\textrm{det}(S)\,\textrm{det}(V\, V^{-1})=\textrm{det}(S)$.
Finally using

\noindent \medskip{}

\noindent 
\begin{equation}
\prod_{k}\textrm{det}(S(k))=\prod_{k>0}\textrm{det}(S(k))\,\textrm{det}(S(-k))=\prod_{k>0}\textrm{det}(S(k))^{2}\,,
\end{equation}

\newpage{}

\noindent we arrive at the final result

\noindent \bigskip{}

\noindent 
\begin{equation}
\int D\varphi(k)D\varphi^{\dagger}(k)\, e^{-\frac{1}{2}\underset{K}{\sum}\varphi^{\dagger}(k)\, S(k)\,\varphi(k)}=\prod_{k}\frac{(2\pi)^{2}}{\sqrt{\textrm{det}(S(k))}}\,.
\end{equation}

\noindent The additional factor of $2\pi$ compared to the case of
real fields $\varphi$ again is a result of the doubled number of
dimensions in the integral and can be absorbed by a normalization
factor $N$ of the path integral. 

~

\subsection{\noindent The stress-energy tensor and current\label{sub:The-stress-energy-tensor-and-current}}

~

\noindent Next, we have to calculate the numerator of equation (\ref{eq:Operator}).
We insert an operator $\hat{A}$ which leads to the expectation value
of $T^{\mu\nu}$ and $j^{\mu}$. To construct this operator and evaluate
$T^{\mu\nu}$ from (\ref{eq:TensorFunc}), we first need to calculate
the propagator in position space . We introduce the shifted complex
fields

\noindent \medskip{}

\noindent 
\begin{equation}
\varphi\rightarrow e^{i\psi(x)}(\frac{1}{\sqrt{2}}\rho+\varphi(x))=\frac{1}{\sqrt{2}}e^{i\psi(x)}(\rho+\varphi_{1}(x)+i\varphi_{2}(x))\,,
\end{equation}

\noindent and insert them into

\noindent \medskip{}

\noindent 
\begin{equation}
\mathcal{L}=\left|\partial_{0}\varphi\right|^{2}-\left|\vec{\nabla}\varphi\right|^{2}-m^{2}\left|\varphi\right|^{2}-\lambda\left|\varphi\right|^{4}\,.
\end{equation}

\noindent This results in

\noindent 
\begin{eqnarray}
\varphi_{1}^{2} & \rightarrow & \frac{1}{2}[\partial_{0}^{2}-\partial_{i}^{2}+(\partial_{0}\psi)^{2}-(\vec{\nabla}\psi)^{2}-m^{2}-3\lambda\rho^{2}]\,,\\
\varphi_{2}^{2} & \rightarrow & \frac{1}{2}[\partial_{0}^{2}-\partial_{i}^{2}+(\partial_{0}\psi)^{2}-(\vec{\nabla}\psi)^{2}-m^{2}-\lambda\rho^{2}]\,,\nonumber \\
\varphi_{1}\varphi_{2} & \rightarrow & (\partial_{0}\psi)\varphi_{1}\partial_{0}\varphi_{2}-(\partial_{0}\psi)\varphi_{2}\partial_{0}\varphi_{1}-\varphi_{1}(\vec{\nabla}\psi\cdot\vec{\nabla}\varphi_{2})-\varphi_{2}(\vec{\nabla}\psi\cdot\vec{\nabla}\varphi_{1})\,.\nonumber 
\end{eqnarray}

\noindent We can thus denote $\mathcal{L}^{(2)}$ (the quadratic terms
in the fluctuations $\varphi(x)$) in the $2\times2$ space spanned
by $\{\varphi_{1}(x),\,\varphi_{2}(x)\}$

\noindent \medskip{}

\noindent 
\begin{equation}
-\frac{1}{2}(\varphi_{1}(x),\,\varphi_{2}(x))\left(\begin{array}{cc}
-\overleftrightarrow{\partial_{\mu}\partial^{\mu}}-\partial_{\mu}\psi\partial^{\mu}\psi+m^{2}+3\lambda\rho^{2} & -2\partial_{\mu}\psi\partial^{\mu}\\
2\partial_{\mu}\psi\partial^{\mu} & -\overleftrightarrow{\partial_{\mu}\partial^{\mu}}-\partial_{\mu}\psi\partial^{\mu}\psi+m^{2}+\lambda\rho^{2}
\end{array}\right)\left(\begin{array}{c}
\varphi_{1}(x)\\
\varphi_{2}(x)
\end{array}\right)
\end{equation}

\noindent where the notation $\varphi_{1}\overleftrightarrow{\partial_{\mu}\partial^{\mu}}\varphi_{1}$
denotes $\partial_{\mu}\varphi_{1}\partial^{\mu}\varphi_{1}$ (alternatively,
using partial integration one can show that it is sufficient to have
all operators act to the right). Partition function and pressure in
position space are given by

\noindent \medskip{}
\begin{equation}
Z=\int D\varphi\,\textrm{exp}[-\frac{1}{2}\int_{x}\varphi^{T}\, S^{-1}\,\varphi]\,,\,\,\,\,\,\,\,\,\Psi=\frac{T}{V}\,\textrm{ln}\, Z\,.
\end{equation}

\noindent We use the pressure as effective action and construct the
stress energy tensor in from

\noindent \medskip{}

\noindent 
\begin{equation}
T^{\mu\nu}=2\frac{\delta\Psi[\psi]}{\delta g_{\mu\nu}}-g^{\mu\nu}\Psi[\psi]\,.
\end{equation}

\noindent This amounts to the calculation of an expectation value
of $T^{\mu\nu}$ 
\begin{eqnarray}
\left\langle T^{\mu\nu}\right\rangle  & = & -\frac{1}{2}Z^{-1}\int D\varphi\,\varphi^{T}\hat{A}^{\mu\nu}\varphi\, e^{-\frac{1}{2}\int_{x}\varphi^{T}\, S^{-1}\,\varphi}\,=\\
 & = & -\frac{1}{2}Z^{-1}\int D\varphi\,\left[\varphi^{T}\left(2\frac{\delta S^{-1}}{\delta g^{\mu\nu}}-g^{\mu\nu}S^{-1}\right)\varphi\right]e^{-\frac{1}{2}\int_{x}\varphi^{T}\, S^{-1}\,\varphi}\,.
\end{eqnarray}

\noindent Explicitly, the inserted operator reads

\noindent \medskip{}
\begin{equation}
\hat{A}^{\mu\nu}=\left(\begin{array}{cc}
2(-\overleftrightarrow{\partial^{\mu}\partial^{\nu}}-\partial^{\mu}\psi\partial^{\nu}\psi)-g^{\mu\nu}S_{11}^{-1} & -2(\partial^{\mu}\psi\partial^{\nu}-\partial^{\nu}\psi\partial^{\mu})-g^{\mu\nu}S_{12}\\
2(\partial^{\mu}\psi\partial^{\nu}-\partial^{\nu}\psi\partial^{\mu})+g^{\mu\nu}S_{12} & 2(-\overleftrightarrow{\partial^{\mu}\partial^{\nu}}-\partial^{\mu}\psi\partial^{\nu}\psi)-g^{\mu\nu}S_{22}^{-1}
\end{array}\right)\,.
\end{equation}

\noindent To perform the actual integration, it is again advantageous
to transform to momentum space 
\begin{eqnarray}
\varphi_{1}(x) & \rightarrow & \frac{1}{\sqrt{TV}}\sum_{p}e^{-ip\cdot x}\varphi_{1}(p)\,,\\
\varphi_{2}(x) & \rightarrow & \frac{1}{\sqrt{TV}}\sum_{q}e^{-iq\cdot x}\varphi_{2}(q)\,.\nonumber 
\end{eqnarray}

\noindent At this point we should remind ourselves that gradients
of the Goldstones fields in the applied approximation are constant

\noindent \medskip{}

\noindent 
\begin{equation}
\partial_{\mu}\psi\partial^{\mu}\psi=\lambda\rho^{2}+m^{2}=const\,,
\end{equation}

\noindent and hence remain unaffected by Fourier transformations.
The Fourier transformed operator thus reads (technically, after Fourier
transformation this matrix is no longer an operator and we shall denote
it 

\noindent \begin{flushleft}
\newpage{}as $A$ instead of $\hat{A}$)
\par\end{flushleft}

\begin{equation}
A^{\mu\nu}=f(\bar{k}\cdot x)\, A^{\mu\nu}(p,\, q)\,.
\end{equation}

\noindent The matrix $A^{\mu\nu}$ is then given by

\noindent 
\begin{equation}
A^{\mu\nu}=\left(\begin{array}{cc}
(p^{\mu}q^{\nu}+p^{\nu}q^{\mu})-2\partial^{\mu}\psi\partial^{\nu}\psi-g^{\mu\nu}m_{1}(p\cdot q) & -2i(\partial^{\mu}\psi\, q^{\nu}-\partial^{\nu}\psi\, q^{\mu})+2iq^{\mu}\\
2i(\partial^{\mu}\psi\, q^{\nu}-\partial^{\nu}\psi\, q^{\mu})-2iq^{\mu} & (p^{\mu}q^{\nu}+p^{\nu}q^{\mu})-2\partial^{\mu}\psi\partial^{\nu}\psi-g^{\mu\nu}m_{2}(p\cdot q)
\end{array}\right)\,,
\end{equation}

\noindent with the following abbreviations:

\noindent 
\begin{eqnarray*}
f(\bar{k}\cdot x) & = & e^{-i(p^{\mu}+q^{\mu})x_{\mu}}\,,\\
m_{1}(p\cdot q) & = & p\cdot q-\sigma^{2}+m^{2}+3\lambda\rho^{2}\,,\\
m_{2}(p\cdot q) & = & p\cdot q-\sigma^{2}+m^{2}+\lambda\rho^{2}\,.
\end{eqnarray*}

\noindent This amounts to the calculation of the following path integral

\noindent \bigskip{}

\noindent 
\begin{equation}
\frac{\int\prod_{k}d\varphi(k)\,\left[\sum_{p,q}(\varphi_{1}(p),\varphi_{2}(p))\, A^{\mu\nu}(p,\, q)\,\left(\begin{array}{c}
\varphi_{1}(q)\\
\varphi_{2}(q)
\end{array}\right)e^{-i(p+q)\cdot x}\right]e^{-\frac{1}{2}\varphi^{T}(-k)\, S^{-1}\varphi(k)}}{\int\prod_{k}d\varphi(k)\, e^{-\frac{1}{2}\varphi^{\dagger}(-k)\, S^{-1}\varphi(k)}}\,.
\end{equation}

\noindent Let us pick out one term in the sum over $p$ and $q$ in
the integrand of the numerator (we shall suppress the Lorentz indices
in the following discussion)

\noindent \medskip{}
\begin{eqnarray}
d\varphi(k_{1})...d\varphi(k_{n})\,\left[A_{11}\,\varphi_{1}(p_{1})\varphi_{1}(q_{1})+A_{22}\,\varphi_{2}(p_{1})\varphi_{2}(q_{1})+...\right]\,\textrm{exp}\left(-\frac{1}{2}\varphi^{\dagger}(-k_{1})\, S^{-1}\varphi(k_{1})\right)\\
...\textrm{exp}\left(-\frac{1}{2}\varphi^{\dagger}(-k_{n})\, S^{-1}\varphi(k_{n})\right) & .\nonumber 
\end{eqnarray}

\noindent The only non-vanishing contributions to the path integral
are obtained by choosing $p_{1}=k_{1}$ and $q_{1}=-k_{1}$. The integration
over all remaining $\varphi(k_{i})$ can be canceled with the corresponding
integrals in the denominator. Repeating this for all $p_{i}$ and
$q_{i}$, we are left with

\noindent \bigskip{}

\noindent 
\begin{equation}
\sum_{k}\frac{\int d\varphi(k)\,\left[(\varphi_{1}(-k),\varphi_{2}(-k))\, A\,(k,\,-k)\,\left(\begin{array}{c}
\varphi_{1}(k)\\
\varphi_{2}(k)
\end{array}\right)\right]e^{-\frac{1}{2}\varphi^{T}(-k)\, S^{-1}\varphi(k)}}{\int d\varphi(k)\, e^{-\frac{1}{2}\phi^{+}(-k)\, S^{-1}\phi(k)}}\,.\label{eq:IntSimple}
\end{equation}

\noindent To restrict $k$ to positive values, we have to consider
$A_{12}(-k)=-A_{12}(k)$ 

\noindent 
\begin{eqnarray}
\sum_{k}\left[A_{11}\varphi_{1}(-k)\varphi_{1}(k)+A_{22}\varphi_{2}(k)\varphi_{2}(-k)\right] & = & 2\sum_{k>0}\left[A_{11}\varphi_{1}(k)\varphi_{1}^{\dagger}(k)+A_{22}\varphi_{2}(k)\varphi_{2}^{\dagger}(k)\right]\,,\\
\sum_{k}\left[A_{12}\varphi_{1}(-k)\varphi_{2}(k)+A_{21}\varphi_{1}(k)\varphi_{2}(-k)\right] & = & 2\sum_{k>0}A_{12}\left[\varphi_{1}(k)\varphi_{2}^{\dagger}(k)-\varphi_{1}^{\dagger}(k)\varphi_{2}(k)\right]\,.\nonumber 
\end{eqnarray}

\noindent Parametrizing $\varphi_{1}$ and $\varphi_{2}$ in exactly
the same way as in the calculation of the denominator, we have

\noindent \medskip{}

\noindent 
\begin{equation}
8\int du\, dv\, dp\, dq\,\left[A_{11}(u^{2}+v^{2})+A_{22}(p^{2}+q^{2})-2A_{12}(vp-uq)\right]e^{-(S_{11}(u^{2}+v^{2})+S_{22}(p^{2}+q^{2})-2S_{12}(vp-uq))}\,.
\end{equation}

\noindent The variables $u,v,p,q$ can now be integrated out resulting
in

\noindent \medskip{}

\noindent 
\begin{equation}
8\int du\, dv\, dp\, dq\,[...]=-\frac{8\pi^{2}}{((S_{12}^{-1})^{2}-S_{11}^{-1}S_{22}^{-1})^{2}}(S_{22}^{-1}A_{11}+S_{11}^{-1}A_{22}-2S_{12}^{-1}A_{12})\,.
\end{equation}

\noindent Note that we have now explicitly included the fact that
the off diagonal elements of $S$ and $A$ are imaginary. All quantities
that appear in the above equation are now real ($S_{12}$ in the formula
below now refers to $\textrm{Im}[S_{12}]$.) To simplify the result
we use

\noindent \medskip{}

\noindent 
\begin{equation}
S=\frac{1}{\textrm{det}\, S^{-1}}\left(\begin{array}{cc}
S_{22}^{-1} & -S_{12}^{-1}\\
-S_{21}^{-1} & S_{11}^{-1}
\end{array}\right)\,,\label{eq:inversS}
\end{equation}

\noindent leading to

\noindent 
\begin{equation}
Tr[S\cdot A]=\frac{1}{\textrm{det}S^{-1}}\left[S_{22}^{-1}A{}_{11}+S_{11}^{-1}A_{22}-2S_{12}^{-1}A_{12}\right]\,,
\end{equation}

\noindent and 

\noindent 
\begin{equation}
8\int du\, dv\, dp\, dq\,[...]=\frac{8\pi^{2}}{detS^{-1}}Tr[SA]\,.
\end{equation}

\noindent In summary, the integral (\ref{eq:IntSimple}) calculates
to

\noindent \medskip{}

\noindent 
\begin{equation}
\frac{\int\prod_{k}d\varphi(k)\,\left[\sum_{p,q}\varphi^{T}(p)\, A\,\varphi(q)e^{-i(p+q)\cdot x}\right]e^{-\frac{1}{2}\varphi^{T}(-k)\, S^{-1}\varphi(k)}}{\int\prod_{k}d\varphi(k)\, e^{-\frac{1}{2}\varphi^{\dagger}(-k)\, S^{-1}\,\varphi(k)}}=2\sum_{k>0}Tr[SA]=\sum_{K}Tr[SA]\,.
\end{equation}

\noindent Inserting the definition of the matrix $A$ , reintroducing
the Lorentz indices and denoting $S^{\prime}=\delta S/\delta g^{\mu\nu}$

\noindent \medskip{}

\noindent 
\begin{equation}
\sum_{k}Tr[SA^{\mu\nu}]=\sum_{k}Tr[2S\cdot(S^{-1})^{\prime\mu\nu}-g^{\mu\nu}S\cdot S^{-1}]=\sum_{k}2Tr[S\cdot(S^{-1})^{\prime}]^{\mu\nu}-2g^{\mu\nu}\,,
\end{equation}

\noindent we finally arrive at the result (\ref{eq:TPath})

\noindent \medskip{}

\noindent 
\begin{equation}
\left\langle T^{\mu\nu}\right\rangle =-\frac{1}{2}\frac{T}{V}\sum_{k}Tr[SA]=-\frac{T}{V}\sum_{k}\left[Tr[S\cdot(S^{-1})^{\prime}]^{\mu\nu}-g^{\mu\nu}\right]\,.
\end{equation}

\noindent One can calculate the charge current in exactly the same
way, 

\noindent \medskip{}

\noindent 
\begin{equation}
\left\langle j^{\mu}\right\rangle =\frac{\delta\Psi[\psi]}{\delta(\partial_{\mu}\psi)}=-\frac{1}{2}Z^{-1}\int D\varphi\left[\varphi^{\dagger}\frac{\delta S^{-1}}{\delta(\partial_{\mu}\psi)}\varphi\right]e^{-\frac{1}{2}\int_{X}\varphi^{T}S^{-1}\varphi}=-\frac{1}{2}\frac{T}{V}\sum_{K}Tr[S\cdot A^{\mu}]\,.
\end{equation}

\noindent This time, the inserted operator is given by

\noindent \medskip{}

\noindent 
\begin{equation}
A^{\mu}=\frac{\partial S^{-1}}{\partial(\partial_{\mu}\psi)}=2\left(\begin{array}{cc}
2\partial^{\mu}\psi & 2ik^{\mu}\\
-2ik^{\mu} & 0
\end{array}\right)\,,
\end{equation}

\noindent resulting in (\ref{eq:JPath}) 
\begin{eqnarray}
Tr[S\cdot A^{\mu}] & = & 2\left[2S_{11}\partial^{\mu}\psi-i(S_{12}-S_{21})k^{\mu}\right]=\frac{2}{\Delta}\left[2S_{22}^{-1}\partial^{\mu}\psi-i(S_{12}-S_{21})k^{\mu}\right]\\
 & = & -\frac{4}{\Delta}\left[k^{2}\partial^{\mu}\psi+2(K\cdot\partial\psi)k^{\mu}\right]\,,\nonumber 
\end{eqnarray}

\noindent with $\Delta=\textrm{det\,}S^{-1}$. 

~

\section{\noindent Renormalization and useful identities for the stress-energy
tensor\label{sec:Renormalization-and-useful-identities}}

~

\noindent With the function $\Psi_{k}$ from (\ref{eq:PsiK}) we can
write the effective action (\ref{eq:effAction}) as 

\noindent \medskip{}

\noindent 
\begin{equation}
\frac{T}{V}\Gamma=-U+\frac{T}{V}\sum_{k}\Psi_{k}\,.\label{eq:PsiKAux}
\end{equation}
 On the other hand, using (\ref{eq:effActionExplicit}), we have 

\noindent \medskip{}

\noindent 
\begin{eqnarray}
\frac{T}{V}\Gamma & = & -U+\frac{1}{3}\frac{T}{V}\sum_{k}\left(C_{k}\vec{k}^{2}-A_{k}\vec{k}\cdot\vec{\nabla}\psi\right)\label{eq:GammaAux}\\
 & = & -U-\frac{1}{3}(g^{\mu\nu}-u^{\mu}u^{\nu})\frac{T}{V}\sum_{k}\left(C_{k}k_{\mu}k_{\nu}+A_{k}k_{\mu}\partial_{\nu}\psi\right)\,,\nonumber 
\end{eqnarray}
 with the four-vector $u^{\mu}=(1,0,0,0)$ and $A_{k}$, $B_{k}$,
$C_{k}$ given in (\ref{eq:AkBkCk}). A useful relation between $A_{k}$,
$B_{k}$, $C_{k}$ can be derived with the help of the explicit form
of the determinant of the inverse tree-level propagator,

\noindent 
\begin{eqnarray}
1 & = & \frac{k^{4}-2k^{2}(\sigma^{2}-m^{2})-4(k\cdot\partial\psi)^{2}}{{\rm det}\, S_{0}^{-1}}\nonumber \\
 & = & -\frac{1}{2}[C_{k}k^{2}+B_{k}\sigma^{2}+2A_{k}(k\cdot\partial\psi)]+\frac{k^{2}m^{2}}{{\rm det}\, S_{0}^{-1}}\,.\label{eq:auxTrace}
\end{eqnarray}
 Next, we rewrite the stress-energy tensor. With 

\noindent \medskip{}

\noindent 
\begin{equation}
-\frac{T}{V}\sum_{k}\textrm{Tr}\left[S_{0}\frac{\partial S_{0}^{-1}}{\partial g_{\mu\nu}}\right]=2\frac{T}{V}\sum_{k}\frac{\partial\Psi_{k}}{\partial g_{\mu\nu}}
\end{equation}
we can write the stress-energy tensor from (\ref{eq:TPath}) as 

\noindent \medskip{}

\noindent 
\begin{equation}
T^{\mu\nu}=-\left(2\frac{\partial U}{\partial g_{\mu\nu}}-g^{\mu\nu}U\right)+\frac{T}{V}\sum_{k}\left[C_{k}k^{\mu}k^{\nu}+B_{k}\partial^{\mu}\psi\partial^{\nu}\psi+A_{k}(k^{\mu}\partial^{\nu}\psi+k^{\nu}\partial^{\nu}\psi)+g^{\mu\nu}+Y^{\mu\nu}\right]\,,\label{eq:TmunuAux}
\end{equation}
 where we have used the definition of $\Psi_{k}$ (\ref{eq:PsiK})
and have added a constant, diagonal tensor $Y^{\mu\nu}$ which has
to be determined such that the conditions $T^{00}=\epsilon$ and $T^{ij}=\delta^{ij}P$
are fulfilled. In order to implement these conditions we now set $\vec{\nabla}\psi=0$
and $\partial_{0}\psi=\mu$. In this case, because of the first line
of (\ref{eq:GammaAux}), the pressure $P=\frac{T}{V}\Gamma$ becomes

\noindent \medskip{}

\noindent 
\begin{equation}
P=-U+\frac{1}{3}\frac{T}{V}\sum_{k}C_{k}\vec{k}^{2}\,.\label{eq:PAux}
\end{equation}
In order to compute the energy density $\epsilon=-P+\mu n+Ts$ we
need

\noindent 
\begin{eqnarray}
n & = & \frac{\partial P}{\partial\mu}=-\frac{\partial U}{\partial\mu}+\frac{T}{V}\sum_{k}(B_{k}\mu+A_{k}k_{0})\,,\\
s & = & \frac{\partial P}{\partial T}=-\frac{\partial U}{\partial T}+\frac{P}{T}+\frac{T}{V}\sum_{k}\left(2+A_{k}\frac{\mu k_{0}}{T}+C_{k}\frac{k_{0}^{2}}{T}\right)\,,\\
\nonumber 
\end{eqnarray}
where we have used the form of the pressure (\ref{eq:PsiKAux}) and
$\partial k_{0}/\partial T=k_{0}/T$ (due to the linear temperature-dependence
of the Matsubara frequencies).

\noindent \newpage{}Consequently, 

\noindent \medskip{}

\noindent 
\begin{equation}
\epsilon=\frac{T}{V}\sum_{k}\left(B_{k}\mu^{2}+2A_{k}k_{0}\mu+C_{k}k_{0}^{2}+2\right)\,.\label{eq:EpxAux}
\end{equation}
On the other hand, the nonzero components of the stress-energy tensor
without superflow are, from (\ref{eq:TmunuAux}), 

\noindent \medskip{}

\noindent 
\begin{equation}
T^{ij}=\frac{(\mu^{2}-m^{2})^{2}}{4\lambda}\delta^{ij}+\frac{T}{V}\sum_{k}\left(C_{k}\frac{\vec{k}^{2}}{3}\delta^{ij}-\delta^{ij}+Y^{ij}\right)\,,\label{eq:TijAux}
\end{equation}
 and

\noindent \medskip{}
\begin{equation}
T^{00}=\frac{(3\mu^{2}+m^{2})(\mu^{2}-m^{2})}{4\lambda}+\frac{T}{V}\sum_{k}\left(B_{k}\mu^{2}+2A_{k}k_{0}\mu+C_{k}k_{0}^{2}+1+Y^{00}\right)\,,\label{eq:T00Aux}
\end{equation}
By comparing (\ref{eq:TijAux}) with (\ref{eq:PAux}) and (\ref{eq:T00Aux})
with (\ref{eq:EpxAux}) we conclude that $Y^{\mu\nu}={\rm diag}(1,1,1,1)$.
Inserting this into (\ref{eq:TmunuAux}), we can write the renormalized
stress-energy tensor as

\noindent \medskip{}

\noindent 
\begin{equation}
T^{\mu\nu}=-\left(2\frac{\partial U}{\partial g_{\mu\nu}}-g^{\mu\nu}U\right)+\frac{T}{V}\sum_{k}\left[C_{k}k^{\mu}k^{\nu}+B_{k}\partial^{\mu}\psi\partial^{\nu}\psi+A_{k}(k^{\mu}\partial^{\nu}\psi+k^{\nu}\partial^{\mu}\psi)+2u^{\mu}u^{\nu}\right]\,,
\end{equation}
 with $u^{\mu}=(1,0,0,0)$.

\noindent ~

\section{\noindent Small-temperature expansion\label{sec:Small-temperature-expansion}}

~

\noindent Here we explain the small-temperature expansion for the
effective action, the stress-energy tensor, and the current density
which is used in section \ref{sub:Explicit-results-in-lowT}. We focus
on the effective action in this appendix, but the other results are
obtained analogously.

\noindent Expanding in powers of the temperature corresponds to expanding
the integrand in powers of $|\vec{k}|$. In order to obtain the result
up to $T^{6}$, we expand the integrand of the momentum integral in
\ref{eq:effectiveAction2} as 

\noindent \medskip{}

\noindent 
\begin{equation}
\frac{F(\epsilon_{1,\vec{k}},\vec{k})}{(\epsilon_{1,\vec{k}}+\epsilon_{1,-\vec{k}})(\epsilon_{1,\vec{k}}+\epsilon_{2,-\vec{k}})(\epsilon_{1,\vec{k}}-\epsilon_{2,\vec{k}})}\simeq a_{1}|\vec{k}|+\frac{a_{2}}{\sigma^{2}}|\vec{k}|^{3}\,,
\end{equation}
\newpage{} and, for the dispersion in the argument of the Bose distribution, 

\noindent \medskip{}

\noindent 
\begin{equation}
\epsilon_{1,\vec{k}}\simeq c_{1}|\vec{k}|+\frac{c_{2}}{\sigma^{2}}|\vec{k}|^{3}\,,
\end{equation}
where $a_{1}$, $a_{2}$, $c_{1}$, $c_{2}$ are angular-dependent,
dimensionless coefficients. Inserting these expansions, introducing
a dimensionless integration variable $y=c_{1}|\vec{k}|/T$, expanding
in $T/\sigma$, and performing the resulting integration over $y$
yields

\noindent \medskip{}

\noindent 
\begin{equation}
\frac{T}{V}\Gamma\simeq\frac{(\sigma^{2}-m^{2})^{2}}{4\lambda}+\frac{2\pi^{2}T^{4}}{45}\int\frac{d\Omega}{4\pi}\left[\frac{a_{1}}{c_{1}^{4}}+\frac{40\pi^{2}}{7c_{1}^{6}}\left(\frac{a_{2}}{3}-\frac{2a_{1}c_{2}}{c_{1}}\right)\frac{T^{2}}{\sigma^{2}}\right]\,,
\end{equation}
 where we have used the integrals 

\noindent \medskip{}

\noindent 
\begin{equation}
\int_{0}^{\infty}dy\frac{y^{3}}{e^{y}-1}=\frac{\pi^{4}}{15}\,,\qquad\int_{0}^{\infty}dy\frac{y^{5}}{e^{y}-1}=\frac{8\pi^{6}}{63}\,,\qquad\int_{0}^{\infty}dy\frac{y^{6}e^{y}}{(e^{y}-1)^{2}}=\frac{16\pi^{6}}{21}\,.
\end{equation}
 For the case without superflow, $\vec{\nabla}\psi=0$, the angular
integral becomes trivial. In this case, with $\partial_{0}\psi=\mu$,
the full dispersions are given by equation (\ref{eq:DispNoSF}), and
we have 

\noindent \medskip{}

\noindent 
\begin{equation}
c_{1}=\sqrt{\frac{\mu^{2}-m^{2}}{3\mu^{2}-m^{2}}}\,,\qquad c_{2}=\frac{\mu^{6}}{\sqrt{\mu^{2}-m^{2}}(3\mu^{2}-m^{2})^{5/2}}\,,
\end{equation}
 and 

\noindent 
\begin{equation}
a_{1}=\frac{c_{1}}{4}\,,\qquad a_{2}=\frac{3c_{1}}{4}\,.
\end{equation}

\noindent We thus find for the pressure

\noindent 
\begin{eqnarray}
P & = & \frac{T}{V}\Gamma\simeq\frac{(\mu^{2}-m^{2})^{2}}{4\lambda}+\frac{\pi^{2}T^{4}}{90c_{1}^{3}}-\frac{4c_{2}\pi^{4}T^{6}}{63\mu^{2}c_{1}^{6}}\nonumber \\
 & = & \frac{(\mu^{2}-m^{2})^{2}}{4\lambda}+\frac{(3\mu^{2}-m^{2})^{3/2}}{(\mu^{2}-m^{2})^{3/2}}\frac{\pi^{2}T^{4}}{90}-\frac{\mu^{6}(3\mu^{2}-m^{2})^{1/2}}{(\mu^{2}-m^{2})^{7/2}}\frac{4\pi^{4}T^{6}}{63\mu^{2}}\,.
\end{eqnarray}
The expressions for the case with superflow are quite lengthy in general,
and we give the final results in the limit $m=0$ in the main text,
see table 2 in section \ref{sub:Explicit-results-in-lowT}. 

\noindent ~

\newpage{}

\section{\noindent Renormalization in 2PI\label{sec:Renormalization-in-2PI}}

~

\noindent In this appendix we discuss the renormalization of the 2PI
approach. The renormalization procedure is done on the level of the
effective action. Its general form (\ref{eq:Action2PI}) can be written
as 

\noindent \medskip{}

\noindent 
\begin{equation}
\Psi=\frac{\rho^{2}}{2}(\mu^{2}-m^{2})-\frac{\lambda}{4}\rho^{4}+J+\frac{M^{2}-m^{2}-2\lambda\rho^{2}}{2}I^{+}+\frac{\delta M^{2}-\lambda\rho^{2}}{2}I^{-}-\frac{\lambda}{2}(I^{+})^{2}-\frac{\lambda}{4}(I^{-})^{2}\,,\label{eq:Reno2PiAction}
\end{equation}
where we have abbreviated

\noindent \medskip{}

\noindent 
\begin{equation}
J\equiv-\frac{1}{2}\frac{T}{V}\sum_{k}\textrm{Tr}\,\ln\frac{S^{-1}}{T^{2}}\,,\qquad I^{\pm}\equiv\frac{T}{V}\sum_{k}[S_{11}(k)\pm S_{22}(k)]\,.\label{eq:RenoIJ}
\end{equation}
 For the $\textrm{Tr}\,[S_{0}^{-1}S-1]$ term we have used the tree-level
propagator (\ref{eq:Propagator}) and the ansatz for the propagator
(\ref{eq:FullProp}), while for $V_{2}$ we have used the definition
(\ref{eq:twoPiPotential}) and the fact that the propagator is antisymmetric,
$S_{12}=-S_{21}$.

\noindent We shall add a counterterm $\delta\Psi$ to $\Psi$, such
that the effective action becomes renormalized at the stationary point.
It is instructive to start with the non-superfluid case where there
is no condensate, then discuss the case with condensate but without
superflow, and then turn to the most complicated case that includes
condensate and superflow.

\subsection{\noindent Uncondensed phase with (spurious) background field \label{sub:Uncondensed-phase-with-sF}}

\noindent First we consider the high-temperature, non-superfluid,
phase. We can formally include a background field $\vec{\nabla}\psi$
also in this phase, although we shall see that the physics will turn
out to be independent of $\vec{\nabla}\psi$. If the condensate vanishes,
there is no need to introduce two different self-consistent masses
$M$ and $\delta M$, and the full propagator is given by 

\noindent \bigskip{}

\noindent 
\begin{equation}
S^{-1}(K)=\left(\begin{array}{cc}
-k^{2}+M^{2}-\sigma^{2} & 2ik_{\mu}\partial^{\mu}\psi\\[2ex]
-2ik_{\mu}\partial^{\mu}\psi & -k^{2}+M^{2}-\sigma^{2}
\end{array}\right)\,.
\end{equation}
 From the poles of the propagator we obtain the dispersion relations 

\noindent \medskip{}

\noindent 
\begin{equation}
\epsilon_{\vec{k}}^{e}=\sqrt{(\vec{k}-e\vec{\nabla}\psi)^{2}+M^{2}}-e\mu\,,\label{eq:RenoDiskNoCond}
\end{equation}
where $\mu=\partial_{0}\psi$. These are simply the usual particle
and anti-particle excitations, carrying one unit of positive and negative
charge, respectively, but with the spatial momentum shifted by $\vec{\nabla}\psi$,
for particles and anti-particles in opposite directions. 

\noindent The effective action in the uncondensed phase is given by
equation (\ref{eq:Reno2PiAction}) with $\rho=0$. Moreover, since
there is only one self-consistent mass $M$, we also have $\delta M=0$
and $I^{-}=0$, 

\noindent \medskip{}

\noindent 
\begin{equation}
\Psi=J+\frac{M^{2}-m^{2}}{2}I^{+}-\frac{\lambda}{2}(I^{+})^{2}\,.\label{eq:RenoPsiNormal}
\end{equation}
 We now add counterterms to the effective action in order to cancel
the infinities in $J$ and $I^{+}$, 

\noindent \medskip{}

\noindent 
\begin{equation}
\delta\Psi=-\frac{\delta m^{2}}{2}I^{+}-\frac{\delta\lambda}{2}(I^{+})^{2}\,.\label{eq:RenoDelPsi}
\end{equation}
The recipe for finding these counterterms is very simple: we add counterterms
$\delta m^{2}$ and $\delta\lambda$ to each mass squared and each
coupling constant that appears in the action (\ref{eq:RenoPsiNormal})
(neither $J$ nor $I^{+}$ depend on $m$ or $\lambda$ explicitly).
This will be a bit less straightforward in the condensed phase, where
we shall need two different counterterms $\delta\lambda_{1}$ and
$\delta\lambda_{2}$, see next subsection. The mass counterterm $\delta m^{2}$
is of order $\lambda$, while $\delta\lambda$ is of order $\lambda^{2}$.
The crucial point will be to show that all divergences can be canceled
with medium independent quantities $\delta m^{2}$ and $\delta\lambda$.
Of course, the total counterterm $\delta\Psi$ does depend on the
medium because $I^{+}$ and $J$ depend on $\mu$, $T$, and $\nabla\psi$.
Let us first discuss the renormalized stationarity equation. In the
uncondensed phase there is only one equation, for the self-consistent
mass $M$,

\noindent 
\begin{equation}
M^{2}=m^{2}+\delta m^{2}+2(\lambda+\delta\lambda)I^{+}\,.\label{eq:RenoStatM}
\end{equation}
 In evaluating integrals like $I^{+}$ we will use a notation where
a subscripted argument indicates subtraction of the function's value
when that argument is zero, 

\noindent \medskip{}

\noindent 
\begin{equation}
I_{x}(A)\equiv I(x,A)-I(0,A)\,.\label{eq:RenoDefineSubs}
\end{equation}
 Using that notation, we split the integral $I^{+}$ into a zero temperature
part that depends on the cutoff $\Lambda$ and a part $I_{T}^{+}$
that depends on $T$ but goes to zero as $T\to0$ and is cutoff-independent, 

\noindent \medskip{}

\noindent 
\begin{equation}
I^{+}(T,\Lambda)=I^{+}(0,\Lambda)+I_{T}^{+}\,,
\end{equation}
 where the dependence on $\mu,\, M,\,\vec{\nabla}\psi$ is not explicitly
shown. Evaluating the Matsubara sum, we find 

\noindent \medskip{}

\noindent 
\begin{equation}
I^{+}(0,\Lambda)=\frac{1}{2}\sum_{e=\pm}\int\frac{d^{3}\vec{k}}{(2\pi)^{3}}\frac{1}{\sqrt{(\vec{k}-e\vec{\nabla}\psi)^{2}+M^{2}}}\,,\qquad I_{T}^{+}\equiv\sum_{e=\pm}\int\frac{d^{3}\vec{k}}{(2\pi)^{3}}\frac{f(\epsilon_{\vec{k}}^{e})}{\sqrt{(\vec{k}-e\vec{\nabla}\psi)^{2}+M^{2}}}\,,
\end{equation}
where $f$ is the Bose distribution function. The terms in the large-momentum
expansion of the integrand that lead to cutoff dependences are shown
in table 4 (with $\delta M=0$ for the uncondensed case).

\noindent \begin{table*}[t] \begin{tabular}{|c|c|c|c|} \hline \rule[-1.5ex]{0em}{6ex} & $I^+$ & $I^-$ & $J$ \\[2ex] \hline \rule[-1.5ex]{0em}{6ex} $\;\;$ \parbox{6em}{UV div.}$\;\;$ &$\;\;$ $\displaystyle{k-\frac{M^2}{2k} }$$\;\;$ &$\;\;$ $\displaystyle{-\frac{\delta M^2}{2k}}$ $\;\;$ & $\;\;$$\displaystyle{-k^3-\left[M^2+\frac{2}{3}(\nabla\psi)^2\right]\frac{k}{2}+\frac{M^4+\delta M^4}{8k} }$$\;\;$ \\[2ex] \hline
\end{tabular} \caption{Ultraviolet divergent contributions to the various integrands of the three-momentum integrals. The contributions are given for the most general case with condensation and superflow. The limit cases discussed in detail in this appendix are obtained by setting $\delta M=0$ (uncondensed case) and $\nabla\psi=0$ (condensed case without superflow). The divergent terms depend implicitly on temperature, chemical potential, and the superfluid velocity, the latter appearing even explicitly in the divergent terms of $J$.} \label{table0} \end{table*} We evaluate the momentum integral $I^{+}(0,\Lambda)$ via proper
time regularization \cite{Schwinger1951}, using the general relation 

\noindent 
\begin{equation}
\frac{1}{x^{a}}=\frac{1}{\Gamma(a)}\int_{0}^{\infty}d\tau\,\tau^{a-1}e^{-\tau x}\,,
\end{equation}
 where, in this case, $x=(\vec{k}-e\vec{\nabla}\psi)^{2}+M^{2}$,
and exchange the order of the $\vec{k}$ and $\tau$ integrals. The
$\vec{k}$ integral is now finite, so we can eliminate $\vec{\nabla}\psi$
because it is simply a shift of the integration variable. The ultraviolet
cutoff $\Lambda$ is implemented by setting the lower limit of the
proper time integral to $1/\Lambda^{2}$. This yields

\noindent 
\begin{eqnarray}
I^{+}(T,\Lambda) & = & \frac{\Lambda^{2}}{8\pi^{2}}-\frac{M^{2}}{8\pi^{2}}\ln\frac{\Lambda^{2}}{\ell^{2}}+I_{{\rm finite}}^{+}(T,\ell)\,,\\
I_{{\rm finite}}^{+}(T,\ell) & = & \frac{M^{2}}{8\pi^{2}}\left(\gamma-1+\ln\frac{M^{2}}{\ell^{2}}\right)+I_{T}^{+}\,,
\end{eqnarray}
where we have introduced the renormalization scale $\ell$, and where
$\gamma\simeq0.5772$ is the Euler-Mascheroni constant. We can now
insert the regularized integral into the stationarity equation (\ref{eq:RenoStatM}),
and separate it in to a cutoff-independent part

\noindent 
\begin{equation}
M^{2}=m^{2}+2\lambda I_{{\rm finite}}^{+}(T,\ell)\,,\label{eq:RenoM}
\end{equation}
and a cutoff-dependent part

\noindent \medskip{}

\noindent 
\begin{equation}
0=\delta m^{2}+\frac{\lambda+\delta\lambda}{4\pi^{2}}\left(\Lambda^{2}-M^{2}\ln\frac{\Lambda^{2}}{\ell^{2}}\right)+2\delta\lambda\, I_{{\rm finite}}^{+}(T,\ell)\,.
\end{equation}
 Note that the ambiguity in performing this separation corresponds
to choosing the renormalization scale $\ell$. In order to determine
$\delta m^{2}$ and $\delta\lambda$, we eliminate $I_{{\rm finite}}^{+}$
with the help of equation \ref{eq:RenoM}. The resulting equation
has two contributions, one of which is medium independent and one
of which is proportional to $M^{2}$. Both contributions have to vanish
separately, and thus we obtain two equations for $\delta m^{2}$ and
$\delta\lambda$ whose solutions are

\noindent \medskip{}

\noindent 
\begin{equation}
\delta\lambda=\frac{\lambda^{2}}{4\pi^{2}}\ln\frac{\Lambda^{2}}{\ell^{2}}\left(1-\frac{\lambda}{4\pi^{2}}\ln\frac{\Lambda^{2}}{\ell^{2}}\right)^{-1}\,,\qquad\delta m^{2}=\delta\lambda\left(\frac{m^{2}}{\lambda}-\frac{\Lambda^{2}}{4\pi^{2}}\right)-\lambda\frac{\Lambda^{2}}{4\pi^{2}}\,.
\end{equation}
If we introduce the bare mass $m_{{\rm bare}}^{2}=m^{2}+\delta m^{2}$
and the bare coupling, $\lambda_{{\rm bare}}=\lambda+\delta\lambda$,
we can write

\noindent \medskip{}

\noindent 
\begin{equation}
\frac{1}{\lambda}=\frac{1}{\lambda_{{\rm bare}}}+\frac{1}{4\pi^{2}}\ln\frac{\Lambda^{2}}{\ell^{2}}\,,\qquad\frac{m^{2}}{\lambda}=\frac{m_{{\rm bare}}^{2}}{\lambda_{{\rm bare}}}+\frac{\Lambda^{2}}{4\pi^{2}}\,.\label{eq:RenoRelations}
\end{equation}
 Next, we need to check whether the same counterterms cancel all divergences
in the pressure $\Psi+\delta\Psi$. Again, we write 

\noindent 
\begin{equation}
J(T,\Lambda)=J(0,\Lambda)+J_{T}\,,
\end{equation}
 where again we do not show the dependence on $\mu,\, M,\,\vec{\nabla}\psi$,
and where, after performing the Matsubara sum and taking the thermodynamic
limit, we have 

\noindent \medskip{}

\noindent 
\begin{equation}
J(0,\Lambda)=-\frac{1}{2}\sum_{e=\pm}\int\frac{d^{3}\vec{k}}{(2\pi)^{3}}\epsilon_{\vec{k}}^{e}\,,\qquad J_{T}=-T\sum_{e=\pm}\int\frac{d^{3}\vec{k}}{(2\pi)^{3}}\ln\left(1-e^{-\epsilon_{{\bf \vec{k}}}^{e}/T}\right)\,.
\end{equation}
 Proper time regularization eliminates the cutoff-dependent term that
depended explicitly on the background field $\vec{\nabla}\psi$, and
we find 

\noindent 
\begin{eqnarray}
J(T,\Lambda) & = & \frac{\Lambda^{4}}{32\pi^{2}}-\frac{M^{2}\Lambda^{2}}{16\pi^{2}}+\frac{M^{4}}{32\pi^{2}}\ln\frac{\Lambda^{2}}{\ell^{2}}+J_{{\rm finite}}(T,\ell)\,,\\
J_{{\rm finite}}(T,\ell) & = & \frac{M^{4}}{32\pi^{2}}\left(\frac{3}{2}-\gamma-\ln\frac{M^{2}}{\ell^{2}}\right)+J_{T}\,.
\end{eqnarray}
 Inserting this into $\Psi+\delta\Psi$, using the stationarity equation
(\ref{eq:RenoStatM}) to eliminate $I^{+}$, and making use of the
relations (\ref{eq:RenoRelations}), we find that indeed all medium
dependent divergences in $\Psi+\delta\Psi$ are canceled. 

\noindent \newpage{}We are left with 

\noindent \medskip{}

\noindent 
\[
\Psi+\delta\Psi=\frac{\Lambda^{4}}{32\pi^{2}}-\frac{m^{4}}{8\lambda}+\frac{m_{{\rm bare}}^{4}}{8\lambda_{{\rm bare}}}+\frac{(M^{2}-m^{2})^{2}}{8\lambda}+J_{{\rm finite}}(T,\ell)\,.
\]
 The first three terms on the right-hand side are independent of the
thermodynamic parameters $\mu$, $T$, and $\vec{\nabla}\psi$, and
hence have no effect on the physics. 

\noindent ~

\subsection{\noindent Condensed phase without superflow\label{sub:Condensed-phase-without-SF}}

\noindent ~

\noindent As a next step, we consider the condensed phase, but first
without supercurrent, $\vec{\nabla}\psi=\vec{0}$. In this case, with
$\mu=\partial_{0}\psi$, the inverse propagator is %
\footnote{Remember that $M^{2}\pm\delta M^{2}$ is just a notation for two different
self-consistent masses, as in reference \cite{CritTKaon}, i.e., $\delta M^{2}$
should not be confused with a counterterm.%
} 

\noindent \bigskip{}

\noindent 
\begin{equation}
S^{-1}(k)=\left(\begin{array}{cc}
-k^{2}+M^{2}+\delta M^{2}-\mu^{2} & 2ik_{0}\mu\\[2ex]
-2ik_{0}\mu & -k^{2}+M^{2}-\delta M^{2}-\mu^{2}
\end{array}\right)\,,\label{eq:RenoPropNoCond}
\end{equation}
 which leads to the dispersion relations 

\noindent \medskip{}

\noindent 
\begin{equation}
\epsilon_{\vec{k}}^{e}=\sqrt{E_{\vec{k}}^{2}+\mu^{2}-e\sqrt{4\mu^{2}E_{\vec{k}}^{2}+\delta M^{4}}}\,,
\end{equation}
 where 

\noindent 
\begin{equation}
E_{\vec{k}}\equiv\sqrt{\vec{k}^{2}+M^{2}}\,.
\end{equation}
 Now, the counterterm (\ref{eq:RenoDelPsi}) is generalized to 

\noindent \medskip{}

\noindent 
\[
\delta\Psi=-\frac{\delta m^{2}}{2}\rho^{2}-\frac{2\delta\lambda_{1}+\delta\lambda_{2}}{4}\rho^{4}-\frac{\delta m^{2}+2\delta\lambda_{1}\rho^{2}}{2}I^{+}-\frac{\delta\lambda_{2}\rho^{2}}{2}I^{-}-\frac{\delta\lambda_{1}}{2}(I^{+})^{2}-\frac{\delta\lambda_{2}}{4}(I^{-})^{2}\,.
\]
 In the condensed phase it is necessary to introduce two different
counterterms $\delta\lambda_{1}$ and $\delta\lambda_{2}$ for the
two structures $I^{+}$ and $I^{-}$ \cite{Fejos2008}%
\footnote{In the notation of \cite{Fejos2008}, $\delta\lambda^{A}\equiv2\delta\lambda_{1}-\delta\lambda_{2}$,
$\delta\lambda^{B}\equiv\delta\lambda_{2}$.%
}. We could have put another different counterterm in front of the
$\rho^{4}$ term, but we have already anticipated the result that
this counterterm is a particular linear combination of $\delta\lambda_{1}$
and $\delta\lambda_{2}$. 

\newpage{}

\noindent The stationarity equations become, in agreement to \cite{Fejos2008},

\noindent 
\begin{eqnarray}
0 & = & \mu^{2}-(m^{2}+\delta m^{2})-(\lambda+2\delta\lambda_{1}+\delta\lambda_{2})\rho^{2}-\left[2(\lambda+\delta\lambda_{1})I^{+}+(\lambda+\delta\lambda_{2})I^{-}\right]\,,\label{eq:RenoStat1}\\
M^{2}+\delta M^{2} & = & m^{2}+\delta m^{2}+(3\lambda+2\delta\lambda_{1}+\delta\lambda_{2})\rho^{2}+2(\lambda+\delta\lambda_{1})I^{+}+(\lambda+\delta\lambda_{2})I^{-}\,,\label{eq:RenoStat2}\\
M^{2}-\delta M^{2} & = & m^{2}+\delta m^{2}+(\lambda+2\delta\lambda_{1}-\delta\lambda_{2})\rho^{2}+2(\lambda+\delta\lambda_{1})I^{+}-(\lambda+\delta\lambda_{2})I^{-}\,,\label{eq:RenoStat3}
\end{eqnarray}
 where the first one is obtained from extremizing the action with
respect to $\rho$ and the second and the third are the two nontrivial
components of the Dyson-Schwinger equation. Inserting (\ref{eq:RenoStat2})
into (\ref{eq:RenoStat1}) as well as adding and subtracting (\ref{eq:RenoStat2})
and (\ref{eq:RenoStat3}) to/from each other yields the simpler system
of equations

\noindent 
\begin{eqnarray}
M^{2}+\delta M^{2} & = & \mu^{2}+2\lambda\rho^{2}\,,\label{eq:RenoStat11}\\
M^{2} & = & m^{2}+\delta m^{2}+2(\lambda+\delta\lambda_{1})(\rho^{2}+I^{+})\,,\label{eq:RenoStat12}\\
\delta M^{2} & = & (\lambda+\delta\lambda_{2})(\rho^{2}+I^{-})\,,\label{eq:RenoStat13}
\end{eqnarray}
where the first equation already has its final, renormalized form.
Using the notation of (\ref{eq:RenoDefineSubs}), we rewrite $I^{\pm}$
by first separating off the $T$-dependent term, and then separating
off the $\mu$-dependence at $T=0$, leaving a $\mu=T=0$ vacuum term
that contains all the cutoff dependence, 

\noindent 
\begin{eqnarray}
I^{\pm}(T,\mu,\Lambda) & = & I^{\pm}(0,\mu,\Lambda)+I_{T}^{\pm}(\mu)\,,\label{eq:RenoDiff1}\\
I^{\pm}(0,\mu,\Lambda) & = & I^{\pm}(0,0,\Lambda)+I_{\mu}^{\pm}(0)\,,\label{eq:RenoDff2}
\end{eqnarray}
 where each quantity has dependence on $(M,\delta M)$ which is not
explicitly shown. As in \cite{Fejos2008}, when we set $\mu$ or $T$
to zero we keep unchanged the mass parameters of the full propagator
$M$ and $\delta M$, even though in reality they depend on $\mu$
and $T$. Evaluating (\ref{eq:RenoDiff1}), (\ref{eq:RenoDff2}) with
the help of (\ref{eq:RenoIJ}) and (\ref{eq:RenoPropNoCond}), the
$T=0$ integrals are 

\noindent 
\begin{eqnarray}
I^{+}(0,\mu,\Lambda) & = & \frac{1}{2}\sum_{e=\pm}\int\frac{d^{3}\vec{k}}{(2\pi)^{3}}\frac{1}{\epsilon_{\vec{k}}^{e}}\left(1-\frac{2e\mu^{2}}{\sqrt{4\mu^{2}E_{\vec{k}}^{2}+\delta M^{4}}}\right)\,,\\
I^{-}(0,\mu,\Lambda) & = & -\frac{1}{2}\sum_{e=\pm}\int\frac{d^{3}\vec{k}}{(2\pi)^{3}}\frac{1}{\epsilon_{\vec{k}}^{e}}\frac{e\delta M^{2}}{\sqrt{4\mu^{2}E_{\vec{k}}^{2}+\delta M^{4}}}\,.
\end{eqnarray}
 The thermal integrals $I_{T}^{\pm}(\mu)$ are simply given by $I^{\pm}(0,\mu,\Lambda)$
with an additional factor $2f(\epsilon_{\vec{k}}^{e})$ in the integrand,
which renders them cutoff-independent.

\noindent \newpage{}The vacuum contribution is 

\noindent \medskip{}

\noindent 
\begin{equation}
I^{\pm}(0,0,\Lambda)=\frac{1}{2}\int\frac{d^{3}\vec{k}}{(2\pi)^{3}}\left(\frac{1}{\sqrt{\vec{k}^{2}+M^{2}+\delta M^{2}}}\pm\frac{1}{\sqrt{\vec{k}^{2}+M^{2}-\delta M^{2}}}\right)\,,\label{eq:RenoAux1}
\end{equation}
 and its cutoff-dependence arises from the terms given in table 4
(after setting $\vec{\nabla}\psi=\vec{0}$). This can be evaluated
using proper-time regularization, 

\noindent 
\begin{eqnarray}
I^{+}(0,0,\Lambda) & = & \frac{\Lambda^{2}}{8\pi^{2}}-\frac{M^{2}}{8\pi^{2}}\ln\frac{\Lambda^{2}}{\ell^{2}}+I_{{\rm vac,finite}}^{+}(\ell)\,,\label{eq:RenoIJreno}\\
I^{-}(0,0,\Lambda) & = & -\frac{\delta M^{2}}{8\pi^{2}}\ln\frac{\Lambda^{2}}{\ell^{2}}+I_{{\rm vac,finite}}^{-}(\ell)\,,\\
I_{{\rm vac,finite}}^{\pm}(\ell) & \equiv & \frac{M^{2}}{8\pi^{2}}(\gamma-1)+\frac{M^{2}+\delta M^{2}}{16\pi^{2}}\ln\frac{M^{2}+\delta M^{2}}{\ell^{2}}\pm\frac{M^{2}-\delta M^{2}}{16\pi^{2}}\ln\frac{M^{2}-\delta M^{2}}{\ell^{2}}\,.\nonumber 
\end{eqnarray}
 The finite parts $I_{{\rm finite}}^{\pm}$ of $I^{\pm}$ are then
given by 

\noindent \smallskip{}

\noindent 
\begin{equation}
I_{{\rm finite}}^{\pm}(T,\mu,\ell)=I_{{\rm vac,finite}}^{\pm}(\ell)+I_{\mu}^{\pm}(0)+I_{T}^{\pm}(\mu)\,,
\end{equation}
 where $I_{\mu}^{\pm}(0)$ is obtained via (\ref{eq:RenoDff2}), by
numerically evaluating $I^{\pm}(0,\mu,\Lambda)-I^{\pm}(0,0,\Lambda)$,
combining them into one cutoff-independent integral. Now we can come
back to the stationarity equations (\ref{eq:RenoStat11}-\ref{eq:RenoStat13}).
The first of these equations does not contain any divergences anymore.
With (\ref{eq:RenoStat12}) and (\ref{eq:RenoStat13}) we proceed
analogously as explained for the uncondensed phase: we insert equations
(\ref{eq:RenoIJreno}) and separate finite and infinite contributions.
The finite contributions are the renormalized equations

\noindent 
\begin{eqnarray}
M^{2} & = & m^{2}+2\lambda[\rho^{2}+I_{{\rm finite}}^{+}(T,\mu,\ell)]\,,\label{eq:RenoMreno}\\
\delta M^{2} & = & \lambda[\rho^{2}+I_{{\rm finite}}^{-}(T,\mu,\ell)]\,.\label{eq:RenoDmreno}
\end{eqnarray}
In the equations for the infinite contributions we first eliminate
$\rho$ and $I_{{\rm finite}}^{\pm}$ with the help of equations (\ref{eq:RenoMreno}),
(\ref{eq:RenoDmreno}) and then separate medium-independent terms
from terms proportional to $M^{2}$ for equation (\ref{eq:RenoStat12})
and $\delta M^{2}$ for (\ref{eq:RenoStat13}). The requirement that
all infinities cancel yields the conditions

\noindent 
\begin{eqnarray}
\delta\lambda_{1} & = & \frac{\lambda^{2}}{4\pi^{2}}\ln\frac{\Lambda^{2}}{\ell^{2}}\left(1-\frac{\lambda}{4\pi^{2}}\ln\frac{\Lambda^{2}}{\ell^{2}}\right)^{-1}\,,\qquad\delta m^{2}=\delta\lambda_{1}\left(\frac{m^{2}}{\lambda}-\frac{\Lambda^{2}}{4\pi^{2}}\right)-\lambda\frac{\Lambda^{2}}{4\pi^{2}}\,,\label{eq:RenoCounterT}\\
\delta\lambda_{2} & = & \frac{\lambda^{2}}{8\pi^{2}}\ln\frac{\Lambda^{2}}{\ell^{2}}\left(1-\frac{\lambda}{8\pi^{2}}\ln\frac{\Lambda^{2}}{\ell^{2}}\right)^{-1}\,,\nonumber 
\end{eqnarray}
 which confirms that $\delta\lambda_{1}$ and $\delta\lambda_{2}$
are indeed different. By introducing the two bare couplings $\lambda_{1/2,{\rm bare}}=\lambda+\delta\lambda_{1/2}$
and the bare mass $m_{{\rm bare}}^{2}=m^{2}+\delta m^{2}$ we can
write this in a more compact way, 

\noindent 
\begin{equation}
\frac{1}{\lambda}=\frac{1}{\lambda_{1,{\rm bare}}}+\frac{1}{4\pi^{2}}\ln\frac{\Lambda^{2}}{\ell^{2}}\,,\qquad\frac{m^{2}}{\lambda}=\frac{m_{{\rm bare}}^{2}}{\lambda_{1,{\rm bare}}}+\frac{\Lambda^{2}}{4\pi^{2}}\,,\qquad\frac{1}{\lambda}=\frac{1}{\lambda_{2,{\rm bare}}}+\frac{1}{8\pi^{2}}\ln\frac{\Lambda^{2}}{\ell^{2}}\,.\label{eq:RenoRelationsreno}
\end{equation}
 Finally, we need to check that all divergences in the pressure cancel.
This requires evaluation of $J$ in (\ref{eq:Reno2PiAction}). In
analogy with our evaluation of $I^{\pm}$, we separate the $T$ and
$\mu$ dependence from the vacuum term, writing 

\noindent 
\begin{eqnarray}
J(T,\mu,\Lambda) & = & J(0,\mu,\Lambda)+J_{T}(\mu)\,,\\
J(0,\mu,\Lambda) & = & J(0,0,\Lambda)+J_{\mu}(0)\,.\nonumber 
\end{eqnarray}
The $T=\mu=0$ ``vacuum'' integral is 

\noindent \medskip{}

\noindent 
\begin{equation}
J(0,0,\Lambda)=-\frac{1}{2}\int\frac{d^{3}\vec{k}}{(2\pi)^{3}}\left(\sqrt{\vec{k}^{2}+M^{2}+\delta M^{2}}+\sqrt{\vec{k}^{2}+M^{2}-\delta M^{2}}\right)\,.
\end{equation}
 Evaluating this using a proper-time regulator we find 

\noindent 
\begin{eqnarray}
J(0,0,\Lambda) & = & \frac{\Lambda^{4}}{32\pi^{2}}-\frac{\Lambda^{2}M^{2}}{16\pi^{2}}+\frac{M^{4}+\delta M^{4}}{32\pi^{2}}\ln\frac{\Lambda^{2}}{\ell^{2}}+J_{{\rm vac,finite}}(\ell)\,,\label{eq:RenoJfinite}\\
J_{{\rm vac,finite}}(\ell) & \equiv & \frac{M^{4}+\delta M^{4}}{64\pi^{2}}(3-2\gamma)-\frac{(M^{2}+\delta M^{2})^{2}}{64\pi^{2}}\ln\frac{M^{2}+\delta M^{2}}{\ell^{2}}-\frac{(M^{2}-\delta M^{2})^{2}}{64\pi^{2}}\ln\frac{M^{2}-\delta M^{2}}{\ell^{2}}\,.\nonumber \\
\end{eqnarray}
The finite part of $J$ is then the finite part of the vacuum contribution
plus the $\mu$ and $T$ dependence,

\noindent \smallskip{}

\noindent 
\begin{equation}
J_{{\rm finite}}(T,\mu,\ell)=J_{{\rm vac,finite}}(\ell)+J_{\mu}(0)+J_{T}(\mu)\,.
\end{equation}
By using equations (\ref{eq:RenoStat12}) and (\ref{eq:RenoStat13})
to eliminate $I^{+}$ and $I^{-}$ we obtain

\noindent \medskip{}

\noindent 
\begin{eqnarray}
\Psi+\delta\Psi & = & \frac{\rho^{2}}{2}(\mu^{2}-m^{2})-\frac{\lambda}{4}\rho^{4}+J-\frac{\delta m^{2}}{2}\rho^{2}-\frac{2\delta\lambda_{1}+\delta\lambda_{2}}{4}\rho^{4}\\
\nonumber \\
 & + & \frac{(M^{2}-m_{{\rm bare}}^{2}-2\lambda_{1,{\rm bare}}\rho^{2})^{2}}{8\lambda_{1,{\rm bare}}}+\frac{(\delta M^{2}-\lambda_{2,{\rm bare}}\rho^{2})^{2}}{4\lambda_{2,{\rm bare}}}\,.\nonumber 
\end{eqnarray}
\newpage{}With the help of (\ref{eq:RenoRelationsreno}) we rewrite
the last two terms of this expression, 

\noindent 
\begin{eqnarray}
\frac{(M^{2}-m_{{\rm bare}}^{2}-2\lambda_{1,{\rm bare}}\rho^{2})^{2}}{8\lambda_{1,{\rm bare}}} & = & \frac{(M^{2}-m^{2}-2\lambda\rho^{2})^{2}}{8\lambda}-\frac{m^{4}}{8\lambda}+\frac{m^{4}}{8\lambda_{1}}+\frac{M^{2}\Lambda^{2}}{16\pi^{2}}\\
 & - & \frac{M^{4}}{32\pi^{2}}\ln\frac{\Lambda^{2}}{\ell^{2}}+\frac{\delta m^{2}}{2}\rho^{2}+\frac{\delta\lambda_{1}}{2}\rho^{4}\,,\nonumber \\
\nonumber \\
\frac{(\delta M^{2}-\lambda_{2,{\rm bare}}\rho^{2})^{2}}{4\lambda_{2,{\rm bare}}} & = & \frac{(\delta M^{2}-\lambda\rho^{2})^{2}}{4\lambda}-\frac{\delta M^{4}}{32\pi^{2}}\ln\frac{\Lambda^{2}}{\ell^{2}}+\frac{\delta\lambda_{2}}{4}\rho^{2}\,.
\end{eqnarray}
 We see that the divergences appearing here cancel all divergences
from $J$ in (\ref{eq:RenoJfinite}) that depend on $M$ and $\delta M$,
and we arrive at the renormalized pressure 

\noindent \medskip{}

\noindent 
\begin{equation}
\Psi+\delta\Psi=\frac{\Lambda^{4}}{32\pi^{2}}-\frac{m^{4}}{8\lambda}+\frac{m_{{\rm bare}}^{4}}{8\lambda_{1,{\rm bare}}}+\frac{\rho^{2}}{2}(\mu^{2}-m^{2})-\frac{\lambda}{4}\rho^{4}+J_{{\rm finite}}(T,\mu,\ell)+\frac{(M^{2}-m^{2}-2\lambda\rho^{2})^{2}}{8\lambda}+\frac{(\delta M^{2}-\lambda\rho^{2})^{2}}{4\lambda}\,.\label{eq:RenoPsidPsi}
\end{equation}
The first three terms on the right-hand side are independent of the
thermodynamic parameters $\mu$ and $T$, and hence have no effect
on the physics; the next two terms are the renormalized tree-level
potential; then, $J_{{\rm finite}}$ is the finite part of the $\textrm{Tr}\,\ln S^{-1}$
term, while the last two terms are the renormalized version of the
combined terms coming from $\textrm{Tr}\,[S_{0}^{-1}S-1]$ and $V_{2}$.

\noindent ~

\subsection{\noindent Renormalization with Goldstone mode\label{sub:Renormalization-with-Goldstone}}

\noindent ~

\noindent As discussed in section \ref{sub:Goldstone-mode}, the present
formalism violates the Goldstone theorem, and since our discussion
of the superfluid properties requires an exact Goldstone mode we need
to consider modified stationarity equations. We thus have to check
how our modification affects the renormalization and what the renormalized
pressure at the new ``Goldstone point'' is (which is slightly off
the ``stationary point''). To this end, we emphasize that the renormalization
procedure explained above is designed to work at the stationary point.
In particular, equation (\ref{eq:RenoPsidPsi}) is the renormalized
pressure at that point because we have used the stationarity equations
(\ref{eq:RenoStat11}-\ref{eq:RenoStat13}) that include finite as
well as infinite parts. It seems we would have to redo our whole analysis
for the ``Goldstone point''. However, we may simply do the modification
in the \textit{finite} part of the stationarity equations, thus preserving
all the results for the counterterms. This amounts to changing equation
(\ref{eq:RenoDmreno}) to

\noindent \medskip{}

\noindent 
\begin{equation}
\delta M^{2}=\lambda\rho^{2}\,,\label{eq:RenoDmGold}
\end{equation}
 but keeping the two other renormalized equations (\ref{eq:RenoStat11})
and (\ref{eq:RenoMreno}) as well as all infinite contributions in
equations (\ref{eq:RenoStat12}) and (\ref{eq:RenoStat13}) as they
are. It is then obvious that the counterterms are still given by (\ref{eq:RenoCounterT}).
All we need to do is compute the finite part of the pressure; by construction,
all infinities in the pressure will still cancel. We can thus simply
replace all integrals in the effective action (\ref{eq:Reno2PiAction})
by their finite parts, and use (\ref{eq:RenoMreno}) and (\ref{eq:RenoDmGold})
to find

\noindent \medskip{}

\noindent 
\begin{equation}
\Psi+\delta\Psi=\frac{\Lambda^{4}}{32\pi^{2}}-\frac{m^{4}}{8\lambda}+\frac{m_{{\rm bare}}^{4}}{8\lambda_{1,{\rm bare}}}+\frac{\rho^{2}}{2}(\mu^{2}-m^{2})-\frac{\lambda}{4}\rho^{4}+J_{{\rm finite}}+\frac{(M^{2}-m^{2}-2\lambda\rho^{2})^{2}}{8\lambda}-\frac{\lambda}{4}(I_{{\rm finite}}^{-})^{2}\,.
\end{equation}

~

\subsection{\noindent Condensed phase with superflow\label{sub:Condensed-phase-with-SF}}

~

\noindent Following the procedure of section \ref{sub:Condensed-phase-without-SF},
we first separate the integrals $I^{\pm}$ and $J$ into their thermal
parts $I_{T}^{\pm}(\mu)$ and $J_{T}(\mu)$ and the cutoff dependent
integrals%
\footnote{For explicit numerical calculations, the identity

\[
(\epsilon_{\vec{k}}^{e}+\epsilon_{-\vec{k}}^{e})(\epsilon_{\vec{k}}^{e}+\epsilon_{-\vec{k}}^{-e})(\epsilon_{\vec{k}}^{e}-\epsilon_{\vec{k}}^{-e})=4\left\{ \epsilon_{\vec{k}}^{e}\left[(\epsilon_{\vec{k}}^{e})^{2}-\vec{k}^{2}-M^{2}-(\partial_{0}\psi)^{2}-(\nabla\psi)^{2}\right]-2\partial_{0}\psi\,\vec{k}\cdot\vec{\nabla}\psi\right\} 
\]
can be useful, the right-hand side being simpler due to the fewer
appearances of the complicated excitation energies.%
}
\begin{eqnarray}
I{}^{+}(0,\mu,\Lambda) & = & 2\sum_{e=\pm}\int\frac{d^{3}\vec{k}}{(2\pi)^{3}}\frac{(\epsilon_{\vec{k}}^{e})^{2}-\vec{k}^{2}-M^{2}+\sigma^{2}}{(\epsilon_{\vec{k}}^{e}+\epsilon_{-\vec{k}}^{e})(\epsilon_{\vec{k}}^{e}+\epsilon_{-\vec{k}}^{-e})(\epsilon_{\vec{k}}^{e}-\epsilon_{\vec{k}}^{-e})}\,,\\
I^{-}(0,\mu,\Lambda) & = & 2\sum_{e=\pm}\int\frac{d^{3}\vec{k}}{(2\pi)^{3}}\frac{\delta M^{2}}{(\epsilon_{\vec{k}}^{e}+\epsilon_{-\vec{k}}^{e})(\epsilon_{\vec{k}}^{e}+\epsilon_{-\vec{k}}^{-e})(\epsilon_{\vec{k}}^{e}-\epsilon_{\vec{k}}^{-e})}\,,
\end{eqnarray}
 where $\epsilon_{\vec{k}}^{e}$ are the positive solutions to (\ref{eq:DispEq}),
which depend on the angle between the momentum of the excitation and
the superflow. Next, we need to regularize $I^{\pm}(0,\mu,\Lambda)$
and $J(0,\mu,\Lambda)$. The divergent contributions of these integrals
are shown in Table 4. The integrals $I^{\pm}(0,\mu,\Lambda)$ show
exactly the same divergences as for the case without superflow discussed
in section \ref{sub:Condensed-phase-without-SF}. In $J(0,\mu,\Lambda)$,
however, there is a divergent contribution that depends explicitly
on $\vec{\nabla}\psi$. This divergence is exactly the same as for
the uncondensed case discussed in section \ref{sub:Uncondensed-phase-with-sF}.
In that case, the $\vec{\nabla}\psi$ dependent divergence in the
pressure was spurious because after regularization with the proper
time method the integrals in pressure and self-energy did not depend
on $\vec{\nabla}\psi$ anymore. 

\noindent One might think that, in order to regularize the divergent
integrals, we should subtract the same integrals at the point $\mu=T=0,\,\vec{\nabla}\psi=\vec{0}$.
However, this procedure would not take care of the $\vec{\nabla}\psi$
dependent divergence. Thus we seem to be forced to subtract the integrals
at the point $\mu=T=0$ with $\vec{\nabla}\psi$ kept fixed, i.e.,
$J(0,\mu,\Lambda)=J(0,0,\Lambda)+J_{\mu}(0)$, which reads

\noindent \medskip{}

\noindent 
\begin{equation}
J(0,\mu,\Lambda)=-\frac{1}{2}\sum_{e=\pm}\int\frac{d^{3}\vec{k}}{(2\pi)^{3}}\epsilon_{\vec{k}}^{e}(\mu=0)-\frac{1}{2}\sum_{e=\pm}\int\frac{d^{3}\vec{k}}{(2\pi)^{3}}[\epsilon_{\vec{k}}^{e}-\epsilon_{\vec{k}}^{e}(\mu=0)]\,,\label{eq:RenoAux4}
\end{equation}
 and analogously for $I^{\pm}(0,\mu,\Lambda)$. The $\mu=0$ dispersion
turns out to be 

\noindent \medskip{}

\noindent 
\begin{equation}
\epsilon_{\vec{k}}^{e}(\mu=0)=\sqrt{\vec{k}^{2}+M^{2}+(\nabla\psi)^{2}\mp\sqrt{4(\vec{\nabla}\psi\cdot\vec{k})^{2}+\delta M^{4}}}\,.\label{eq:RenoAux5}
\end{equation}
 The presence of the two square roots in this expression renders a
straightforward application of the proper time regularization very
complicated and one would have to proceed numerically.

\noindent We notice, however, that there is another way to treat the
ultraviolet divergences, using the same proper time regularization.
Since the structure of the divergences is a simple combination of
the divergences of the cases discussed above, it is easy to ``guess''
a generalization of the subtraction terms to the present case,

\noindent 
\begin{eqnarray}
I^{+}(0,\mu,\Lambda) & = & \frac{1}{2}\sum_{e=\pm}\int\frac{d^{3}\vec{k}}{(2\pi)^{3}}\frac{1}{\omega_{\vec{k}}^{e}}+\sum_{e=\pm}\int\frac{d^{3}\vec{k}}{(2\pi)^{3}}\left[\frac{2\left[(\epsilon_{\vec{k}}^{e})^{2}-\vec{k}^{2}-M^{2}+\sigma^{2}\right]}{(\epsilon_{\vec{k}}^{e}+\epsilon_{-\vec{k}}^{e})(\epsilon_{\vec{k}}^{e}+\epsilon_{-\vec{k}}^{-e})(\epsilon_{\vec{k}}^{e}-\epsilon_{\vec{k}}^{-e})}-\frac{1}{2\omega_{\vec{k}}^{e}}\right]\label{eq:RenoAux2}\\
I^{-}(0,\mu,\Lambda) & = & \frac{1}{2}\sum_{e=\pm}\int\frac{d^{3}\vec{k}}{(2\pi)^{3}}\frac{e}{\omega_{\vec{k}}^{e}}+\sum_{e=\pm}\int\frac{d^{3}\vec{k}}{(2\pi)^{3}}\left[\frac{2\delta M^{2}}{(\epsilon_{\vec{k}}^{e}+\epsilon_{-\vec{k}}^{e})(\epsilon_{\vec{k}}^{e}+\epsilon_{-\vec{k}}^{-e})(\epsilon_{\vec{k}}^{e}-\epsilon_{\vec{k}}^{-e})}-\frac{e}{2\omega_{\vec{k}}^{e}}\right]\label{eq:RenoAux3}
\end{eqnarray}
 and 

\noindent 
\begin{equation}
J(0,\mu,\Lambda)=-\frac{1}{2}\sum_{e=\pm}\int\frac{d^{3}\vec{k}}{(2\pi)^{3}}\omega_{\vec{k}}^{e}-\frac{1}{2}\sum_{e=\pm}\int\frac{d^{3}\vec{k}}{(2\pi)^{3}}(\epsilon_{\vec{k}}^{e}-\omega_{\vec{k}}^{e})\,.\label{eq:RenoAux6}
\end{equation}
 Here, 

\noindent 
\begin{equation}
\omega_{\vec{k}}^{e}\equiv\sqrt{(\vec{k}+e\vec{\nabla}\psi)^{2}+M^{2}+e\delta M^{2}}\label{eq:RenoAux7}
\end{equation}
 is simply the $\mu=0$ dispersion of the uncondensed phase in the
presence of a $\vec{\nabla}\psi$, see equation (\ref{eq:RenoDiskNoCond}),
generalized to two different mass parameters $M^{2}+\delta M^{2}$
and $M^{2}-\delta M^{2}$. It is also the $\mu=0$ dispersion of the
condensed phase without $\vec{\nabla}\psi$, see equation (\ref{eq:RenoAux1}),
with $\vec{\nabla}\psi$ added as a simple shift of the three-momentum.
According to the structure of the divergences, it is clear that the
second integrals on the right-hand sides of equations (\ref{eq:RenoAux2})
and (\ref{eq:RenoAux3}) are finite. And, the first integrals can
be regularized with the proper time method just as in the previous
subsections: the $\vec{\nabla}\psi$ dependence drops out since the
proper time integrals ``ignore'' this dependence, and the resulting
cutoff-dependent terms together with the finite parts $I^{\pm}(\ell)_{{\rm vac,finite}}$,
$J(\ell)_{{\rm vac,finite}}$ are exactly the same as in section \ref{sub:Condensed-phase-without-SF}.
Therefore, the renormalization works as above, with exactly the same
medium independent counterterms as given in (\ref{eq:RenoCounterT}).

\noindent The choice of the subtraction term corresponds to a renormalization
condition, and usually this term is the vacuum contribution. The appearance
of the superflow in the divergent contributions appears to make the
choice ambiguous, and it is not a priori clear whether using (\ref{eq:RenoAux4})-(\ref{eq:RenoAux5})
or (\ref{eq:RenoAux2})-(\ref{eq:RenoAux7}) is the correct physical
choice. We are rather led to the conclusion that the very existence
of the $\vec{\nabla}\psi$ dependent divergence is problematic, because
we seem to have found two renormalization conditions that differ in
their predictions of how physical observables depend on the superflow.
Here we only point out this problem, and leave its solution to further
studies. It will not affect our physical results because we shall
restrict ourselves to weak coupling strengths where these ambiguous
terms are negligibly small, see discussion in section \ref{sub:Critical-temperature,-condensate-critV-forallT}.

\noindent In the main part we summarize the results of the renormalization
procedure using equations (\ref{eq:RenoAux2})-(\ref{eq:RenoAux7}),
see equations (\ref{eq:stat1reno})-(\ref{eq:stat3reno}) for the
stationarity equation and (\ref{eq:PsiStat}) for the pressure. 

~

\section{\noindent Calculation of sound velocities\label{sec:Calculation-of-sound-vel}}

~

\subsection{\noindent Derivation of the wave equations\label{sub:Derivation-of-the-waveEQ}}

\noindent We start from the hydrodynamic equations

\noindent 
\begin{eqnarray}
0 & = & \partial_{\mu}j^{\mu}\,,\label{eq:ContJ}\\
0 & = & \partial_{\mu}s^{\mu}\,,\label{eq:ContS}\\
0 & = & s_{\mu}(\partial^{\mu}\Theta^{\nu}-\partial^{\nu}\Theta^{\mu})\,.\label{eq:ContVort}
\end{eqnarray}
Before we evaluate them, we collect some useful relations. We denote
$P\equiv P_{n}+P_{s}=\Psi$ and thus can write with (\ref{eq:VariationalMaster})

\noindent 
\begin{eqnarray}
dP & = & j^{\mu}d(\partial_{\mu}\psi)+s^{\mu}d\Theta_{\mu}\label{eq:dpAux}\\
 & = & nd\mu+sdT-\frac{n_{s}}{\sigma}\vec{\nabla}\psi\cdot d\vec{\nabla}\psi+\frac{n_{n}}{s}\vec{s}\cdot d\vec{\nabla}\psi-\vec{s}\cdot d\left(\frac{n_{n}}{s}\vec{\nabla}\psi\right)-\vec{s}\cdot d\left(\frac{w}{s^{2}}\vec{s}\right)\,,\nonumber 
\end{eqnarray}
\newpage{}where $j^{0}=n$, $s^{0}=s$, $\partial^{0}\psi=\mu$,
$\Theta^{0}=T$, and we have eliminated $\vec{j}$ and ${\bf \Theta}$
by using 

\noindent \medskip{}

\noindent 
\begin{equation}
j^{\mu}=n_{n}u^{\mu}+n_{s}\frac{\partial^{\mu}\psi}{\sigma}\,,\qquad\Theta^{\mu}=-\frac{n_{n}}{s}\partial^{\mu}\psi+\frac{w}{s}u^{\mu}\,,\label{eq:JThetaAux}
\end{equation}
 where $w\equiv\epsilon_{n}+P_{n}=\mu n_{n}+sT$ is the enthalpy density
of the normal fluid. In the linear approximation, $\vec{s}$ times
a space-time derivative is negligible, because $\vec{s}=s^{0}\vec{v}_{n}$
and we neglect products of $\vec{v}_{n}$ with space-time derivatives.
Therefore, we may approximate

\noindent \medskip{}

\noindent 
\begin{equation}
dP\simeq nd\mu+sdT-\frac{n_{s}}{\sigma}\vec{\nabla}\psi\cdot d\vec{\nabla}\psi\,.
\end{equation}
This relation is needed to express derivatives of any thermodynamic
quantity in terms of derivatives of $T$, $\mu$, and $\vec{\nabla}\psi$.
For instance, we can write $\partial_{0}n=\partial_{0}\frac{\partial P}{\partial\mu}=\frac{\partial}{\partial\mu}\partial_{0}P$
etc. and obtain the following useful identities,

\noindent 
\begin{eqnarray}
\partial_{0}n & = & \frac{\partial n}{\partial\mu}\partial_{0}\mu+\frac{\partial s}{\partial\mu}\partial_{0}T-\frac{\partial(n_{s}/\sigma)}{\partial\mu}\vec{\nabla}\psi\cdot\vec{\nabla}\mu\,,\label{eq:d0n}\\
\partial_{0}s & = & \frac{\partial n}{\partial T}\partial_{0}\mu+\frac{\partial s}{\partial T}\partial_{0}T-\frac{\partial(n_{s}/\sigma)}{\partial T}\vec{\nabla}\psi\cdot\vec{\nabla}\mu\,,\label{eq:d0s}\\
\partial_{0}\left(\frac{n_{s}}{\sigma}\partial_{i}\psi\right) & = & -\frac{\partial n}{\partial(\partial_{i}\psi)}\partial_{0}\mu-\frac{\partial s}{\partial(\partial_{i}\psi)}\partial_{0}T+\frac{\partial(n_{s}/\sigma)}{\partial(\partial_{i}\psi)}\vec{\nabla}\psi\cdot\vec{\nabla}\mu+\frac{n_{s}}{\sigma}\partial_{i}\mu\,,\label{eq:d0ns}\\
\vec{\nabla}\cdot\left(\frac{n_{s}}{\sigma}\vec{\nabla}\psi\right) & = & -\frac{\partial n}{\partial(\partial_{i}\psi)}\partial_{i}\mu-\frac{\partial s}{\partial(\partial_{i}\psi)}\partial_{i}T+\frac{\partial(n_{s}/\sigma)}{\partial(\partial_{i}\psi)}\partial_{j}\psi\partial_{i}\partial_{j}\psi+\frac{n_{s}}{\sigma}\Delta\psi\,,\label{eq:dins}
\end{eqnarray}
 where $\partial_{0}\psi=\mu$ has been used. With these preparations
we can discuss the hydrodynamic equations. The current conservation
(\ref{eq:ContJ}) obviously becomes 

\noindent \medskip{}

\noindent 
\begin{equation}
0\simeq\partial_{0}n+n_{n}\vec{\nabla}\cdot\vec{v}_{n}-\vec{\nabla}\cdot\left(\frac{n_{s}}{\sigma}\vec{\nabla}\psi\right)\,,
\end{equation}
 where we have used $\vec{u}\simeq\vec{v}_{n}$ and $\vec{\nabla}\cdot(n_{n}\vec{v}_{n})\simeq n_{n}\vec{\nabla}\cdot\vec{v}_{n}$.
Inserting (\ref{eq:d0n}) and (\ref{eq:dins}) into this equation,
taking the time derivative of the result, and multiplying the whole
equation by $\mu$ yields 

\noindent 
\begin{eqnarray}
0 & \simeq & \mu\frac{\partial n}{\partial\mu}\partial_{0}^{2}\mu+\mu\frac{\partial s}{\partial\mu}\partial_{0}^{2}T-n_{s}\frac{\mu}{\sigma}\Delta\mu+\mu n_{n}\vec{\nabla}\cdot\partial_{0}\vec{v}_{n}-\mu\frac{\partial(n_{s}/\sigma)}{\partial\mu}\vec{\nabla}\psi\cdot\vec{\nabla}\partial_{0}\mu\label{eq:CurrentConAux}\\
 & + & \mu\frac{\partial n}{\partial(\partial_{i}\psi)}\partial_{0}\partial_{i}\mu+\mu\frac{\partial s}{\partial(\partial_{i}\psi)}\partial_{0}\partial_{i}T-\mu\frac{\partial(n_{s}/\sigma)}{\partial(\partial_{i}\psi)}\partial_{j}\psi\partial_{i}\partial_{j}\mu\,.\nonumber 
\end{eqnarray}
 Due to the linear approximation, all expressions have the form (equilibrium
quantity) $\times$ (second space-time derivative), since products
of two first space-time derivatives are of higher order. 

\noindent The entropy conservation (\ref{eq:ContS}) reads

\noindent \smallskip{}

\noindent 
\begin{equation}
0\simeq\partial_{0}s+s\vec{\nabla}\cdot\vec{v}_{n}\,.
\end{equation}
 Inserting (\ref{eq:d0s}), taking the time derivative of the result
and multiplying the whole equation by $T$ yields 

\noindent \smallskip{}

\noindent 
\begin{equation}
0\simeq T\frac{\partial n}{\partial T}\partial_{0}^{2}\mu+T\frac{\partial s}{\partial T}\partial_{0}^{2}T-T\frac{\partial(n_{s}/\sigma)}{\partial T}\vec{\nabla}\psi\cdot\vec{\nabla}\partial_{0}\mu+sT\vec{\nabla}\cdot\partial_{0}\vec{v}_{n}\,.\label{eq:EntropyConAux}
\end{equation}
 It is convenient for the following to add (\ref{eq:CurrentConAux})
and (\ref{eq:EntropyConAux})

\noindent \medskip{}

\noindent 
\begin{eqnarray}
0 & \simeq & \left(\mu\frac{\partial n}{\partial\mu}+T\frac{\partial n}{\partial T}\right)\partial_{0}^{2}\mu+\left(\mu\frac{\partial s}{\partial\mu}+T\frac{\partial s}{\partial T}\right)\partial_{0}^{2}T-\left[\mu\frac{\partial(n_{s}/\sigma)}{\partial\mu}+T\frac{\partial(n_{s}/\sigma)}{\partial T}\right]\vec{\nabla}\psi\cdot\vec{\nabla}\partial_{0}\mu\label{eq:SummAux}\\
 & + & w\vec{\nabla}\cdot\partial_{0}\vec{v}_{n}+2\mu\frac{\partial n}{\partial(\nabla\psi)^{2}}\vec{\nabla}\psi\cdot\vec{\nabla}\partial_{0}\mu+2\mu\frac{\partial s}{\partial(\nabla\psi)^{2}}\vec{\nabla}\psi\cdot\vec{\nabla}\partial_{0}T-2\mu\frac{\partial(n_{s}/\sigma)}{\partial(\nabla\psi)^{2}}(\vec{\nabla}\psi\cdot\vec{\nabla})^{2}\mu-n_{s}\frac{\mu}{\sigma}\Delta\mu\,,\nonumber 
\end{eqnarray}
 where we have rewritten the derivative with respect to $\partial_{i}\psi$
in terms of the derivative with respect to $(\vec{\nabla}\psi)^{2}$.
Finally, we need the vorticity equation (\ref{eq:ContVort}). The
temporal component becomes

\noindent \smallskip{}

\noindent 
\begin{equation}
0=\vec{s}\cdot(\vec{\nabla}T+\partial_{0}\vec{\Theta})\,.\label{eq:VortSpatial}
\end{equation}
 We can neglect this equation completely, since in both terms the
normal-fluid velocity is multiplied with a space-time derivative.
The spatial components are

\noindent \medskip{}

\noindent 
\begin{equation}
0\simeq s(\partial_{0}\vec{\Theta}+\vec{\nabla}T)\simeq s\vec{\nabla}T+w\partial_{0}\vec{v}_{n}+s\partial_{0}\left(\frac{n_{n}}{s}\vec{\nabla}\psi\right)\,,
\end{equation}
 where $\vec{\Theta}$ from equations (\ref{eq:JThetaAux}) has been
used. The last term needs some rearrangements, 

\noindent \medskip{}

\noindent 
\begin{eqnarray}
s\partial_{0}\left(\frac{n_{n}}{s}\vec{\nabla}\psi\right) & = & -\frac{n_{n}}{s}\vec{\nabla}\psi\,\partial_{0}s+\partial_{0}(n_{n}\vec{\nabla}\psi)=-\frac{n_{n}}{s}\vec{\nabla}\psi\,\partial_{0}s+\partial_{0}(n\vec{\nabla}\psi)-\partial_{0}\left(n_{s}\frac{\mu}{\sigma}\vec{\nabla}\psi\right)\\
 & = & -\frac{n_{n}}{s}\vec{\nabla}\psi\,\partial_{0}s+\vec{\nabla}\psi\partial_{0}n+n\vec{\nabla}\mu-\frac{n_{s}}{\sigma}\vec{\nabla}\psi\partial_{0}\mu-\mu\partial_{0}\left(\frac{n_{s}}{\sigma}\vec{\nabla}\psi\right)\,.\nonumber 
\end{eqnarray}
 Inserting (\ref{eq:d0n}), (\ref{eq:d0s}) and (\ref{eq:d0ns}) into
this relation, the result into (\ref{eq:VortSpatial}) and taking
the divergence of

\noindent \newpage{}the resulting equation, we arrive at 

\noindent 
\begin{eqnarray}
0 & \simeq & n_{n}\Delta\mu+s\Delta T+w\vec{\nabla}\cdot\partial_{0}\vec{v}_{n}-\frac{n_{s}}{\sigma}\vec{\nabla}\psi\cdot\vec{\nabla}\partial_{0}\mu-\left[\frac{n_{n}}{s}\frac{\partial n}{\partial T}-\frac{\partial n}{\partial\mu}-2\mu\frac{\partial n}{\partial(\vec{\nabla}\psi)^{2}}\right]\vec{\nabla}\psi\cdot\vec{\nabla}\partial_{0}\mu\\
 & - & \left[\frac{n_{n}}{s}\frac{\partial s}{\partial T}-\frac{\partial s}{\partial\mu}-2\mu\frac{\partial s}{\partial(\vec{\nabla}\psi)^{2}}\right]\vec{\nabla}\psi\cdot\vec{\nabla}\partial_{0}T+\left[\frac{n_{n}}{s}\frac{\partial(n_{s}/\sigma)}{\partial T}-\frac{\partial(n_{s}/\sigma)}{\partial\mu}-2\mu\frac{\partial(n_{s}/\sigma)}{\partial(\vec{\nabla}\psi)^{2}}\right](\vec{\nabla}\psi\cdot\vec{\nabla})^{2}\mu\,.\nonumber 
\end{eqnarray}
The normal-fluid velocity can now be eliminated by solving this relation
for $\vec{\nabla}\cdot\partial_{0}\vec{v}_{n}$ and inserting the
result into the other two equations: inserting it into \ref{eq:EntropyConAux}
yields, after multiplying the whole equation with $w/(Ts)$, (\ref{eq:soundEq1}),
while inserting it into (\ref{eq:SummAux}) yields (\ref{eq:soundEq2}).
These are the sound wave equations from which the sound velocities
are computed as follows.

\subsection{\noindent Solution of sound wave equations and low-temperature approximation\label{sub:Solution-of-waveEQ}}

\noindent Replacing the chemical potential and the temperature in
all space-time derivatives of the sound wave equations (\ref{eq:soundEq1}),
(\ref{eq:soundEq2}) by $\delta\mu=\delta\mu_{0}e^{i(\omega t-\vec{k}\cdot\vec{x})}$
and $\delta T=\delta T_{0}e^{i(\omega t-\vec{k}\cdot\vec{x})}$, the
sound wave equations become

\noindent 
\begin{eqnarray}
0 & \simeq & \left[a_{1}\tilde{\omega}^{2}+(a_{2}+a_{4}\mu^{2}\vec{v}_{s}^{2}\cos^{2}\theta)+a_{3}\mu|\vec{v}_{s}|\tilde{\omega}\cos\theta\right]\delta\mu_{0}\nonumber \\
 & + & (b_{1}\tilde{\omega}^{2}+b_{2}+b_{3}\mu|\vec{v}_{s}|\tilde{\omega}\cos\theta)\,\delta T_{0}\,,\label{eq:SolSoundAux1}\\
0 & \simeq & \left[A_{1}\tilde{\omega}^{2}+(A_{2}+A_{4}\mu^{2}\vec{v}_{s}^{2}\cos^{2}\theta)+A_{3}\mu|\vec{v}_{s}|\tilde{\omega}\cos\theta\right]\delta\mu_{0}\label{eq:SolSoundAux2}\\
 & + & (B_{1}\tilde{\omega}^{2}+B_{2}+B_{3}\mu|\vec{v}_{s}|\tilde{\omega}\cos\theta)\,\delta T_{0}\,,\nonumber 
\end{eqnarray}
 where $\cos\theta\equiv\hat{k}\cdot\hat{v_{s}}$, $\tilde{\omega}\equiv\omega/|\vec{k}|$,
and we have abbreviated 

\noindent 
\begin{eqnarray}
a_{1} & \equiv & \frac{w}{s}\frac{\partial n}{\partial T}\,,\qquad a_{2}\equiv-n_{n}\,,\qquad a_{3}\equiv\frac{n_{s}}{\sigma}-\frac{w}{s}\frac{\partial(n_{s}/\sigma)}{\partial T}+\frac{n_{n}}{s}\frac{\partial n}{\partial T}-\frac{\partial n}{\partial\mu}-2\mu\frac{\partial n}{\partial(\nabla\psi)^{2}}\,,\\
a_{4} & \equiv & -\left[\frac{n_{n}}{s}\frac{\partial(n_{s}/\sigma)}{\partial T}-\frac{\partial(n_{s}/\sigma)}{\partial\mu}-2\mu\frac{\partial(n_{s}/\sigma)}{\partial(\nabla\psi)^{2}}\right]\,,\qquad b_{1}\equiv\frac{w}{s}\frac{\partial s}{\partial T}\,,\qquad b_{2}\equiv-s\,,\nonumber \\
b_{3} & \equiv & \frac{n_{n}}{s}\frac{\partial s}{\partial T}-\frac{\partial s}{\partial\mu}-2\mu\frac{\partial s}{\partial(\nabla\psi)^{2}}\,,\nonumber \\
A_{1} & \equiv & \mu\frac{\partial n}{\partial\mu}+T\frac{\partial n}{\partial T}\,,\qquad A_{2}\equiv-n\,,\qquad A_{3}\equiv\frac{n_{s}}{\sigma}-\mu\frac{\partial(n_{s}/\sigma)}{\partial\mu}-T\frac{\partial(n_{s}/\sigma)}{\partial T}+\frac{n_{n}}{s}\frac{\partial n}{\partial T}-\frac{\partial n}{\partial\mu}\,,\nonumber \\
A_{4} & \equiv & -\left[\frac{n_{n}}{s}\frac{\partial(n_{s}/\sigma)}{\partial T}-\frac{\partial(n_{s}/\sigma)}{\partial\mu}\right]\,,\qquad B_{1}\equiv\mu\frac{\partial s}{\partial\mu}+T\frac{\partial s}{\partial T}\,,\qquad B_{2}\equiv-s\,,\qquad\nonumber \\
B_{3} & \equiv & \frac{n_{n}}{s}\frac{\partial s}{\partial T}-\frac{\partial s}{\partial\mu}\,.\nonumber 
\end{eqnarray}
In general, the determinant of the coefficient matrix of the system
of two equations (\ref{eq:SolSoundAux1}), (\ref{eq:SolSoundAux2})
yields a complicated quartic equation for $\tilde{\omega}$. However,
we can simplify the result in the low-temperature approximation as
follows. First one can check, for instance by explicit calculation,
that the temperature dependence of the various coefficients is

\noindent 
\begin{eqnarray}
a_{i} & = & a_{i}^{(4)}T^{4}+a_{i}^{(6)}T^{6}\,,\quad A_{i}=A_{i}^{(0)}+A_{i}^{(4)}T^{4}+A_{i}^{(6)}T^{6}\,,\\
b_{j} & = & b_{j}^{(3)}T^{3}+b_{j}^{(5)}T^{5}\,,\quad b_{j}=b_{j}^{(3)}T^{3}+b_{j}^{(5)}T^{5}\,,\quad B_{j}=B_{j}^{(3)}T^{3}+B_{j}^{(5)}T^{5}\,.\nonumber 
\end{eqnarray}
where $i=1,2,3,4$, $j=1,2,3$, and where the prefactors in front
of the various powers of $T$ depend on the superfluid velocity $\vec{v}_{s}$.
Since we have computed the pressure up to order $T^{6}$, all terms
of order $T^{7}$ and higher must be neglected in these expressions.
We see in particular that only the $A_{i}$'s contribute in the limit
$T=0$. Now, for the determinant we encounter two kinds of products,
namely $A_{i}b_{j}$ and $a_{i}B_{j}$, 

\noindent 
\begin{eqnarray}
A_{i}b_{j} & = & T^{3}[A_{i}^{(0)}+A_{i}^{(4)}T^{4}+A_{i}^{(6)}T^{6}][b_{j}^{(3)}+b_{j}^{(5)}T^{2}]=T^{3}A_{i}^{(0)}[b_{j}^{(3)}+b_{j}^{(5)}T^{2}]+{\cal O}(T^{7})\,,\\
a_{i}B_{j} & = & T^{7}[a_{i}^{(4)}+a_{i}^{(6)}T^{2}][B_{j}^{(3)}+B_{j}^{(5)}T^{2}]={\cal O}(T^{7})\,.\nonumber 
\end{eqnarray}
 The first line shows that the ${\cal O}(T^{7})$ terms are unknown
in our expansion because a $b_{j}^{(7)}T^{7}$ term in $b_{j}$ would
give rise to a $T^{7}$ term in $A_{i}b_{j}$, but we have not computed
$b_{j}^{(7)}$. Therefore, we must neglect all products of the form
given in the second line since they are all of order $T^{7}$ and
higher. In other words, it is consistent with our approximation to
set $a_{i}\simeq B_{i}\simeq0$ and use the $T=0$ results for $A_{i}$.
In this case the quartic equation for $\omega$ factorizes into two
quadratic equations, 

\noindent 
\begin{eqnarray}
0 & \simeq & A_{1}^{(0)}\tilde{\omega}^{2}+(A_{2}^{(0)}+A_{4}^{(0)}\mu^{2}\vec{v}_{s}^{2}\cos^{2}\theta)+A_{3}^{(0)}\mu|\vec{v}_{s}|\tilde{\omega}\cos\theta\,,\label{eq:Quadsol1}\\
\nonumber \\
0 & \simeq & [b_{1}^{(3)}+b_{1}^{(5)}T^{2}]\tilde{\omega}^{2}+[b_{2}^{(3)}+b_{2}^{(5)}T^{2}]+[b_{3}^{(3)}+b_{3}^{(5)}T^{2}]\mu|\vec{v}_{s}|\tilde{\omega}\cos\theta\,.\label{eq:quadsol2}
\end{eqnarray}
 After dividing out the overall factor $T^{3}$ of the quartic equation,
the highest remaining power of temperature is 2, i.e., our approximation
allows us to reliably compute the sound velocities up to $T^{2}$.
From the first equation we see that one of the solutions has no $T^{2}$
correction. This is the velocity of first sound. The second equation
yields the velocity of second sound which does have a $T^{2}$ correction.
We see that the coefficients $b_{i}^{(5)}$ are needed to compute
this correction. These coefficients arise from the $T^{5}$ terms
in the entropy, i.e., from the $T^{6}$ terms in the pressure. Had
we truncated our expansion of the pressure at order $T^{4}$, the
velocity of second sound would have turned out to be independent of
temperature. There are two physical solutions of (\ref{eq:Quadsol1}),
(\ref{eq:quadsol2}), $\omega=u_{1,2}|\vec{k}|$. For the explicit
calculation of the two sound velocities $u_{1}$, $u_{2}$ we need
$n_{s}$ and $n_{n}$ from (\ref{eq:nsLowT}) (\ref{eq:nnlowT}) and
the entropy $s$ (obtained by taking the derivative with respect to
temperature of $\Psi=T_{\perp}$ from table 2). The results are shown
given in (\ref{eq:solLOWT1}), (\ref{eq:solLOWT2}) .

~

\subsection{\noindent Sound velocities at low temperatures for arbitrary m\label{sub:Sound-velocities-at-arbitraryM}}

\noindent In this appendix we derive the result (\ref{eq:uLowTwithM})
for the sound velocities in the low-temperature approximation and
in the limit of vanishing superflow. As input, we need the tree level
(and $m\neq0$) expression of the condensate 

\noindent \medskip{}

\noindent 
\begin{equation}
\rho^{2}\simeq\frac{\sigma^{2}-m^{2}}{\lambda}\,,\label{eq:RenoRho}
\end{equation}
 and the low-temperature approximation for the pressure 

\noindent \medskip{}

\noindent 
\begin{equation}
\Psi\simeq\frac{(\sigma^{2}-m^{2})^{2}}{4\lambda}-T\int\frac{d^{3}\vec{k}}{(2\pi)^{3}}\ln\left(1-e^{-\epsilon_{\vec{k}}/T}\right)\,,
\end{equation}
 where $\epsilon_{\vec{k}}$ is the dispersion relation of the Goldstone
mode, containing the superflow; the massive mode only becomes relevant
at higher temperatures and can be neglected. With the help of relation
(\ref{eq:NsSFaprox}) we write

\noindent \medskip{}

\noindent 
\begin{equation}
n_{s}=\mu\frac{\mu^{2}-m^{2}}{\lambda}-\mu\int\frac{d^{3}\vec{k}}{(2\pi)^{3}}\left\{ \left(\frac{\partial\epsilon_{\vec{k}}}{\partial|\vec{\nabla}\psi|}\right)^{2}\frac{f(\epsilon_{\vec{k}})[1+f(\epsilon_{\vec{k}})]}{T}-\frac{\partial^{2}\epsilon_{\vec{k}}}{\partial|\vec{\nabla}\psi|^{2}}f(\epsilon_{\vec{k}})\right\} _{|\vec{\nabla}\psi|\to0}\,.\label{eq:RenoNs}
\end{equation}
Using that the dispersion $\epsilon_{\vec{k}}$ is given by the zeros
of the determinant of the inverse free tree-level propagator $S_{0}$, 

\noindent 
\begin{equation}
S_{0}^{-1}=-k^{2}[-k^{2}+2(\sigma^{2}-m^{2})]-4(k_{\mu}\partial^{\mu}\psi)^{2}\,,
\end{equation}
 where (\ref{eq:RenoRho}) has been used, we find 

\noindent 
\begin{eqnarray*}
\left.\frac{\partial\epsilon_{\vec{k}}}{\partial|\vec{\nabla}\psi|}\right|_{\left|\vec{\nabla}\psi\right|=0} & = & -\frac{2\mu k_{\parallel}}{\sqrt{4\mu^{2}\vec{k}^{2}+(3\mu^{2}-m^{2})^{2}}}\,,\\
\left.\frac{\partial^{2}\epsilon_{\vec{k}}}{\partial|\vec{\nabla}\psi|^{2}}\right|_{\left|\vec{\nabla}\psi\right|=0} & = & \frac{\epsilon_{k}^{2}-2k_{\parallel}^{2}-k^{2}}{\epsilon_{k}\sqrt{4\mu^{2}\vec{k}^{2}+(3\mu^{2}-m^{2})^{2}}}+\frac{8\mu^{2}k_{\parallel}^{2}}{\epsilon_{k}[4\mu^{2}\vec{k}^{2}+(3\mu^{2}-m^{2})^{2}]}+\frac{4\mu^{2}k_{\parallel}^{2}(3\epsilon_{k}^{2}-\vec{k}^{2}-3\mu^{2}+m^{2})}{\epsilon_{k}[4\mu^{2}\vec{k}^{2}+(3\mu^{2}-m^{2})^{2}]^{3/2}}\,.
\end{eqnarray*}
 Here, $k_{\parallel}$ is the longitudinal component of the momentum
with respect to the superflow, and 

\noindent \medskip{}

\noindent 
\begin{equation}
\epsilon_{\vec{k}}=\sqrt{\vec{k}^{2}+3\mu^{2}-m^{2}-\sqrt{4\mu^{2}\vec{k}^{2}+(3\mu^{2}-m^{2})^{2}}}
\end{equation}
 is the dispersion of the Goldstone mode at vanishing superflow. Now
the only nontrivial angular integration is the one over $k_{\parallel}^{2}$, 

\noindent 
\[
k_{\parallel}^{2}=\frac{|\vec{k}|^{2}}{3}\,.
\]
For low temperatures, we can expand the integrand in (\ref{eq:RenoNs})
for small $k$. We find

\noindent \smallskip{}

\[
\mu^{2}\int\frac{d\Omega}{4\pi}\left(\frac{\partial\epsilon_{\vec{k}}}{\partial|\vec{\nabla}\psi|}\right)_{\left|\vec{\nabla}\psi\right|=0}^{2}\simeq q_{1}|\vec{k}|^{2}+\frac{q_{2}}{\mu^{2}}|\vec{k}|^{4}\,,\qquad\mu^{2}\int\frac{d\Omega}{4\pi}\left(\frac{\partial^{2}\epsilon_{\vec{k}}}{\partial|\nabla\psi|^{2}}\right)_{\left|\vec{\nabla}\psi\right|=0}\simeq p_{1}|\vec{k}|+\frac{p_{2}}{\mu^{2}}|\vec{k}|^{3}\,,
\]

\noindent with

\begin{equation}
\epsilon_{\vec{k}}\simeq c_{1}|\vec{k}|+\frac{c_{2}^{2}}{\mu^{2}}|\vec{k}|^{3}\,,
\end{equation}

\noindent and with the dimensionless coefficients

\noindent \medskip{}

\noindent 
\begin{eqnarray}
q_{1} & = & \frac{4\mu^{4}}{3(3\mu^{2}-m^{2})^{2}}\,,\qquad q_{2}=-\frac{16\mu^{8}}{3(3\mu^{2}-m^{2})^{4}}\,,\label{eq:CoeffDef}\\
p_{1} & = & -\frac{2\mu^{2}(4\mu^{2}-m^{2})}{3(\mu^{2}-m^{2})^{1/2}(3\mu^{2}-m^{2})^{3/2}}\,,\qquad p_{2}=\frac{2\mu^{6}(5\mu^{4}-6\mu^{2}m^{2}+2m^{4})}{(\mu^{2}-m^{2})^{3/2}(3\mu^{2}-m^{2})^{7/2}}\,,\nonumber \\
c_{1} & = & \frac{(\mu^{2}-m^{2})^{1/2}}{(3\mu^{2}-m^{2})^{1/2}}\,,\qquad c_{2}=\frac{\mu^{6}}{(\mu^{2}-m^{2})^{1/2}(3\mu^{2}-m^{2})^{5/2}}\,.\nonumber 
\end{eqnarray}
This yields

\noindent \medskip{}

\noindent 
\begin{equation}
n_{s}\simeq\mu\frac{\mu^{2}-m^{2}}{\lambda}-\frac{\pi^{2}T^{4}}{6\mu c_{1}^{4}}\left[\frac{1}{5}\left(\frac{4q_{1}}{c_{1}}-p_{1}\right)+\frac{8\pi^{2}T^{2}}{\mu^{2}c_{1}^{2}}\left(\frac{2}{7}\frac{q_{2}+p_{1}c_{2}}{c_{1}}-\frac{2q_{1}c_{2}}{c_{1}^{2}}-\frac{p_{2}}{21}\right)\right]\,,\label{eq:NswithM}
\end{equation}
\newpage{}where the integrals

\noindent \medskip{}

\noindent 
\begin{eqnarray}
\int_{0}^{\infty}dy\,\frac{y^{3}}{e^{y}-1} & = & \frac{\pi^{4}}{15}\,,\qquad\int_{0}^{\infty}dy\,\frac{y^{4}e^{y}}{(e^{y}-1)^{2}}=\frac{4\pi^{4}}{15}\,,\\
\int_{0}^{\infty}dy\,\frac{y^{5}}{e^{y}-1} & = & \frac{8\pi^{6}}{63}\,,\qquad\int_{0}^{\infty}dy\,\frac{y^{6}e^{y}}{(e^{y}-1)^{2}}=\frac{16\pi^{6}}{21}\,,\qquad\int_{0}^{\infty}dy\,\frac{y^{7}e^{y}(e^{y}+1)}{(e^{y}-1)^{3}}=\frac{16\pi^{6}}{3}\nonumber 
\end{eqnarray}
 for the dimensionless variable $y=c_{1}k/T$ have been used. Inserting
the coefficients from (\ref{eq:CoeffDef}) into (\ref{eq:NswithM})
yields the result for the superfluid density 

\noindent \medskip{}

\noindent 
\begin{eqnarray}
n_{s} & \simeq & \mu\frac{\mu^{2}-m^{2}}{\lambda}-\frac{\pi^{2}T^{4}}{9\mu}\left[\frac{\mu^{2}(12\mu^{2}-m^{2})(3\mu^{2}-m^{2})^{1/2}}{5(\mu^{2}-m^{2})^{5/2}}-\frac{8}{7}\left(\frac{\pi T}{\mu}\right)^{2}\frac{\mu^{6}(57\mu^{4}-24\mu^{2}m^{2}+2m^{4})}{(\mu^{2}-m^{2})^{9/2}(3\mu^{2}-m^{2})^{1/2}}\right]\,.\nonumber \\
\end{eqnarray}
 The pressure (\ref{eq:PressureLowT}), evaluated at $|\vec{\nabla}\psi|=0$
becomes 

\noindent \medskip{}

\noindent 
\begin{equation}
\Psi\simeq\frac{(\mu^{2}-m^{2})^{2}}{4\lambda}+\frac{\pi^{2}T^{4}}{90}\left[\frac{(3\mu^{2}-m^{2})^{3/2}}{(\mu^{2}-m^{2})^{3/2}}-\frac{40}{7}\left(\frac{\pi T}{\mu}\right)^{2}\frac{\mu^{6}(3\mu^{2}-m^{2})^{1/2}}{(\mu^{2}-m^{2})^{7/2}}\right]\,.
\end{equation}
This is all we need to compute the sound velocities: we can now straightforwardly
take all relevant derivatives of the pressure, compute the normal-fluid
density via $n_{n}=n-n_{s}$, and insert the results into the wave
equation for the sound velocities. 

~

\section{\noindent Chemical potential in a system of two coupled superfluids\label{sec:Chemical-potential-in-2SF}}

~

\noindent In this part of the appendix, we shall prove that the usual
way in which a chemical potential is introduced in a field theory
(i.e. similar to the zeroth component of a gauge field) remains intact
in the presence of a gradient coupling. To this end, we start from
the Lagrangian 
\begin{equation}
\mathcal{L}=\partial_{\mu}\varphi_{1}\partial^{\mu}\varphi_{1}^{*}-m_{1}^{2}\left|\varphi_{1}\right|^{2}-\lambda_{1}\left|\varphi_{1}\right|^{4}+\partial_{\mu}\varphi_{2}\partial^{\mu}\varphi_{2}^{*}-m_{2}^{2}\left|\varphi_{2}\right|^{2}-\lambda_{2}\left|\varphi_{2}\right|^{4}+\mathcal{L}_{int}\,,\label{eq:Laux}
\end{equation}

\newpage{}

\noindent with 

\noindent 
\begin{equation}
\mathcal{L}_{int}=-\frac{1}{2}\lambda_{12}\varphi_{1}\varphi_{2}^{*}\partial_{\mu}\varphi_{1}^{*}\partial^{\mu}\varphi_{2}+\, h.c.\,,
\end{equation}
and decompose the complex fields $\varphi_{i}$ in the usual way

\noindent 
\begin{eqnarray}
\varphi_{1} & = & \frac{1}{\sqrt{2}}\left(\varphi_{11}+i\varphi_{12}\right)\,,\\
\varphi_{2} & = & \frac{1}{\sqrt{2}}\left(\varphi_{21}+i\varphi_{22}\right)\,.
\end{eqnarray}

\noindent This results in the following kinetic terms

\noindent 
\begin{eqnarray}
\mathcal{L}_{kin} & = & \frac{1}{2}\left(\partial_{\mu}\varphi_{11}\partial^{\mu}\varphi_{11}+\partial_{\mu}\varphi_{12}\partial^{\mu}\varphi_{12}+\partial_{\mu}\varphi_{21}\partial^{\mu}\varphi_{21}+\partial_{\mu}\varphi_{22}\partial^{\mu}\varphi_{22}\right)\label{eq:KinAux}\\
\nonumber 
\end{eqnarray}

\noindent and interaction terms

\noindent 
\begin{eqnarray}
\mathcal{L}_{int} & = & -\left(A+A^{*}\right)\left(\partial_{\mu}\varphi_{11}\partial^{\mu}\varphi_{21}+\partial_{\mu}\varphi_{12}\partial^{\mu}\varphi_{22}\right)-i\left(A-A^{*}\right)\left(\partial_{\mu}\varphi_{11}\partial^{\mu}\varphi_{22}-\partial_{\mu}\varphi_{12}\partial^{\mu}\varphi_{21}\right)\label{eq:IntAux}\\
 &  & -2Re\left(A\right)\left(\partial_{\mu}\varphi_{11}\partial^{\mu}\varphi_{21}+\partial_{\mu}\varphi_{12}\partial^{\mu}\varphi_{22}\right)+2Im\left(A\right)\left(\partial_{\mu}\varphi_{11}\partial^{\mu}\varphi_{22}-\partial_{\mu}\varphi_{12}\partial^{\mu}\varphi_{21}\right)\nonumber 
\end{eqnarray}

\noindent with the abbreviation

\noindent 
\begin{equation}
A=\frac{1}{8}\lambda_{12}\left[\left(\varphi_{11}+i\varphi_{12}\right)\left(\varphi_{21}-i\varphi_{22}\right)\right]\,.
\end{equation}

\noindent All remaining terms in (\ref{eq:Laux}) are irrelevant for
this analysis. The canonically conjugate momenta $\pi_{ij}=\partial\mathcal{L}/\partial(\partial_{0}\varphi_{ij})$
are given by:

\noindent \medskip{}

\noindent 
\begin{equation}
\left(\begin{array}{c}
\pi_{11}\\
\pi_{12}\\
\pi_{21}\\
\pi_{22}
\end{array}\right)=\left(\begin{array}{cccc}
1 & 0 & -2R & 2I\\
0 & 1 & -2I & -2R\\
-2R & -2I & 1 & 0\\
2I & -2R & 0 & 1
\end{array}\right)\left(\begin{array}{c}
\partial_{0}\varphi_{11}\\
\partial_{0}\varphi_{12}\\
\partial_{0}\varphi_{21}\\
\partial_{0}\varphi_{22}
\end{array}\right)\,,\label{eq:PID0Rel}
\end{equation}

\noindent where 
\begin{eqnarray}
2R & = & 2Re(A)=\frac{1}{4}\lambda_{12}\left(\varphi_{11}\varphi_{21}+\varphi_{12}\varphi_{22}\right)\,,\\
2I & = & 2Im(A)=\frac{1}{4}\lambda_{12}\left(\varphi_{12}\varphi_{21}-\varphi_{11}\varphi_{22}\right)\,.
\end{eqnarray}

\newpage{}

\noindent This relation can easily be inverted to

\noindent \medskip{}

\noindent 
\begin{equation}
\left(\begin{array}{c}
\partial_{0}\varphi_{11}\\
\partial_{0}\varphi_{12}\\
\partial_{0}\varphi_{21}\\
\partial_{0}\varphi_{22}
\end{array}\right)=\mathcal{N}\left(\begin{array}{cccc}
1 & 0 & 2R & -2I\\
0 & 1 & 2I & 2R\\
2R & 2I & 1 & 0\\
-2I & 2R & 0 & 1
\end{array}\right)\left(\begin{array}{c}
\pi_{11}\\
\pi_{12}\\
\pi_{21}\\
\pi_{22}
\end{array}\right)\label{eq:inverse}
\end{equation}

\noindent with

\noindent 
\[
\mathcal{N}=\frac{1}{1-4\left|A\right|^{2}}=\frac{1}{1-\left(\frac{\lambda_{12}}{4}\right)^{2}\left(\varphi_{11}^{2}+\varphi_{12}^{2}\right)\left(\varphi_{21}^{2}+\varphi_{22}^{2}\right)}\,.
\]

\noindent The currents are given by

\noindent 
\begin{eqnarray*}
j_{1}^{\mu} & = & i\left(\phi_{1}^{*}\partial^{\mu}\phi_{1}-\phi_{1}\partial^{\mu}\phi_{1}^{*}\right)-i\frac{1}{2}\lambda_{12}\left|\phi_{1}\right|^{2}\left(\phi_{2}^{*}\partial^{\mu}\phi_{2}-\phi_{2}\partial^{\mu}\phi_{2}^{*}\right)\\
 & = & \left(\varphi_{12}\partial^{\mu}\varphi_{11}-\varphi_{11}\partial^{\mu}\varphi_{12}\right)-\frac{1}{4}\lambda_{12}\left(\varphi_{11}^{2}+\varphi_{12}^{2}\right)\left(\varphi_{22}\partial^{\mu}\varphi_{21}-\varphi_{21}\partial^{\mu}\varphi_{22}\right)\,,
\end{eqnarray*}

\begin{eqnarray*}
j_{2}^{\mu} & = & i\left(\phi_{2}^{*}\partial^{\mu}\phi_{2}-\phi_{2}\partial^{\mu}\phi_{2}^{*}\right)-i\frac{1}{2}\lambda_{12}\left|\phi_{2}\right|^{2}\left(\phi_{1}^{*}\partial^{\mu}\phi_{1}-\phi_{1}\partial^{\mu}\phi_{1}^{*}\right)\\
 & = & \left(\varphi_{22}\partial^{\mu}\varphi_{21}-\varphi_{21}\partial^{\mu}\varphi_{22}\right)-\frac{1}{4}\lambda_{12}\left(\varphi_{21}^{2}+\varphi_{22}^{2}\right)\left(\varphi_{12}\partial^{\mu}\varphi_{11}-\varphi_{11}\partial^{\mu}\varphi_{12}\right)\,.
\end{eqnarray*}

\noindent Note that after plugging in the conjugate momenta from (\ref{eq:inverse}),
the zeroth components of these currents simply read: 

\noindent 
\begin{eqnarray}
j_{1}^{0} & = & \pi_{11}\varphi_{12}-\pi_{12}\varphi_{11}\,,\label{eq:AuxCur}\\
j_{2}^{0} & = & \pi_{21}\varphi_{22}-\pi_{22}\varphi_{21}\,.
\end{eqnarray}

\noindent To eliminate the dependence of $\mathcal{L}$ on $\partial_{0}\varphi_{ij}$
in favor of the momenta $\pi_{ij}$, we need calculate the factor
required for the Legendre transform of $\mathcal{L}$\medskip{}

\noindent 
\begin{eqnarray}
\pi_{11}\partial_{0}\phi_{11}+\pi_{12}\partial_{0}\phi_{12}+\pi_{21}\partial_{0}\phi_{21}+\pi_{22}\partial_{0}\phi_{22} & = & \mathcal{N}\left[\pi_{11}^{2}+\pi_{12}^{2}+\pi_{21}^{2}+\pi_{22}^{2}\right.\label{eq:AuxSum}\\
 &  & \,\,\,\left.+4R\left(\pi_{11}\pi_{21}+\pi_{12}\pi_{22}\right)+4I\left(\pi_{12}\pi_{21}-\pi_{11}\pi_{22}\right)\right]\nonumber 
\end{eqnarray}

\noindent Observe that for $\lambda_{12}\rightarrow0$ this reduces
to $\pi_{11}^{2}+\pi_{12}^{2}+\pi_{21}^{2}+\pi_{22}^{2}$. From (\ref{eq:KinAux})
and (\ref{eq:IntAux}) we need all contributions which include time
derivatives of $\varphi_{ij}$. From these, we calculate all contributions
to $\mathcal{H}=\sum_{ij}(\pi_{ij}\partial_{0}\varphi_{ij})-\mathcal{L}$
which involve time derivatives (we could carry along terms in $\vec{\nabla}\varphi_{ij}$
through the whole calculation but they are irrelevant), eliminate
them in favor of $\pi_{ij}$ and obtain \medskip{}

\noindent 
\begin{eqnarray}
\mathcal{H}-\mu_{1}j_{1}^{0}-\mu_{2}j_{2}^{0} & = & \frac{1}{2}\mathcal{N}\left(\pi_{11}^{2}+\pi_{12}^{2}+\pi_{21}^{2}+\pi_{22}^{2}\right)-\mu_{1}\left(\pi_{11}\varphi_{12}-\pi_{12}\varphi_{11}\right)\label{eq:AuxH}\\
 &  & -\mu_{2}\left(\pi_{21}\varphi_{22}-\pi_{22}\varphi_{21}\right)+2\mathcal{N}\left[R\left(\pi_{11}\pi_{21}+\pi_{12}\pi_{22}\right)+I\left(\pi_{12}\pi_{21}-\pi_{11}\pi_{22}\right)\right]\,.\nonumber 
\end{eqnarray}

\noindent This is not yet the expression which appears in the path
integral of the partition function (see equation (\ref{eq:Partition})).
We need \medskip{}

\noindent 
\begin{equation}
\pi_{11}\partial_{0}\varphi_{11}+\pi_{12}\partial_{0}\varphi_{12}+\pi_{21}\partial_{0}\varphi_{21}+\pi_{22}\partial_{0}\varphi_{22}-\mathcal{H}+\mu_{1}j_{1}^{0}+\mu_{2}j_{2}^{0}\,.\label{eq:AuxHelp}
\end{equation}

\noindent Plugging in equations (\ref{eq:AuxCur}), (\ref{eq:AuxSum})
and (\ref{eq:AuxH}), we obtain a result which can be summarized as\bigskip{}

\noindent 
\[
\frac{1}{2}\Pi^{T}\, A\,\Pi+\xi^{T}\Pi\,,
\]

\noindent \begin{flushleft}
where
\par\end{flushleft}

\noindent 
\begin{equation}
A=-\mathcal{N}\left(\begin{array}{cccc}
1 & 0 & 2R & -2I\\
0 & 1 & 2I & 2R\\
2R & 2I & 1 & 0\\
-2I & 2R & 0 & 1
\end{array}\right)\,,\,\,\,\,\,\,\,\,\,\,\xi=\left(\begin{array}{c}
\partial_{0}\varphi_{11}+\mu_{1}\varphi_{12}\\
\partial_{0}\varphi_{12}-\mu_{1}\varphi_{11}\\
\partial_{0}\varphi_{21}+\mu_{2}\varphi_{22}\\
\partial_{0}\varphi_{22}-\mu_{2}\varphi_{21}
\end{array}\right),\,\,\,\,\,\,\,\,\,\,\,\Pi=\left(\begin{array}{c}
\pi_{11}\\
\pi_{12}\\
\pi_{21}\\
\pi_{22}
\end{array}\right)\,.
\end{equation}

\noindent $A$ is the inverse of the matrix defined in (\ref{eq:PID0Rel})
with an additional factor $(-1)$ because $-\mathcal{H}$ appears
in formula (\ref{eq:AuxHelp}). We can now use\medskip{}

\noindent 
\begin{equation}
\frac{1}{2}\Pi^{T}\, A\,\Pi+\xi^{T}\Pi=-\frac{1}{2}\xi^{T}A^{-1}\xi+\frac{1}{2}\Pi^{\prime T}A\,\Pi^{\prime}\label{eq:AuxShift}
\end{equation}

\noindent where we have introduced the shifted fields $\Pi^{\prime}$\medskip{}

\noindent 
\begin{equation}
\Pi^{\prime}=\Pi+A^{-1}\xi\,.
\end{equation}

\noindent The second term on the right hand side of equation (\ref{eq:AuxShift})
is now quadratic in the shifted momenta and can easily be integrated
out (the linear shift doesn\textquoteright t affect the integration
measure $\mathcal{D}\pi$). The first term on the right hand side
represents the new shifted Lagrangian and evaluates to\medskip{}

\noindent 
\begin{eqnarray}
\mathcal{L}_{kin} & = & \left(\partial_{0}\varphi_{11}\right)^{2}+\left(\partial_{0}\varphi_{12}\right)^{2}+\left(\partial_{0}\varphi_{21}\right)^{2}+\left(\partial_{0}\varphi_{22}\right)^{2}+\mu_{1}^{2}\left(\varphi_{11}^{2}+\varphi_{12}^{2}\right)+\mu_{2}^{2}\left(\varphi_{21}^{2}+\varphi_{22}^{2}\right)\nonumber \\
 &  & +2\mu_{1}\left(\varphi_{12}\partial_{0}\varphi_{11}-\varphi_{11}\partial_{0}\varphi_{12}\right)+2\mu_{2}\left(\varphi_{22}\partial_{0}\varphi_{21}-\varphi_{21}\partial_{0}\varphi_{22}\right)\,,
\end{eqnarray}

\newpage{}

\noindent and

\noindent 
\begin{eqnarray}
\mathcal{L}_{int} & = & -2R\left[\partial_{0}\varphi_{11}\partial_{0}\varphi_{21}+\partial_{0}\varphi_{12}\partial_{0}\varphi_{22}+\mu_{1}\left(\varphi_{12}\partial_{0}\varphi_{21}-\varphi_{11}\partial_{0}\varphi_{22}\right)\right]\nonumber \\
 & + & 2R\left[\mu_{2}\left(\varphi_{21}\partial_{0}\varphi_{12}-\varphi_{22}\partial_{0}\varphi_{11}\right)-\mu_{1}\mu_{2}\left(\varphi_{11}\varphi_{21}+\varphi_{21}\varphi_{22}\right)\right]\nonumber \\
 & + & 2I\left[\partial_{0}\varphi_{11}\partial_{0}\varphi_{22}-\partial_{0}\varphi_{12}\partial_{0}\varphi_{21}+\mu_{1}\left(\varphi_{11}\partial_{0}\varphi_{21}+\varphi_{12}\partial_{0}\varphi_{22}\right)\right]\nonumber \\
 & - & 2I\left[\mu_{2}\left(\varphi_{21}\partial_{0}\varphi_{11}+\varphi_{22}\partial_{0}\varphi_{12}\right)+\mu_{1}\mu_{2}\left(\varphi_{12}\varphi_{21}-\varphi_{11}\varphi_{22}\right)\right]
\end{eqnarray}

\noindent This corresponds precisely to\medskip{}

\noindent 
\[
\mathcal{L}_{kin}=\left[\left(\partial_{\mu}-iA_{1\mu}\right)\varphi_{1}\right]\left[\left(\partial^{\mu}+iA_{1}^{\mu}\right)\varphi_{1}^{*}\right]+\left[\left(\partial_{\mu}-iA_{2\mu}\right)\varphi_{2}\right]\left[\left(\partial^{\mu}+iA_{2}^{\mu}\right)\varphi_{2}^{*}\right]
\]

\noindent and

\noindent 
\[
\mathcal{L}_{int}=-\frac{1}{2}\lambda_{12}\varphi_{1}\varphi_{2}^{*}\left[\left(\partial_{\mu}+iA_{1\mu}\right)\varphi_{1}^{*}\left(\partial^{\mu}-iA_{2}^{\mu}\right)\varphi_{2}\right]+c.c.
\]

\noindent with 
\[
A_{1}^{\mu}=(\mu_{1},\vec{0}),\,\,\,\,\,\,\, A_{2}^{\mu}=(\mu_{2},\vec{0})\,.
\]

\noindent This completes the proof. 

\newpage{}

\end{document}